\documentclass[12pt]{article}

\usepackage{natbib}
\usepackage[T1]{fontenc}
\usepackage{palatino}
\usepackage{graphics}
\usepackage[demo]{graphicx}
\usepackage{booktabs,caption}
\usepackage{threeparttable}
\usepackage{adjustbox}
\usepackage[format=plain, labelsep=period, labelfont=bf, font=footnotesize, singlelinecheck=false,skip=5pt,justification=justified]{caption}
\usepackage{tabularx}
\usepackage{array}
\usepackage{amsthm}
\usepackage{lscape}
\usepackage{appendix}
\usepackage{hyperref}
\usepackage{subcaption}
\usepackage{bm}
\usepackage{physics}
\usepackage[thinc]{esdiff}
\usepackage{epstopdf}
\usepackage{amsmath}
\usepackage{multirow}
\usepackage{setspace}
\usepackage{amssymb}
\usepackage{textcomp}
\usepackage{amsfonts}
\usepackage{soul}
\usepackage{bbm}
\usepackage{multirow}
\usepackage{multicol}
\usepackage{verbatim}
\usepackage{mdwlist}
\usepackage{color,soul}
\usepackage{eurosym}
\usepackage{rotating}
\usepackage{lmodern}
\usepackage{color}
\usepackage{longtable}
\usepackage{etoolbox}
\usepackage{epigraph}
\usepackage{xargs}
\usepackage[dvipsnames]{xcolor}
\usepackage{lipsum}
\usepackage{todonotes}
\usepackage[left=2cm, right=2cm, bottom=2cm, top=2cm]{geometry}

\newcolumntype{H}{>{\setbox0=\hbox\bgroup}c<{\egroup}@{}}

\makeatletter
\newlength\epitextskip
\patchcmd{\epigraph}{\@epitext{#1}}{\itshape\@epitext{#1}}{}{}
\makeatother

\DeclareMathOperator*{\argmin}{arg\,min}

\title{Winners and losers of immigration}

\author{Davide Fiaschi\thanks{ University of Pisa, Dipartimento di Economia e Management, Via Ridolfi 10, 56124 Pisa (Italy), Phone: +39 0502216208, Email: davide.fiaschi@unipi.it. }  \and Cristina Tealdi\thanks{Corresponding author. Heriot-Watt University, EH14 4AS Edinburgh (UK) and IZA Institute of Labor, Phone: +44 0131 4513208, Email: c.tealdi@hw.ac.uk. }}

\date{\today}

\begin{document}

\maketitle

%\epigraph{
%	``... immigration has consequences, and these consequences generally imply that some people lose while others benefit.''}{--- \textup{George Borjas (2014)}}
 
\begin{abstract}
We study the impact of low-skilled immigration in a general equilibrium search and matching model, with heterogeneous workers producing intermediate goods, which are used in the production of two final goods. In addition to complementarity/substitution between native and non-native workers, we explore how immigration affects the relative prices of final goods and wages.  An application to Italy reveals a positive contribution of immigrants to GDP, public revenues, and the per capita provision of public goods. Employers and employees in the high-skilled-intensive market are winners, while losers are employers in the low-skilled-intensive market. The effects on low-skilled employees are instead inconclusive.

%We study the impact of low-skilled immigration on wages, profits and welfare in a general equilibrium search and matching model, in which low and high-skilled workers produce intermediate goods, which are in turn used in the production of two final goods. Unemployment benefits and public goods are financed by taxes on wages, profits and production. By changing the labour force composition and  the relative supply  of intermediate goods, immigration leads to a decrease in both the relative price of the low-skill intensive final good and  the relative wages of low-skilled employees.
%An application to Italy, for the period 2008-2017, points to a large positive contribution of immigrants to GDP, government revenues, and the per capita furniture of public goods. Employees and employers in the high-skill intensive market are  the winners,  and employers in the low-skill intensive market are the losers. For employees in the low-skill intensive market the effects are less conclusive, as the decline in prices and wages is offset by the increase in the provision of public goods.
\end{abstract}


\clearpage

\section{Introduction}

Immigration is a multifaceted phenomenon that keeps shaping the world. Between July 2021 and July 2022 net international migration added more than a million people to the U.S. population (US Census Bureau), and in 2022 about 1.46 million to the German population. Nevertheless, the question of how immigration flows affect inequality within the receiving economies is still open and largely debated \citep{EUParliament}.
To answer this question, we develop a  general equilibrium search and matching model, in which two non-tradable final goods are developed using two intermediate goods, which, in turn, are produced by high-skilled and low-skilled labour (either native or non-native). Both final goods are consumed by all employees and employers. The provision of public goods and unemployment benefits are financed by a progressive taxation on wages and profits. 
We highlight that on top of the standard channel of complementarity/substitution between native and non-native workers,  immigration crucially affects both the wages and the prices of goods and, therefore, the profits of employers and the welfare of employees. 

A large inflow of low-skilled employees has two major direct effects on the economy: first, it changes the skill composition of the workforce, with an ambiguous impact on the real wages of low-skilled employees, depending on the elasticity of the low-skilled wage to the labour market tightness, which could overturn the direct negative effect of the increased supply of low-skilled employees. Second, by increasing the supply of the low-skill intensive good and by raising the demand for both goods, it leads to a drop in the price of the low-skill intensive good and to an increase in the price of the high-skill intensive good, causing the real wage of high-skilled employees to rise.
The adverse effect on the price of the low-skill intensive good might be so large to induce a loss for employers who operate in the low-skill intensive market, even when the wage of the low-skilled employees is lower. Finally, the inflow of low-skilled employees increases GDP, government revenues, and social contributions, while the effect on GDP per worker and the per capita provision of  public goods is ambiguous.

We estimate the model using data from Italy for each year in the period 2008-2017. We find that, compared to a counter-factual scenario with no non-natives, in 2017 total production was higher by approximately 16\%. Both indirect and direct taxes, which are proportional to total production were 16\% higher, and so did social security contributions, corresponding to an overall increase in the government revenues of approximately 75 billions \euro. GDP per worker and the per capita provision of public goods were both higher by approximately 5\% and 8\%, respectively. The wage of low-skilled employees was lower by approximately 5\%, while the wage of high-skilled employees was higher by approximately 15\%. On the other hand, the effect on unemployment rates was minimal. The overall effect of the presence of non-natives on the employees' lifetime utilities was positive for high-skilled employees (+10\%) and null for low-skilled employees. At the same time, the value of a filled vacancy, which we interpret as a proxy for the employers' expected profits, was 10\% higher for firms operating in the high-skill intensive market and 5\% lower for firms operating in the low-skill intensive market, due to the increased (decreased) real price of the high-skill intensive (low-skill intensive) good.
A sensitivity analysis on the parameters not estimated, i.e., the elasticity of substitution between the two final goods and the congestion in the provision of public goods, supports the robustness of these findings.

In Appendix \ref{app:SMO} we discuss an extension of the model to a small open economy where the high-skill intensive good is tradable; however, we argue that the main focus should be on the case of non-tradable goods for two reasons. First, non-tradable activities, typically services, are the ones which absorb the largest share of low-skilled immigrants (see Section \ref{sec:ilm} for Italy). Moreover, on average the share of tradable activities in an economy is very small. \citet{blinder2013alternative} show that in the US only 25\% of activities (tasks) are offshorable and offshorability does not have systematic effects on either wages or the probability of layoff.
 
The contribution of this paper to the existing literature is multi-fold.
First,  while in most of the existing research the effects of immigration have been studied through the factor ratios within firms \citep{dustmann2012, borjas2003, JEEA:JEEA1052}, in line with \citet{monras2020immigration} and \citet{ burstein2020tradability} this paper explores the \textit{price transmission} through the inclusion of two goods, whose equilibrium prices strongly depend on the skill composition of migrants, while maintaining some degree of socio-economic heterogeneity between native and non-native employees.\footnote{Among others, job finding and job exit rates of employees within the same labour market differ according to the country of origin, hence natives and non-natives are not perfect substitute, although their value added is the same.} This is motivated by the evidence provided by \cite{cortes2008effect}, who empirically shows that in the US an increase in the share of low-skilled immigrants in the labour force significantly decreased the price of immigrant-intensive services. Appendix \ref{HS_LS_prices} reports anecdotal evidence for Italy about the differential changes in the prices of high and low-skill intensive goods in the 2004-2017 period in which the percentage of foreign workforce increased from 6\% to 15\%. Different from \cite{monras2020immigration} and \cite{burstein2020tradability},  we use a search and matching framework, instead of perfect competition, since the impact of a low-skilled immigration flow on the wage of low-skilled workers is determined not only by the elasticity of substitution between different types of workers, but also by the profitability of operating in different labour markets. Moreover, in such framework a change in the size and composition of the workforce admits asymmetric effects on the unemployment rates of different types of workers, as found in our empirical application. Finally, it allows for a comparison of the impact of immigration across countries with different employment protection legislation (EPL).

Second, we provide a general equilibrium framework to assess the impact of immigration on public finances, which is crucial for formulating a comprehensive evaluation of the immigration phenomenon \citep{national2017economic}.  As natives and non-natives have the same access to unemployment benefits and the provision of public goods (i.e., the fiscal costs are the same for all employees) and the impact of immigration on unemployment is negligible, we find for Italy a \textit{net fiscal benefit} of immigration as measured by the increased per capita provision of public goods.  

Third, we advance our understanding of the impact of immigration on profits and the welfare of employers, which are outcomes generally overlooked in the literature which uses competitive equilibrium frameworks \citep{monras2020immigration, burstein2020tradability, caliendo2021goods}. This is particularly important as the mass of employers is generally not negligible; in 2019, it spanned from 10\% of total employment in the USA (CPS) to 23\% in Italy and 32\% in Greece (OECD). Our model points out that employers in the low-skill intensive market might be at risk of being negatively affected by low-skilled immigration as the benefit from the lower wage of low-skilled employees might be overtaken by the lower revenues deriving from the drop of the price of the low-skill intensive good. A similar risk is faced by employers in the high-skill intensive market due to the parallel increase in the wage of high-skilled employees  and the higher price of the high-skill intensive good. Our empirical analysis for Italy suggests a loss for employers in the low-skill intensive market and a gain for employers in the high-skill intensive market. 

Moreover, we are the first to estimate the impact of immigration in Italy using a micro-founded general equilibrium model. Italy has experienced large immigration inflows in the last decade and several analysis on the fiscal consequences of immigration have been conducted by policy-focused groups \citep{moressa2019rapporto, INPS2017}. This research however presents weaknesses typical of partial equilibrium analysis and the lack of micro-foundation of individual decisions. Not surprisingly, although we find qualitatively similar (but quantitatively different) results  in terms of increased social security contributions \citep{INPS2017}, we offer a more balanced view on the costs of immigration, specifically for low-skilled workers. 
Finally, in our estimate we match a much larger set of moments with respect to the estimated parameters with the aim of assessing the ability of the model to reproduce the dynamics of the most important variables, while in the literature the estimation of general equilibrium models is usually performed using perfect identification, i.e. the number of parameters estimated is equal to the number of moments \citep{flinn2015labor}.

This paper is organized as follows. Section \ref{sec:literatureReview} presents a brief literature review; in Section \ref{sec:model} we describe the search and matching model with non-tradable goods, while in Section \ref{sec:equilibrium} we study its equilibrium. In Section \ref{EmpiricalEvidence} we provide labour market statistics for Italy in the 2008-2017 period and illustrate the methodology used to estimate the model. Section \ref{sec:winnersLosersMigration} shows the results of the model estimation and presents two counter-factual scenarios. Section \ref{sec:conclusions} concludes the paper.

\section{Literature review \label{sec:literatureReview}}

Traditionally, the literature on the economics of immigration has focused on three major issues: the determinants of the size and  composition of immigrant flows, the integration of immigrants into the host country, and the impact of immigrants on the host country's economy \citep{borjas1989economic}.
Two different approaches have been used to address these issues, one based on the estimation/calibration of theoretical models (e.g., \citealp{borjas2003}) and one based on the exploitation of exogenous shocks (e.g., \citealp{card2001}).
Theoretical studies on the effects of immigration on native labour market outcomes have populated the literature since the 1980s \citep{borjas1987economic,greenwood1986factor, borjas2003}. The neoclassical  theory suggests that a natural starting point for this type of analysis is the specification of a production technology, which describes how immigrants, natives and capital interact in the production process \citep{borjas1989economic}.\footnote{See \cite{bluekanh2012} for a survey of the literature.} A number of papers take a medium/long-run approach, by assuming that over time firms are able to adjust the composition of factors (capital/labour ratio) \citep{JEEA:JEEA1052}, while another strand using a short-term approach represents immigration as an increase in labour supply for a given capital stock \citep{borjas2003}. The large majority of papers discusses partial equilibrium models with a production technology that distinguishes between high-skilled and low-skilled labour, and assumes that immigrants are perfect substitutes with their corresponding native skill category. Within this framework, a sudden inflow of low-skilled migrants, changing the skill composition, leads to an excess supply of low-skilled employees and therefore a downward pressure on wages of all low-skilled employees, at least in the short run \citep{ dustmann2012, Moreno2016, Ortega2000}. On the contrary, whenever immigrants and natives are only imperfect substitutes within the same (observable) skill group, the increase in immigration primarily affects the wages of immigrants already living in the host country \citep{manacorda2012, d2010labor}.

Scant is instead the literature on the use of a general equilibrium search and matching model to quantify the impact of immigration. Some papers  concentrate only on the production side of the economy and assume perfect competition in the labour market \citep{JEEA:JEEA1052, borjas2003}. More recently, a number of papers have analysed the issue within a general equilibrium model with imperfect competition, but include only one sector and assume perfect substitutability between immigrants and natives \citep{albert2018immigration,amior2020monopsony}. 
Generally, within these models consumption goods are produced by a nested CES production function, where employees of different skill levels enter as complementary factors, while less emphasis is placed on the equilibrium in the goods market \citep{Chassamboulli2014, battisti2017, Iftikhar2016, monras2020immigration}. Another recent approach involves the inclusion of tradable and non-tradable activities (tasks), within a general equilibrium model, but with the assumption of  perfect competition \citep{burstein2020tradability}. Within a similar framework, \cite{caliendo2021goods} develop a multi-country model to study the economic effects of the 2004 European Union enlargement, where labour mobility and the trade of goods and services are subject to institutional barriers and tariffs. 

\section{The model \label{sec:model}}

In the economy two non-tradable final goods are produced using two intermediate goods, which in turn are produced by  high-skilled and low-skilled labour (either native or non-native). Both final goods are then consumed by all employees and employers. The labour market displays imperfect competition, frictions, and the inflows and outflows of employees. Finally, the government collects resources through a progressive taxation on wages and profits and provides public goods and unemployment benefits. We start the description of the model from the labour market.

\subsection{The labour market \label{sec.labourMarket}}

Consider a continuous time infinite-horizon economy with constant stocks of physical and human capital and technology.
There is a continuum of mass $\sigma$ of employees. All employees supply labour inelastically, are risk neutral and discount the future at constant rate $r$. Employees differ according to their skill level and  their country of origin. We distinguish between \textit{natives} $N$, who are born in the home country and \textit{non-natives} $I$, who are born in a foreign country.  Each individual is either high-skilled, $h$, or low-skilled, $l$. The total measure of resident employees is therefore the sum of the four different categories of employees, i.e. $\sigma=\sigma_{l,N}+\sigma_{l,I}+\sigma_{h,N}+\sigma_{h,I}$.  \textit{Native employees} can be either employed or unemployed, hence  $\sigma_{i,N}=e_{i,N}+u_{i,N}$, where $i \in \{l,h\}$. The stock of \textit{non-native employees} of type $i$ ($L_{i,I}$) instead includes residents in the country, who can be either employed or unemployed, and those living in the foreign country ($FC_i$), i.e.,  $L_{i,I}=\sigma_{i,I}+FC_{i}= e_{i,I}+u_{i,I}+FC_{i}$. 
Every worker supplies labour in one of the two intermediate good sectors: high-skill intensive (\textit{HS}) and low-skill intensive (\textit{LS}). 
The economy is also populated by a measure $\chi$ of \textit{employers}.
Employers are ex-ante homogeneous and post skill-specific vacancies, which are open to both natives and non-natives.  From the match between an employer and an employee two types of intermediate goods, $h$ and $l$ are produced, using labour as sole input. 

Employers and employees come together via a standard matching function  $m\left(v_i,u_i\right)$, where $u_i$ is the measure of unemployment  (natives and non-natives, $u_i=u_{i,N}+u_{i,I}$)  and $v_i$ is the measure of vacancies, and $\theta_i \equiv v_i/u_i $  is defined as the labour market tightness. The function $m\left(v_i,u_i\right)$ is twice differentiable, increasing in its arguments, and exhibits constant returns to scale. Each vacancy is skill-specific, but open to both native and non-native employees. Hence, there are two labour markets, one for high-skilled and one for low-skilled employees. Within each market, the probability that an employer with an open  vacancy meets either a native or an non-native employee may be defined as $q\left(\theta_i\right) \equiv m\left(u_{i},v_i\right)/v_i $. The probability that an unemployed worker, either native or non-native, meets an employer may be defined as $\theta_i q\left(\theta_i\right)=m\left(u_{i},v_i\right)/u_{i}$.  It is assumed that $q(\theta_i) \rightarrow 1$ and $\theta_i q(\theta_i) \rightarrow 0$ as $\theta_i \rightarrow 0$, and
$q(\theta_i) \rightarrow 0$ and $\theta_i q(\theta_i) \rightarrow 1$ as $\theta_i \rightarrow \infty$.

The probability that a vacancy is filled with a worker is equal to $\kappa_{i,j} q\left(\theta_i\right)$, which is the product of the probability that an employer meets an employee, $q\left(\theta_i\right)$, and the probability that the job offer is signed, $\kappa_{i,j}$. Similarly, the probability that an unemployed worker finds a job is equal to the product of the probability that an employee meets an employer $\theta_i q\left(\theta_i\right)$ and the probability that the job offer is signed $\kappa_{i,j}$, i.e., $\kappa_{i,j} \theta_i q\left(\theta_i\right)$. While the probability for an employee to meet an employer $q(\theta_i)$ and the probability for an employer to fill a vacancy $\theta_iq(\theta_i)$ are identical for all employers and employees searching in a given labour market (either high-skilled or low-skilled), the probability to sign a job offer $\kappa_{i,j}$ depends on both the country of origin and the skill level of the employee. The heterogeneity, which is discussed in the literature \citep{Iftikhar2016} and confirmed in the Italian data (Appendix \ref{JobCreationandJobDestruction}), can be explained by the \textit{compatibility} factor: observable skills (e.g., language proficiency), but also unobservable characteristics may favour the match formation of natives compared to non-natives once a contact happens. On the other hand, non-natives might have different outside options compared to natives \citep{chassamboulli2020immigration}, they are on average younger, less tied to a specific geographic location, face lower moving costs \citep{schundeln2014immigrants} and  are more in need to find a job \citep{fullin_2011}: all these features make them more mobile and flexible, with an impact on their probability to form a match. Finally, the heterogeneity among employees is ex-ante and does not imply an ex-post heterogeneity in the new matches, as ``non-compatible'' matches are not formed \citep{rogerson_survey}. 

The exogenous destruction rate  $\delta_{i,j}$ is specific to the type of worker to match the data (Appendix \ref{JobCreationandJobDestruction}): when the shock hits the match, the employee becomes unemployed and the employer is left with an open vacancy. In this circumstance, as in \cite{GaribaldiViolante2005}, the employer is required to pay firing costs $F$, which include two components: a share $\phi$ of the cost is transferred to the employee as a severance payment, while the share $1-\phi$ is a dead-weight loss (red-tape cost). 
Non-native employees coming from abroad join the labour market as unemployed at exogenous rate $\eta$. They also leave the labour market and go abroad at exogenous rate $\lambda$. Both $\lambda$ and $\eta$ are assumed to be the same among employees with different skill levels as they are influenced by socio-political phenomena, which affect all employees similarly.

\subsection{The final good sector}

As in \cite{acemoglu2001good}, the two sub-markets with different skill requirements produce two non-storable intermediate goods that are then sold in a competitive market and then used to produce the final good in each of the two production sectors ($HS$ and  $LS$). The technology of production for the final goods $q_a$, with $a \in \{HS,LS\}$, is:
 \begin{eqnarray}
q_a&=&[\alpha_a q_{h,a}^{\rho_a}+(1-\alpha_a) q_{l,a}^{\rho_a}]^{1/\rho_a},
\label{eq:productionFinalGoods}
\end{eqnarray} 
where $q_{h,a}$ is the quantity of the high-skilled intermediate good $h$ and  $q_{l,a}$ is the quantity of the low-skilled intermediate good $l$ used in the production of final good $a$. Final good $HS$ is assumed to be high-skill intensive, while final good $LS$ low-skill intensive, i.e. $\alpha_{HS} > \alpha_{LS}$. The term $1/(1-\rho_a)$ is the elasticity of substitution between high-skill and low-skill intensive goods in sector $a$, with $\rho_a \in (-\infty,1]$.

\subsection{The intermediate good sector}

The technology of production for the intermediate goods is \textit{Leontief}, i.e., when matched with a firm, a worker produces $x_i$ ($i \in \{h,l\}$) unit of intermediate good. The total production of the intermediate good  $h$ is $q_h=x_h e_h$ and the total production of the intermediate good  $l$ is $q_l = x_l e_l$, where $e_h$ is the employment of high-skilled workers and $e_l$ is the employment of low-skilled workers.
Each intermediate good is then used in the final good sector (Equation (\ref{eq:productionFinalGoods})), i.e. $q_{h,HS}+q_{h,LS}=q_h$ and $q_{l,HS}+q_{l,LS}=q_l$. The goods $h$ and $l$ are sold at price $p_h$ and $p_l$, respectively.

\subsection{Employees}\label{sec:employees}

All employees consume both final goods $q_{HS}$ and $q_{LS}$ and benefit from the provision of public goods $\nu$. As they consume all the income they earn and do not save, their utility function reads as follows:
\begin{equation}\label{eq:CESutility}
Z_{i,j}=\left[\gamma d^{\rho}_{HS,i,j}+(1-\gamma) d^{\rho}_{LS,i,j}\right]^{1/\rho} + \iota \nu,
\end{equation}
where $d_{HS,i,j}$ and $d_{LS,i,j}$ are the quantities demanded and consumed of product $HS$ and $LS$, respectively, by an individual with skill level $i \in \{h,l\}$ and country of birth $j \in \{N,I\}$. The parameter $\rho$ determines the \textit{elasticity of substitution between the two goods} $\epsilon \equiv 1/\left(1-\rho\right) \geq 0$. Whenever $\epsilon \to \infty$ ($\rho \to 1$) the two goods are perfect substitutes, whenever $\epsilon \to 0$ ($\rho \to -\infty$) the two goods are perfect complements, and, finally, whenever $\epsilon = 1$ ($\rho =0$), we have the baseline case of a Cobb-Douglas utility function. The utility $\iota \nu$ comes from the \textit{public goods} provided by the government, where $\iota$ is a proxy for the marginal rate of substitution between private and public goods. We assume that $\gamma \in \left(0,1\right)$ and $\iota>0$.\footnote{We include the term $\iota \nu$ as an additive factor, rather than multiplicative, for tractability purposes.}

Employees maximize their utility function, subject to the following budget constraint:
\begin{equation}\label{budgetConstraint}
p_{HS} d_{HS,i,j} + p_{LS} d_{LS,i,j} = y_{i,j}=
\begin{cases}
\left(1-t\right)\left(w_{i,j}+\tau \right) + m & \text{if the worker $(i,j)$ is employed,}  \\
b\left(1-t\right)\left(w_{i,j}+\tau \right) + m & \text{if the worker $(i,j)$ is unemployed.} 
\end{cases}
\end{equation}
Employees earn a wage $w_{i,j}$, gross of social security contributions, which is taxed at proportional rate $t$.\footnote{In the model (and in real life), the wages paid by the firms ($w_{i,j}$) are gross of taxes and social security contributions, and are  commonly defined as gross welfare wages; gross fiscal wages $(w^{gf}_{i,j})$ instead refer to the amount on which income taxes are computed, i.e., wages gross of taxes but net of social security contributions $(w^{gf}_{i,j}=(1-ssc)w_{i,j})$. Finally, in the data we observe net wages $(w^{n}_{i,j})$ which refer to wages net of income taxes and social security contributions $(w^{n}_{i,j}=(1-t)(1-ssc)w_{i,j})$. See Appendix \ref{sec:Wages} for  Italian data on net wages.}
In order to introduce the possibility of progressive taxation, we assume that all employees receive a \textit{tax subsidy} $\tau$.\footnote{As in \cite{pissarides2000equilibrium}, employees receive a tax subsidy $\tau$, and then are taxed on total labour earnings, including the subsidy, at the proportional tax rate $t$. Hence, the net taxation paid by the employee is $T(w_{i,j}) = t\left(w_{i,j}+\tau\right)- \tau=t w_{i,j}-(1-t)\tau$.} Employees who are unemployed instead receive \textit{unemployment benefits}, which are a proportion $b$ of their net wage. Finally, all employees receive a \textit{lump-sum transfer} from the government equal to $m$.

From the  utility maximization, we get the optimal quantities of goods $HS$ and $LS$ demanded by each individual, depending on her skill level, country of origin	 and employment status. Specifically, if the worker is employed she will demand the following quantities of good $HS$ and good $LS$:

\begin{eqnarray}\label{eq:demandGoodH}
d_{HS,i,j}=\left[\frac{\left(\frac{p_{HS}}{\gamma}\right)^{1/(\rho-1)}}{p_{LS}(\frac{p_{LS}}{1-\gamma})^{1/(\rho-1)}+p_{HS}(\frac{p_{HS}}{\gamma})^{1/(\rho-1)}}\right]y_{i,j} \text{ and} \\\label{eq:demandGoodL}
d_{LS,i,j}=\left[ \frac{\left(\frac{p_{LS}}{1-\gamma}\right)^{1/(\rho-1)}}{p_{LS}(\frac{p_{LS}}{1-\gamma})^{1/(\rho-1)}+p_{HS}(\frac{p_{HS}}{\gamma})^{1/(\rho-1)}}\right]y_{i,j},
\end{eqnarray}  
where $y_{i,j}$ is the work income as defined in Equation (\ref{budgetConstraint}).

%\begin{equation}
%Y^e_i= \left[\gamma \frac{(1-t)(w_i+\tau \mathbbm{1}_i)}{p_h}\right]^\gamma \left[(1-\gamma) \frac{(1-t)(w_i+\tau \mathbbm{1}_i)}{p_l}\right]^{(1-\gamma)}+\nu,
%\end{equation}
%while for unemployed workers:
%\begin{equation}
%Y^u_i= \left[\gamma \frac{b[(1-t)(w_i+\tau \mathbbm{1}_i)]}{p_h}\right]^\gamma \left[(1-\gamma) \frac{b[(1-t)(w_i+\tau \mathbbm{1}_i)]}{p_l}\right]^{(1-\gamma)}+\nu.
%\end{equation}
%Thus, they can be expressed as:
By plugging the optimal demands of the two goods (Equations (\ref{eq:demandGoodH}) and (\ref{eq:demandGoodL})) into Equation (\ref{eq:CESutility}), we can  compute the indirect utility of employed and unemployed employees as:
\begin{equation}
Z_{i,j}=\frac{y_{i,j}}{P}+ \iota \nu,
\end{equation}
where P is the \textit{aggregate price index of final goods} defined as:
%\begin{equation}
%P=\left\{\gamma^{1/(1-\rho)}p_h^{\rho/(\rho-1)}+(1-\gamma)^{1/(1-\rho)}p_l^{\rho/(\rho-1)}\right\}^{(\rho-1)/\rho}.
%\end{equation}
\begin{equation}\label{priceIndexCES}
P \equiv \left[p_{HS}\left(\frac{p_{HS}}{\gamma}\right)^{1/(\rho-1)}+p_{LS}\left(\frac{p_{LS}}{1-\gamma}\right)^{1/(\rho-1)}\right]^{(\rho-1)/\rho}.
\end{equation}
Finally, we can  express the demanded quantities of good $HS$ and good $LS$ as:
\begin{eqnarray}\label{dh}
d_{HS,i,j}=\left(\frac{\tilde{p}_{HS}}{\gamma}\right)^{1/(\rho-1)}\tilde{y}_{i,j} \text{ and}\\\label{dl}	d_{LS,i,j}=\left(\frac{\tilde{p}_{LS}}{1-\gamma}\right)^{1/(\rho-1)}\tilde{y}_{i,j},
\end{eqnarray}
where $\tilde{p}_a \equiv p_a/P$ is the \textit{real price} of good  $a \in \{HS,LS\}$ and $\tilde{y}_{i,j} \equiv y_{i,j}/P$ is the \textit{real income}. From Equations (\ref{dh}) and (\ref{dl}),  we can then calculate the shares of expenditure per each good:
\begin{eqnarray} \label{eq:shareExpenditureHighSkilledGood}
\dfrac{\tilde{p}_{HS} d_{HS,i,j}}{\tilde{y}_{i,j}}=\gamma^{\epsilon} \tilde{p}_{HS}^{1-\epsilon}, \text{ and} \\	
\dfrac{\tilde{p}_{LS} d_{LS,i,j}}{\tilde{y}_{i,j}}=\left(1-\gamma\right)^{\epsilon} \tilde{p}_{LS}^{1-\epsilon};	 \label{eq:shareExpenditureLowSkilledGood}
\end{eqnarray}
for $\epsilon=1$ such shares are independent of the prices level, while for $\epsilon \neq 1$ the rate of change of the shares is equal to the rate of change of the prices multiplied by $1-\epsilon$.\footnote{In Appendix \ref{app:elasticityOfSubstitutionFinalGood} we exploit Eqq. (\ref{eq:shareExpenditureHighSkilledGood}) and (\ref{eq:shareExpenditureLowSkilledGood}) to argue that $\epsilon$ falls in the range $[1,2]$ for the case of the Italian economy.}

\subsection{Employers}

Employers can be active on both markets and open multiple vacancies simultaneously.
  The utility of employer $q$ reads as:
%\begin{equation}
%Z_q= d_{h,q}^\gamma d_{l,q}^{(1-\gamma)}+ \iota \nu.
%\end{equation}
\begin{equation}
Z_{q}=\left[\gamma d^{\rho}_{HS,q}+(1-\gamma) d^{\rho}_{LS,q}\right]^{1/\rho} + \iota \nu.
\end{equation}
Employers consume both final goods and receive the provision of public goods $\nu$, as the employees. We assume that also employers do not save. Employers maximize their utility subject to the following budget constraint:
\begin{equation}
p_{HS} d_{HS,q} + p_{LS} d_{LS,q} =\Pi_q +m,
%\label{bc_empl}
\end{equation}
where $\Pi_q$ are the employer's $q$ net profits and $m$ is the lump-sum transfer from the government.
%The optimal real quantities of goods $h$ and $l$ demanded by employer $q$ are:
%%\begin{eqnarray} \nonumber
%%d_{h,q} &=& \frac{\gamma }{p_h} \text{ and}\\
%%d_{l,q} &=& \frac{\left(1-\gamma\right)\Pi_q}{p_l}.
%%\label{eq:demandGoodEntrepreneur}
%%\end{eqnarray}
%\begin{eqnarray}
%d_{h,q}=\frac{\left(\frac{p_h}{\gamma}\right)^{1/(\rho-1)}}{p_l(\frac{p_l}{1-\gamma})^{1/(\rho-1)}+p_h(\frac{p_h}{\gamma})^{1/(\rho-1)}}\Pi_q\equiv B_h\left(\frac{p_h}{p_l}\right) \frac{\Pi_q}{p_h},\\
%d_{l,q}=\frac{\left(\frac{p_l}{1-\gamma}\right)^{1/(\rho-1)}}{p_l(\frac{p_l}{1-\gamma})^{1/(\rho-1)}+p_h(\frac{p_h}{\gamma})^{1/(\rho-1)}}\Pi_q\equiv B_l\left(\frac{p_h}{p_l}\right) \frac{\Pi_q}{p_l},
%\end{eqnarray}  
%where $\Pi_q$ is the work income as defined in Equation (\ref{budgetConstraint}).
We can then express the  demanded quantities of good $HS$ and good $LS$ in real terms, using the price index P (Equation (\ref{priceIndexCES})) as:
\begin{eqnarray}\label{demandHSEmployer}
d_{HS,q}&=&\left(\frac{\tilde{p}_{HS}}{\gamma}\right)^{1/(\rho-1)}\tilde{\Pi}_q \text{ and}\\\label{demandLSEmployer}	d_{LS,q}&=&\left(\frac{\tilde{p}_{LS}}{1-\gamma}\right)^{1/(\rho-1)}\tilde{\Pi}_q ,
\end{eqnarray}
where $\tilde{\Pi}_q \equiv (\Pi_q+m)/P$ denotes the \textit{real net profits} plus the government transfer.

%The indirect utility of employer $q$ operating in the market $i$ ($i \in \{h,l\}$) hiring a employee of type $j$ ($j \in \{N,I\}$) is:
%\begin{eqnarray}\label{y_empl}
%Z^f_{q,i,j}&=& \left(1-t\right)\left(\tilde{p}_i x_i-\tilde{w}_{i,j}\right) + \iota \nu, \label{y_employers}
%\end{eqnarray}
%while the indirect utility of employers with an open vacancy reads as 
%\begin{eqnarray}\label{y_empl}
%Z^v_{q,i}&=&  \iota \nu -c\tilde{p}_i x_i  \label{y_employers}
%\end{eqnarray}

 \subsection{Bellman's equations of employees and employers}
 
 Let $W^{e}_{i,j}$ be the present discounted value of the utility of an employee with skill level $i$ and country of origin $j$ who is currently employed.  Following \cite{pissarides2000equilibrium}, the corresponding Bellman's equation reads:
 \begin{eqnarray}
 rW^{e}_{i,j}  &=&  Z^e_{i,j} + \delta_{i,j}\left(W^u_{i,j} + \phi \tilde{p}_i x_i F- W^e_{i,j}\right).  \label{eq:BellmanEmployedNative} 
 %rW^{e}_{i,I}  &=&  Z^e_{i,I}+ \delta_{i,I}\left(W^u_{i,I} + \phi\tilde{p}_i x_i F- W^e_{i,I}\right). \label{eq:BellmanEmployedImmigrant}  
 \end{eqnarray}
 Employees who are employed, both natives and non-natives, enjoy the indirect utility of being employed $Z_{i,j}^e$. At rate $\delta_{i,j}$ the match is exogenously destroyed  and the employees become unemployed: in these circumstances, the employees receive a transfer, which is a share $\phi$ of the total firing costs $\tilde{p}_i x_i F$ paid by the employers.
 
 The Bellman's equations for employees who are unemployed  read:
 \begin{eqnarray}
 rW^{u}_{i,N} &=& Z^u_{i,N}+\kappa_{i,N}\theta_i q\left(\theta_i\right)\left(W^e_{i,N}- W^u_{i,N}\right)
 \label{eq:BellmanUnemployedNative} \text{ and}\\
 rW^{u}_{i,I} &=& Z^u_{i,I}+\kappa_{i,I} \theta_i q\left(\theta_i\right)\left(W^e_{i,I}- W^u_{i,I}\right) + \lambda\left(W_{i,FC} - W^u_{i,I}\right).
 \label{eq:BellmanUnemployedImmigrant}
 \end{eqnarray}
 While both natives and non-native employees find a job with probability $\kappa_{i,j}\theta_i q(\theta_i)$, non-native employees have the additional outside option of leaving the  country at rate $\lambda$,\footnote{The model could be extended by allowing native employees to leave the country too. However, in Italy the share of native unemployed who emigrate is very low (about 2.5\%, Appendix \ref{app:inflowOutflowNatives}). Nevertheless, our main theoretical and empirical findings would be substantially unchanged.} and enjoying utility $W_{i,FC}$ elsewhere. We can interpret $\lambda$  as either the individual decision to go back home or as a government policy, which forces non-native unemployed  to be expelled.\footnote{The rate at which non-natives exit the country $\lambda$ is  determined by socio-political factors in origin countries, by migration policies in host countries, and by the personal choice of non-natives. The first two factors appear to be the most important for several countries,  suggesting that the rate at which non-natives exit the country is  mainly independent of the skill level of non-natives. Nevertheless, introducing skill-specific $\lambda$ would not change our main findings, while substantially increasing the algebraic computation.}
 
 Let $J_{i,j}$ be the present discounted value of the utility of an employer hiring an employee with skill level $i$ and of country of origin $j$. The employer's Bellman's equation for a filled position reads:\footnote{See in particular Equation (9.9) in \cite{pissarides2000equilibrium}.}
 %\todo[author=Davide,inline]{L'imprenditore quando decide che fare sembra considerare un profitto al netto delle tasse, ma che non include il tranferimenti e i beni pubblici. Effettivamente queste cose rientrano nel reddito netto che lui, ricevendo i profitti tassati, dovrà calcolare per decideere il suo consumo. E per arrivare all'utilità dovremo considerare anche i beni pubblci. Quindi forse è improprio parlare di utilità nel testo. }
 \begin{eqnarray} 
 r J_{i,j} &=& \left(1-t\right)\left(\tilde{p}_i x_i-\tilde{w}_{i,j}\right) + \delta_{i,j}\left(V_i-J_{i,j}- \tilde{p}_i x_iF \right), 
 \label{eq:filledjobNative}
 %r J_{i,I} &=& (1-t)(\tilde{p}_i x_i-\tilde{w}_{i,I})+ \delta_{i,I}\left(V_i-J_{i,I}- \tilde{p}_i x_iF \right),
 %\label{eq:filledjobImmigrant}
 \end{eqnarray}
 where $i \in \{l,h\}$ and $\tilde{p}_ix_i$ is the real value added of each employee in the labour market $i$. The  employer value function in Equation (\ref{eq:filledjobNative}) takes into account the presence of \textit{firing costs}. In particular, every time an exogenous  shock $\delta_{i,j}$  destroys a match, the employer is required to pay firing costs $F$, which are proportional to the value added of the employee, as in \citet{pissarides2000equilibrium}. 
 The firm's Bellman's equation for hiring an employee with skill level $i$, i.e. the value of a skill-specific vacancy, is given by:
 \begin{equation}
 rV_i=  -c\tilde{p}_i x_i + \pi_{i,N} \kappa_{i,N}q \left(\theta_i\right) (J_{i,N} -V_i)+\left(1-\pi_{i,N}\right) \kappa_{i,I}q\left(\theta_i\right) (J_{i,I} -V_i),
 \label{eq:vacancy}
 \end{equation}
 where $c\tilde{p}_i x_i$ is the vacancy cost which is proportional to the value added of the employee, $\kappa_{i,j}q\left(\theta_i\right)$ is the rate at which a vacancy is filled, and $\pi_{i,N} \equiv u_{i,N}/\left(u_{i,N}+u_{i,I}\right)$ is the probability for an employer operating in the good market $i$ to meet a native employee, which is computed as the  share of unemployed natives on the total number of unemployed.

\subsection{Wage bargaining}

The  wages of native and non-native employees are chosen to maximize the surplus of the match between employer and employee:
\begin{eqnarray}
\left(W^{e}_{i,j}-\phi \tilde{p}_i x_i F-W^{u}_{i,j}\right)^{\beta_j}\left(J_{i,j}+ \tilde{p}_i x_i F-V_i\right)^{1-\beta_j},
%\label{eq:NashProductNative}
\end{eqnarray}
where $\beta_j$ is the bargaining power of the employees and $1-\beta_j$ is the bargaining power of the employers. The bargaining power $\beta_j$ is different for employees with different country of origin, and specifically we expect $\beta_N\geq\beta_I$, for several reasons \citep{muthoo2001economics}. In particular, he points to three main features affecting the worker's bargaining power: the degree of impatience, the risk of breakdown due to exogenous and uncontrollable factors and  inside and outside options. All these features concur to lower the bargaining power of non-natives.
The maximization with respect to wages amounts to maximize the Nash product with respect to $\tilde{w}_i$, being the price index $P$ taken as given in competitive markets. Employers and employees also take $V_i$, $W^{u}_{i,N}$ and $W^{u}_{i,I}$ as given in the bargaining process. Firing costs $F$ enter into the maximization as employers internalize the cost they will have to pay in case of match destruction. The share of the firing costs $\phi$ which is transferred to the employees  enters into the maximization as it is part of the outside option of the employees. Hence, the first order condition for the maximization of the Nash product reads:
\begin{equation}
(1-\beta_j)(W^{e}_{i,j}-\phi \tilde{p}_i x_iF-W^{u}_{i,j}) = \beta_j  \left( J_{i,j} +  \tilde{p}_i x_i F - V_i \right).
\label{eq:NashProductNative}
\end{equation}

\subsection{The government}

To conclude, we describe the government activities. In particular, the government collects the following amount of \textit{net direct taxes} (NDT) :
\begin{eqnarray} \label{eq:NDT}\nonumber
NDT &\equiv & t\left[\tilde{p}_{HS} q_{HS} +  \tilde{p}_{LS} q_{LS} + b\left( \tilde{w}_{h,N}u_{h,N} + \tilde{w}_{h,I}u_{h,I} + \tilde{w}_{l,N}u_{l,N} + \tilde{w}_{l,I}u_{l,I}\right) \right] + \\ \notag
&-& \left(1-t\right)\tilde{\tau}\left[e_{l,N}+e_{l,I}+e_{h,N}+e_{h,I} + b\left(u_{l,N}+u_{l,I}+u_{h,N}+u_{h,I}\right) \right] + \\
&-& \left( \sigma_{h,N}+\sigma_{l,N}+\sigma_{h,I}+\sigma_{l,I} + \chi \right) \tilde{m}
, \label{eq:DT}
\end{eqnarray}
where $\tilde{m}\equiv m/P$. The first term of Equation \ref{eq:NDT} represents the taxation of income and profits, while the second and third terms represent the tax subsidy and the lumps-sum transfer, respectively.
The government also collects the following amount of \textit{indirect taxes} (IT):\footnote{The amount of indirect taxes depends on the ratio between $t_p$ and $1-t_p$ as $\tilde{p}_ax_a$ is the production \textit{net} of the indirect tax and production subsidies.}
\begin{eqnarray}
IT \equiv \left(\frac{t_p}{1-t_p}\right)\left(\tilde{p}_{HS} q_{HS} + \tilde{p}_{LS} q_{LS}\right),
\label{eq:IT}
\end{eqnarray}
where $t_p$ is the indirect tax rate net of production subsidies.

Finally, we assume that the government expenditure in public goods is equal to the share $g_{HS}$ and $g_{LS}$ of the aggregate real value added (GVA) of $HS$ and $LS$ goods. Thus, the \textit{total government expenditure} (TGE) amounts to:
\begin{eqnarray} 
TGE &\equiv &  b\left[\tilde{w}_{h,N}u_{h,N} + \tilde{w}_{h,I}u_{h,I} + \tilde{w}_{l,N}u_{l,N} + \tilde{w}_{l,I}u_{l,I}\right] + g_{HS} \tilde{p}_{HS}  q_{HS} + g_{LS} \tilde{p}_{LS} q_{LS},
\label{eq:TGE}
\end{eqnarray}
where the first term is the provision of unemployment benefits and the remaining two terms capture government expenditure in public goods. In equilibrium, the government primary surplus is equal to zero, i.e., $TGE=NDT+IT$ as in any closed economy with no investments and savings, independent on any fiscal parameter (see Equation (\ref{app:demandsupplygoods}) in  Appendix \ref{app:equilibriumGoodMarket}). Therefore, an increase in the lump-sum transfer $\tilde{m}$ would imply both a decrease in NDT and a decrease in TGE.

The per capita provision of public goods is subject to congestion, as measured by the parameter $\zeta \geq 0$ ($\zeta = 0$ is the case of pure public goods):
\begin{eqnarray}
\nu & = & \dfrac{ g_{HS}\tilde{p}_{HS} q_{HS} + g_{LS}\tilde{p}_{LS} q_{LS}}{\left(\sigma_{h,N}+\sigma_{l,N}+\sigma_{h,I}+\sigma_{l,I}+ \chi +IP\right)^\zeta},\label{budget}
\end{eqnarray}
where the denominator of Equation (\ref{budget}) is the total population, which includes the labour force (employers and employees, employed and unemployed) and the rest of the population $IP$ (inactive individuals and people who are not in the working age). Thus, immigrants affect the per capita provision of public goods, and in turn welfare, through their impact on aggregate productivity, via a compositional effect. In particular, since immigrants are mostly low-skilled, they will have less than proportional  impact on the aggregate value added, in turn reducing the per capita provision of public goods. This effect could be counterbalanced by a low rate of congestion in the provision of public goods, as measured by $\zeta$ (a higher $\zeta$ implies a higher rate of congestion).\footnote{For example, public health and education would  call for a  $\zeta>0$, while defense would call for a  $\zeta=0$.}

Finally, a gap between the amount of taxes that the government should collect and the actual amount of taxes paid is present, the so-called tax gap ($t_g$), due to tax policies and (low) tax compliance (see Appendix \ref{app:TaxGap}). 
The lump-sum transfer $\tilde{m}$ provided by the government to all employees and employers is proportional to this tax gap:
\begin{eqnarray}
\tilde{m} &=&   \left(\dfrac{t_g}{\sigma_{h,N}+\sigma_{l,N}+\sigma_{h,I}+\sigma_{l,I}+ \chi}\right)\bigg\{  t\left[\tilde{p}_{HS} q_{HS} +  \tilde{p}_{LS} q_{LS} + b\left( \tilde{w}_{h,N}u_{h,N} + \tilde{w}_{h,I}u_{h,I} \right. \right. \\ \notag&+& \left.\left. \tilde{w}_{l,N}u_{l,N} + \tilde{w}_{l,I}u_{l,I}\right) \right] - \left(1-t\right)\tilde{\tau}\left[e_{l,N}+e_{l,I}+e_{h,N}+e_{h,I} + b\left(u_{l,N}+u_{l,I}+u_{h,N}+u_{h,I}\right) \right]
\bigg\},
\label{eq:lumpSumTrasfer}
\end{eqnarray}
or, more intuitively, the net direct taxes can be expressed as:
\begin{eqnarray}
NDT &=& (1-t_g)  \big\{  t\left[\tilde{p}_{HS} q_{HS} +  \tilde{p}_{LS} q_{LS} + b\left( \tilde{w}_{h,N}u_{h,N} + \tilde{w}_{h,I}u_{h,I} + \tilde{w}_{l,N}u_{l,N} + \tilde{w}_{l,I}u_{l,I}\right) \right]  \notag\\&-&  \left(1-t\right)\tilde{\tau}\left[e_{l,N}+e_{l,I}+e_{h,N}+e_{h,I} + b\left(u_{l,N}+u_{l,I}+u_{h,N}+u_{h,I}\right) \right]
\big\}.
\label{eq:taxGapNDT}
\end{eqnarray}

\section{The equilibrium \label{sec:equilibrium}}

In the equilibrium all markets clear, i.e., an equilibrium is reached in both labour markets (high-skilled and low-skilled), and in both intermediate and final goods markets.

\subsection{The equilibrium in the labour markets}

The free-entry condition in each labour market implies that the value of a vacancy in each of the two markets is equal to zero, that is $V_i= 0$. Hence, in equilibrium  employers are indifferent whether to open a low-skilled or a high-skilled vacancy.
We derive the job-creation curve, which is market-specific, by substituting Equation (\ref{eq:filledjobNative}) in  Equation (\ref{eq:vacancy}) and applying the free-entry condition:
\begin{eqnarray}
&&\nonumber \pi_{i,N}\kappa_{i,N}q(\theta_i)\left[\frac{(1-t)(\tilde{p}_ix_i-\tilde{w}_{i,N})-\delta_{i,N}\tilde{p}_ix_iF}{r+\delta_{i,N}}\right]+\\&+& 
(1-\pi_{i,N})\kappa_{i,I}q(\theta_i)\left[\frac{(1-t)(\tilde{p}_ix_i-\tilde{w}_{i,I})-\delta_{i,I}\tilde{p}_ix_iF}{r+\delta_{i,I}}\right]= c \tilde{p}_i x_i.
\label{eq:JCC}
\end{eqnarray}
Equation (\ref{eq:JCC}) states that the expected benefit for an employer with a filled vacancy, which is given by the weighted average of the benefit received in case the employer hires a native or a non-native worker, appropriately discounted at rate $r+ \delta_{i,j}$ must be equal to the cost of creating the vacancy (the right hand side of Equation (\ref{eq:JCC})). The benefit is a positive function of the employee's value added $\tilde{p}_i x_{i}$ and a negative function of the wage paid $w_{i,j}$ and of the firing cost to be paid in case of dismissal $\delta_{ij}\tilde{p}_ix_iF$. 

To compute the equilibrium wages of native and non-native employees in each market we subtract Equation (\ref{eq:BellmanUnemployedNative}) from Equation (\ref{eq:BellmanEmployedNative}), and 
%\begin{eqnarray}
%W^{e}_{i,N}-W^{u}_{i,N}&=&\dfrac{(1-b)\left(1-t\right)(\tilde{w}_{i,j}+\tilde{\tau} + \delta_{i,N}\phi \tilde{p}_i x_i F)}{r+\delta_{i,N}+ \kappa_{i,N} \theta_i q\left(\theta_i\right)} \text{ and} \label{we-wu_native1}\\
%W^{e}_{i,I}-W^{u}_{i,I}&=&\dfrac{(1-b)\left(1-t\right)\left(\tilde{w}_{i,j}+\tilde{\tau} \right)-\lambda\left(W_{i,FC} - W^u_{i,I}\right)+\delta_{i,I}\phi \tilde{p}_i x_i F}{r+\delta_{i,I}+ \kappa_{i,I} \theta_i q\left(\theta_i\right)}.\label{we-wu_immigrant1}
%\end{eqnarray}
by plugging the resulting equation into the Nash bargaining Equation  (\ref{eq:NashProductNative}), we get an expression for the wages of native employees in each market, as a function of the parameters of the model:\footnote{See Appendix \ref{app:equilibriumwages} for the details of calculations.}
\begin{eqnarray}
\tilde{w}_{i,N}&=&\underbrace{A_{i,N}(\theta_i)}_{>0}\tilde{p}_i x_i- \underbrace{B_{i,N}(\theta_i)}_{>0} \tilde{\tau} +  \underbrace{C_{i,N}(\theta_i)}_{>0}\tilde{p}_i x_i F, \label{wageNshort}
\end{eqnarray}
where $A_{i,N}(\theta_i)$, $B_{i,N}(\theta_i)$ and $C_{i,N}(\theta_i)$ are positive and the first derivatives with respect to $\theta_i$ are positive for $A_{i,N}(\theta_i)$ and $C_{i,N}(\theta_i)$ and negative for $B_{i,N}(\theta_i)$.
Similarly, we get an expression for the wages of immigrant employees in each market, as a function of the parameters of the model:\footnote{See Appendix \ref{app:equilibriumwages} for the details of calculations.}
\begin{eqnarray}
\tilde{w}_{i,I}&=& \underbrace{D_{i,I}(\theta_i)}_{>0}\tilde{p}_i x_i - \underbrace{E_{i,I}(\theta_i)}_{>0}\tilde{\tau} +
\underbrace{G_{i,I}(\theta_i)}_{>0}\tilde{p}_i x_i F+ \underbrace{K_{i,I}(\theta_i)}_{>0} W_{i,FC}  - \underbrace{H_{i,I}(\theta_i)}_{>0} (\iota\nu + \tilde{m}),\label{wageIshort}
\end{eqnarray}
where $D_{i,I}(\theta_i)$, $E_{i,I}(\theta_i)$, $G_{i,I}(\theta_i)$, $K_{i,I}(\theta_i)$ and $H_{i,I}(\theta_i)$ are positive and the first derivatives with respect to $\theta_i$ are positive for $D_{i,I}(\theta_i)$ and $G_{i,I}(\theta_i)$ and negative for  $E_{i,I}(\theta_i)$, $K_{i,I}(\theta_i)$ and $H_{i,I}(\theta_i)$.

Equations (\ref{wageNshort}) and (\ref{wageIshort}) show that the real wages of  both native and non-native employees are positive functions of the employees' real value added $\tilde{p}_i x_i$ and of the firing cost $\tilde{p}_i x_i F$ paid by the employer in case of match destruction, as in the standard Diamond-Mortensen-Pissarides (DMP) model, because employers internalize the cost of dismissal. Instead, the real wages are negative functions of the real tax subsidy $\tilde{\tau}$ provided by the government to all employees. As the possibility of leaving the country at rate $\lambda$ affects the outside option of non-native employees, their wage includes also additional factors. Specifically, their real wage is a positive function of the utility the employees would get if deciding to leave the country $W_{i,FC}$, and a negative function of public goods $\nu$ and lump-sum transfer $\tilde{m}$, which they will lose in case of emigration.
Finally, as in the standard DMP, wages are positively related to the market tightness $\theta_i$, with the partial exception of the term depending on the outside option of non-native employees as when the number of vacancies is higher the opportunity cost of leaving is higher.

Employment and unemployment for native employees in equilibrium can be computed as:
\begin{eqnarray}
e_{i,N} &=& \sigma_{i,N} \left[\dfrac{\kappa_{i,N}\theta_i q\left(\theta_i\right)}{\delta_{i,N} + \kappa_{i,N} \theta_i q\left(\theta_i\right)}\right] \text{ and}
\label{eq:equilibriumEmploymentNatives}\\
u_{i,N} &=& \sigma_{i,N} \left[\dfrac{\delta_{i,N}}{\delta_{i,N} + \kappa_{i,N}\theta_i q\left(\theta_i\right)}\right],
\label{eq:equilibriumUnemploymentNatives}
\end{eqnarray}
while non-natives employed and unemployed in equilibrium are given by:
\begin{eqnarray}
e_{i,I}&=&\sigma_{i,I} \left\{ \dfrac{\kappa_{i,I}\theta_i q\left(\theta_i\right)}{ \delta_{i,I}+\kappa_{i,I}\theta_i q\left(\theta_i\right)} \right\} \text{ and}
\label{eq:equilibriumEmploymentImmigrants}\\
u_{i,I}&=&\sigma_{i,I} \left\{  \dfrac{ \delta_{i,I}}{\delta_{i,I}+\kappa_{i,I}\theta_i q\left(\theta_i\right)} \right\}.
\label{eq:equilibriumUnemploymentImmigrants}
\end{eqnarray}
Finally, the measure of non-natives in the foreign country in equilibrium is:
\begin{equation}
FC_{i,I}=\sigma_{i,I} \left\{ \dfrac{\lambda \delta_{i,I}}{\eta\left[\kappa_{i,I}\theta_i q_{i,I}\left(\theta_i\right)+\delta_{i,I}\right]} \right\}.
\label{eq:equilibriumResidentRSoL}
\end{equation}

\subsection{The equilibrium in the final good markets}

Using Equation (\ref{priceIndexCES}), we can express the \textit{real} price of good $HS$ as a function of the real price of good $LS$ as:
\begin{equation}
\tilde{p}_{HS}=\left[\frac{1-\tilde{p}_{LS}^{\rho/(\rho-1)}\left(1-\gamma\right)^{1/(1-\rho)}}{\gamma^{1/(1-\rho)}}\right]^{(\rho-1)/\rho}.
\label{phslequation1}
\end{equation}
By equating demand of good $HS$ and good $LS$ (Equations (\ref{dh})-(\ref{dl}) and (\ref{demandHSEmployer})-(\ref{demandLSEmployer})) and supply of good $HS$ and good $LS$ and using Equation (\ref{phslequation1}), we get the following expressions for the equilibrium prices:
\begin{eqnarray}\label{priceHpriceL}
\tilde{p}_{HS}&=&\left\{\gamma^{1/(1-\rho)}+(1-\gamma)\gamma^{\rho/(1-\rho)}\left[\frac{(1-g_{LS})q_{LS}}{(1-g_{HS})q_{HS}}\right]^{\rho}\right\}^{(1-\rho)/\rho} \text{ and}\\\label{priceL} \tilde{p}_{LS}&=&\left\{(1-\gamma)^{1/(1-\rho)}+\gamma(1-\gamma)^{\rho/(1-\rho)}\left[\frac{(1-g_{HS})q_{HS}}{(1-g_{LS})q_{LS}}\right]^{\rho}\right\}^{(1-\rho)/\rho}.
\end{eqnarray}

\subsection{The equilibrium in the intermediate good markets}

The two intermediate goods are sold in competitive markets, thus their prices are equal to their marginal productivity. Moreover, the same type of intermediate good sold in the two markets ($HS$ and $LS$) must have the same price. Therefore, in equilibrium:
\begin{eqnarray}
	p_{h}&=&p_{HS}\alpha_{HS} q_{h,HS}^{\rho_{HS}-1} q_{HS}^{1-\rho_{HS}}=p_{LS}\alpha_{LS} q_{h,LS}^{\rho_{LS}-1} q_{LS}^{1-\rho_{LS}}; \text{and}\\
	p_{l}&=&p_{HS}(1-\alpha_{HS}) q_{l,HS}^{\rho_{HS}-1} q_{HS}^{1-\rho_{HS}}=p_{LS}(1-\alpha_{LS}) q_{l,LS}^{\rho_{LS}-1} q_{LS}^{1-\rho_{LS}},
	\label{eq:intermediateGoodPrices}
\end{eqnarray}
where $q_{i,a}$, $i \in \{h, l \}$ and $a \in \{HS, LS \}$  is the quantity of the intermediate good $i$ used in sector $a$.

\subsection{General equilibrium effects of an immigration shock}\label{Generalequilibriumeffects}

A complete analytical study of the equilibrium properties appears as a difficult task; however,  an intuition about the effects of an immigration shock on the labour and goods markets can help clarify the results of the counter-factual scenarios, discussed in Section \ref{sec:calibration}. 
A sudden increase in the stock of low-skilled non-native employees produces a change in the equilibrium value of a firm (see Equation (\ref{eq:filledjobNative}), setting $V_i=0$) which is operating in the low-skill intensive market equal to:
\begin{eqnarray}
\frac{\partial J_{l,j}}{\partial \sigma_{l,I}}=\frac{(1-t-\delta_{l,j}F)x_l}{r+\delta_{l,j}}\frac{\partial \tilde{p}_l}{\partial \sigma_{l,j}} -\frac{1-t}{r+\delta_{l,j}}\frac{\partial \tilde{w}_{l,j}}{\partial \sigma_{l,I}},\label{eq:partailjsigma}
\end{eqnarray}
which implies that:
\begin{eqnarray}
\frac{\partial J_{l,j}}{\partial \sigma_{l,I}}>0 \iff \epsilon_{\tilde{w}_{l,j},\sigma_{l,I}} <\epsilon_{\tilde{p}_{l},\sigma_{l,I}}\frac{(1-t-\delta_{l,j}F)\tilde{p}_l x_i}{(1-t)\tilde{w}_{l,j}},\label{eq:elasticities}
\end{eqnarray}
where $\epsilon_{\tilde{p}_{l},\sigma_{l,I}}$ is the elasticity of the price of the low-skill intensive good with respect to the stock of low-skilled non-natives and $\epsilon_{\tilde{w}_{l,j},\sigma_{l,I}}$ is the elasticity of the wages of the low-skilled employees with respect to the stock of low-skilled non-natives. 
Hence, the sign of the impact of an immigration shock is undetermined a priori, depending on the magnitude of the two elasticities $\epsilon_{\tilde{w}_{l,j},\sigma_{l,I}}$ and $\epsilon_{\tilde{p}_{l},\sigma_{l,I}}$. While the latter is always negative  (or zero in case of perfect substitutability of the two goods) (Equation (\ref{priceL})), the former depends on the elasticity of the low-skilled wage  with respect to the market tightness, which is ambiguous, and on the elasticity of the price of the low-skill intensive good with respect to $\sigma_{l,I}$, which is negative.\footnote{In particular, assuming $B_{i,N}=0$ in Equation (\ref{wageNshort}), $ \epsilon_{\tilde{w}_{l,N},\sigma_{l,I}}= \epsilon_{{A}_{l,N}+{C}_{l,N},\theta_{l}}\times \epsilon_{{\theta}_{l},\sigma_{l,I}}+\epsilon_{{p}_{l},\sigma_{l,I}}$, where the first term is positive and the last is negative. However, the sign of the second term is ambiguous, depending on the values of the probability that the job offer is signed, $\kappa_{i,j}$.} Thus, the overall effect can be only determined numerically. In other words, employers operating in the low-skill intensive market can be winners or losers in case of a low-skilled immigration shock depending on the functioning of the goods and labour markets.

In a similar fashion, the change in the value of a firm which operates in the high-skilled sector following a low-skilled immigration shock reads:
\begin{eqnarray}
\frac{\partial J_{h,j}}{\partial \sigma_{l,I}}>0 \iff
\epsilon_{\tilde{w}_{h,j},\sigma_{l,j}} <\epsilon_{\tilde{p}_{h},\sigma_{l,I}}\frac{(1-t-\delta_{h,j}F)\tilde{p}_h x_h}{(1-t)\tilde{w}_{h,j}},\label{eq:elasticitiessigmah}
\end{eqnarray}
where $\epsilon_{\tilde{p}_{h},\sigma_{l,I}}$ is the elasticity of the price of the high-skill intensive good with respect to the stock of low-skilled non-natives and $\epsilon_{\tilde{w}_{h,j},\sigma_{l,I}}$ is the elasticity of the wage of high-skilled employees with respect to the stock of low-skilled non-natives.
Once again, the impact of an immigration shock on the value of a firm which operates in the high-skill intensive market cannot be determined analytically, but only numerically. In fact, the elasticity of the price of the high-skill intensive good with respect to the stock of low-skilled non-natives  $\sigma_{l,I}$ is positive (Equation (\ref{priceHpriceL})), however the sign of the elasticity of the wage of high-skilled employees with respect to the stock of low-skilled non-natives is ambiguous, depending on the sign of the elasticity of the wage of high-skilled employees with respect to the market tightness.

The change in the equilibrium lifetime utility of employed employees (Equation (\ref{eq:BellmanEmployedNative})) as a consequence of an increase in the stock of low-skilled non-natives reads:
\begin{eqnarray}
\frac{\partial W^e_{i,j}}{\partial \sigma_{l,j}}=\frac{\partial W^e_{i,j}}{\partial \tilde{w}_{i,j}}\frac{\partial \tilde{w}_{i,j}}{\partial \sigma_{l,j}}+\frac{\partial W^e_{i,j}}{\partial \nu}\frac{\partial \nu}{\partial \sigma_{l,j}}+\frac{\partial W^e_{i,j}}{\partial \tilde{p}_{i}}\frac{\partial \tilde{p}_{i}}{\partial \sigma_{l,j}}\label{eq:partialwpartialsigma}.
\end{eqnarray}
 The price of the high (low)-skilled good positively  (negatively) depends on $\sigma_{l,j}$ (Equation (\ref{priceHpriceL})), and hence the last term is always positive (negative) for high (low)-skilled employees. However, the signs of the first and second terms are undetermined, as the wage of high (low)-skilled employees and the degree of congestion in the provision of public goods ambiguously depend on $\sigma_{l,j}$ (Equation (\ref{wageNshort}) and Equation (\ref{budget})). Hence, also the impact of an immigration shock on the lifetime utility of employed employees can be calculated only numerically. Finally, an increase in the stock of non-natives raises total GDP as well as government revenues and social contributions. However, the impact on GDP per worker is analytically ambiguous, depending on the value of the elasticities of several variables with respect to the stock of non-natives. Among those, crucial is the magnitude of the  elasticities of the goods prices. In Section \ref{sec:calibration} we will discuss evidence for Italy for all the variables.

\section{Model estimation \label{EmpiricalEvidence}}

In this section we bring the model to the data to quantify the impact of immigration  in Italy.\footnote{Appendix \ref{app:ItalianLabourMarket} contains some figures about the immigration and the labour market in Italy.} The model is estimated year by year over the period 2008-2017, i.e., we consider each year separately assuming that the economy is near its equilibrium.
We classify individuals in two categories, high-skilled (h) and low-skilled (l),  following the ILO classification: high-skilled employees are those with skill levels 3 or 4, e.g. managers, professionals and technicians; while, low-skilled employees are those with skill levels 1 or 2, e.g. clerks, sales employees, craft employees, plant and machine operators.
For those for whom we do not observe the occupation, we use the occupation in their last job. For those for whom no information is available, we use the education level and correct for the issue of mismatch.\footnote{Details are reported in Appendix \ref{workerclassification}. The classification by education level is in line with a group of studies \citep{card2001can,card2009immigration,goldin2009race}  which has argued that the most relevant partition across employees by education groups is between people with at least some college education and people with a high school degree or less, i.e., "college-educated" and "non-college-educated" \citep{peri2016immigrants}.}
Sectors are classified in the two categories of low-skilled intensive (LS) and high-skilled intensive (HS), according to the skill composition of their workers. In particular, we identify fifteen sectors as HS and forty-eight as LS on the basis of the high-skilled/low-skilled employee share (if $\geq1$ the sector is classified as HS) (see Appendix \ref{app:sectors}).
In the following, in Section \ref{sec:calibration} we discuss the calibration of some of the model’s parameters, while in Section \ref{sec:estimation} we explain in detail the estimation method.\footnote{Data (only the part publicly available) and codes are available at \url{https:/https://people.unipi.it/davide_fiaschi/ricerca/}.}

\subsection{Model's parameters \label{sec:calibration}} 

Table \ref{tab:parameterCalibrationDes} reports the list of the model's parameters.
We take the value of three parameters, i.e., the discount rate, the share of firing costs transferred to employees and the elasticity of the matching function with respect to unemployment, from the literature. Specifically, \citet{paserman2008job} discusses how the discount rate $r$ is not identifiable, and needs to be set exogenously. In standard search models  future choices are discounted exponentially, however according to the hyperbolic discounting theory \citep{laibson1997golden, Cohen2020}, individual choices are time inconsistent, as people discount much more the close future rather the long run, i.e., $r$ is generally set to a value which is too low. In our calibration, we set the discount rate $r$ to a conservative level of 0.01, which corresponds to an annual discount rate of  12\%, in line with the average annual  discount rate estimated in the literature \citep{paserman2008job, dellavigna2005job}. We also perform a sensitivity analysis with $r$ equal to 0.005 and 0.015, which correspond respectively to annual discount rates of 5\% and 18\% (the lower and upper bounds suggested in the literature), and we get
approximately the same quantitative results. The share of firing costs transferred to employees $\phi$ is taken from the literature \citep{GaribaldiViolante2005}. As standard in the literature (\citealp{pissarides2000equilibrium}), we assume that the matching function is shaped as a Cobb-Douglas, $m(v_i,u_i)= u_i^\alpha v_i^{1-\alpha}$,  with $\alpha$ defined as the elasticity with respect to unemployment.  The parameter $\alpha$ is set to a conservative value of 0.4, which is in the range suggested  in the literature and in line with the evidence reported by \citet{petrongolo_looking_2001}.\footnote{In an alternative baseline specification, we set the parameter $\alpha$ free to match the moments described in Section \ref{sec:estimation}, and as a result the value of $\alpha$ shows fluctuations around the value of 0.4.}

\begin{table}[!htbp]
	\scriptsize
	\caption{Description of model's parameters and data sources.}
	\label{tab:parameterCalibrationDes}
	\centering
	\begin{tabular}{c|l| l| c}
		\hline 
		\hline
		\textbf{Parameter}	 & \textbf{Description}	 & \textbf{Source}&\textbf{Time }\\
		&&&\textbf{varying}\\  
		\hline 
		$r$ & Discount rate &\cite{dellavigna2005job}& no\\
		$\phi$ & Share of firing costs transferred to employees &\cite{GaribaldiViolante2005}&no	 \\
		$\alpha$ & Elasticity of matching function w.r.t. unemployment & Petrongolo and Pissarides (2001) &no \\
		$c$ &Vacancy cost & Our calculation (Appendix \ref{vacancycost})&yes\\
		$F$ &Firing cost & Our calculation (Appendix \ref{firingcost})&yes\\
		$\delta_{h,N}$ & Job destruction rate of h native employees& Our calculation (Appendix \ref{app:jfrjer})&yes\\
		$\delta_{l,N}$ & Job destruction rate of l native employees& Our calculation (Appendix \ref{app:jfrjer})& yes\\
		$\delta_{h,I}$ & Job destruction rate of h non-native employees & Our calculation (Appendix \ref{app:jfrjer})& yes\\
		$\delta_{l,I}$ & Job destruction rate of l non-native employees& Our calculation (Appendix \ref{app:jfrjer})&yes\\
		$\chi$ & Mass of employers& Italian Labour Force Survey (RCFL)&yes\\
		$\sigma_{h,N}$ &Mass of h native employees & Italian Labour Force Survey (RCFL)& yes\\
		$\sigma_{l,N}$ &Mass of l native employees& Italian Labour Force Survey (RCFL) &yes\\
		$\sigma_{h,I}$ &Mass of  h non-native employees& Italian Labour Force Survey (RCFL)& yes\\
		$\sigma_{l,I}$ &Mass of  l non-native employees & Italian Labour Force Survey (RCFL)&yes\\ 
		$g_{HS}$ & Government expenditure in the HS final good  ($\%$ GVA) & Our calculation (Appendix \ref{govExp})& yes\\
		$g_{LS}$ & Government expenditure in the LS final good  ($\%$ GVA) & Our calculation (Appendix \ref{govExp})& yes\\
		%$g_l$ & Government expenditure in LS public goods as $\%$ of GVA & Our calculation (Appendix \ref{govExp})& yes\\
		$b$&Unemployment benefits& Our calculation (Appendix \ref{unBenefits})&yes\\
		$t_p$ & Indirect tax rate & Our calculation (Appendix \ref{indirectTax})&yes\\
							$t$ & Direct tax rate (income and profits) & Our calculation (Appendix \ref{indirectTax}) & no\\
								$ssc$ & Social security contribution rate & Our calculation (Appendix \ref{indirectTax})&yes\\ 
		$\eta$ &Rate at which non-natives enter the country& Our calculation (Appendix \ref{app:ExitRateImmigrant})& yes\\
		$\lambda$ & Rate at which non-natives exit the country& Our calculation (Appendix \ref{app:ExitRateImmigrant})&	yes\\
						$\iota$ &Degree of substitutability between public and private goods & Set to 1 (perfect substitutability)&no\\
					$\zeta$& Degree of congestion in accessing the public good & Set to 1 in baseline model$^1$ &no\\
						$\rho$& Parameter of the utility function & Set to 0 in baseline model$^2$ &no\\
					$\tau$& Tax subsidy& Estimated by matching moments&yes\\%$m$&Lump-sum transfer from the government&Estimated by matching moments&yes\\
						$t_g$& Tax gap& Estimated by matching moments&yes\\
			$\beta_N$ &Bargaining power of native employees& Estimated by matching moments&yes\\
		$\beta_I$ &Bargaining power of non-native employees& Estimated by matching moments&yes\\
		$\gamma$& Preference for the h final good& Estimated by matching moments&yes\\
		$x_{h}$& Productivity of h employees in intermediate good sectors & Estimated by matching moments&yes\\ 
		$x_{l}$& Productivity in l employees in intermediate good sectors  & Estimated by matching moments&yes\\
		$\kappa_{h,N}$&Hiring chances of h native employees& Estimated by matching moments&yes\\
		$\kappa_{l,N}$&Hiring chances of l native employees & Estimated by matching moments&yes\\
		$\kappa_{h,I}$&Hiring chances of h non-native employees& Estimated by matching moments&yes\\
		$\kappa_{l,I}$&Hiring chances of l non-native employees& Estimated by matching moments&yes\\ 
		$W_{h,FC}$&Utility of h employees abroad & Estimated by matching moments&yes\\
		$W_{l,FC}$&Utility of l employees abroad& Estimated by matching moments&yes\\
		$\alpha_{HS}$& Parameter of the production function of HS sector & Estimated by matching moments &yes\\
		$\alpha_{LS}$& Parameter of the production function of LS sector & Estimated by matching moments &yes\\		
		$\rho_{HS}$& Parameter of the production function of HS sector & Estimated by matching moments&yes\\
		$\rho_{LS}$& Parameter of the production function of LS sector & Estimated by matching moments&yes\\
		%\bottomrule
		\hline \hline
		\multicolumn{3}{l}{\scriptsize{\textit{Note}: l=low-skilled, h=high-skilled, LS=low-skilled intensity, and HS=high-skilled intensity.}}&\\
		\multicolumn{3}{l}{\scriptsize{$^1$ $\zeta=1$ indicates medium congestion.¸}}&\\
		\multicolumn{3}{l}{\scriptsize{$^2$ $\rho=1$ indicates an elasticity of substitution $=$ 1.}}&
	\end{tabular}
\end{table}

Some parameters are instead the results of our calculations. In particular, the vacancy cost $c$ which is  proportional to the employee's value added, is set equal to the estimated total start-up costs (as a percentage of per capita income)  \citep{djankov2002}, converted to a fraction of the value added as in \citet{boeri_burda}, using data from The World Bank (Appendix \ref{vacancycost}). The firing cost, which is proportional to the employee's value added, is computed according to the Italian regulations, using data on trials as provided by the Ministry of Justice (Appendix \ref{firingcost}).\footnote{As explained in details in Appendix \ref{firingcost}, two thirds of employer-initiated separations do not end up in court, while one third do, with additional costs, which can be potentially very large, depending on the length of the trial. We provide a calculation of the expected firing costs for firms, taking into account the probability to go to court, the average trial length and the related additional disburse.}  The values of the job exit rates $\delta_{h,N}$, $\delta_{h,I}$, $\delta_{l,N}$, and $\delta_{l,I}$ are estimated following \cite{SHIMER2012127} (Appendix \ref{JobCreationandJobDestruction}). The mass of employers $\chi$  is taken from the Italian Labour Force Survey (\textit{Rilevazione Continua sulle Forze di Lavoro}) and corresponds to the number of self-employed in the country. The mass of different types of employees defined by their skill level and country of origin are taken from the Italian Labour Force Survey (\textit{Rilevazione Continua sulle Forze di Lavoro}) and refer to the number of natives and non-natives individuals in the workforce with high or low skill levels, in line with  the definition reported in Appendix \ref{workerclassification}.   The government expenditure in public goods as percentage of GVA, $g_{HS}$ and $g_{LS}$, is calculated using Eurostat data (Appendix \ref{govExp}). The unemployment benefits are calculated as the ratio of total government expenditure for unemployment benefits and the number of unemployed in the economy, converted as a percentage of the employees' value added, using data from Eurostat (Appendix \ref{unBenefits}). The indirect tax rate $t_p$  is set  by dividing the total revenues from indirect taxation by the GDP at current prices, as reported by the Italian Ministry of Economics and Finance and the  Italian  Institute of Statistics (ISTAT) (Appendix \ref{indirectTax}). The parameters $\eta$ and $\lambda$, which refer to the rate at which non-native employees enter and leave the country,  respectively, are set using inflow and outflow rates, as reported by the OECD statistics (Appendix \ref{app:ExitRateImmigrant}).\footnote{We use OECD statistics instead of data from the Italian Institute of Statistics as foreign employees in the OECD statistics are defined by country of birth and not by nationality.}

Within this setup, and given the available set of information, we cannot identify  $\iota$, $\zeta$ and $\rho$. The parameter $\iota$, which defines the degree of substitutability between public and private goods, is set equal to 1, implying \textit{perfect substitutability} between private and public goods.
The parameter $\zeta$, which measures the congestion rate in the furniture of the public goods, is set equal to 1. This value implies a medium degree of congestion, i.e., the per capita furniture of public goods would be \textit{independent} on the size of the economy, whenever the labour force is homogeneous and equal to the total population.
The parameter $\rho$ is set equal to 0,  corresponding to a unit elasticity of substitution in consumption between the high-skilled and low-skilled intensive final goods. 
In Section \ref{sec:estimationresultsRobustness} we report robustness exercises perturbing the parameters  $\rho$ and  $\zeta$.
Appendix \ref{app:calibratedParameters} reports all the twenty-two calibrated parameters for each year of the analysis. The remaining seventeen parameters are estimated, as described in Section \ref{sec:estimation}.

\subsection{Estimation methodology \label{sec:estimation}}

The seventeen model's parameters are estimated using the \textit{method of simulated moments} \citep{gourieroux1996simulation}, year by year in the period 2008-2017, assuming that the bargaining power is the same for natives and non-natives, i.e $\beta=\beta_N=\beta_I$, since both $\beta$s cannot be identified simultaneously. In particular, the estimator is given by:
\begin{equation}
\hat{\bm{\omega}} = \argmin_{\bm{\omega} \in \Omega}  (\textbf{M} - \textbf{M}^s(\bm{\omega}))'\textbf{W}(\textbf{\textbf{M}} - \textbf{M}^s(\bm{\omega})),
\label{eq:matchingMoments}
\end{equation}
where $\hat{\bm{\omega}}$ is the vector of the estimated model's parameters, $\textbf{M}$ the $m-$vector of moments used in the estimate, $\textbf{M}^s(\bm{\omega})$ the $m-$vector of simulated moments calculated from the model taking the $m-$vector of  parameters $\bm{\omega}$, $\textbf{W}$ the $m \times m$ weighting matrix, and $\Omega$ the parameter space. The weighting matrix $\textbf{W}$ has to be positive-definite to guarantee that the moment-based estimator is consistent \citep[p. 380]{flinn2015labor}. Given the comparable magnitude of the scale of each moment, in our estimation we set $\textbf{W}$ to be an identity matrix.\footnote{In \citet{flinn2015labor} the diagonal elements of the matrix $\textbf{W}$  are set equal to the inverse of the variance of the corresponding element of the matrix $\textbf{M}$. However, in our analysis some moments to be matched have a negligible variance (they are taken from national accounts and from large national surveys), hence we take a conservative approach and use weights equal to one for all moments, while \citet{flinn2015labor} deal with these cases by setting an ``extremely large weight''.}

\subsection{Matched moments and estimated parameters \label{sec:estimationresults}}
 
 We match twenty-six moments of our theoretical model listed in Table \ref{tab:ListofMomentsToBeMatched}, which are filtered by a Hodrick-Prescott procedure to minimize the impact of the business cycle fluctuations on the estimated parameters.
 The model is able to match all observed moments very well, with few divergences in 2008 (Appendix \ref{app:matchedMoments}). Specifically, the average distance expressed in \% between the observed and the model's moments (i.e. $(1/26)\sum_{m=1}^{26}(\textbf{M}(m)-\textbf{M}^s(m))/\textbf{M}^s(m)$, see Equation (\ref{eq:matchingMoments})) is 2.2, with a standard deviation of 5.3; the average absolute distance is 3.4, with a standard deviation of 4.6; and, finally, the maximum distance is lower than 20, while the median distance is about 0.2. These figures appear to be remarkable, considering that we only have sixteen free parameters to match twenty-six moments.
 
 The estimated parameters are smooth over time and fluctuate in a range of plausible values, with a partial exception in 2008 (Appendix \ref{app:matchedParameters}). The tax subsidy $\tau$ and the tax gap $t_g$ are both slightly increasing over time in the period considered, with the latter ranging between 62\% and 64\% and matching remarkably well the Italian data reported in Appendix \ref{app:TaxGap}.
 The bargaining power $\beta$ is estimated to be in the range between 0.11 and 0.14, which is rather far from the level of 0.5 usually chosen in the literature (\citealp{pissarides2000equilibrium}). Nevertheless, the accurate matching of the share of labour income on total gross value added supports the robustness of this estimate (Appendix \ref{app:matchedMoments}).
 
\begin{table}[!htbp]
	\scriptsize
	\caption{Description of twenty-six moments to be matched in the estimate.}
	\label{tab:ListofMomentsToBeMatched}
	\centering
	\begin{tabular}{l|c}
		\hline 
		\hline
		  \textbf{Description}&	\textbf{Model parameter}\\	 
		\hline 
	Wage of h natives&$w_{h,N}$\\\\[-1.8ex]
 Wage of h non-natives&	$w_{h,I}$\\\\[-1.8ex]
	 Wage of l natives&$w_{l,N}$\\\\[-1.8ex]
		 Wage of l natives&$w_{l,I}$\\\\[-1.8ex]
			 Unemployment rate of h natives&$u_{h,N}$\\\\[-1.8ex]
			 Unemployment rate of h non-natives&	$u_{h,I}$\\\\[-1.8ex]
		 Unemployment rate of l natives&$u_{l,N}$\\\\[-1.8ex]
		 Unemployment rate of l non-natives&$u_{l,I}$\\\\[-1.8ex]
		 Job finding rate  of h natives&	$\kappa_{h,N} q(\theta_h)$\\\\[-1.8ex]
	 Job finding rate  of h non-natives&	$\kappa_{h,I} q(\theta_h)$\\\\[-1.8ex]
		Job finding rate of l natives&$\kappa_{l,N} q(\theta_l)$\\\\[-1.8ex]
	 Job finding rate of l non-natives&$\kappa_{l,I} q(\theta_l)$\\\\[-1.8ex]
h workers in HS sector&	$e_{h,HS}$\\\\[-1.8ex]
	h workers in LS sector&$e_{h,LS}$\\\\[-1.8ex]
	l workers in HS sector&$e_{l,HS}$\\\\[-1.8ex]
	h workers in LS sector&$e_{l,LS}$\\\\[-1.8ex]
		Share of h native unemployed&$\pi_{h,N}$\\\\[-1.8ex]
	Share of l native unemployed&$\pi_{l,N}$	\\\\[-1.8ex]
		Share of labour income on total GVA&$\frac{(w_{h,N}e_{h,N}+w_{h,I}e_{h,I+w_{l,N}e_{l,N}+w_{l,I}e_{l,I}})}{(\tilde{p}_{HS}q_{HS}+\tilde{p}_{LS}q_{LS})}$\\\\[-1.8ex]
			 	Share of expenditure in HS final good on GVA&$\frac{(\tilde{p}_{HS}q_{HS})}{(\tilde{p}_{HS}q_{HS}+\tilde{p}_{LS}q_{LS})}$\\\\[-1.8ex]
		Share of expenditure in  LS final good on GVA&$\frac{(\tilde{p}_{LS}q_{LS})}{(\tilde{p}_{HS}q_{HS}+\tilde{p}_{LS}q_{LS})}$\\\\[-1.8ex]
		Total realTechnicians GDP&$(1-t_p)(\tilde{p}_{HS}q_{HS}+\tilde{p}_{LS}q_{LS})$\\\\[-1.8ex]
		Total real GDP per worker& $\frac{(1-t_p)(\tilde{p}_{HS}q_{HS}+\tilde{p}_{LS}q_{LS})}{(e_{h,N}+e_{h,I}+e_{l,N}+e_{l,I})}$\\\\[-1.8ex]
		Net direct taxes&$NDT$\\\\[-1.8ex]
		Indirect taxes&$IT$\\\\[-1.8ex]
	Social security contributions&	$SSC$\\
\hline \hline 
\multicolumn{2}{l}{\scriptsize{\textit{Note}: l=low-skilled, h=high-skilled, LS=low-skilled intensity, and HS=high-skilled intensity.}}
	\end{tabular}
\end{table}

The preference parameter for high-skill intensive final goods ($\gamma$) fluctuates around 0.23.
The quantities of goods produced by high-skilled and low-skilled employees ($x_h$ and $x_l$) alone are scarcely informative; they need to be paired with the corresponding prices ($\tilde{p}_h$ and $\tilde{p}_l$) to be able to evaluate the value added of firms  ($\tilde{p}_h x_h$ and $\tilde{p}_l x_l$). The latter is estimated to be decreasing over time for both high-skilled and low-skilled firms, reflecting a declining GDP per worker (Appendix \ref{app:relevantEndoVariables}).Technicians
The hiring chances ($\kappa_{i,j}$) smoothly decrease over time for all types of employees, pointing to a declining matching efficiency. 
The non-native expected utilities of living abroad $W_{FC}$ for high-skilled and low-skilled employees are decreasing over time, reflecting changes in the socio-economic conditions in foreign countries.\footnote{This might reflect for instance, among others, the ongoing civil wars and persecutions in African and Middle Eastern countries, which are the sources of large immigration flows to Italy in the period of our analysis.} 
The parameters of the CES production function of  final good sectors $\alpha_{HS}$ and $\alpha_{LS}$ are estimated to be above $65\%$ and below $15\%$, respectively, which matches our proposed partition between high-skill versus low skill intensive sectors. The parameters $\rho_{HS}$ and $\rho_{LS}$ are both estimated to be negative, with corresponding average elasticities of about 0.77 and 0.33 for $LS$ and $HS$ final good sectors respectively. 
Finally, the per-capita level of public goods $\nu$ has been almost constant in the period considered, except for 2008 and 2009 (Figure \ref{PublicGood} in Appendix \ref{app:relevantEndoVariables}). 

\section{Winners and losers of immigration \label{sec:winnersLosersMigration}}

In this section, we run two counter-factual scenarios to identify the winners and losers of immigration. The first scenario envisions an economy with no non-natives, while the second one considers the case with a sudden inflow of 165 thousands low-skilled non-natives, which is about the forecasted/planned net migration for Italy in 2025 according to Italian Government. In particular, these exercises are performed by setting the number of non-natives and calculating the new equilibrium in the labour and goods markets, keeping the parameters at their original estimated values, while allowing all endogenous variables such as wages, prices, etc. to adjust to the new equilibrium conditions.

\subsection{No non-natives} \label{CFnononnatives}

\begin{figure}[!htbp]
 	\vspace{-1cm}
 	\caption{The main variables in the counter-factual scenario with no non-natives.}
 	\label{CFnonatives}
 	\begin{subfigure}{0.48\textwidth}
 		\centering
 		\includegraphics[width=0.8\linewidth]{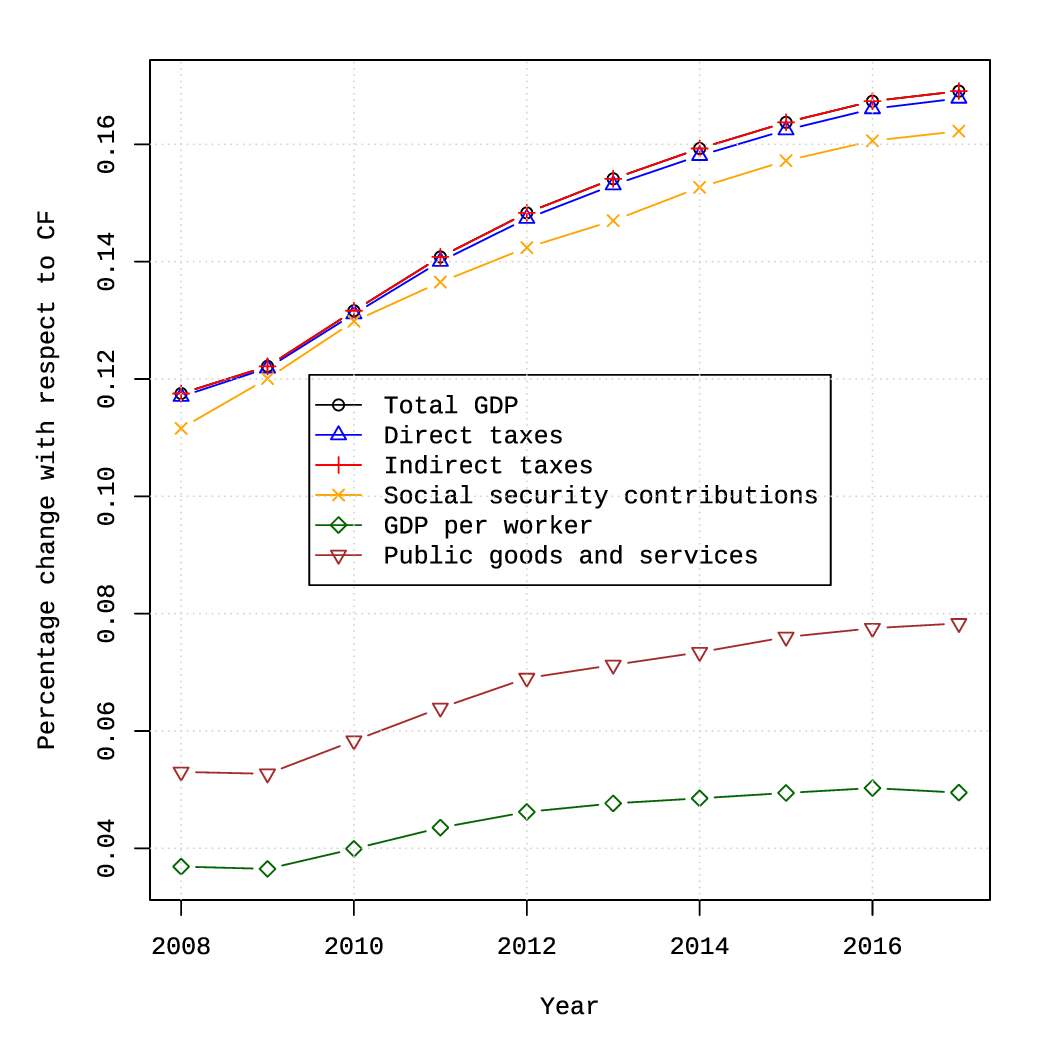}
 		\vspace{-0.2cm}
 		\caption{Percentage changes of aggregate variables}
 		\label{fig:totalProductionAV}
 	\end{subfigure}
  	\begin{subfigure}{0.48\textwidth}
 	\centering
 	\includegraphics[width=0.8\linewidth]{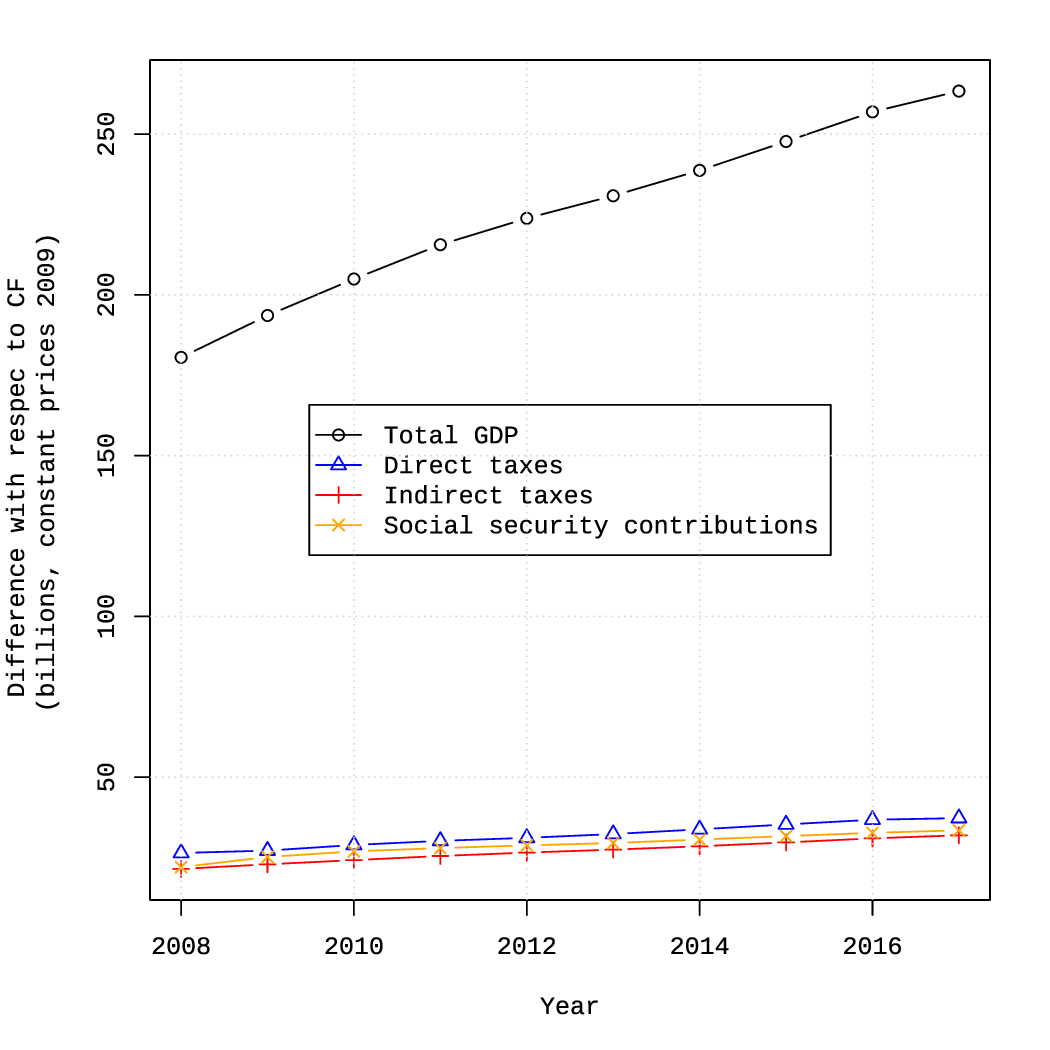}
 	 		\vspace{-0.2cm}
 	\caption{Absolute changes of aggregate variables}
 	\label{fig:totalProduction}
 \end{subfigure}
 	\begin{subfigure}{0.48\textwidth}
	\centering
	\includegraphics[width=0.8\linewidth]{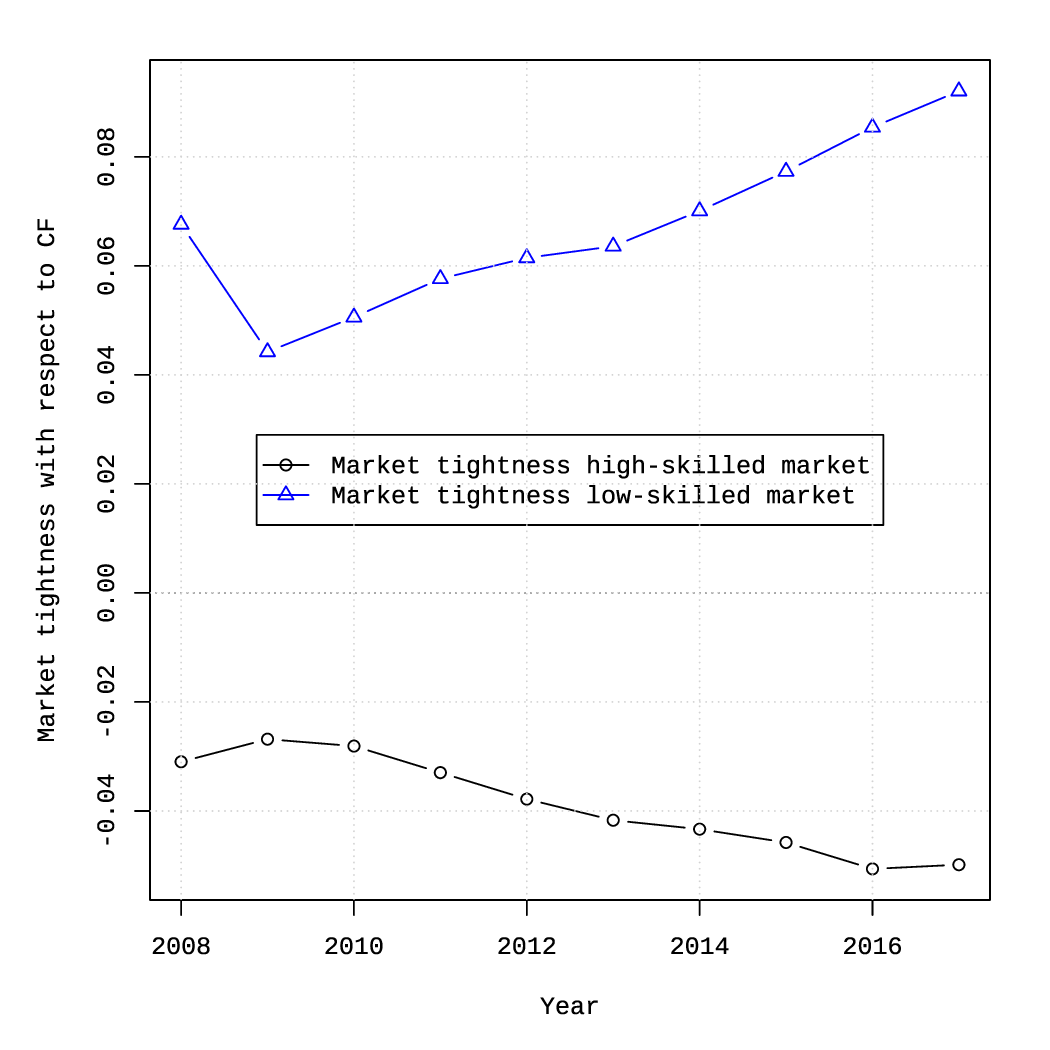}
	\vspace{-0.2cm}
	\caption{Market tightness}
	\label{fig:marketTightness}
\end{subfigure}
 	\begin{subfigure}{0.48\textwidth}
 		\centering
 		\includegraphics[width=0.8\linewidth]{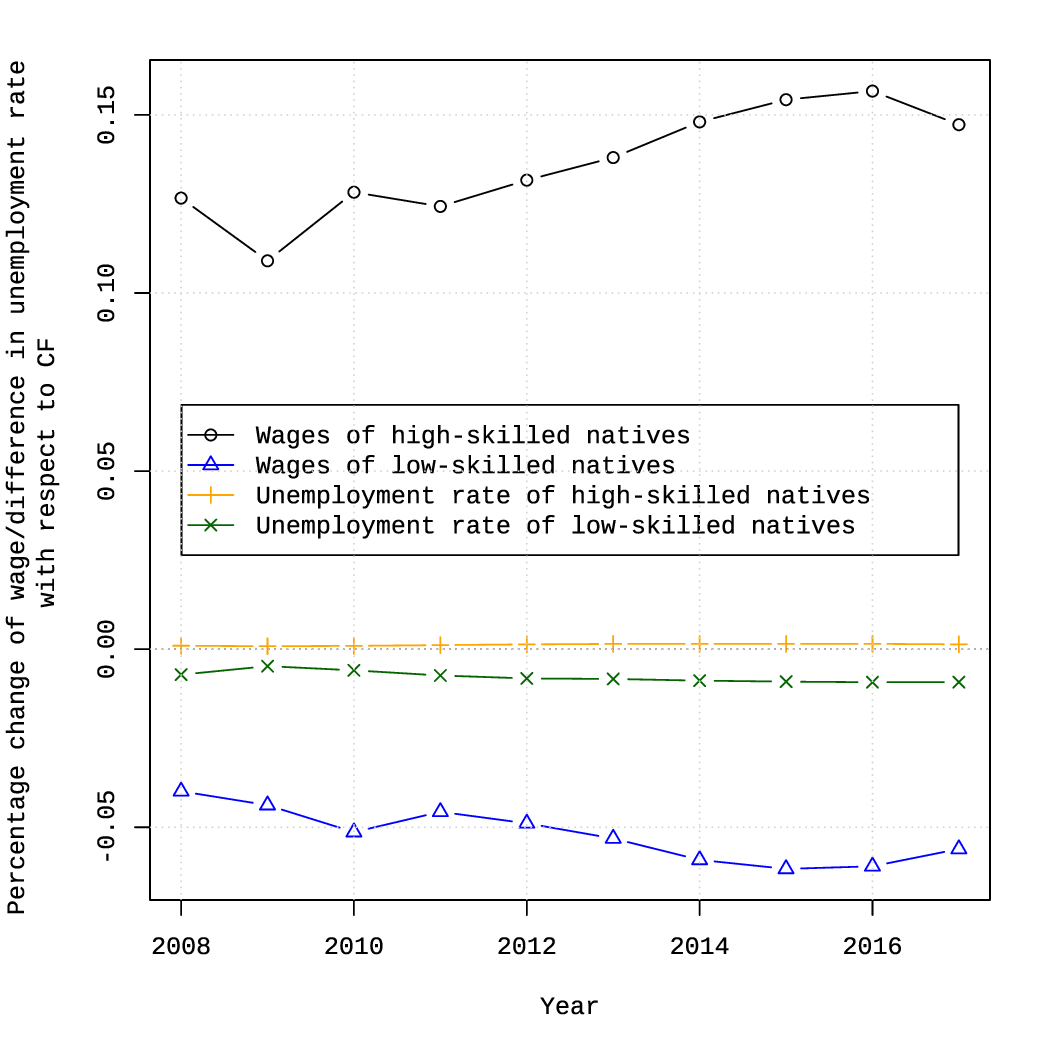}
 		 		\vspace{-0.2cm}
 		\caption{Wages and unemployment rates}
 		\label{fig:labourmarket}
 	\end{subfigure}
 	\begin{subfigure}{0.48\textwidth}
 		\centering
 		\includegraphics[width=0.8\linewidth]{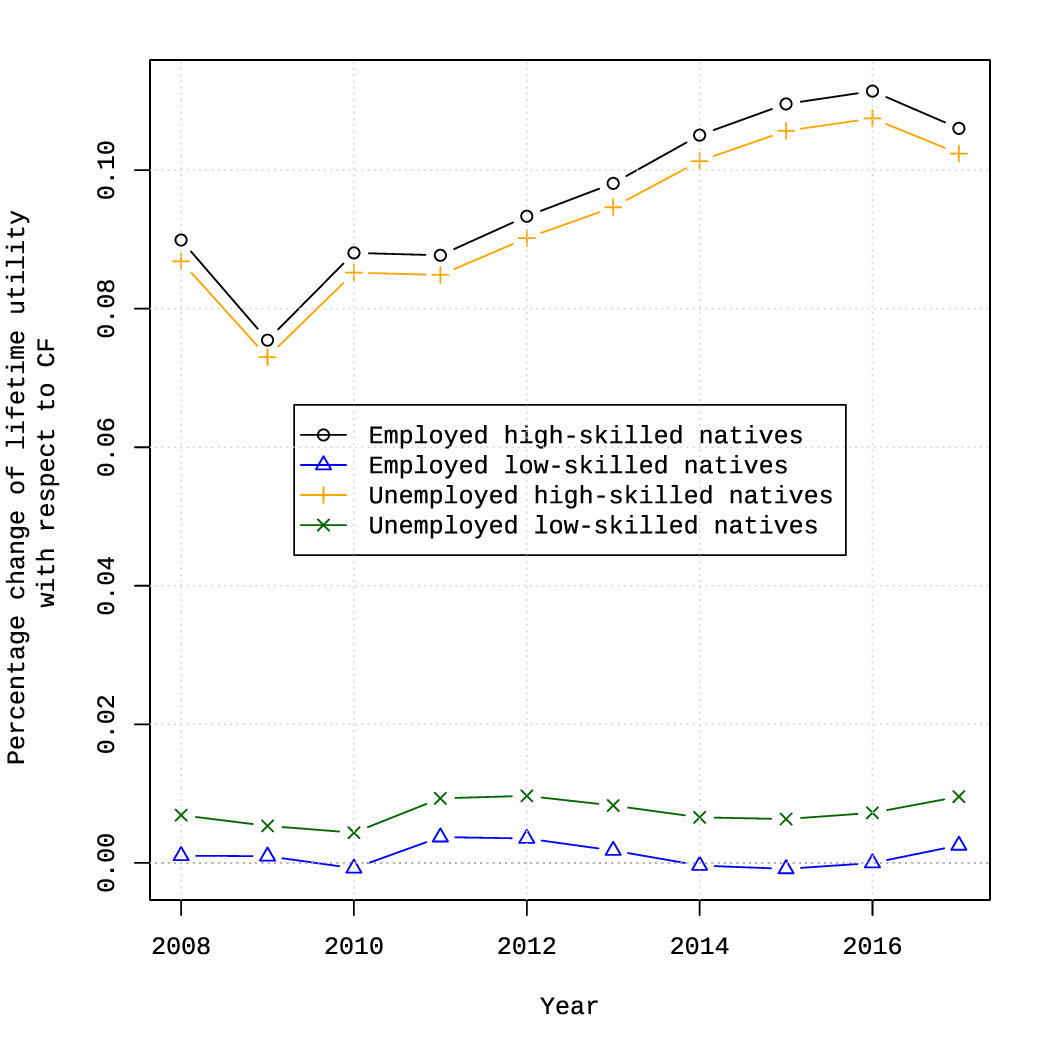}
 		 		\vspace{-0.2cm}
 		\caption{Employees' lifetime utility}
 		\label{fig:WLU}
 	\end{subfigure}
 	\begin{subfigure}{0.48\textwidth}
 		\centering
 		\includegraphics[width=0.8\linewidth]{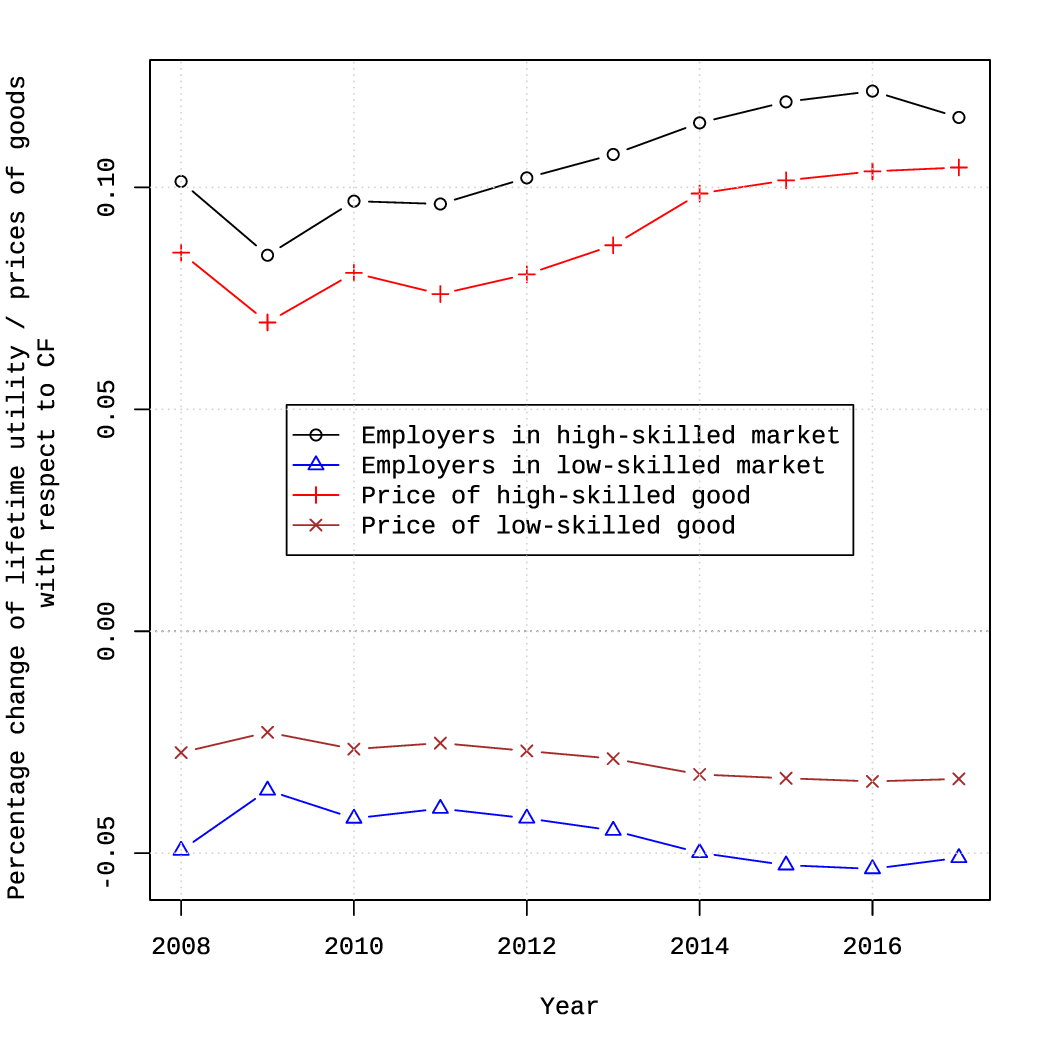}
 		 		\vspace{-0.2cm}
 		\caption{Employers' lifetime utility and real prices}
 		\label{fig:ELU}
 	\end{subfigure}
 \vspace{0.2cm}
 		\caption*{\scriptsize{\textit{Note}: The lines show the percentage/absolute change in the main variables in the counter-factual scenario in which there are no non-natives compared to the equilibrium in each year between 2008 and 2017. Only for the case of unemployment, we report the difference between the unemployment rate in the counter-factual scenario and in equilibrium, by skill level and country of origin. In Figure (e) the employees' lifetime utility is represented by the present discounted value of having a job $W_{i,j}$ (Equation \ref{eq:BellmanEmployedNative}), while the employers' lifetime utility is represented by the present discounted value of a filled vacancy $J_{i,j}$ (Equation \ref{eq:filledjobNative}). }}
 \end{figure}

Figures \ref{CFnonatives} reports the changes in the main aggregated variables relative to the counter-factual scenario with no non-natives. The presence of non-natives, which peaked at 15\% of the workforce in 2017, has lead to an increase in the gross domestic product (GDP) of almost 12\% in 2008 (approximately 180 billions \euro) and more than 16\% in 2017 (approximately 250 billions \euro)(Figure \ref{fig:totalProductionAV}). The quasi-proportional increase of GDP is driven by the increased labour force as well as the higher job finding rate of non-native low-skilled workers (see Figure \ref{JobCreationandJobDestruction}). Revenues from direct and indirect taxes, which are proportional to the GDP, increased by 12\% and 16\% in 2008 and 2017, respectively,  corresponding to an increase of approximately 25 billions \euro. Social security contributions have followed a similar pattern. The per-capita provision of public goods has increased in the presence of non-natives employees, by almost 8\% in 2017. Finally, the GDP per worker has increased by about 5\%, reflecting both the considerable differential impact on the prices of the two goods (approximately +10\%  for the high-skill intensive good versus -3.5\% for the low-skill intensive good, see Figure \ref{fig:ELU}) and the change in the labour force composition. 

With the presence of non-natives, the market tightness in the low-skill intensive market is higher by approximately 9\%, while the market tightness in the high-skill intensive market is lower and ranges between -2\% in 2008 and -5\% in 2017 (Figure  \ref{fig:marketTightness}). The positive effect of immigration of low-skilled employees on the job creation, i.e., the job creation effect, has been discussed in the literature, although for the case of illegal immigrants \cite{chassamboulli2020immigration}.
With the presence of non-natives, the real wage of low-skilled employees is lower by approximately $5\%$ in 2017, while the real wage of high-skilled employees is higher on average by approximately $15\%$ (Figure \ref{fig:labourmarket}). This is the result of two complementary effects: the increased supply of non-native low-skilled employees has on one side pushed down the real wage of low-skilled employees, while also expanding the supply of the low-skill intensive good, driving its price down. In 2017 the real price of the high-skill intensive good $\tilde{p}_{HS}$ has increased by approximately $10\%$, while the price of the low-skill intensive good $\tilde{p}_l$ has decreases by approximately $5\%$ (Figure \ref{fig:ELU}). These findings are in line with our conjecture that (real) wages and prices are positively related and therefore show similar patterns. In support of these results Appendix \ref{HS_LS_prices} provides anecdotal evidence of the divergent behaviour of the prices of some goods produced by low-skilled and high-skilled employees in Italy during the period considered.

The effects on unemployment rates are minimal (Figure \ref{fig:labourmarket}). We observe a slight decrease in the unemployment rate of low-skilled employees, ascribable to the asymmetries in the job finding and job exit rates between natives and non-natives. Specifically, both the job finding and the job exit rates are higher among non-natives, but the former prevails. In other words, the presence of non-natives is a source of efficiency for the Italian labour market.

We find that the lifetime utility of low-skilled employees is unchanged compared to a scenario with no non-natives (Figure \ref{fig:WLU}). This is due to the fact that their wages are lower, but the higher per capita provision of public goods fully compensates this negative effect. When non-natives are present in the economy, high-skilled employees are instead better off: their lifetime utility is higher by more than 10\% in 2017, due to their higher wages, which is on top of the increased provision of public goods. The lifetime utility of unemployed employees, both high-skilled and low-skilled, follows a similar pattern: the lifetime utility is slightly higher for low-skilled ($+1\%$) and remarkably higher for high-skilled unemployed employees ($+10\%$). Employers in the low-skill intensive market are worse off in the presence of non-natives: while both the salaries and the value added are lower due to the  decline in the price of the low-skill intensive good (Figure \ref{fig:ELU}). Their lifetime utility is therefore approximately 5\% lower when non-natives are present. On the other hand, although the salaries of high-skilled employees are higher, employers in the high-skill intensive market are better off in the presence of non-natives, as they take advantage of the higher value added due to the increase in the price of the high-skill intensive good. Their lifetime utility is between 10\% and 11\% higher in the presence of non-natives.

\subsubsection{Robustness analysis}\label{sec:estimationresultsRobustness}

The first robustness check considers two alternative counter-factual scenarios with no non-natives when $\rho$ is such that the elasticity of substitution between the two final goods $\epsilon$ is equal to 0.5 and 2, which represent the lower and upper bounds of its plausible range (Appendix \ref{app:elasticityOfSubstitutionFinalGood}).\footnote{We also explore an extreme value of 10 for $\epsilon$.}
In all scenarios, the results are consistent with the baseline specification (Appendix \ref{app:elasticityOfSubstitution}). In particular, the magnitude of the immigration's impacts on GDP, GDP per worker, taxes, and social contributions is about the same as for the baseline case with $\epsilon=1$. Instead, the polarization of wages and lifetime utilities is decreasing with $\epsilon$ due to the differential response of the prices of final goods to the change in the composition of the labour force. When $\epsilon=0.5$ (the final goods are less substitute), the real wage of high-skilled employees and the price of the high-skill intensive good are higher and the real wage of low-skilled employees and the price of the low-skill intensive good are lower in the presence of non-natives, with respect to the baseline scenario with $\epsilon=1$. This is then reflected in more polarized lifetime utilities for both employers and employees. Specifically, for the case of low-skilled employees, the lifetime utility turns slightly negative as the increase in the provision of public goods is not enough to fully compensate  for the decrease in wages. On the other hand, the lifetime utility of high-skilled employees is now higher than 12\% compared to the 10\% estimated for the case of $\epsilon=1$. Regarding the employers, the lifetime utility of those operating in the high(low)-skill intensive market is approximately 2 percentage points higher(lower) compared to the baseline scenario. The opposite is true for the case with $\epsilon=2$ (the final goods are more substitutes). Interestingly, in the extreme case of $\epsilon=10$, the wage of low-skilled employees is almost unchanged in the presence of non-natives, while the wage of high-skilled employees is on average 3.5\% higher, leading the lifetime utility of low-skilled employees to be between  3\% and 4\% higher, and that one of low-skilled employees to be between 2\% and 3\% higher. However, the lifetime utility of employers operating in the low-skill intensive market is still on average 1\% lower, while the lifetime utility of employers operating in the high-skill intensive market is between 4\% and 5\% higher.

The second robustness check considers a counter-factual scenario with no non-natives when we set $\zeta=0.5$, i.e., lower congestion in the provision of public goods, and $\zeta=1.5$, i.e., higher congestion (Appendix \ref{app:congestion}).
The provision of public goods is 4\% higher when $\zeta=0.5$ and 1.5\% lower when $\zeta=1.5$, compared to the baseline model. Specifically, the provision of public goods is between 10\% and 12\% higher compared to the scenario with no non-natives with less congestion, and 2\% higher compared to the scenario with no non-natives with more congestion. Overall, the presence of immigrants seems to increase the per capita provision of public goods even in the worst scenario of high congestion.
The higher (lower) provision of public goods is directly reflected in the lifetime utility of low-skilled employees. This is between 5\% and 6\%  higher compared to a scenario with no non-natives, when $\zeta$ is $0.5$, and 1\% lower compared to a scenario with no non-natives, when $\zeta$ is $1.5$. In the baseline scenario of $\zeta=1$, the lifetime utility of low-skilled workers is unchanged in the absence of non-natives.

\subsection{Immigration shock} 

In this second counter-factual analysis, we investigate the impact of an increase in the stock of non-native individuals, using the net migration forecast for Italy in 2023. While we expect the effects on native employees to be similar to the ones estimated within the previous counter-factual analysis, we are particularly interested in the effects on the outcomes of non-natives employees \citep{manacorda2012}.
Table \ref{CFshockAV} reports the impact of the counter-factual increase of 139 thousands additional working age non-natives  and 26 thousands non-working age non-natives on aggregate variables, compared to the 2017 equilibrium. Increased immigration  leads to a GDP increase of approximately 0.67\%, corresponding to an additional 10.46 billions \euro. Revenues from indirect and direct taxes increased by 0.65\%, corresponding to an increase of approximately 1.02 and 3.38 billions \euro, respectively. Social security contributions increased by 0.60\%, corresponding to 1.34 billions \euro.

\begin{table}[!htbp]
	\centering
	\caption{Counter-factual scenario: the effect of an increase in 165 thousands of non-natives individuals with respect to 2017.} 
	\label{CFshockAV}
	\scriptsize
	\begin{tabular}{lcc} 
		%\toprule
		\hline \hline 
		\\[-1.8ex]
		\textbf{Variable}&\textbf{\% change}&\textbf{Absolute change} \\
	&&	\textbf{(in billions of \euro)}\\
	\\[-1.8ex]
		\hline
		Monthly GDP	& 0.67 & 10.46\\
		Direct taxes&	0.65 &	3.38\\
		Indirect taxes&	0.65 & 1.02\\
		Social security contributions&	0.60 & 1.34\\
				%\bottomrule
				\hline \hline 
			\end{tabular}
		\end{table}
	
	\begin{table}[!htbp]
		\centering
		\caption{Counter-factual scenario: the effect of an increase in 165 thousands of non-natives individuals with respect to 2017.} 
		\label{CFshockIV}
		\scriptsize
		\begin{tabular}{lc} 
			%\toprule
			\hline \hline 
			\\[-1.8ex]
			\textbf{Variable}&\textbf{\% change}\\
			\\[-1.8ex]
			\hline
		Real GDP per worker&	0.15\\
		Public goods &	0.15 \\
		Wages of high-skilled natives&	0.83 \\
		Wages of high-skilled non-natives&	0.48 \\
		Wages of low-skilled natives&	-0.30 \\
		Wages of low-skilled non-natives&	-0.36 \\
		Unemployment rate of high-skilled natives (absolute change)&	0.000009 \\
		Unemployment rate of high-skilled non-natives  (absolute change)&	0.00001 \\
		Unemployment rate of low-skilled natives  (absolute change)&	-0.00033 \\
		Unemployment rate of low-skilled non-natives  (absolute change) &	-0.00032 \\
		Lifetime utility of employed high-skilled natives&	0.56 \\
		Lifetime utility of employed high-skilled non-natives	& 0.33 \\
		Lifetime utility of employed low-skilled natives &	-0.036 \\
		Lifetime utility of employed low-skilled non-natives &	-0.036\\
		Lifetime utility of unemployed high-skilled natives	& 0.54 \\
		Lifetime utility of unemployed high-skilled non-natives & 0.28 \\
		Lifetime utility of unemployed low-skilled natives&	-0.007\\
		Lifetime utility of unemployed low-skilled non-natives&	-0.006 \\
		Lifetime utility of employers hiring high-skilled natives&	0.54 \\
		Lifetime utility of employers hiring high-skilled non-natives&	0.78 \\
		Lifetime utility of employers hiring low-skilled natives&	-0.25 \\
		Lifetime utility of employers hiring low-skilled non-natives&	-0.23 \\
		Price of high-skill intensive good&	0.55 \\
		Price of low-skill intensive good&	-0.16 \\
		%\bottomrule
		\hline \hline 
	\end{tabular}
\end{table}

Looking at the effect of the increased stock of non-natives on individual variables (Table \ref{CFshockIV}), GDP per worker is higher in the presence of an increased number of non-native people by 0.15\% as well as the per-capita provision of public goods. The wage of low-skilled native and non-native employees is lower by approximately 0.30\% and 0.36\%, respectively. On the other hand, the wage of high-skilled native employees is higher on average by approximately 0.83\%, while the wage of high-skilled non-native employees is higher on average by approximately 0.48\%. The effects on unemployment rates are negligible. Accordingly, we find that both low-skilled native and non-native employees are slightly worse off compared to a situation of fewer non-natives: their lifetime utility is lower by approximately 0.036\%, as the result of lower wages almost fully compensated by a higher provision of public goods. High-skilled native and non-native employees are instead better off: their lifetime utility is higher by 0.56\% and 0.33\%, respectively. In this case, native employees are the ones gaining the most, as the wage increase is higher for natives compared to non-natives. Finally, employers operating in the low-skill intensive market are worse off: while the salaries are lower, also the value added is lower due to the fall in the price of the low-skill intensive good. Their lifetime utility is lower by approximately 0.23\%/0.25\% if natives or non-natives. On the other hand, although the salary of high-skilled employees is higher, employers in the high-skill intensive market are better off, as they take advantage of the higher value added of their production due to the increase in the price of the high-skill intensive good. Their lifetime utility is higher by 0.54\% if they hire a native employee and 0.78\% if they hire a non-native employee, due to the larger wage increase among native employees.

\section{Conclusions}\label{sec:conclusions}

We describe the asymmetric effects of low-skilled immigration, within a  framework which includes key elements relevant for policy makers aiming at developing an "effective, fair and robust  migration policy" \citep{EUParliament}, such as the elasticity of substitution between intermediate goods produced by high-skilled and low-skilled workers, the elasticity of substitution between final goods, a frictional labour market, the taxation of wages and profits, and the provision of public goods and unemployment benefits. 

Our model suggests that a positive shock to the stock of low-skilled immigrants has a direct impact on the labour market, pushing down the wages of low-skilled employees and increasing the relative supply of the low-skilled intermediate good. In turn, this induces an increase in the relative supply of the low-skilled intensity final good, with a fall in its relative price. This decrease, finally, reduces the profitability of being an employer in the low-skilled intermediate good sector.
Overall, the change in the relative prices in final good sector identifies a \textit{price channel}, which works in favour of high-skilled employees (more vacancies and higher real wages) and employers hiring them (more profits), while hurting low-skilled employees (less vacancies and lower real wages) and employers (less profits).
This works in parallel to the change in relative wages, identifying the traditional \textit{substitution channel}, which tends to dampen the direct impact on the labour market, working in favour of low-skilled employees (more vacancies) and employers hiring them (more profits).
These two channels operate in opposite directions, with the final outcome crucially depending on the elasticity of substitution between intermediate goods produced by high-skilled and low-skilled workers, and on the elasticity of substitution between final goods.
Finally, the critical role of the labour force composition in shaping aggregate dynamics can be explained by the heterogeneity in the bargaining power, the outside opportunity, and the job finding and exit rates between natives and non-natives.

Our empirical analysis on Italy suggests that immigration  had sizable and largely asymmetric effects on employees and employers, but a positive net fiscal impact, and led to an increased provision of public goods. Interestingly, non-natives have proven to be a source of efficiency for the Italian labour market.
The losers from the recent inflows of migrants are employers operating in the low-skill intensive market, while the winners are  employers and employees in the high-skill intensive market. The lifetime utility of employees operating in the low-skill intensive market is instead unchanged as a result of the higher provision of public goods which compensates their lower real wage. Among the winners we also include the Italian government, which benefited from the substantial  additional direct and indirect taxes, and social security contributions.
As the opening of the Italian labour market to foreign low-skilled workers contributed to income inequality both among employees and employers, it is not surprising that social tensions arise in the debate on the effects of immigration, also in light of the possible incomplete perception of the positive impact of immigrants on the provision of public goods \citep{10.1162/rest.91.2.295,EC2018}.

On the basis of our analysis we argue that when policy makers face the choice between a labour market with a limited participation of non-native low-skilled workers, e.g., the Brexit strategy, and a labour market open to non-native workers, the latter alternative is preferable, if it is implemented  together with interventions to compensate the losers. The effective design and implementation of these policies represents the direction of our future research.

\clearpage
	
\bibliographystyle{chicago}
\bibliography{MyLibrary}

\clearpage

\appendix

\textbf{\Large{Appendix}}

\section{Small open economy \label{app:SMO}}

In the main text, we have considered a closed economy. In this section, we  extend the model to a small open economy, assuming that  good $LS$ is tradable, while good $HS$ is non-tradable. The price of the non-tradable good $HS$, will still be determined  by equating demand and supply in the domestic market. The price of the tradable  good $LS$, instead  is equal to the product of the price of the good $LS$ expressed in foreign currency in the international market $p_{LS}^{\$}$ and the \textit{nominal exchange rate} $e$, i.e., 
\begin{eqnarray}\label{internationalprice}
	p_{LS}=e p_{LS}^{\$}.
\end{eqnarray}
The equilibrium in the goods markets $LS$ and $HS$ is defined as:
\begin{eqnarray} \notag
	DD_{LS}+NX&=&DS_{LS} \text{ and}\\ \notag
	DD_{HS}&=&DS_{HS}, 
\end{eqnarray}
where $DD_a$ and $DS_a$ are the domestic demand and supply of good $a \in \{HS,LS\}$, respectively, and $NX$ is the \textit{net export} of good $LS$. 
Let  $C$ aggregate consumption,  $D$ aggregate demand, $Y$ aggregate production and $DY$ aggregate disposable income  ($T$ and $TR$ are the collected taxes and transfers, respectively); therefore, in equilibrium:
\begin{eqnarray}
	\begin{cases} \notag
		D=DD_{LS}+NX+DD_{HS}=C+I+G+NX;\\ \notag
		DY=Y-T+TR=C+S;\\ \notag
		Y=DS_{LS}+DS_{HS};   \\ \notag
		D = Y, \notag
	\end{cases}
\end{eqnarray}
from which
\begin{eqnarray} \notag
	Y &=& Y-T+TR-S+I+G+NX,
\end{eqnarray}
that is:
\begin{eqnarray} \notag
	S-I+T-TR-G & = & NX.
\end{eqnarray}	
In our model, $S=I=0$ and hence in equilibrium
\begin{eqnarray}\label{psnx}
	T-TR-G=NX.
\end{eqnarray}
In a  \textit{flexible exchange rate regime} or a \textit{fixed exchange rate regime} with flexible price of the non-tradable good $p_{HS}$, in the absence of international financial markets, $NX$ should be equal to $0$ to guarantee the excess demand of national currency to be equal to zero in equilibrium. As a further consequence, in our economy with $S=I=0$, the government runs a balanced primary budget, i.e. $T-TR-G=0$. In other words, the \textit{real} prices of the  goods in equilibrium are not affected by the fact that the \textit{nominal} price of good $LS$ is set in the international good market (Equation (\ref{internationalprice})); that is the price of the non-tradable good $p_{HS}$ adjusts to guarantee that the real prices of both goods satisfy Equation (\ref{psnx}).
As a consequence, in an economy with two goods, in equilibrium an immigration shock  has exactly the same impact  in a closed economy or in a small open economy with one tradable good.

An alternative assumption would be that, at a given \textit{nominal} price of the tradable good, it is possible to produce any quantity of the tradable good, which is then absorbed by the international market \citep{Iftikhar2016}. As such $NX$ can get any value compatible with the decisions taken by the supply side of the economy. However, this assumption would imply that the equilibrium in the trade balance and in the government budget is not guaranteed.

\section{Prices of goods produced by high-skilled and low-skilled employees}\label{HS_LS_prices}

\begin{figure}[!htbp]
	\caption{Prices of goods produced by high-skilled and low-skilled employees in Italy in the period 2010-2017.}
	\label{fig:sectoralPrices}
	\vspace{-1cm}
 	\begin{subfigure}[t]{0.49\textwidth}
		\includegraphics[width=1\linewidth]{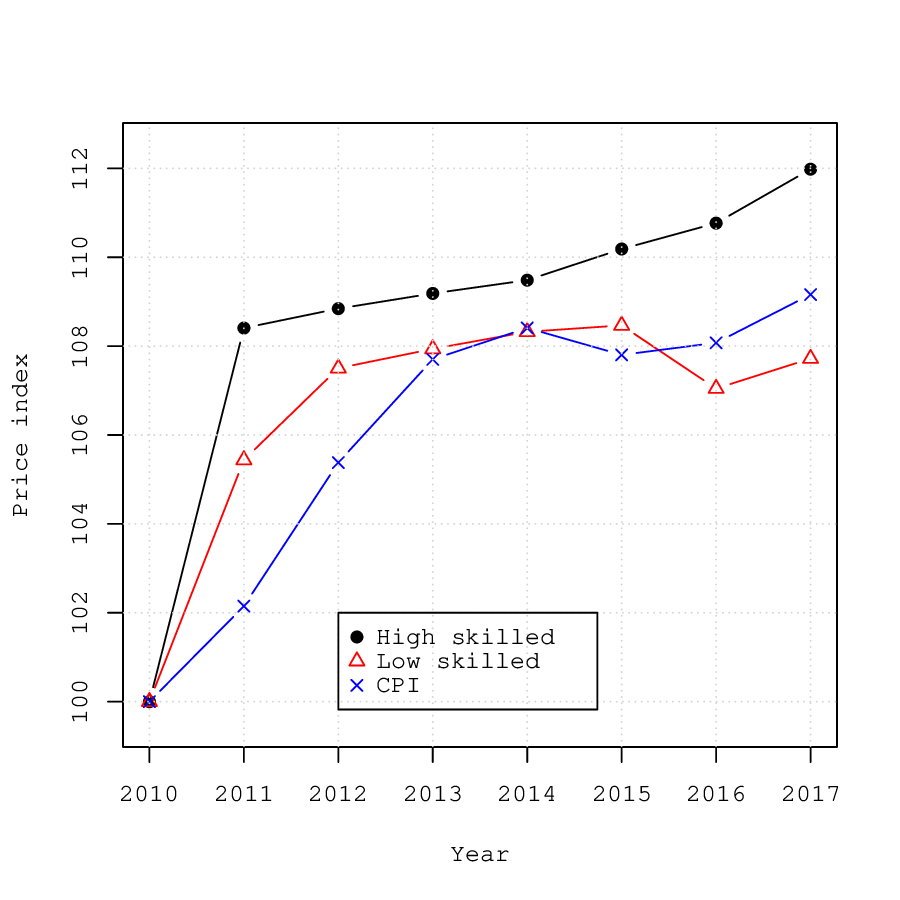}
	\caption{Price index for high-skill and low-skill intensive goods.\\ \bigskip
		\textit{Note}: We include among high-skilled sectors pharmaceuticals, computer programming and information services, legal and accounting activities, and architectural and engineering activities. Among low-skilled sectors we include textiles, wood, paper, glass, basic metals, constructions, transports, postal and courier activities, security and investigation and office administrative activities. The CPI index is reported for reference. \textit{Source}: Italian Institute of Statistics.}
		\label{priceindex}
	\end{subfigure}
	\begin{subfigure}[t]{0.49\textwidth}
	\includegraphics[width=1\linewidth]{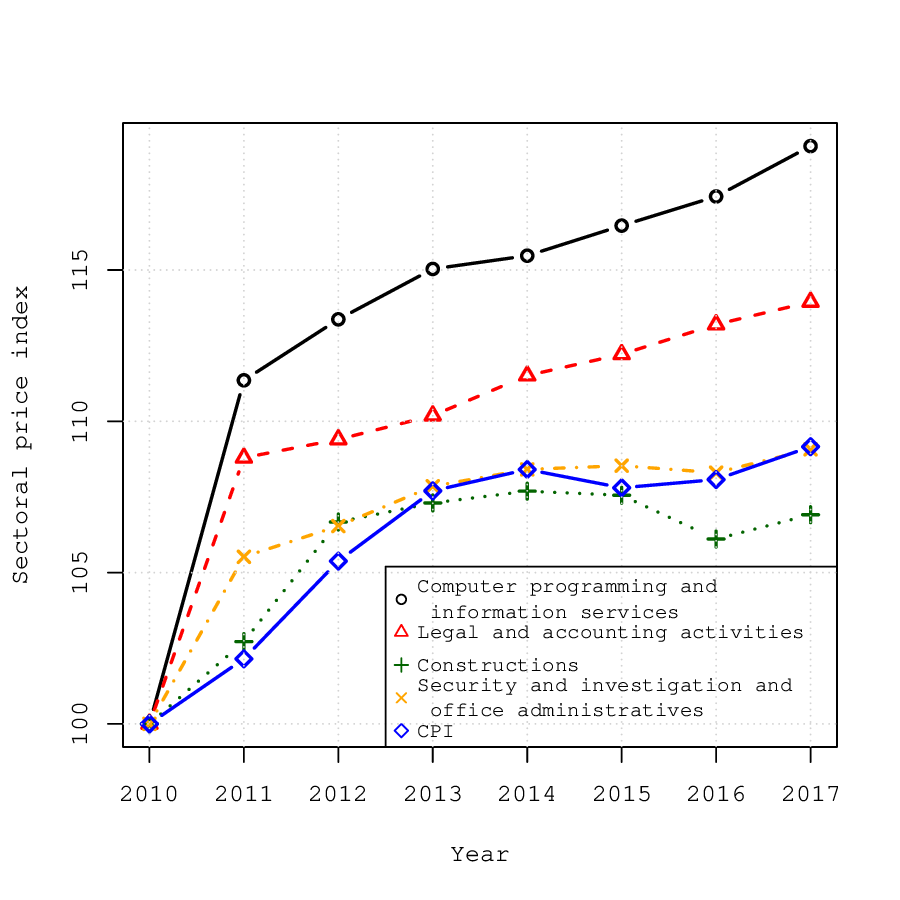}
			\caption{Price index of selected sectors. \\
				\bigskip
				\textit{Note}: we include as high-skilled sectors computer programming and information services and legal and accounting activities. We include as low-skilled sectors constructions, and security and investigation and office administrative activities. The CPI index is reported for reference. \textit{Source}: Italian Institute of Statistics.}
			\label{fig:sectoralpricesnorm}
	\end{subfigure}
\end{figure}

We document the evolution of the prices of goods in sectors in which mainly high-skilled or low-skilled employees are employed. Given that the availability of Italian data on production prices by sector is limited, we construct an index of prices for low-skill and high-skill intensive goods (Figure \ref{priceindex}) by considering a number of sectors in which the share of high-skilled and low-skilled employees is higher than 50\% and 85\%, respectively. The index is built by weighting the price of each sector by the share of workers, either high-skilled or low-skilled, employed in the sector. This corresponds to the 29.3\% of total high-skilled employees and 54.4\% of total low-skilled employees in the Italian economy.

In addition, we select four sectors for which the prices of goods and services should be less influenced by changes in technology (e.g., technological shocks), by external factors and import/export dynamics (e.g., oil shocks), and by the price of raw materials (e.g., commodities). For these four sectors, we report the producer prices (base year 2010) from 2010 to 2017 (Figure \ref{fig:sectoralpricesnorm}). In sectors which are mainly high-skilled, the prices have been increasing much more relative to the prices in sectors which are mainly low-skilled.  
Employment in the two low-skilled sectors considered amounts to 28\% of the total low-skilled employees, and the employment in the two high-skilled sectors considered amounts to 21\% of the total high-skilled employees.
Overall, this anecdotal evidence provides support to our findings that as a consequence of an increase in the supply of low-skilled labour in Italy in the period considered, prices in the low-skill intensive market have \textit{relatively} declined with respect to  prices in the high-skill intensive market.
 
 \newpage
 
\section{The elasticity of substitution between final goods}\label{app:elasticityOfSubstitutionFinalGood}

In this section we discuss the range of plausible values for the  elasticity of substitution between final goods $\epsilon\equiv 1/(1-\rho)$ in the spirit of \citep{ACEMOGLU20111043}. Aggregating the demand for final goods $HS$ and $LS$ from Equation (\ref{demandHSEmployer}) and Equation (\ref{demandLSEmployer}), we get:
\begin{eqnarray} \label{eq:shareHS}
	\dfrac{\tilde{p}_{HS} d_{HS}}{\tilde{y}}=\gamma^{\epsilon} \tilde{p}_{HS}^{1-\epsilon}, \text{ and} \\	\label{eq:shareLS}
	\dfrac{\tilde{p}_{LS} d_{LS}}{\tilde{y}}=\left(1-\gamma\right)^{\epsilon} \tilde{p}_{LS}^{1-\epsilon},	  
\end{eqnarray}
and therefore :
\begin{equation}
	\dfrac{\tilde{p}_{HS}}{\tilde{p}_{LS}} = \left( \dfrac{1-\gamma}{\gamma}\right)^{\epsilon/(1-\epsilon)} \left(\dfrac{\tilde{p}_{HS}d_{HS}}{\tilde{p}_{LS}d_{LS}}\right)^{1/(1-\epsilon)}.
	\label{eq:realtionRatioFinalGoodPricesRatioFinalGFoodsExpenditure}
\end{equation}
Equation (\ref{eq:realtionRatioFinalGoodPricesRatioFinalGFoodsExpenditure}) links the relative price of the two final goods with (i) the relative expenditure for the two goods $\tilde{p}_{HS}d_{HS}/\tilde{p}_{LS}d_{LS}$ and (ii) their elasticity of substitution $\epsilon$.
Unfortunately, in the data we have only the relative expenditure, but no information on the relative prices of the two goods $\tilde{p}_{HS}/\tilde{p}_{LS}$. To circumvent this limitation, we conjecture that the relative price is proportional to the relative wage of high and low skilled workers, i.e.:
\begin{equation}
	\dfrac{w_h}{w_l} \propto \dfrac{p_{HS}}{p_{LS}} = \left( \dfrac{1-\gamma}{\gamma}\right)^{\epsilon/(1-\epsilon)} \left(\dfrac{p_{HS}d_{HS}}{p_{LS}d_{LS}}\right)^{1/(1-\epsilon)}.
	\label{eq:realtionRatioFinalGoodPricesRatioFinalGFoodsExpenditureRatioWages}
\end{equation}
Hence, from Equation (\ref{eq:realtionRatioFinalGoodPricesRatioFinalGFoodsExpenditureRatioWages}):
\begin{equation}
	\dfrac{\mathring{w_h/w_l}}{w_h/w_l} = \left(\dfrac{1}{1-\epsilon}\right) 
	\left(
	\dfrac{\mathring{p_{HS}d_{HS}/p_{LS}d_{LS}}}{p_{HS}d_{HS}/p_{LS}d_{LS}}
	\right),
	\label{eq:wagesVersusRelativeExpenditure}
\end{equation}
and the estimate of $\epsilon$ is directly derived by the estimate of Equation (\ref{eq:wagesVersusRelativeExpenditure}). The estimated coefficient reported  in Column 1 of Table \ref{tab:estimateElasticityOfSubstitutionFinalGoods} suggests a value of $\epsilon$ not different from one, i.e. a value such that prices (and wages) cannot affect the share of expenditure in the two goods (see Equations (\ref{eq:shareHS}) and (\ref{eq:shareLS})). When we control for possible supply side effects, by including the growth of the relative size of the high-skilled and low-skilled workforce $\sigma_h/\sigma_l$ (Column 2 of Table \ref{tab:estimateElasticityOfSubstitutionFinalGoods}), the estimate of $\epsilon$ jumps close to 2 ((1+1.124)/1.124=1.89). Therefore, the reasonable range for $\epsilon$ appears to be $[1,2]$, i.e., a change of 1\% in the relative prices determines a change in the relative demanded quantities in the range $[0.5\%,1\%]$. 

% Table created by stargazer v.5.2.3 by Marek Hlavac, Social Policy Institute. E-mail: marek.hlavac at gmail.com
% Date and time: mar, ott 10, 2023 - 17:33:16
\begin{table}[t] \centering 
	\footnotesize
	\caption{The estimates of the elasticity of substitution between the two final goods.} 
	\label{tab:estimateElasticityOfSubstitutionFinalGoods} 
	\begin{tabular}{@{\extracolsep{5pt}}lccH} 
		\hline 
		\hline
		\\[-1.8ex] 
		 & \multicolumn{2}{c}{Growth rate of } \\
		 & \multicolumn{2}{c}{relative wages $w_h/w_l$} \\
		& (1) & (2) & 3 \\
		\hline \\[-1.8ex] 
		Growth rate of relative & $-$0.194 &$-$1.124$^{**}$ & \\ 
		expenditure $p_{HS}d_{HS}/p_{LS}d_{LS}$ & (0.606)& (0.449)&\\ 
		\\[-1.8ex]
			Growth rate of relative &&$-$1.056$^{**}$ &$-$0.635$^{*}$ \\
			workforce $\sigma_h/\sigma_l$&&(0.293)&(0.317)\\
			\\[-1.8ex]
		Constant & $-$0.001& $-$0.019$^{*}$ &$-$0.006\\ 
		& (0.010)& (0.008)& (0.008)\\ 
		\hline \\[-1.8ex] 
		Observations &9& 9& 9\\ 
		R$^{2}$ & 0.014& 0.689&0.364\\ 
		\hline 
		\hline \\[-1.8ex] 
		   \multicolumn{4}{l}{\scriptsize{\textit{Note:} $^{*}$p$<$0.1; $^{**}$p$<$0.05; $^{***}$p$<$0.01.} } \\  
		   \multicolumn{4}{l}{\scriptsize{\textit{Source}: Own calculation based on the Italian Labour Force}}\\\multicolumn{4}{l}{\scriptsize{ Survey (RCFL) for the period 2008-2017. }}  
	\end{tabular} 
\end{table} 

\clearpage

%\textbf{\LARGE{Appendix only for the referees}}

\section{Sensitivity analysis on the elasticity of substitution \label{app:elasticityOfSubstitution}}
\begin{figure}[!htbp]
	\caption{Counter-factual variables - no non-natives with CES parameter $\epsilon=0.5$.}
	\label{CFnonativesCES05}
	\begin{subfigure}[!htbp]{0.49\textwidth}
		\centering
		\includegraphics[width=0.70\linewidth]{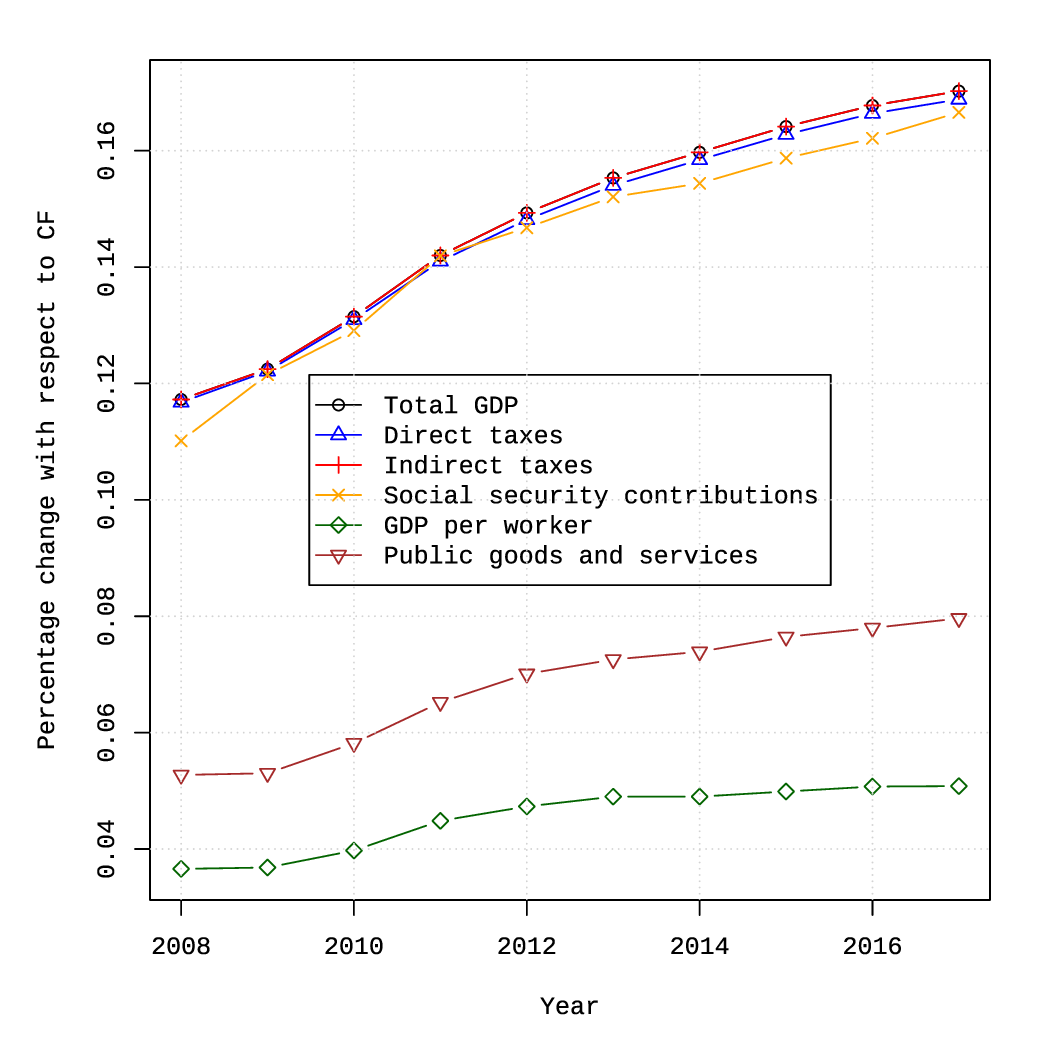}
		\vspace{-0.2cm}
		\caption{Percentage changes of aggregate variables}
		\label{fig:totalProductionAVCES05}
	\end{subfigure}
	\begin{subfigure}[!htbp]{0.49\textwidth}
		\centering
		\includegraphics[width=0.70\linewidth]{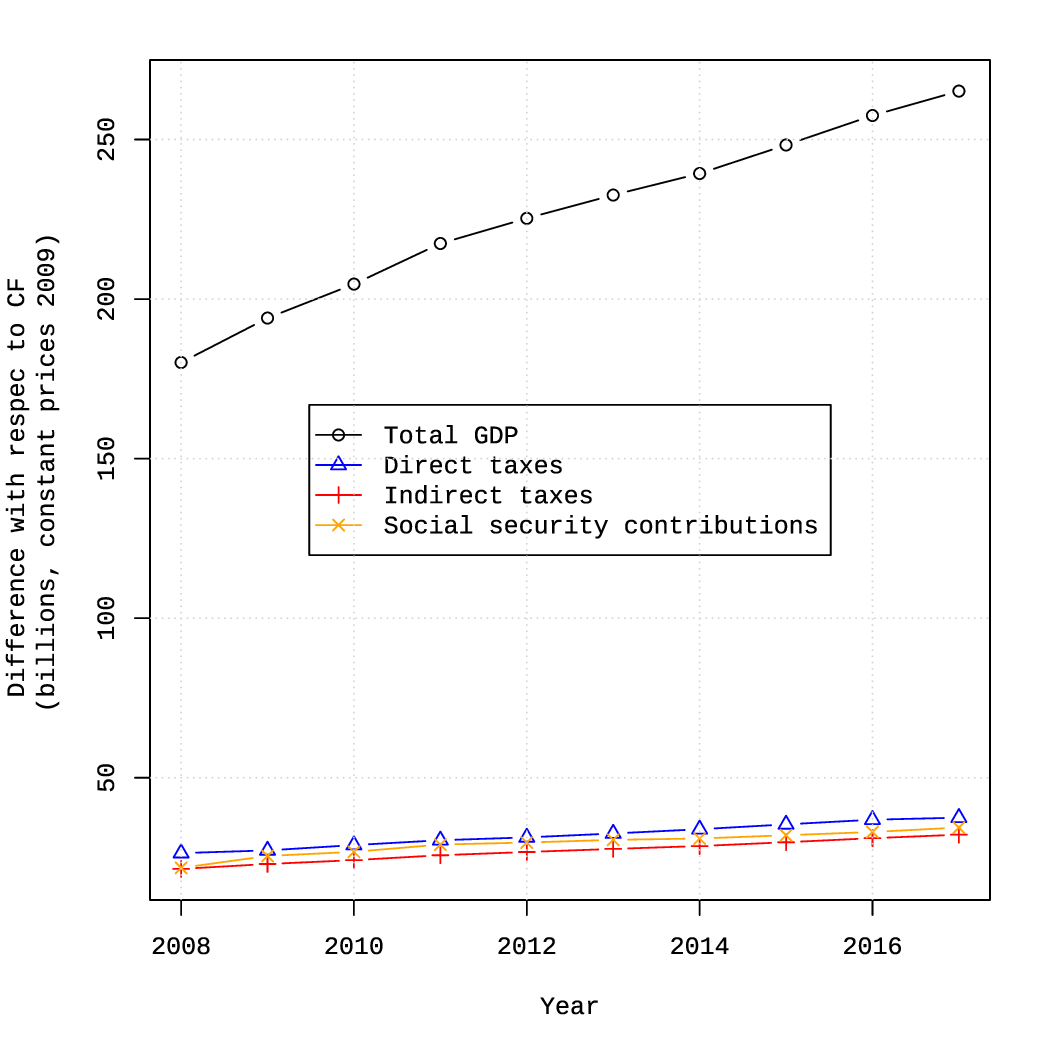}
		\vspace{-0.2cm}
		\caption{Changes of aggregate variables}
		\label{fig:totalProductionCES05}
	\end{subfigure}
	\begin{subfigure}{0.49\textwidth}
	\centering
	\includegraphics[width=0.70\linewidth]{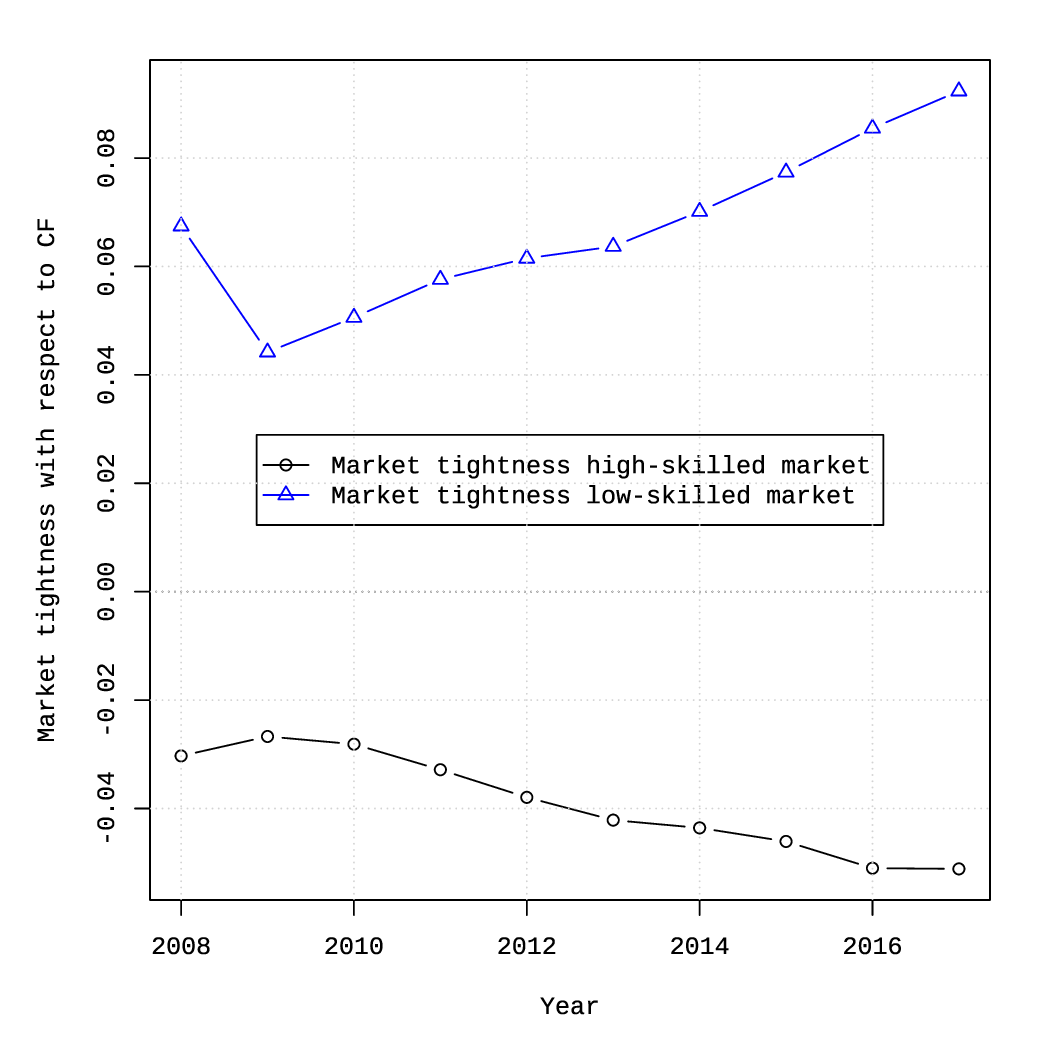}
	\vspace{-0.2cm}
	\caption{Market tightness}
	\label{fig:tightnessCES05}
\end{subfigure}
	\begin{subfigure}{0.49\textwidth}
		\centering
		\includegraphics[width=0.70\linewidth]{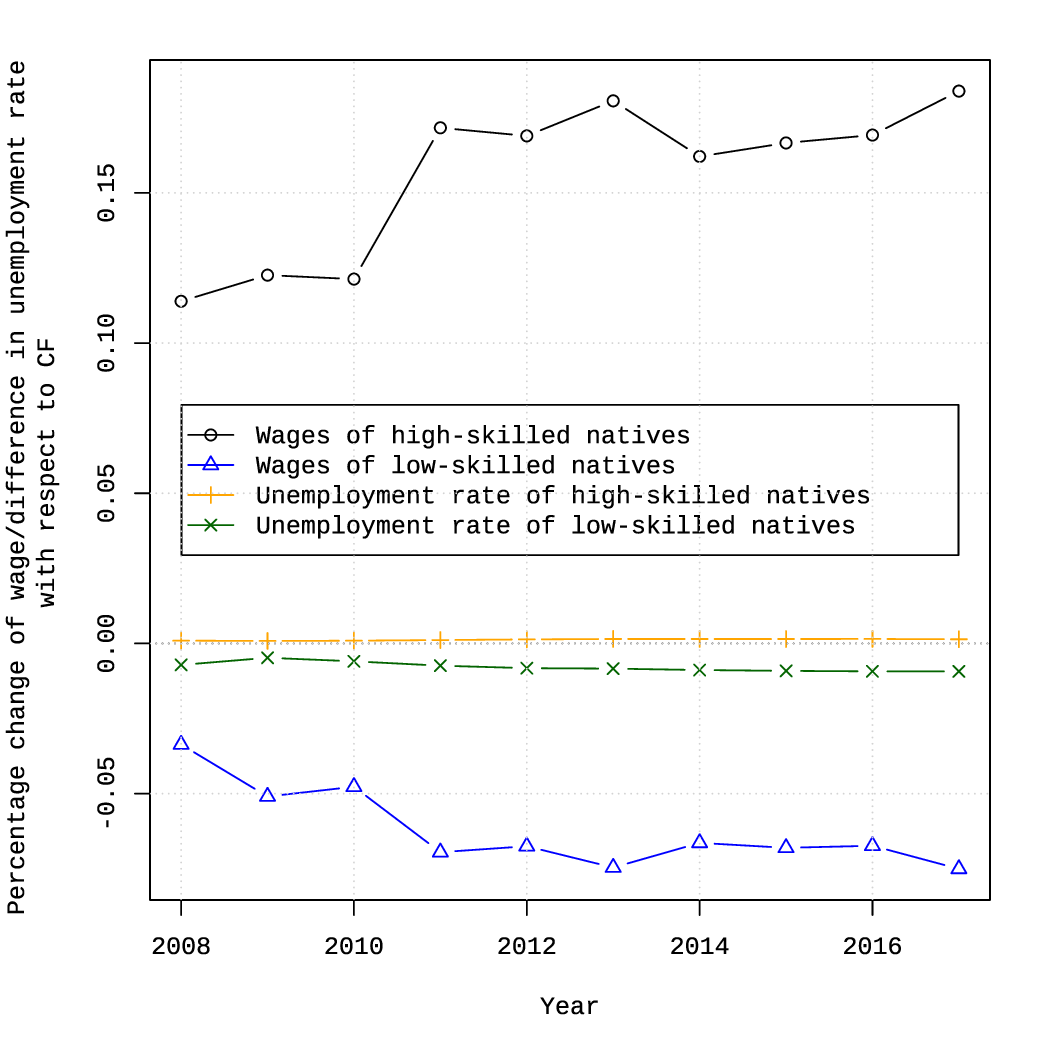}
		\vspace{-0.2cm}
		\caption{Wages and unemployment rates}
		\label{fig:labourmarketCES05}
	\end{subfigure}
	\begin{subfigure}{0.49\textwidth}
		\centering
		\includegraphics[width=0.70\linewidth]{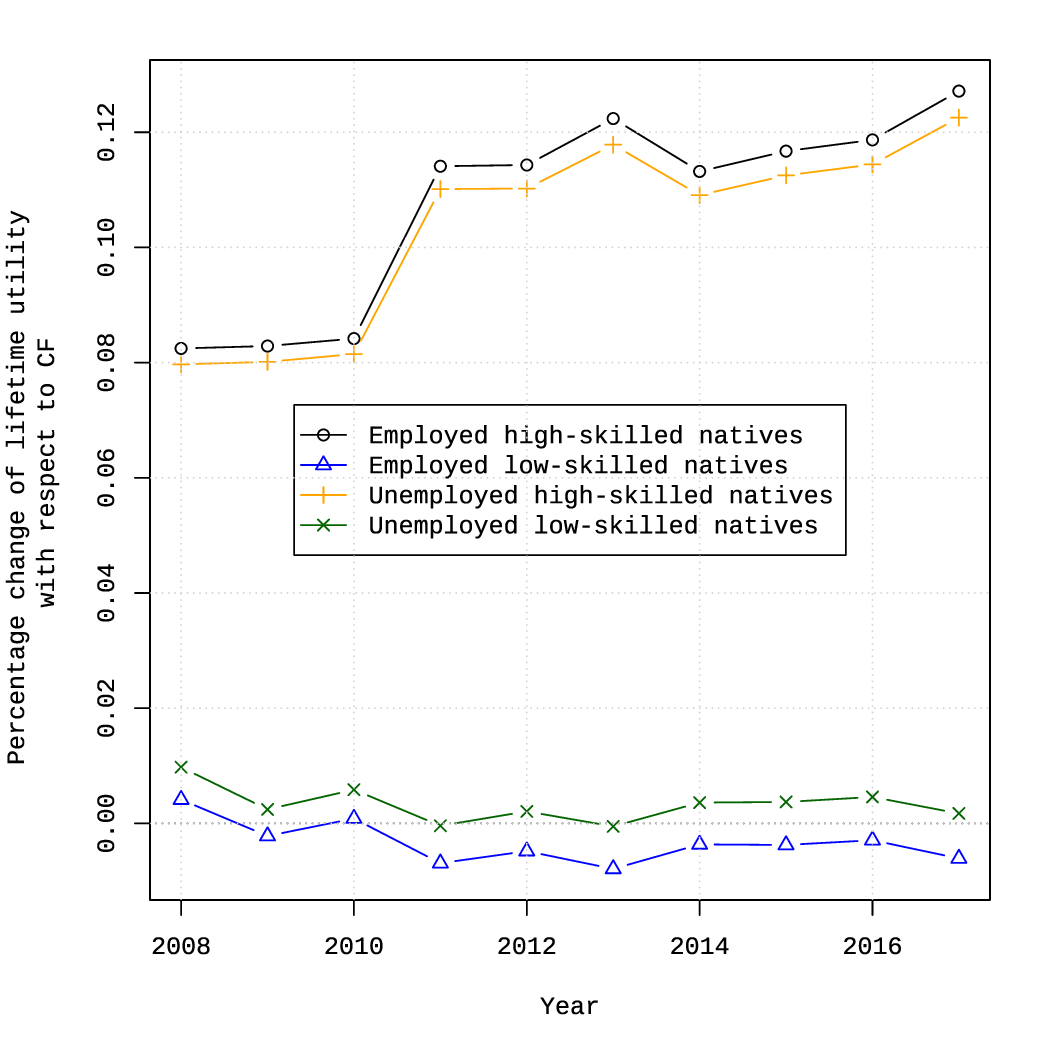}
		\vspace{-0.2cm}
		\caption{Employees' lifetime utility}
		\label{fig:WLUCES05}
	\end{subfigure}
	\begin{subfigure}{0.49\textwidth}
		\centering
		\includegraphics[width=0.70\linewidth]{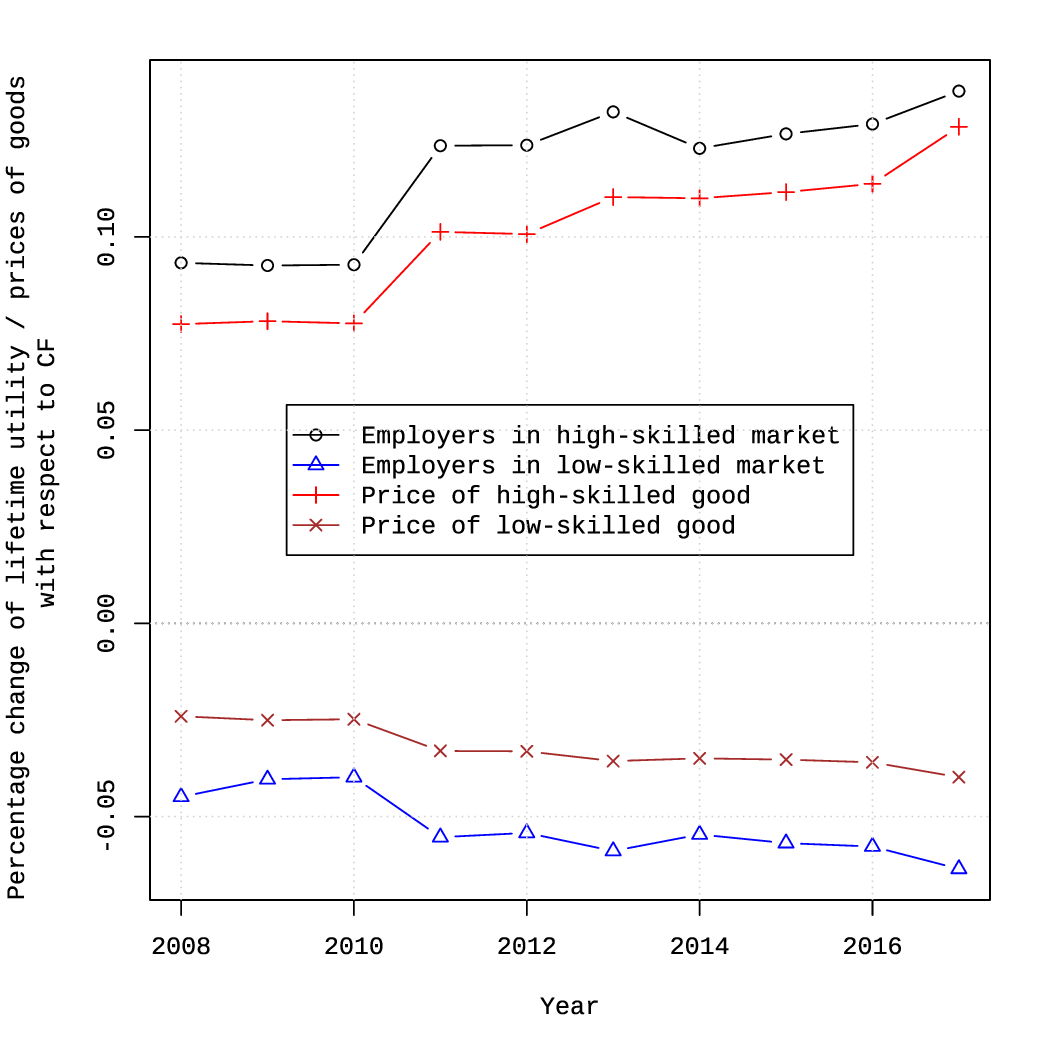}
		\vspace{-0.2cm}
		\caption{Employers' lifetime utility and real prices}
		\label{fig:ELUCES05}
	\end{subfigure}
	\vspace{0.1cm}
	\caption*{\scriptsize{\textit{Note}: The lines show the percentage/absolute change in the variables in the counter-factual scenario in which there are no non-natives compared to the equilibrium in each year between 2008 and 2017. Only for the case of unemployment, we report the difference between the unemployment rate in the counter-factual scenario and in equilibrium, by skill level and country of origin. We define by $CF$ the counter-factual with no non-natives. In Figure (e) the employees' lifetime utility is represented by the present discounted value of having a job $W_{i,j}$ (Equation \ref{eq:BellmanEmployedNative}). In Figure (e) the employers' lifetime utility is represented by the present discounted value of a filled vacancy $J_{i,j}$ (Equation \ref{eq:filledjobNative}).}}
\end{figure}

\begin{figure}[!htbp]
	%\vspace{-1cm}
	\caption{Counter-factual variables - no non-natives with CES parameter $\epsilon=2$.}
	\label{CFnonativesCES20}
	\begin{subfigure}{0.49\textwidth}
		\centering
		\includegraphics[width=0.70\linewidth]{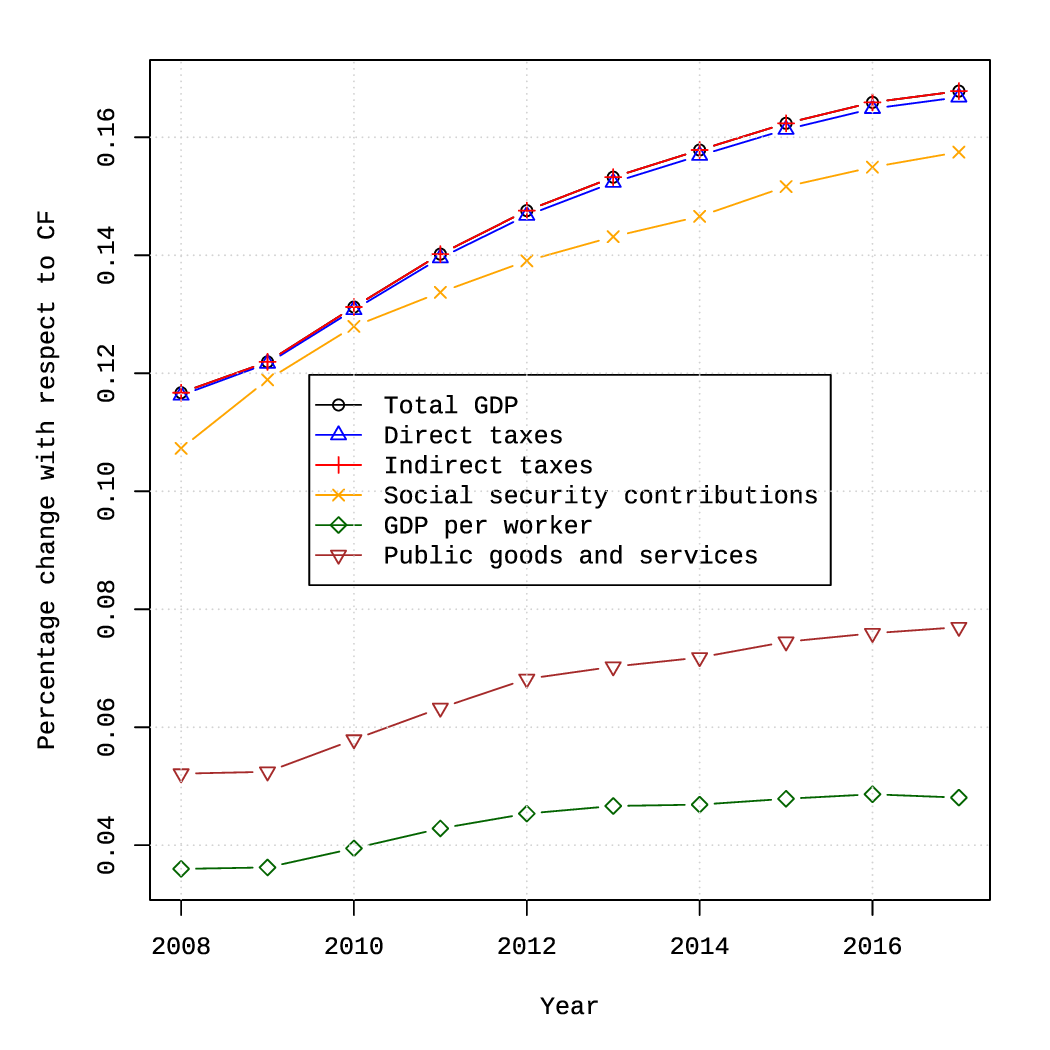}
		\vspace{-0.2cm}
		\caption{Percentage changes of aggregate variables}
		\label{fig:totalProductionAVCES20}
	\end{subfigure}
	\begin{subfigure}{0.49\textwidth}
		\centering
		\includegraphics[width=0.70\linewidth]{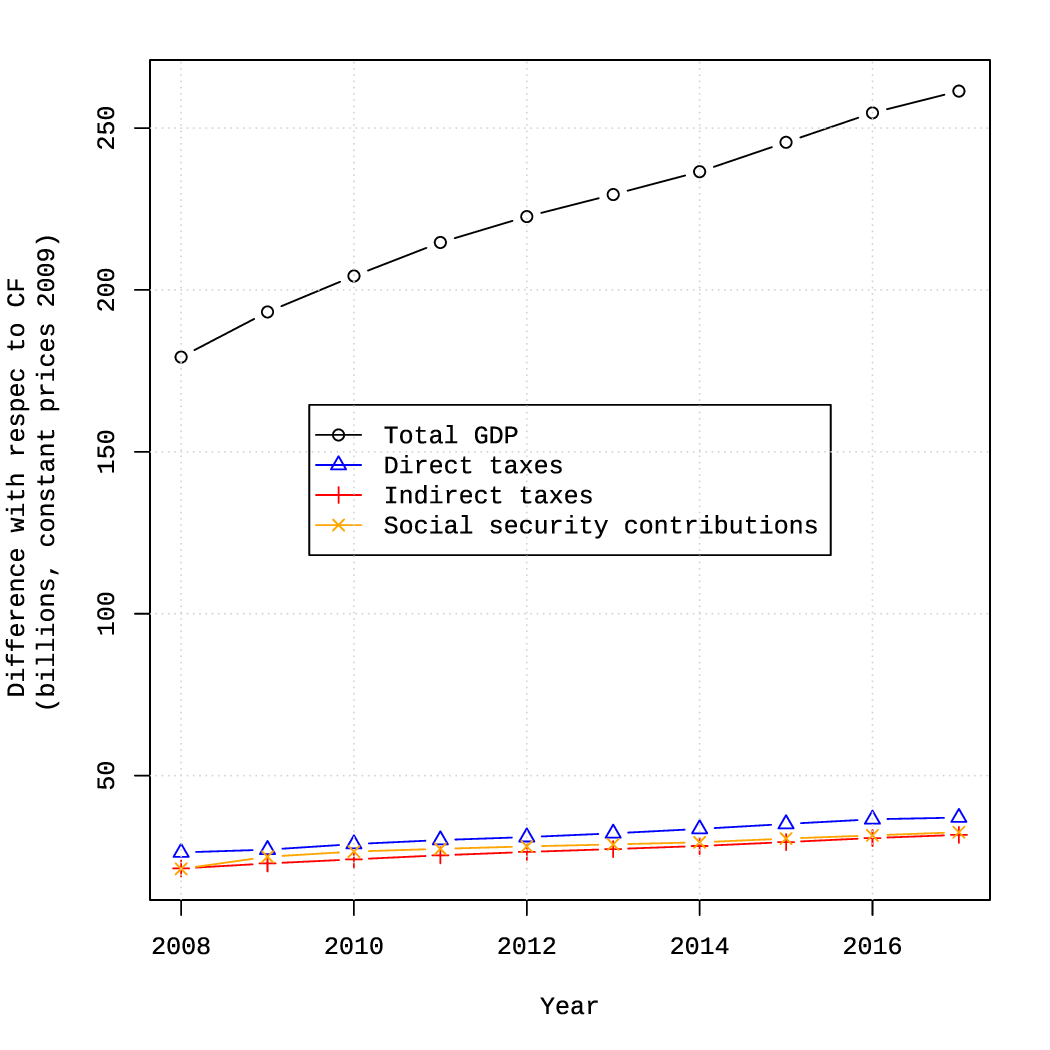}
		\vspace{-0.2cm}
		\caption{Changes of aggregate variables}
		\label{fig:totalProductionCES20}
	\end{subfigure}
	\begin{subfigure}{0.49\textwidth}
	\centering
	\includegraphics[width=0.70\linewidth]{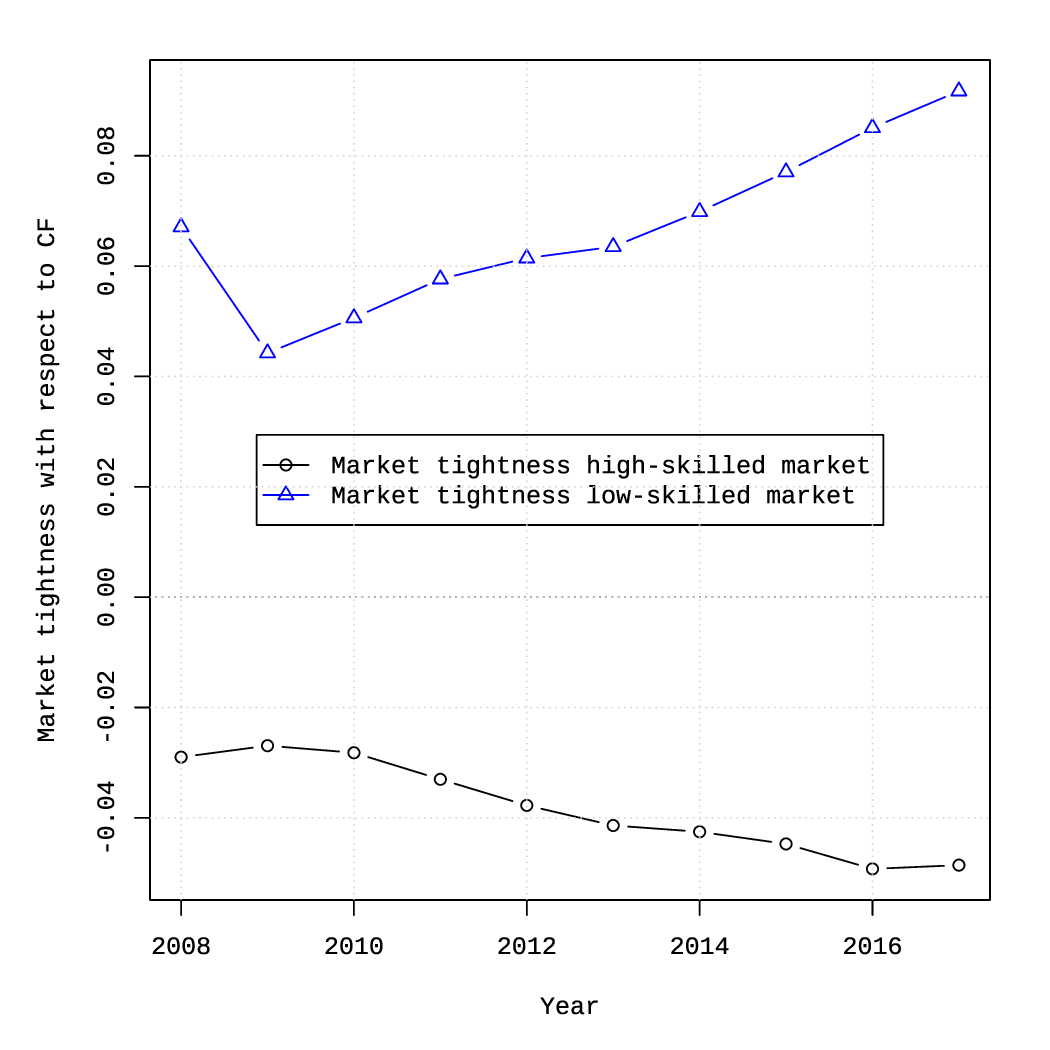}
	\vspace{-0.2cm}
	\caption{Market tightness}
	\label{fig:tightnessCES20}
\end{subfigure}
	\begin{subfigure}{0.49\textwidth}
		\centering
		\includegraphics[width=0.70\linewidth]{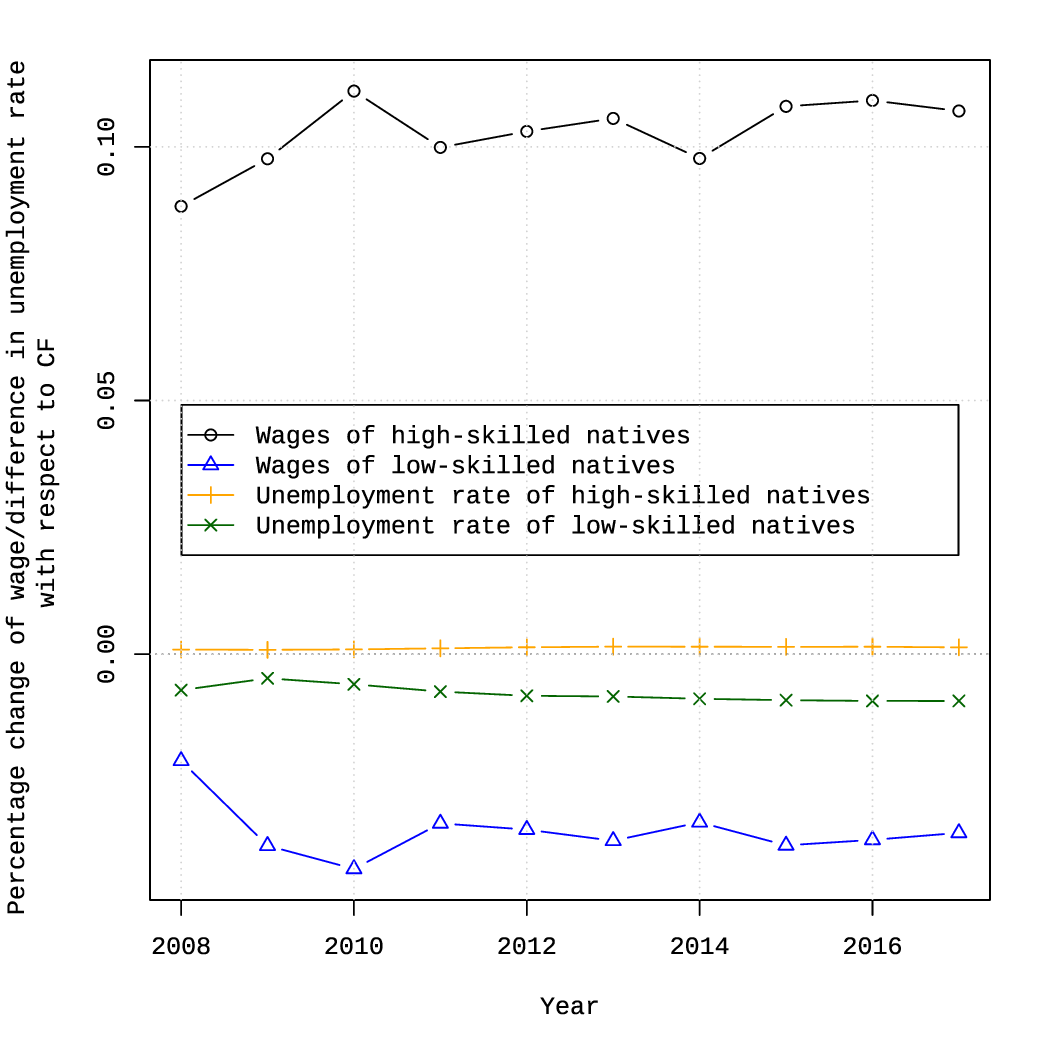}
		\vspace{-0.2cm}
		\caption{Wages and unemployment rates}
		\label{fig:labourmarketCES20}
	\end{subfigure}
	\begin{subfigure}{0.49\textwidth}
		\centering
		\includegraphics[width=0.70\linewidth]{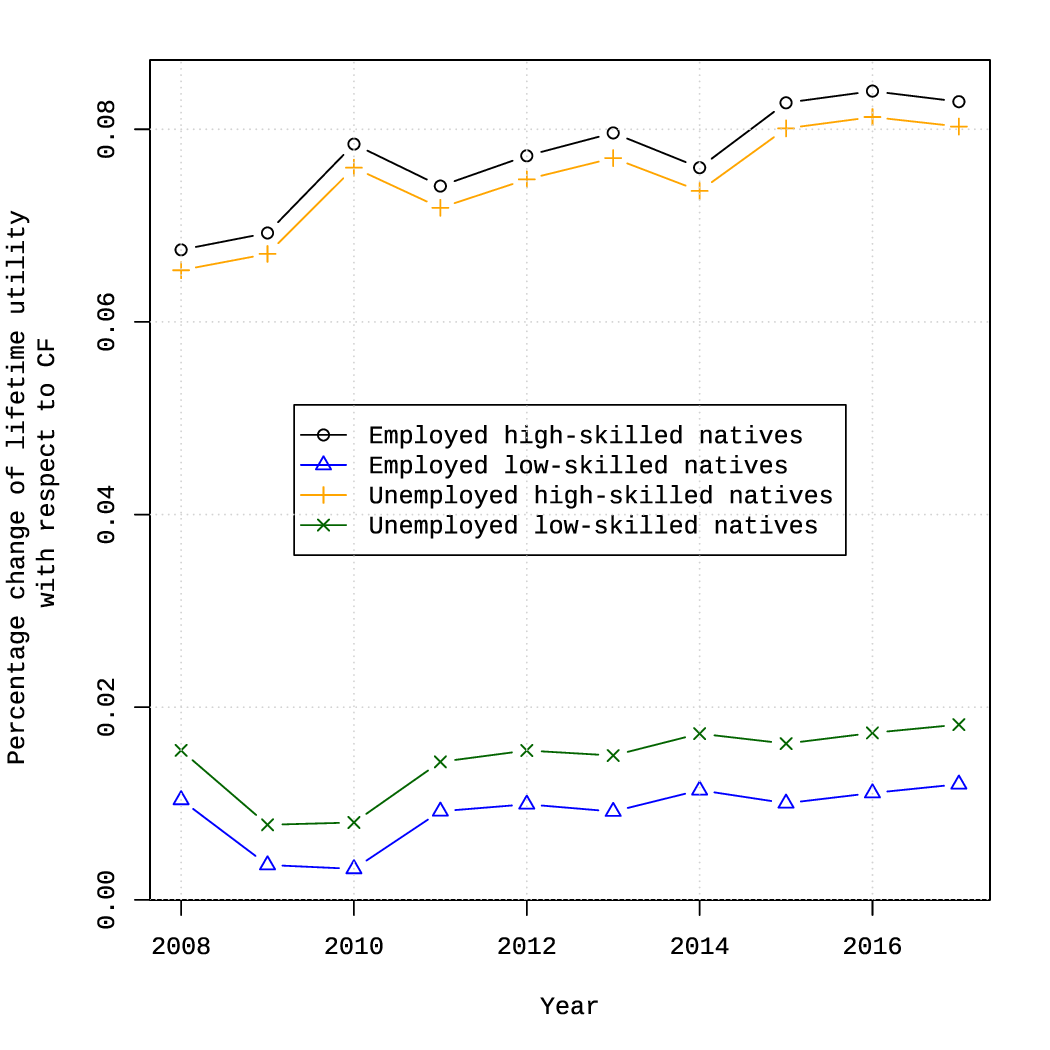}
		\vspace{-0.2cm}
		\caption{Employees' lifetime utility}
		\label{fig:WLUCES40}
	\end{subfigure}
	\begin{subfigure}{0.49\textwidth}
		\centering
		\includegraphics[width=0.70\linewidth]{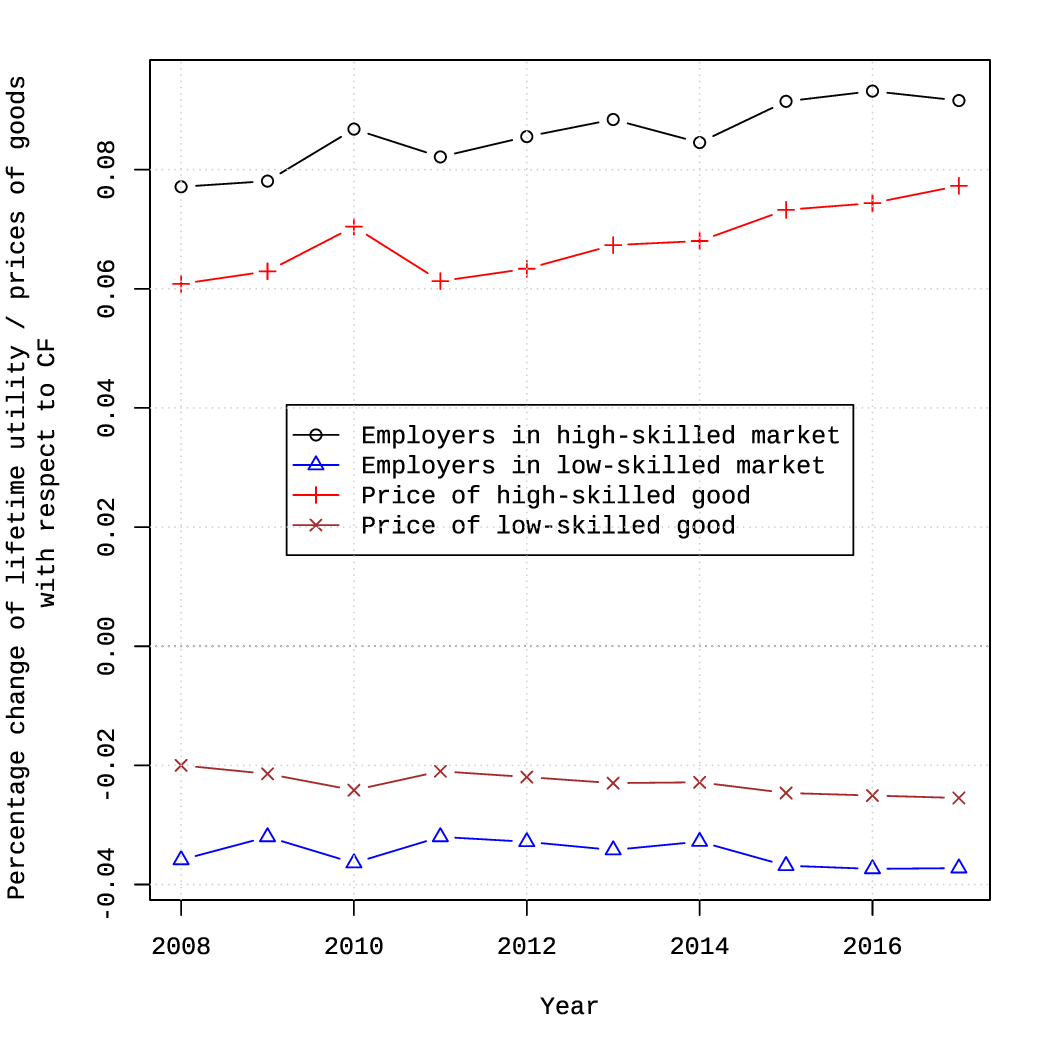}
		\vspace{-0.2cm}
		\caption{Employers' lifetime utility and real prices}
		\label{fig:ELUCES20}
	\end{subfigure}
	\vspace{0.2cm}
	\caption*{\scriptsize{\textit{Note}: The lines show the percentage/absolute change in the variables in the counter-factual scenario in which there are no non-natives compared to the equilibrium in each year between 2008 and 2017. Only for the case of unemployment, we report the difference between the unemployment rate in the counter-factual scenario and in equilibrium, by skill level and country of origin. We define by $CF$ the counter-factual. We define by $CF$ the counter-factual with no non-natives. In Figure (e) the employees' lifetime utility is represented by the present discounted value of having a job $W_{i,j}$ (Equation \ref{eq:BellmanEmployedNative}). In Figure (e) the employers' lifetime utility is represented by the present discounted value of a filled vacancy $J_{i,j}$ (Equation \ref{eq:filledjobNative}).}}
\end{figure}

\clearpage

\begin{figure}[!htbp]
	\vspace{-1cm}
	\caption{Counter-factual variables - no non-natives with CES parameter $\epsilon=10$.}
	\label{CFnonativesCES09}
	\begin{subfigure}{0.49\textwidth}
		\centering
		\includegraphics[width=0.70\linewidth]{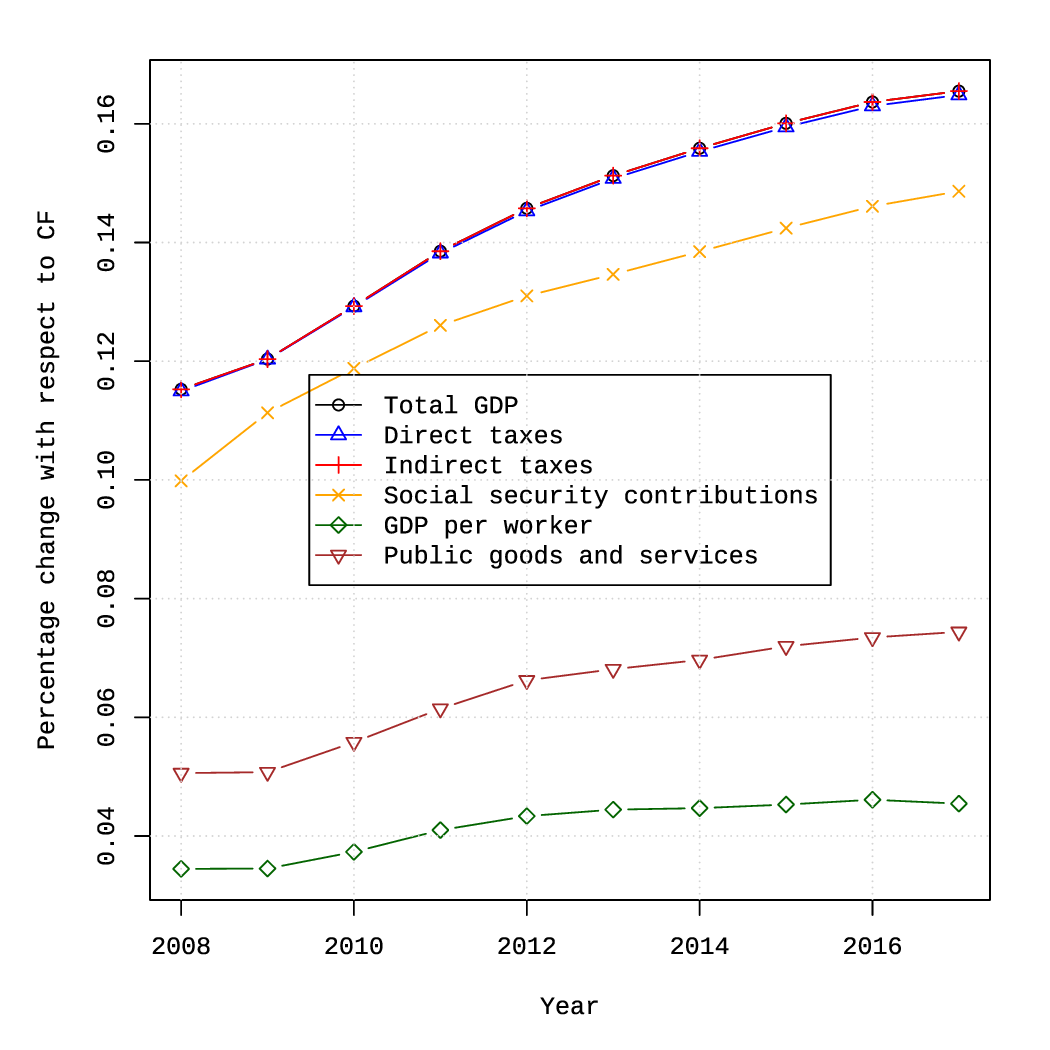}
		\vspace{-0.2cm}
		\caption{Percentage changes of aggregate variables}
		\label{fig:counterfactualWithoutAnyNotNativesAggregateVariables_II_CES10}
	\end{subfigure}
	\begin{subfigure}{0.49\textwidth}
		\centering
		\includegraphics[width=0.70\linewidth]{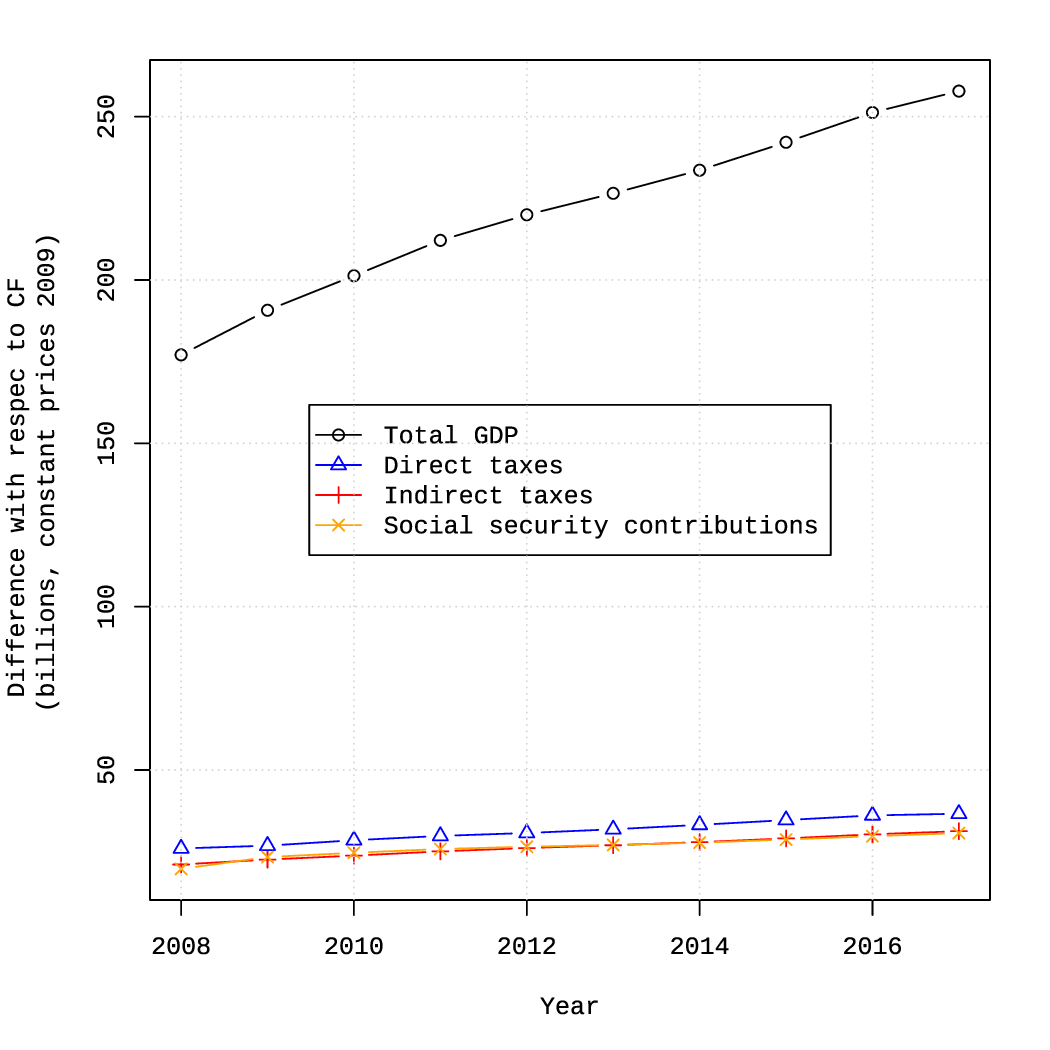}
		\vspace{-0.2cm}
		\caption{Changes of aggregate variables}
		\label{fig:counterfactualWithoutAnyNotNativesAggregateVariablesAbsDiff_II_CES10}
	\end{subfigure}
	\begin{subfigure}{0.49\textwidth}
	\centering
	\includegraphics[width=0.70\linewidth]{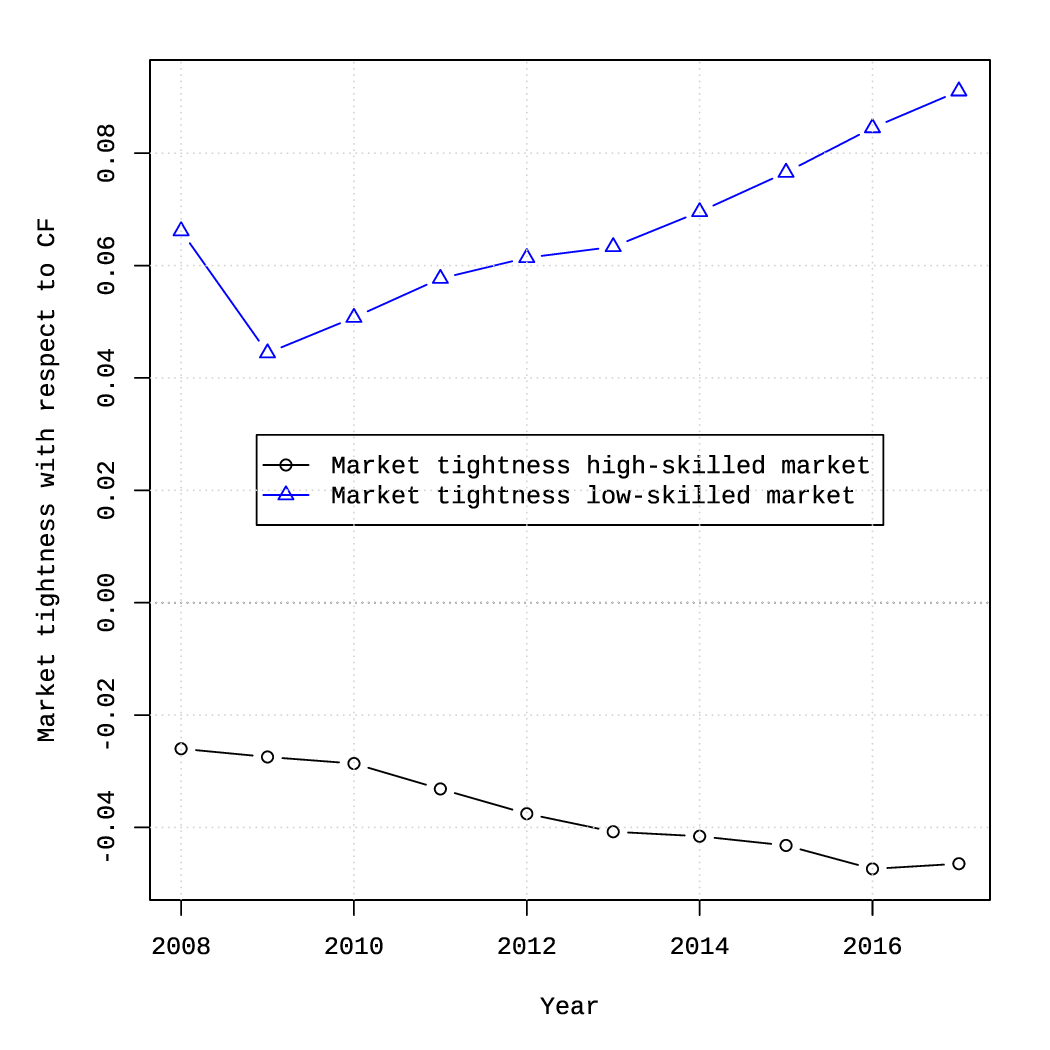}
	\vspace{-0.2cm}
	\caption{Market tightness}
	\label{fig:counterfactualWithoutAnyNotNativesMarketTightness_II_CES10}
\end{subfigure}
	\begin{subfigure}{0.49\textwidth}
		\centering
		\includegraphics[width=0.70\linewidth]{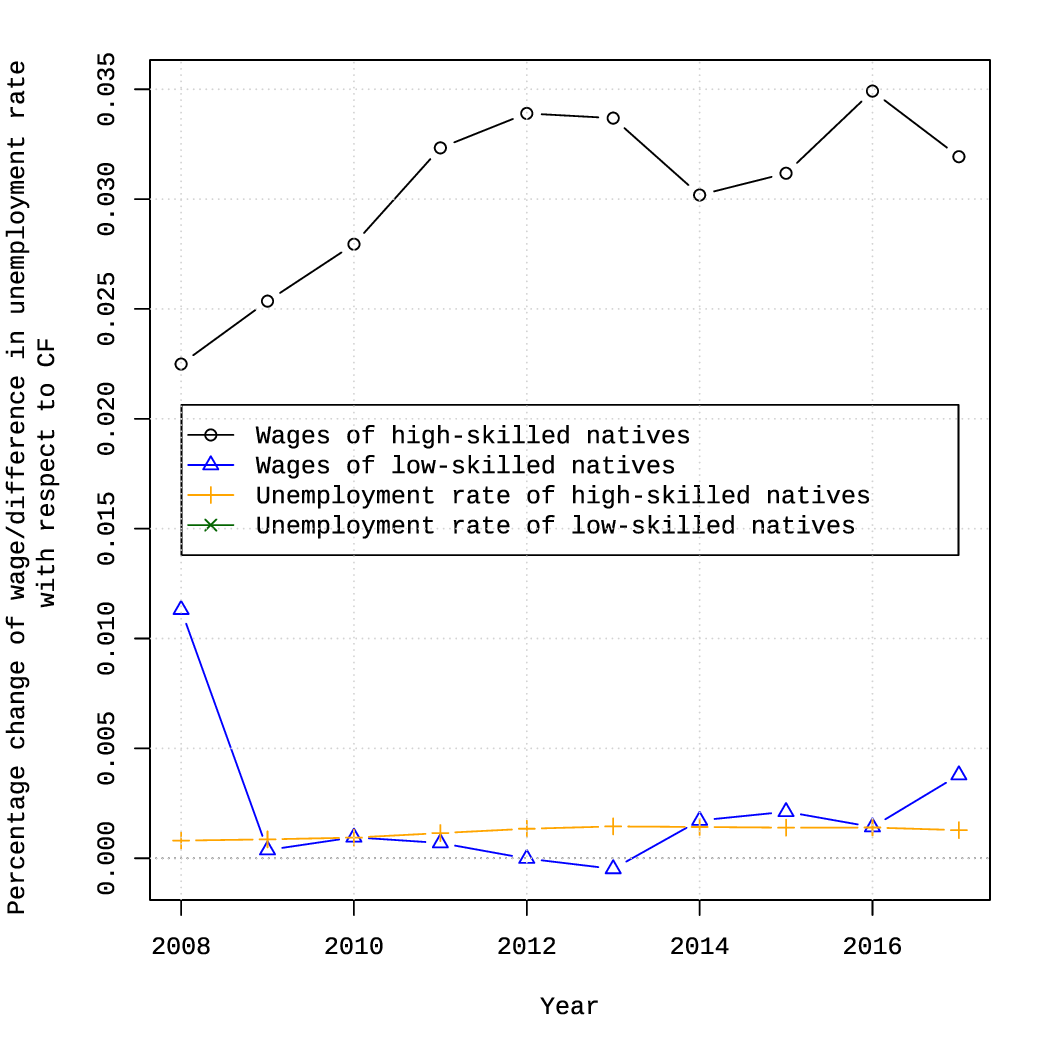}
		\vspace{-0.2cm}
		\caption{Wages and unemployment rates}
		\label{fig:counterfactualWithoutAnyNotNativesLabourMarkets_II_CES10}
	\end{subfigure}
	\begin{subfigure}{0.49\textwidth}
		\centering
		\includegraphics[width=0.70\linewidth]{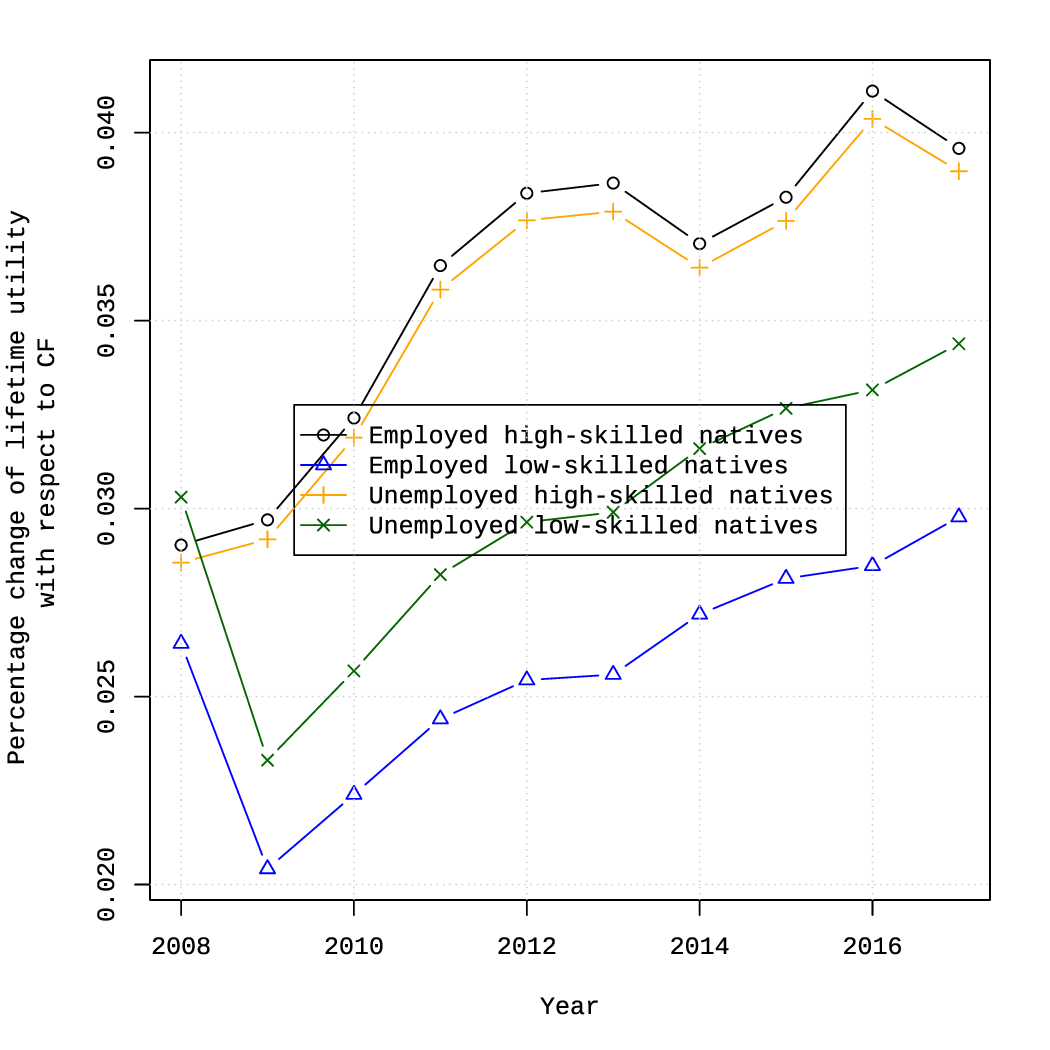}
		\vspace{-0.2cm}
		\caption{Employees' lifetime utility}
		\label{fig:counterfactualWithoutAnyNotNativesLifetimeUtilities_II_CES10}
	\end{subfigure}
	\begin{subfigure}{0.49\textwidth}
		\centering
		\includegraphics[width=0.70\linewidth]{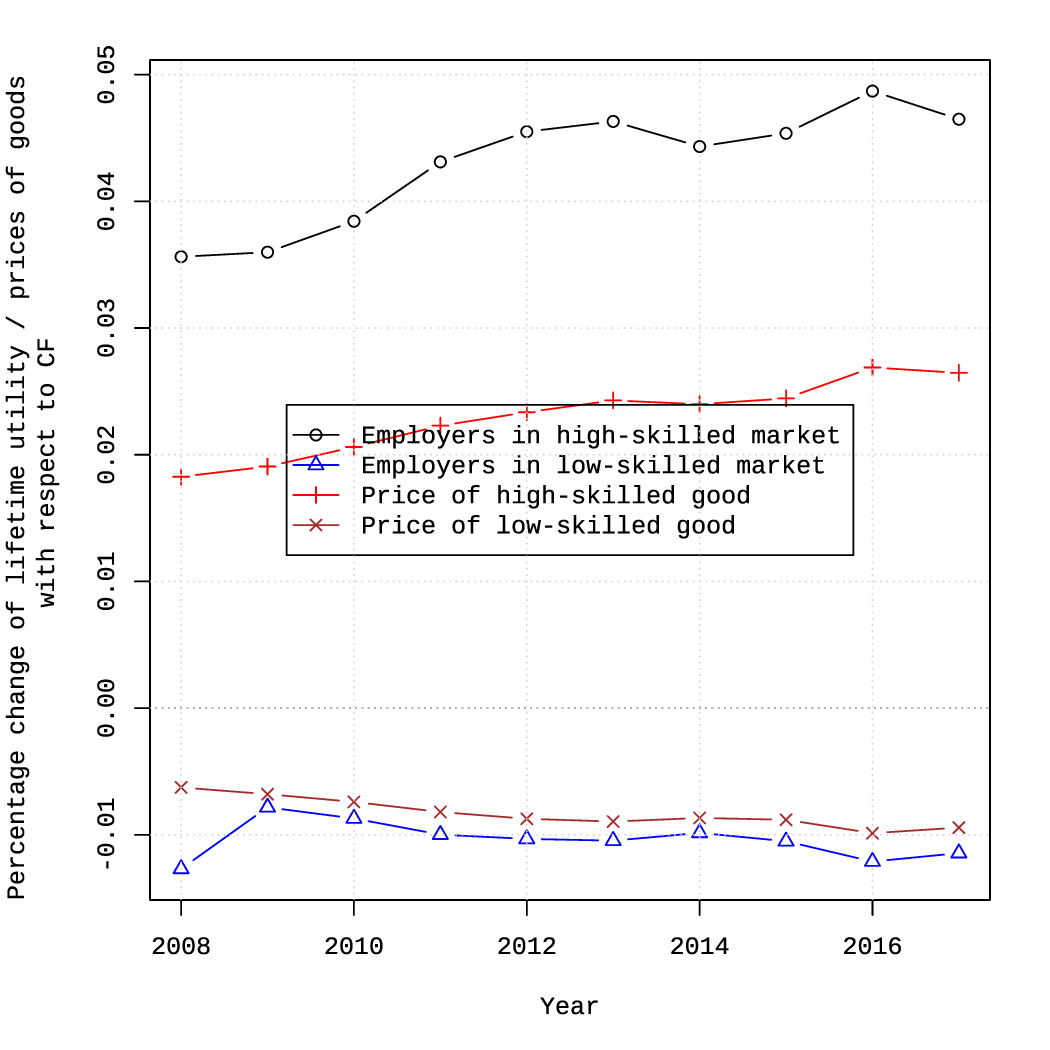}
		\vspace{-0.2cm}
		\caption{Employers' lifetime utility and real prices}
		\label{fig:counterfactualWithoutAnyNotNativesLifetimeUtilitiesEmployers_II_CES10}
	\end{subfigure}
	\vspace{0.2cm}
	\caption*{\scriptsize{\textit{Note}: The lines show the percentage/absolute change in the variables in the counter-factual scenario in which there are no non-natives compared to the equilibrium in each year between 2008 and 2017. Only for the case of unemployment, we report the difference between the unemployment rate in the counter-factual scenario and in equilibrium, by skill level and country of origin. We define by $CF$ the counter-factual. We define by $CF$ the counter-factual with no non-natives. In Figure (e) the employees' lifetime utility is represented by the present discounted value of having a job $W_{i,j}$ (Equation \ref{eq:BellmanEmployedNative}). In Figure (e) the employers' lifetime utility is represented by the present discounted value of a filled vacancy $J_{i,j}$ (Equation \ref{eq:filledjobNative}).}}
\end{figure}

\clearpage

\section{Sensitivity analysis on the congestion level \label{app:congestion}}
\begin{figure}[!htbp]
	\caption{Counter-factual variables - no non-natives with congestion parameter $\zeta=0.5$.}
	\label{fig:CFnonativesCongestion09}
	\begin{subfigure}{0.49\textwidth}
		\centering
		\includegraphics[width=0.70\linewidth]{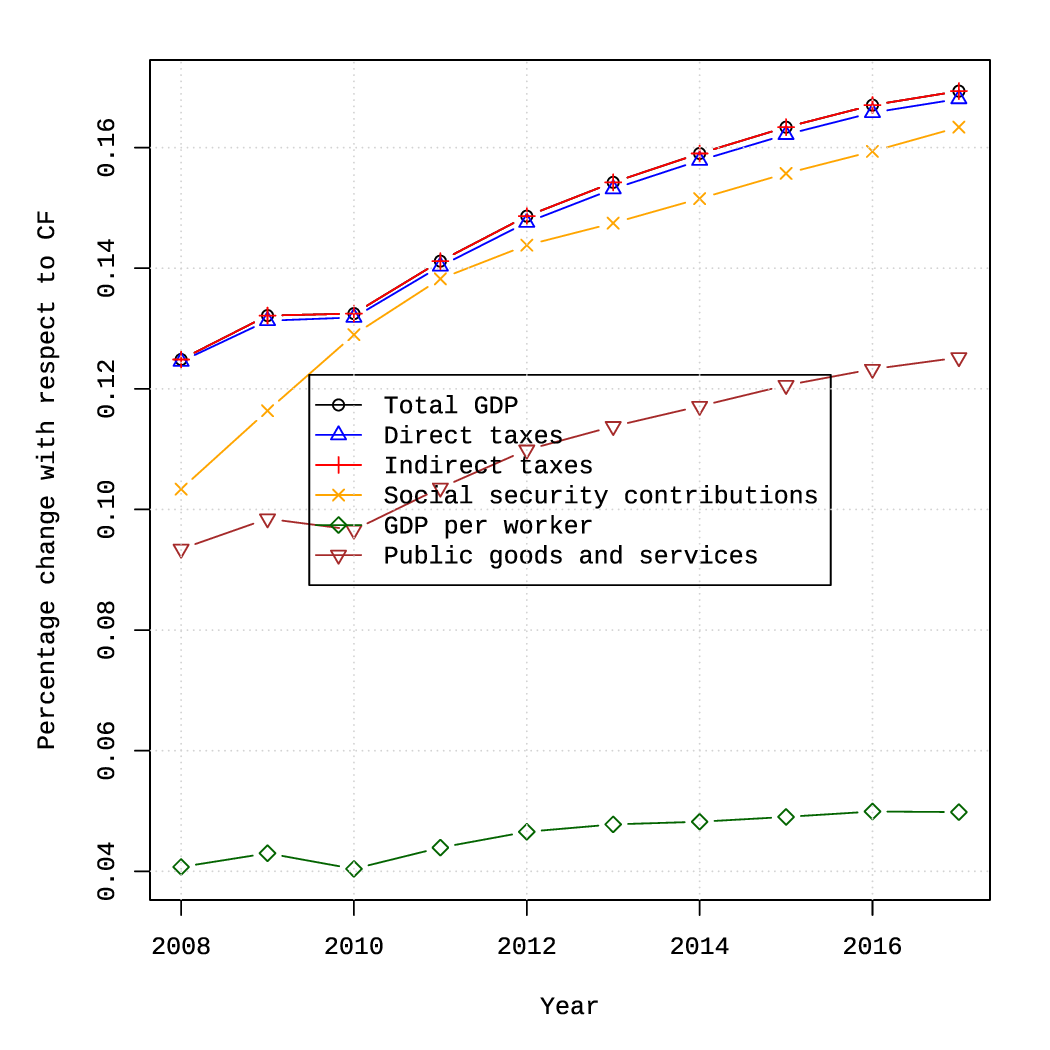}
		\vspace{-0.2cm}
		\caption{Percentage changes of aggregate variables}
		\label{fig:counterfactualWithoutAnyNotNativesAggregateVariables_II_congestion_parameter_09}
	\end{subfigure}
	\begin{subfigure}{0.49\textwidth}
		\centering
		\includegraphics[width=0.70\linewidth]{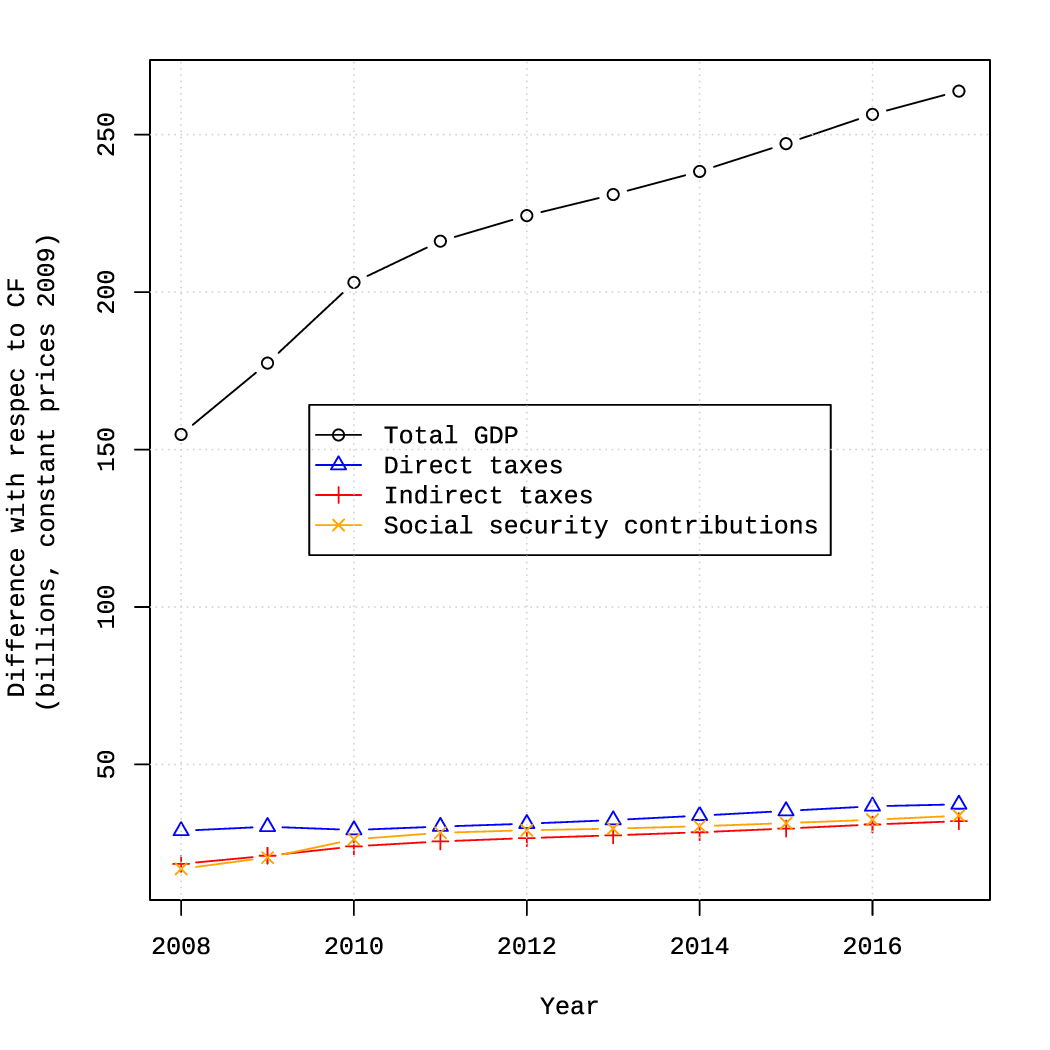}
		\vspace{-0.2cm}
		\caption{Changes of aggregate variables}
		\label{fig:counterfactualWithoutAnyNotNativesAggregateVariablesAbsDiff_II_congestion_parameter_09}
	\end{subfigure}
	\begin{subfigure}{0.49\textwidth}
		\centering
		\includegraphics[width=0.70\linewidth]{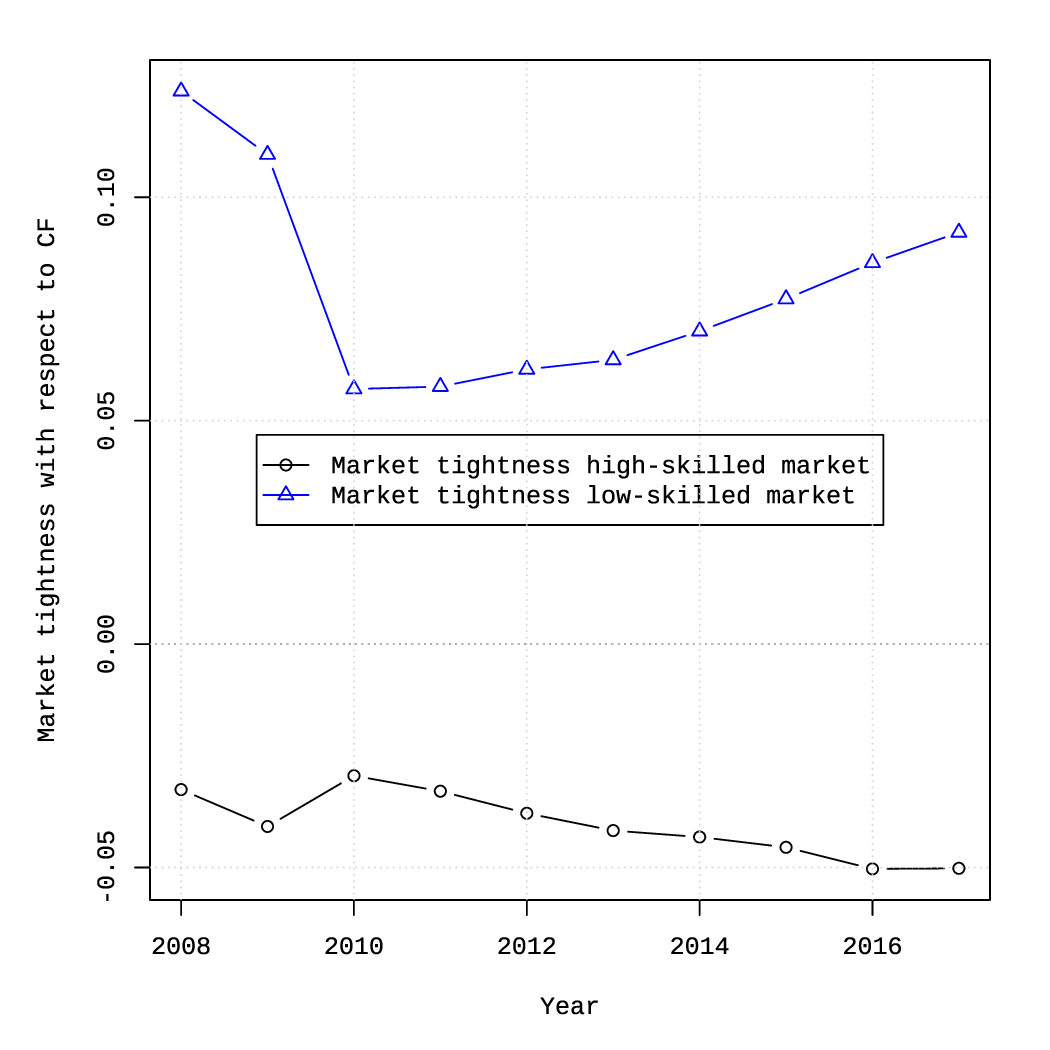}
		\vspace{-0.2cm}
		\caption{Market tightness}
		\label{fig:counterfactualWithoutAnyNotNativesMarketTightness_II_congestion_parameter_09}
	\end{subfigure}
	\begin{subfigure}{0.49\textwidth}
		\centering
		\includegraphics[width=0.70\linewidth]{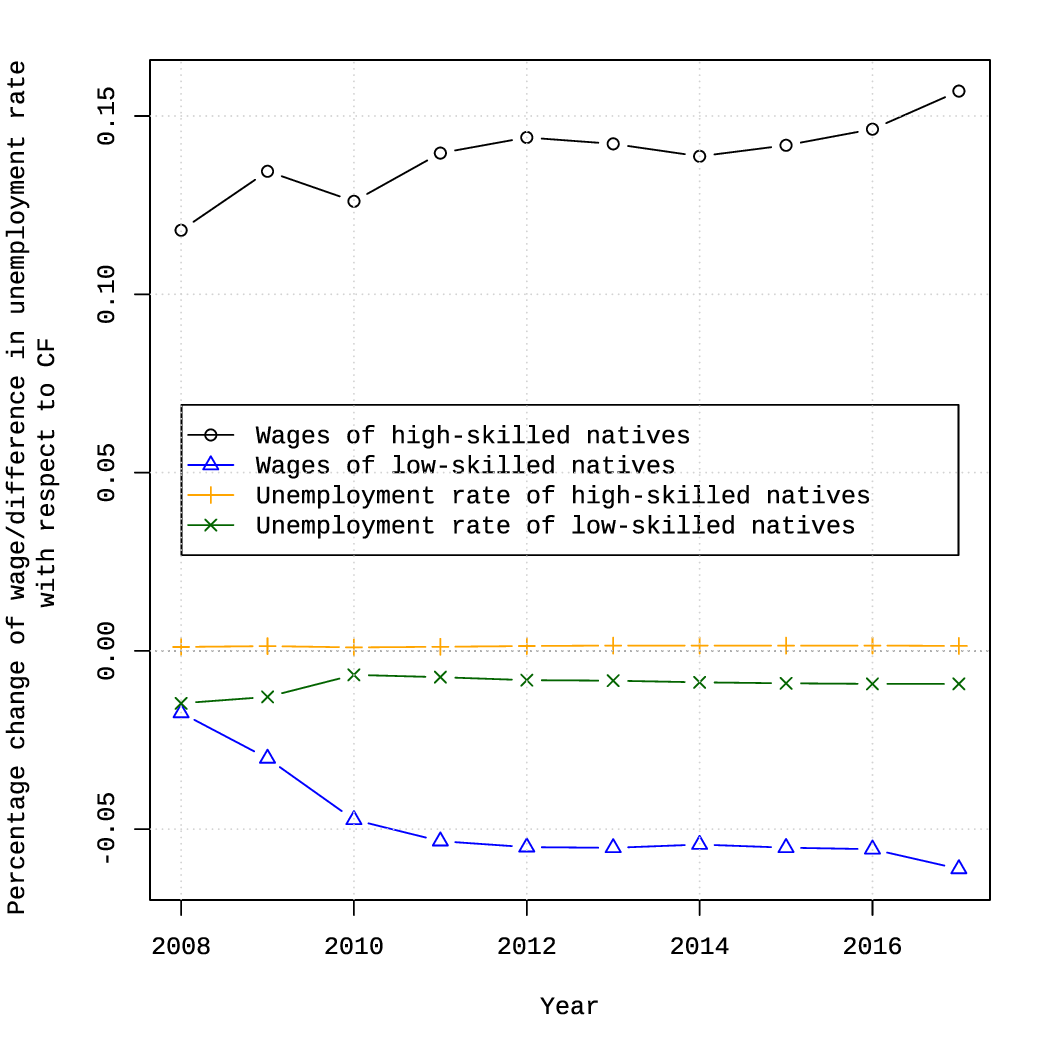}
		\vspace{-0.2cm}
		\caption{Wages and unemployment rates}
		\label{fig:counterfactualWithoutAnyNotNativesLabourMarkets_II_congestion_parameter_09}
	\end{subfigure}
	\begin{subfigure}{0.49\textwidth}
		\centering
		\includegraphics[width=0.70\linewidth]{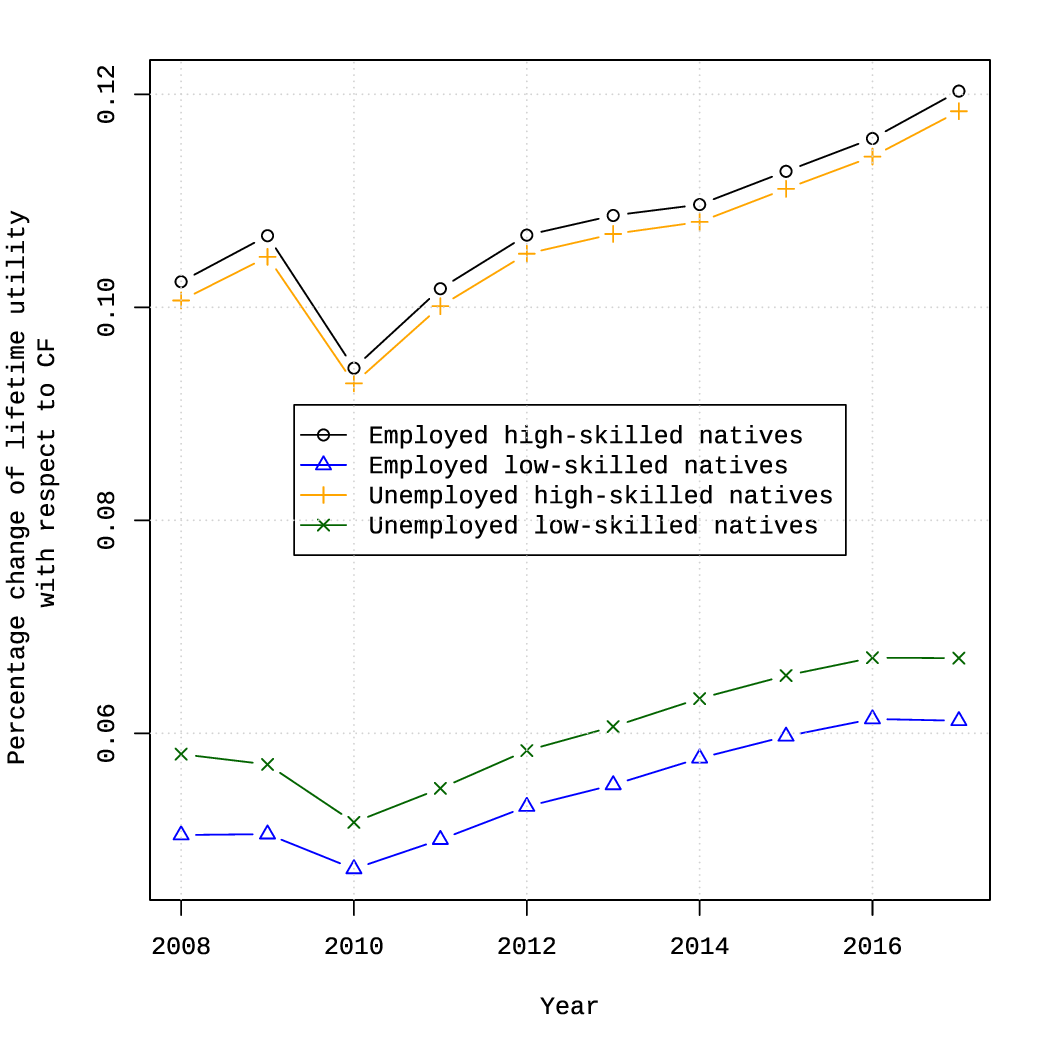}
		\vspace{-0.2cm}
		\caption{Employees' lifetime utility}
		\label{fig:counterfactualWithoutAnyNotNativesLifetimeUtilities_II_congestion_parameter_09}
	\end{subfigure}
	\begin{subfigure}{0.49\textwidth}
		\centering
		\includegraphics[width=0.70\linewidth]{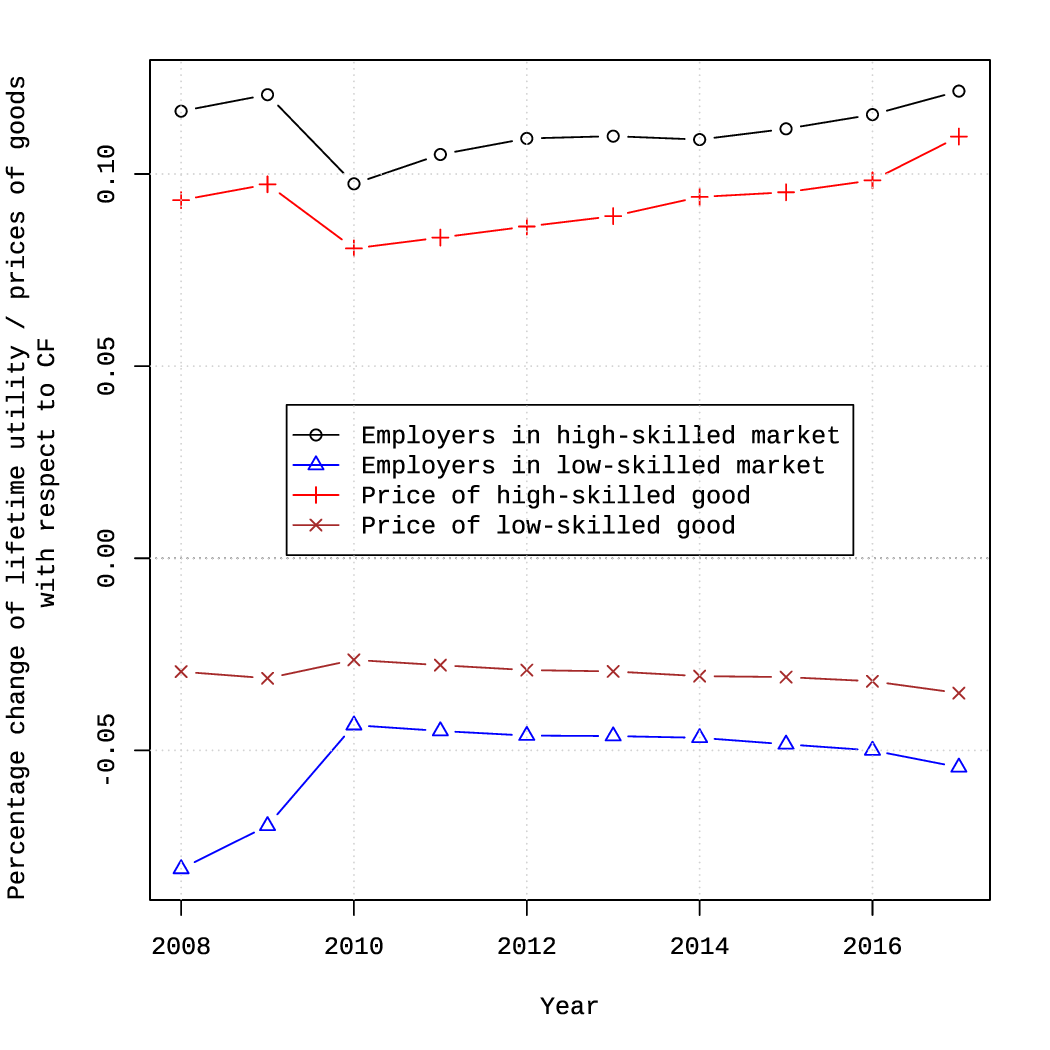}
		\vspace{-0.2cm}
		\caption{Employers' lifetime utility and real prices}
		\label{fig:counterfactualWithoutAnyNotNativesLifetimeUtilitiesEmployers_II_congestion_parameter_09}
	\end{subfigure}
	\vspace{0.1cm}
	\caption*{\scriptsize{\textit{Note}: The lines show the percentage/absolute change in the variables in the counter-factual scenario in which there are no non-natives compared to the equilibrium in each year between 2008 and 2017. Only for the case of unemployment, we report the difference between the unemployment rate in the counter-factual scenario and in equilibrium, by skill level and country of origin. We define by $CF$ the counter-factual. We define by $CF$ the counter-factual with no non-natives. In Figure (e) the employees' lifetime utility is represented by the present discounted value of having a job $W_{i,j}$ (Equation \ref{eq:BellmanEmployedNative}). In Figure (e) the employers' lifetime utility is represented by the present discounted value of a filled vacancy $J_{i,j}$ (Equation \ref{eq:filledjobNative}).}}
\end{figure}

\begin{figure}[!htbp]
	\caption{Counter-factual variables - no non-natives with congestion parameter $\zeta=1.5$.}
	\label{fig:CFnonativesCongestion11}
	\begin{subfigure}{0.49\textwidth}
		\centering
		\includegraphics[width=0.70\linewidth]{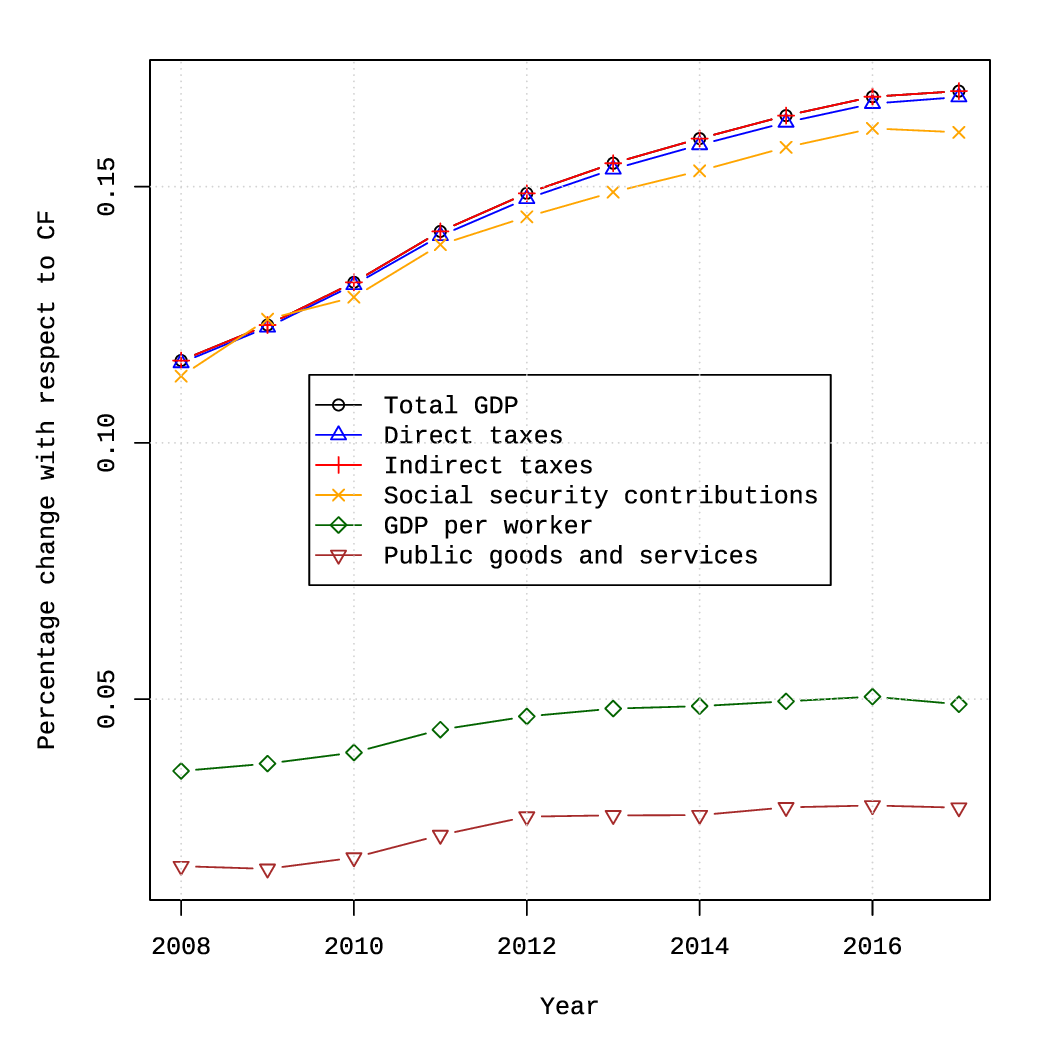}
		\vspace{-0.2cm}
		\caption{Percentage changes of aggregate variables}
		\label{fig:counterfactualWithoutAnyNotNativesAggregateVariables_II_congestion_parameter_11}
	\end{subfigure}
	\begin{subfigure}{0.49\textwidth}
		\centering
		\includegraphics[width=0.70\linewidth]{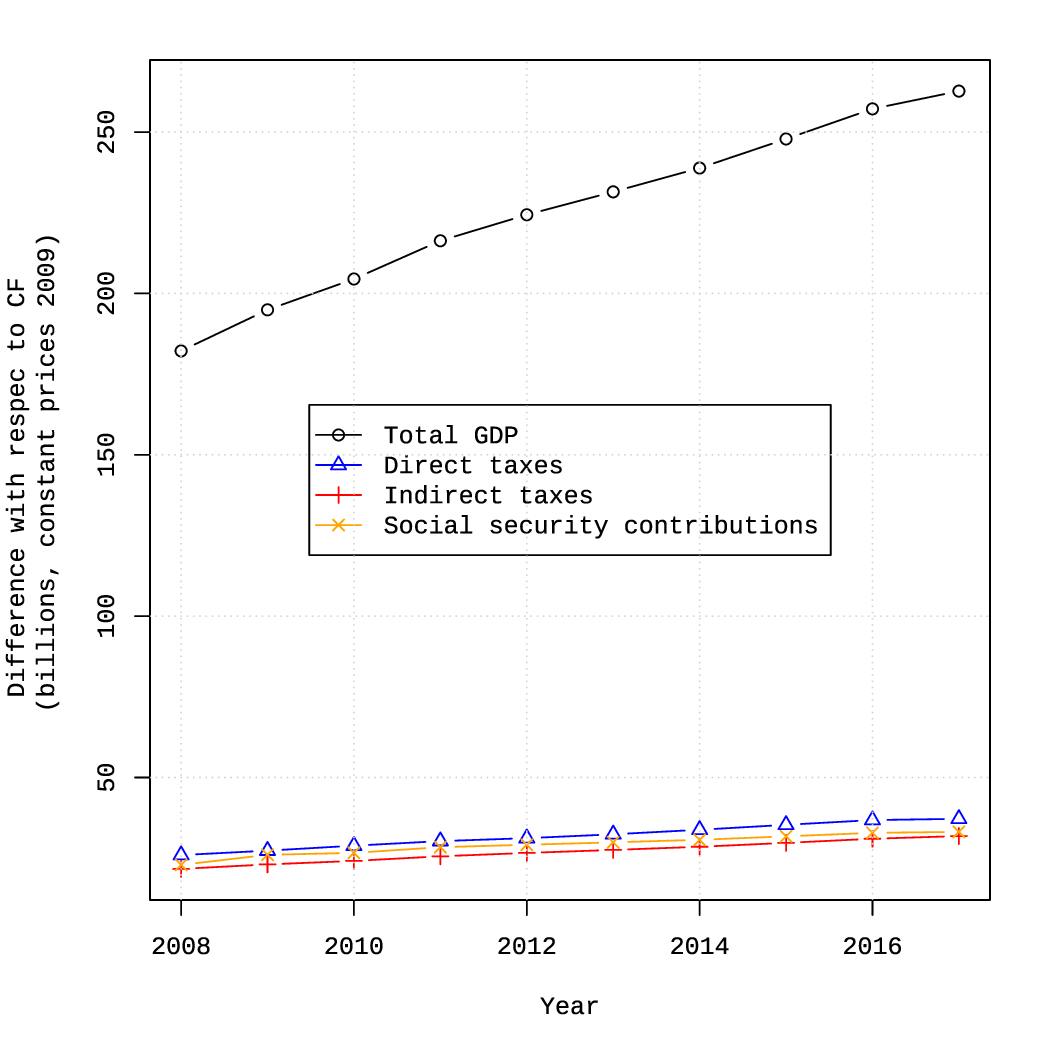}
		\vspace{-0.2cm}
		\caption{Changes of aggregate variables}
		\label{fig:counterfactualWithoutAnyNotNativesAggregateVariablesAbsDiff_II_congestion_parameter_11}
	\end{subfigure}
	\begin{subfigure}{0.49\textwidth}
		\centering
		\includegraphics[width=0.70\linewidth]{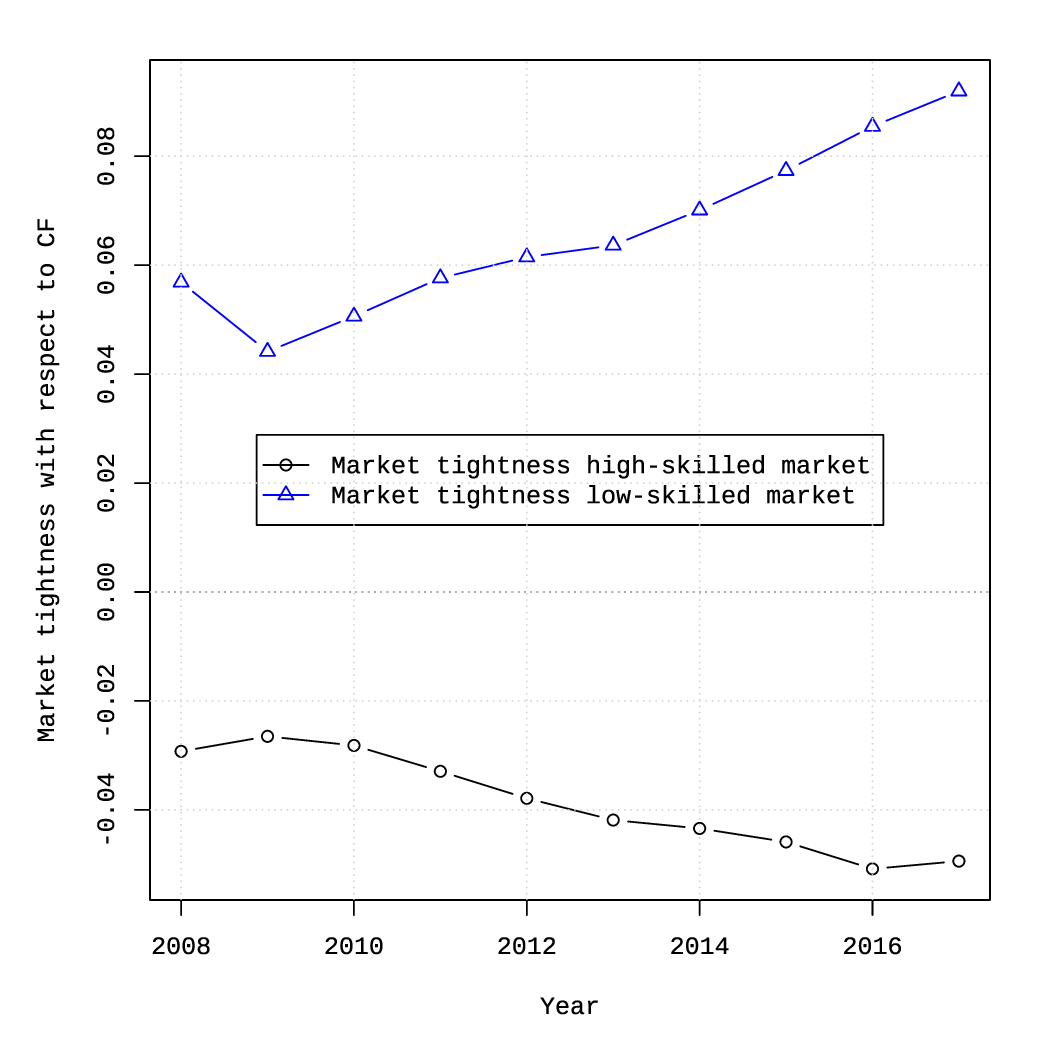}
		\vspace{-0.2cm}
		\caption{Market tightness}
		\label{fig:counterfactualWithoutAnyNotNativesMarketTightness_II_congestion_parameter_11}
	\end{subfigure}
	\begin{subfigure}{0.49\textwidth}
		\centering
		\includegraphics[width=0.70\linewidth]{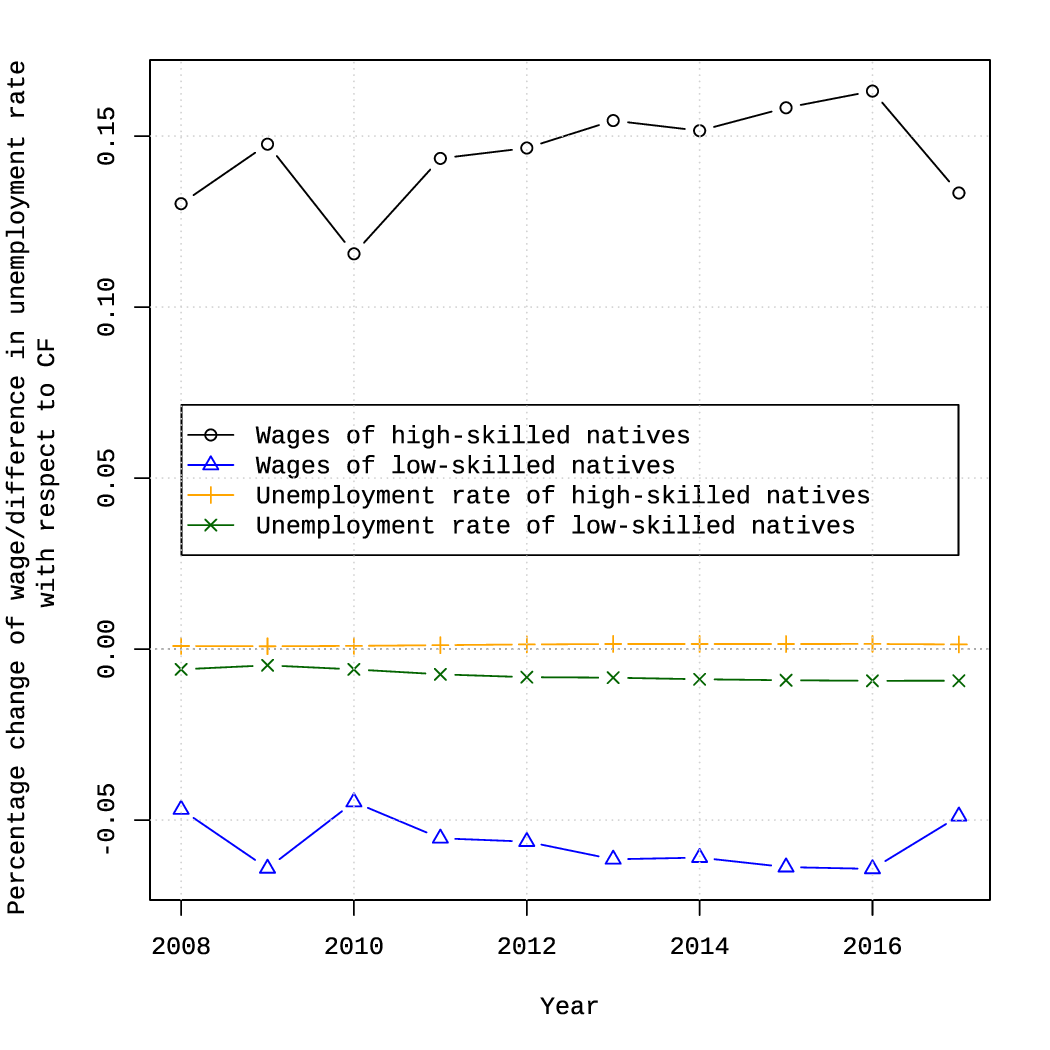}
		\vspace{-0.2cm}
		\caption{Wages and unemployment rates}
		\label{fig:counterfactualWithoutAnyNotNativesLabourMarkets_II_congestion_parameter_11}
	\end{subfigure}
	\begin{subfigure}{0.49\textwidth}
		\centering
		\includegraphics[width=0.70\linewidth]{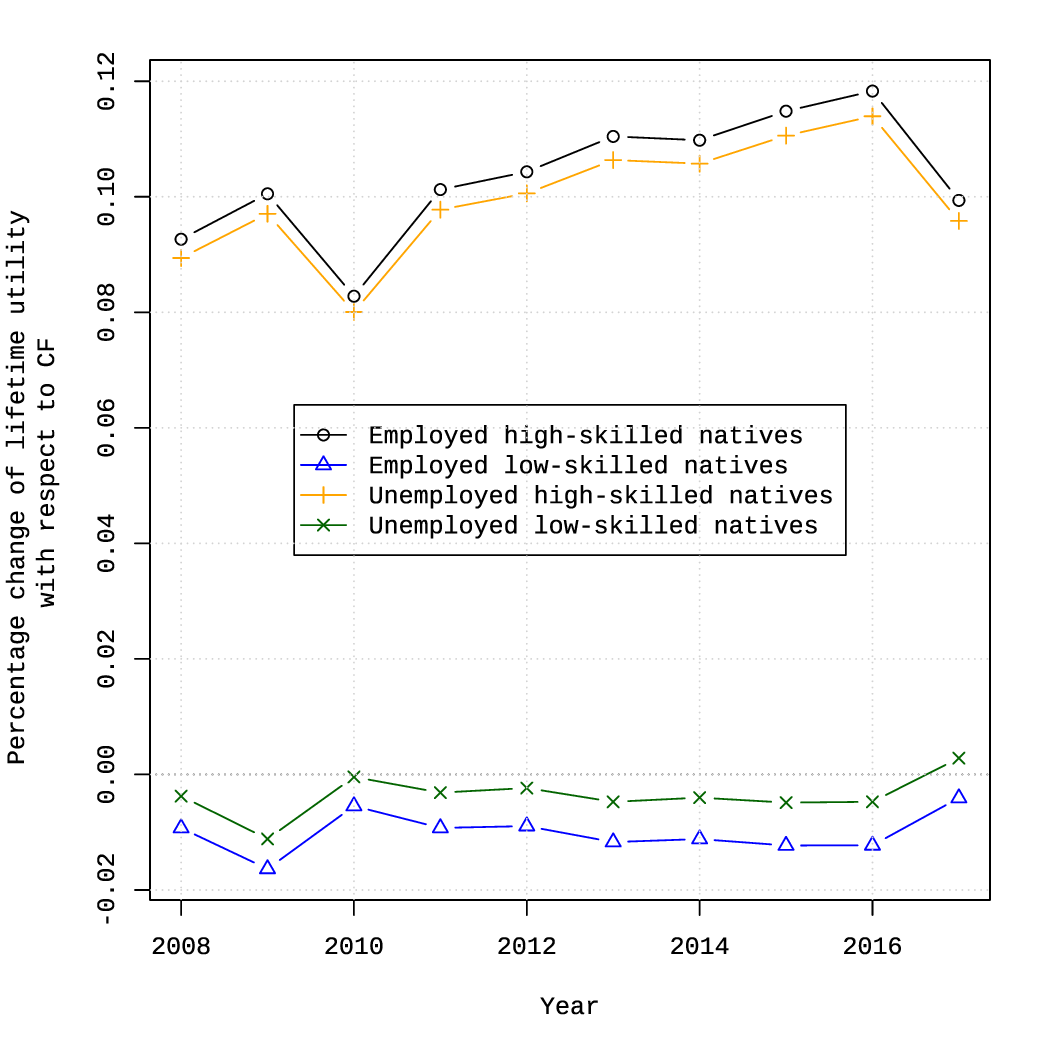}
		\vspace{-0.2cm}
		\caption{Employees' lifetime utility}
		\label{fig:counterfactualWithoutAnyNotNativesLifetimeUtilities_II_congestion_parameter_11}
	\end{subfigure}
	\begin{subfigure}{0.49\textwidth}
		\centering
		\includegraphics[width=0.70\linewidth]{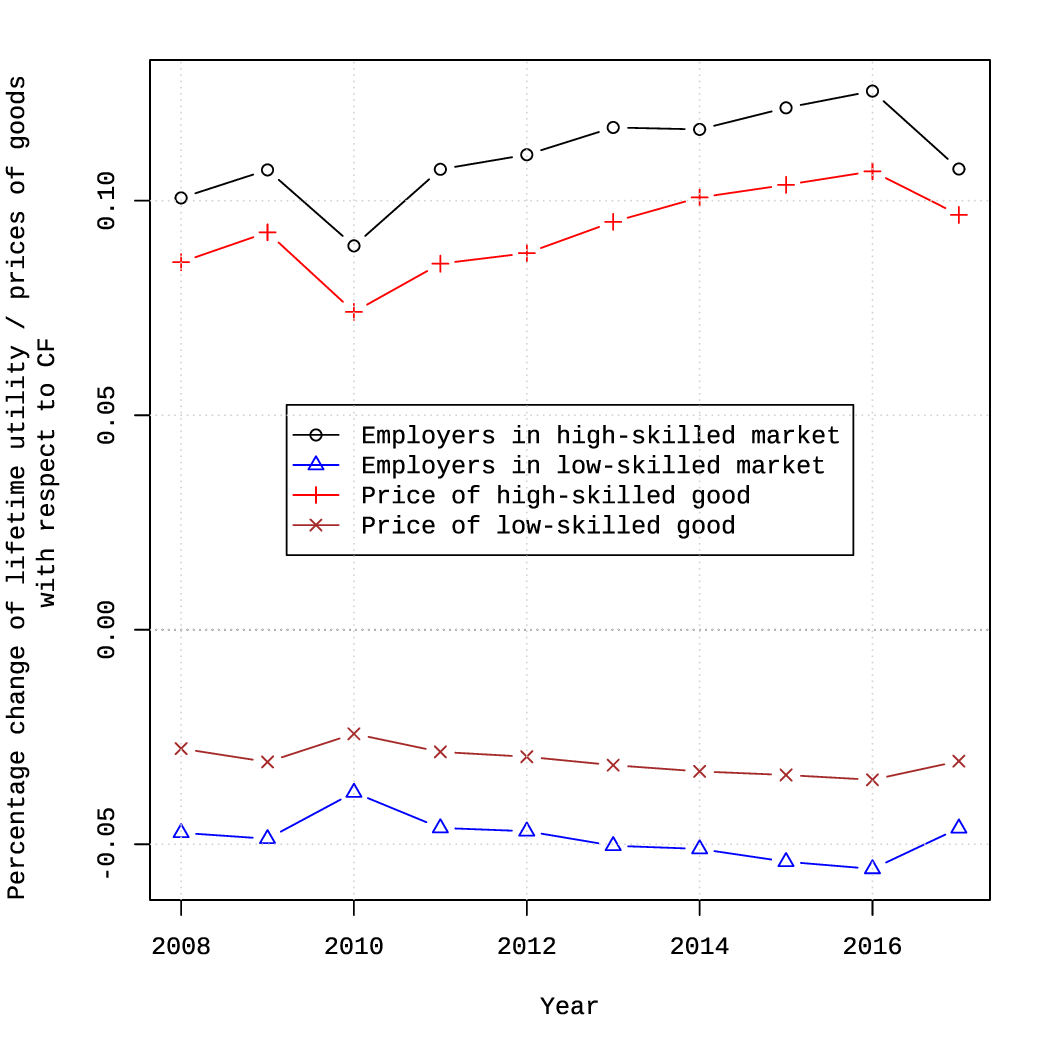}
		\vspace{-0.2cm}
		\caption{Employers' lifetime utility and real prices}
		\label{fig:counterfactualWithoutAnyNotNativesLifetimeUtilitiesEmployers_II_congestion_parameter_11}
	\end{subfigure}
	\vspace{0.1cm}
	\caption*{\scriptsize{\textit{Note}: The lines show the percentage/absolute change in the variables in the counter-factual scenario in which there are no non-natives compared to the equilibrium in each year between 2008 and 2017. Only for the case of unemployment, we report the difference between the unemployment rate in the counter-factual scenario and in equilibrium, by skill level and country of origin. We define by $CF$ the counter-factual. We define by $CF$ the counter-factual with no non-natives. In Figure (e) the employees' lifetime utility is represented by the present discounted value of having a job $W_{i,j}$ (Equation \ref{eq:BellmanEmployedNative}). In Figure (e) the employers' lifetime utility is represented by the present discounted value of a filled vacancy $J_{i,j}$ (Equation \ref{eq:filledjobNative}).}}
\end{figure}

\clearpage

\section{The equilibrium in the goods market \label{app:equilibriumGoodMarket}}

Using Equation (\ref{priceIndexCES}), we can express the real price of good $HS$ as a function of the real price of good $LS$ as:
\begin{equation}\label{app:phplequation1}
	\tilde{p}_{HS}=\left[\frac{1-\tilde{p}_{LS}^{\rho/(\rho-1)}\left(1-\gamma\right)^{1/(1-\rho)}}{\gamma^{1/(1-\rho)}}\right]^{(\rho-1)/\rho}.
\end{equation}
Moreover, by equating demand and supply of good $HS$ and good $LS$ and dividing member by member we get:
\begin{equation}\label{app:phplequation2} \frac{\tilde{p}_{HS}}{\tilde{p}_{LS}}=\left(\frac{\gamma}{1-\gamma}\right)\left[\frac{(1-g_{LS})q_{LS}}{(1-g_{HS})q_{HS}}\right]^{1-\rho}.
\end{equation}
Solving the system of two equations (Equations (\ref{app:phplequation1}) and (\ref{app:phplequation2})) in two unknowns ($\tilde{p}_{HS}$ and $\tilde{p}_{LS}$), we can derive the following equations for the two prices, as function of the parameters of the model: 
%\begin{eqnarray}\label{app:priceHpriceL}
%\tilde{p}_h&=&\left\{\frac{\gamma^{\rho/(\rho-1)}\left[(1-g_h)x_h(e_{h,N}+e_{h,I})\right]^{\rho}}{\gamma\left[(1-g_h)x_l(e_{h,N}+e_{h,I})\right]^{\rho}+(1-\gamma)\left[(1-g_l)x_l(e_{l,N}+e_{l,I})\right]^{\rho}}\right\}^{(\rho-1)/\rho}\\ \tilde{p}_l&=&\left\{\frac{(1-\gamma)^{\rho/(\rho-1)}\left[(1-g_l)x_l(e_{l,N}+e_{l,I})\right]^{\rho}}{\gamma\left[(1-g_h)x_l(e_{h,N}+e_{h,I})\right]^{\rho}+(1-\gamma)\left[(1-g_l)x_l(e_{l,N}+e_{l,I})\right]^{\rho}}\right\}^{(\rho-1)/\rho}
%\end{eqnarray}
%
%
%\begin{eqnarray}\label{app:priceHpriceL}
%\tilde{p}_h&=&\left\{\frac{\gamma\left[(1-g_h)x_l(e_{h,N}+e_{h,I})\right]^{\rho}+(1-\gamma)\left[(1-g_l)x_l(e_{l,N}+e_{l,I})\right]^{\rho}}{\gamma^{\rho/(\rho-1)}\left[(1-g_h)x_h(e_{h,N}+e_{h,I})\right]^{\rho}}\right\}^{(1-\rho)/\rho}\\ \tilde{p}_l&=&\left\{\frac{\gamma\left[(1-g_h)x_l(e_{h,N}+e_{h,I})\right]^{\rho}+(1-\gamma)\left[(1-g_l)x_l(e_{l,N}+e_{l,I})\right]^{\rho}}{(1-\gamma)^{\rho/(\rho-1)}\left[(1-g_l)x_l(e_{l,N}+e_{l,I})\right]^{\rho}}\right\}^{(1-\rho)/\rho}
%\end{eqnarray}
\begin{eqnarray}\label{app:priceHpriceL}
\tilde{p}_{HS}&=&\left\{\gamma^{1/(1-\rho)}+(1-\gamma)\gamma^{\rho/(1-\rho)}\left[\frac{(1-g_{LS})q_{LS}}{(1-g_{HS})q_{HS}}\right]^{\rho}\right\}^{(1-\rho)/\rho}, \text{ and}\\ \tilde{p}_{LS}&=&\left\{(1-\gamma)^{1/(1-\rho)}+\gamma(1-\gamma)^{\rho/(1-\rho)}\left[\frac{(1-g_{HS})q_{HS}}{(1-g_{LS})q_{LS}}\right]^{\rho}\right\}^{(1-\rho)/\rho}.
\end{eqnarray}

At aggregate level, by equating demand (left hand side) and supply (right hand side) of good $HS$, we get:
\begin{eqnarray}\label{app:demandsupplygoodh}
&&\notag \tilde{p}_{HS}\left(\frac{\tilde{p}_{HS}}{\gamma}\right)^{1/(\rho-1)} \left\{ (1-t) \left\{ b(\tilde{w}_{h,N}u_{h,N}+\tilde{w}_{h,I}u_{h,I}+\tilde{w}_{l,N}u_{l,N}+\tilde{w}_{l,I}u_{l,I})+\left[e_{h,N}+e_{h,I}+e_{l,N}+e_{l,I}+ \right. \right. \right.\\ \notag
 &+& \left. \left. \left. b(u_{h,N}+u_{h,I}+u_{l,N}+u_{l,I}) \right]\tilde{\tau} + \tilde{p}_{LS} q_{LS}+ \tilde{p}_{HS}q_{HS}\right\} + \left( \sigma_{h,N}+\sigma_{l,N}+\sigma_{h,I}+\sigma_{l,I} + \chi\right) \tilde{m} \right\} +\\
&+& g_{HS} \left( \frac{\tilde{p}_{HS}q_{HS}}{1-t_p}\right)= \frac{\tilde{p}_{HS}q_{HS}}{1-t_p},
\end{eqnarray}
where $g_{HS}$ is the proportion of good $HS$ which is demanded by the government and $\tilde{p}_{HS}q_{HS}/(1-t_p)$ is the total production of  good $HS$ before indirect taxes. Similarly,  equating demand (left hand side) and supply (right hand side) of good $LS$, we get:
\begin{eqnarray}\label{app:demandsupplygoodl}
&&\notag \tilde{p}_{LS}\left(\frac{\tilde{p}_{LS}}{1-\gamma}\right)^{1/(\rho-1)} \left\{ (1-t) \left\{ b(\tilde{w}_{h,N}u_{h,N}+\tilde{w}_{h,I}u_{h,I}+\tilde{w}_{l,N}u_{l,N}+\tilde{w}_{l,I}u_{l,I})+\left[e_{h,N}+e_{h,I}+e_{l,N}+e_{l,I}+ \right. \right. \right.\\ \notag
 &+& \left. \left. \left. b(u_{h,N}+u_{h,I}+u_{l,N}+u_{l,I}) \right]\tilde{\tau}  +\tilde{p}_{LS} q_{LS} + \tilde{p}_{HS}q_{HS}\right\} + \left( \sigma_{h,N}+\sigma_{l,N}+\sigma_{h,I}+\sigma_{l,I} + \chi\right) \tilde{m} \right\} + \\
&+& g_{LS} \left( \frac{\tilde{p}_{LS}q_{LS}}{1-t_p} \right)= \frac{\tilde{p}_{LS}q_{LS}}{1-t_p},
\end{eqnarray}
where $g_a, a \in \{HS,LS\}$, is the proportion of good $a$ which is demanded by the government and $\tilde{p}_aq_a/(1-t_p)$ is the total production of  good $a$ before indirect taxes.

Summing up side by side Equations (\ref{app:demandsupplygoodh}) and (\ref{app:demandsupplygoodl}) and using Equation \ref{app:phplequation1}, we get:
\begin{eqnarray}\label{app:demandsupplygoods}
&&\notag   b(\tilde{w}_{h,N}u_{h,N}+\tilde{w}_{h,I}u_{h,I}+\tilde{w}_{l,N}u_{l,N}+\tilde{w}_{l,I}u_{l,I})+\left[e_{h,N}+e_{h,I}+e_{l,N}+e_{l,I}+ \right. \\ \notag &+&  \left. b(u_{h,N}+u_{h,I}+u_{l,N}+u_{l,I}) \right]\tilde{\tau} +\left( \sigma_{h,N}+\sigma_{l,N}+\sigma_{h,I}+\sigma_{l,I} + \chi\right) \tilde{m}  + \frac{ g_{LS} \tilde{p}_{LS}q_{LS}+g_{HS}\tilde{p}_{HS}q_{HS}}{1-t_p} =\\ \notag
&=& \frac{t_p}{1-t_p}[\tilde{p}_{LS}q_{LS}+\tilde{p}_{HS}q_{HS}]+\\ \notag 
&+&t\left\{ b(\tilde{w}_{h,N}u_{h,N}+\tilde{w}_{h,I}u_{h,I}+\tilde{w}_{l,N}u_{l,N}+\tilde{w}_{l,I}u_{l,I})+\left[e_{h,N}+e_{h,I}+e_{l,N}+e_{l,I}+ \right. \right.\\ 
&+& \left. \left. b(u_{h,N}+u_{h,I}+u_{l,N}+u_{l,I}) \right]\tilde{\tau}  +\tilde{p}_{LS} q_{LS} + \tilde{p}_{HS}q_{HS}\right\}.
\end{eqnarray}
Equation (\ref{app:demandsupplygoods}) states that the sum of government expenditures on unemployment benefits, tax rebates, public transfers, and consumption of public goods is equal to  the government revenues in the form of direct and indirect taxes, i.e., $TGE=IT+NDT$. In other words, as in all closed economy macroeconomic models with zero private savings and investment, the government primary surplus is equal to zero in equilibrium (see also Appendix \ref{app:SMO}).

\newpage

\section{Equilibrium wages}\label{app:equilibriumwages}

To compute the equilibrium wages of native and non-native employees in each market we subtract Equation (\ref{eq:BellmanUnemployedNative}) from Equation (\ref{eq:BellmanEmployedNative}), and Equation (\ref{eq:BellmanUnemployedImmigrant}) from Equation (\ref{eq:BellmanEmployedNative}), we get:
\begin{eqnarray}
W^{e}_{i,N}-W^{u}_{i,N}&=&\dfrac{(1-b)\left(1-t\right)(\tilde{w}_{i,j}+\tilde{\tau} + \delta_{i,N}\phi \tilde{p}_i x_i F)}{r+\delta_{i,N}+ \kappa_{i,N} \theta_i q\left(\theta_i\right)} \text{ and} \label{we-wu_native1}\\
W^{e}_{i,I}-W^{u}_{i,I}&=&\dfrac{(1-b)\left(1-t\right)\left(\tilde{w}_{i,j}+\tilde{\tau} \right)-\lambda\left(W_{i,FC} - W^u_{i,I}\right)+\delta_{i,I}\phi \tilde{p}_i x_i F}{r+\delta_{i,I}+ \kappa_{i,I} \theta_i q\left(\theta_i\right)}.\label{we-wu_immigrant1}
\end{eqnarray}
By plugging Equation (\ref{we-wu_native1}) into the Nash bargaining Equation  (\ref{eq:NashProductNative}), we get an expression for the wages of native employees in each market, as a function of the parameters of the model:
\begin{eqnarray}\notag
\tilde{w}_{i,N}&=&\notag\beta_N (1-t)\left\{ \dfrac{r+\delta_{i,N}+ \kappa_{i,N} \theta_i q\left(\theta_i\right)  }{(1-t)[(r+\delta_{i,N})[1-b(1-\beta_N)]+\beta_N\kappa_{i,N} \theta_i q\left(\theta_i\right)]}\right\}\tilde{p}_i x_i+\\\notag &-&  \left\{ \dfrac{\left(1-t\right)(1-\beta_N)(r+\delta_{i,N})(1-b) }{(1-t)[(r+\delta_{i,N})[1-b(1-\beta_N)]+\beta_j\kappa_{i,N} \theta_i q\left(\theta_i\right)]} \right\} \tilde{\tau} +\\ &+&  \left\{\dfrac{ \beta_N  r[r+\delta_{i,N}+ \kappa_{i,N} \theta_i q\left(\theta_i\right)]    + \phi   (r+\delta_{i,N})(1-\beta_N)[r+ \kappa_{i,N} \theta_i q\left(\theta_i\right)] }{(1-t)[(r+\delta_{i,N})[1-b(1-\beta_N)]+\beta_N\kappa_{i,N} \theta_i q\left(\theta_i\right)]} \right\}\tilde{p}_i x_i F. \label{wageN}
\end{eqnarray}

Similarly, by plugging Equation (\ref{we-wu_immigrant1}) into the Nash bargaining Equation (\ref{eq:NashProductNative}) and substituting into Equation (\ref{eq:BellmanUnemployedImmigrant}), and reshuffling, we get an expression for the wages of non-native employees:
\begin{eqnarray}\nonumber
\tilde{w}_{i,I}\notag&=& \beta_I   (1-t)\left\{\dfrac{(r + \lambda) (r+\delta_{i,I})+ r \kappa_{i,I} \theta_i q\left(\theta_i\right)}{\left(1-t\right)\left\{ (r+\delta_{i,I})\left[(\lambda+r)-rb(1-\beta)\right]+ \beta_Ir\kappa_{i,I} \theta_i q\left(\theta_i\right)  \right\}}\right\}\tilde{p}_i x_i +\\&-&\notag \left\{ \dfrac{(1-\beta_I)(r+\delta_{i,I})\left(1-t\right)\left[(1-b)r+\lambda\right]}{\left(1-t\right)\left\{ (r+\delta_{i,I})\left[(\lambda+r)-rb(1-\beta)\right]+ \beta_Ir\kappa_{i,I} \theta_i q\left(\theta_i\right)  \right\}} \right\}  \tilde{\tau} +\\\notag&+& \left\{ \dfrac{\lambda r(1-\beta_I) (r+\delta_{i,I})}{\left(1-t\right)\left\{ (r+\delta_{i,I})\left[(\lambda+r)-rb(1-\beta)\right]+ \beta_Ir\kappa_{i,I} \theta_i q\left(\theta_i\right)  \right\}} \right\} W_{i,FC}  + \\&+&\notag\left\{\dfrac{\phi (r+\delta_{i,I})(1-\beta_I)(r + \lambda) \left( r+ \kappa_{i,I} \theta_i q\left(\theta_i\right)- \lambda\right)}{\left(1-t\right)\left\{ (r+\delta_{i,I})\left[(\lambda+r)-rb(1-\beta)\right]+ \beta_Ir\kappa_{i,I} \theta_i q\left(\theta_i\right)  \right\}}  \right.+\\&+&\notag\left. \dfrac{\beta_I   r\left[(r+\delta_{i,I})(r + \lambda)+ r\kappa_{i,I} \theta_i q\left(\theta_i\right)\right]}{\left(1-t\right)\left\{ (r+\delta_{i,I})\left[(\lambda+r)-rb(1-\beta)\right]+ \beta_Ir\kappa_{i,I} \theta_i q\left(\theta_i\right)  \right\}}\right\}\tilde{p}_i x_i F+\\
&-& \left\{ \dfrac{\lambda(1-\beta_I)(r+\delta_{i,I})  }{\left(1-t\right)\left\{ (r+\delta_{i,I})\left[(\lambda+r)-rb(1-\beta)\right]+ \beta_Ir\kappa_{i,I} \theta_i q\left(\theta_i\right)  \right\}} \right\} \left(\iota\nu +\tilde{m} \right).\label{wageI}
\end{eqnarray}

We can rewrite Equation (\ref{wageN}) as:
\begin{eqnarray}\notag
\tilde{w}_{i,N}&=&\notag\underbrace{A_{i,N}(\theta_i)}_{>0}\tilde{p}_i x_i- \underbrace{B_{i,N}(\theta_i)}_{>0} \tilde{\tau} +  \underbrace{C_{i,N}(\theta_i)}_{>0}\tilde{p}_i x_i F, \label{wageNshort2}
\end{eqnarray}
where
\begin{eqnarray}
A_{i,N}(\theta_i)= \dfrac{\beta_N[r+\delta_{i,N}+ \kappa_{i,N} \theta_i q\left(\theta_i\right)]  }{[(r+\delta_{i,N})[1-b(1-\beta_N)]+\beta_N\kappa_{i,N} \theta_i q\left(\theta_i\right)]},
\end{eqnarray}
\begin{eqnarray}
B_{i,N}(\theta_i)=\dfrac{(1-\beta_N)(r+\delta_{i,N})(1-b) }{[(r+\delta_{i,N})[1-b(1-\beta_N)]+\beta_N\kappa_{i,N} \theta_i q\left(\theta_i\right)]},
\end{eqnarray}
\begin{eqnarray}
C_{i,N}(\theta_i)=\dfrac{ \beta_N  r[r+\delta_{i,N}+ \kappa_{i,N} \theta_i q\left(\theta_i\right)]    + \phi   (r+\delta_{i,N})(1-\beta_N)[r+ \kappa_{i,N} \theta_i q\left(\theta_i\right)] }{(1-t)[(r+\delta_{i,N})[1-b(1-\beta_N)]+\beta_N\kappa_{i,N} \theta_i q\left(\theta_i\right)]}.
\end{eqnarray}

Similarly, we can rewrite the equation for the wages of non-native employees (Equation (\ref{wageI})) as:
\begin{eqnarray}
\tilde{w}_{i,I}&=& \underbrace{D_{i,I}(\theta_i)}_{>0}\tilde{p}_i x_i - \underbrace{E_{i,I}(\theta_i)}_{>0}\tilde{\tau} +
\underbrace{G_{i,I}(\theta_i)}_{>0}\tilde{p}_i x_i F + \underbrace{K_{i,I}(\theta_i)}_{>0} W_{i,FC}  - \underbrace{H_{i,I}(\theta_i)}_{>0} \left(\iota\nu +\tilde{m} \right),\label{wageIshort2}
\end{eqnarray}
where
\begin{eqnarray}\nonumber
D_{i,I}(\theta_i)=  \dfrac{\beta_I \left[(r + \lambda) (r+\delta_{i,I})+ r \kappa_{i,I} \theta_i q\left(\theta_i\right)\right]}{\left\{ (r+\delta_{i,I})\left[(\lambda+r)-rb(1-\beta)\right]+ \beta_Ir\kappa_{i,I} \theta_i q\left(\theta_i\right)  \right\}}.
\end{eqnarray}
\begin{eqnarray}\nonumber
E_{i,I}(\theta_i)=\dfrac{(1-\beta_I)(r+\delta_{i,I})\left[(1-b)r+\lambda\right]}{ (r+\delta_{i,I})\left[(\lambda+r)-rb(1-\beta)\right]+ \beta_Ir\kappa_{i,I} \theta_i q\left(\theta_i\right) }. 
\end{eqnarray}
\begin{eqnarray}\nonumber
K_{i,I}(\theta_i)= \dfrac{\lambda r(1-\beta_I) (r+\delta_{i,I})}{\left(1-t\right)\left\{ (r+\delta_{i,I})\left[(\lambda+r)-rb(1-\beta)\right]+ \beta_Ir\kappa_{i,I} \theta_i q\left(\theta_i\right)  \right\}}.
\end{eqnarray}
\begin{eqnarray}\nonumber
G_{i,I}(\theta_i)&=&\dfrac{\phi (r+\delta_{i,I})(1-\beta_I)(r + \lambda) \left( r+ \kappa_{i,I} \theta_i q\left(\theta_i\right)- \lambda\right)}{\left(1-t\right)\left\{ (r+\delta_{i,I})\left[(\lambda+r)-rb(1-\beta)\right]+ \beta_Ir\kappa_{i,I} \theta_i q\left(\theta_i\right)  \right\}}  +\\&+&\notag \dfrac{\beta_I   r\left[(r+\delta_{i,I})(r + \lambda)+ r\kappa_{i,I} \theta_i q\left(\theta_i\right)\right]}{\left(1-t\right)\left\{ (r+\delta_{i,I})\left[(\lambda+r)-rb(1-\beta)\right]+ \beta_Ir\kappa_{i,I} \theta_i q\left(\theta_i\right)  \right\}}.
\end{eqnarray}
\begin{eqnarray}\nonumber
H_{i,I}(\theta_i)= \dfrac{\lambda(1-\beta_I)(r+\delta_{i,I})  }{\left(1-t\right)\left\{ (r+\delta_{i,I})\left[(\lambda+r)-rb(1-\beta)\right]+ \beta_Ir\kappa_{i,I} \theta_i q\left(\theta_i\right)  \right\}}.
\end{eqnarray}

\newpage

\section{Calculation of job finding and job exit rates \label{app:jfrjer}}

To compute the probability for a worker to find a job as well as the probability for a worker to lose her job we follow \cite{SHIMER2012127}. Specifically, to calculate the job finding probability for the unemployed $Q_t\in [0,1]$ and the exit probability for the employed  $\Delta_t\in [0,1]$ in Italy in the period 2004-2014 we use publicly available data from the Italian Labour Force Survey (LFS).  We do not consider transitions in and out of the labour force, but we focus on the employees' transitions between employment and unemployment. We also assume that all the unemployed find a job with probability $Q_t $ and all the employed lose a job with probability $\Delta_t$ during period $t$, ignoring any heterogeneity or duration dependence that makes some unemployed more likely to find and some employed less likely to lose a job within the period. 

For $t \in \{0, 1, 2, . . .\}$, we refer to the interval $[t,t+1]$ as period $t$.  We assume that during period $t$, all unemployed  find a job according to a Poisson process with arrival rate $q_t \equiv -log(1-Q_t ) >0$ and all employed  lose their job according to a Poisson process with arrival rate $\delta_t \equiv -log(1-\Delta_t ) >0$. Hence, $q_t$ and $\delta_t$ represent the job finding and employment exit rates and $Q_t$ and $\Delta_t$ are the corresponding probabilities.
By fixing $t \in \{0, 1, 2, . . .\}$ and letting $\tau \in [0, 1]$ be the time elapsed since the last measurement date, we can define $e_{t+\tau}$ as the number of employed  at time $t +\tau$, $u_{t+\tau}$ as the number of unemployed  at time $t +\tau$, and $u^s_{t}(\tau)$ denote "short term unemployment", that is employees who are unemployed at time $t +\tau$, but were employed at some time $t' \in [t, t +\tau]$. Note that $u^s_{t}(0)=0$ for all $t$. 
Therefore, the law of motion for unemployment at time $t +\tau$ reads:
\begin{equation}\label{Eq0}
\dot{u}_{t+\tau} =e_{t+\tau} \delta_t-u_{t+\tau} q_t.
\end{equation}
The number of unemployed  at date $[t +1]$ is then equal to the number of unemployed  at date $t$ who do not find a job (a fraction $1-Q_t = e^{-qt}$ ) plus the $u^s_{t+1}$ short-term unemployed, i.e., those who are unemployed at date $[t +1]$ but held a job at some point during period $t$:
\begin{equation}\label{Eq1}
u_{t + 1} = (1-Q_t)u_t + u_{t+ 1}^s.
\end{equation}
By inverting Equation \ref{Eq1}, we find an expression for the job finding probability as a function of unemployment and short term unemployment:
\begin{equation}
Q_t = 1 - \frac{u_{t+ 1}-u^s_{t+1}}{u_t}.
\end{equation}

As in \cite[p.130]{shimer_cyclical_2005}, an implicit equation for the employment exit rate can be obtained by solving Equation \ref{Eq0} :
\begin{equation}
u_{t+ 1}=\frac{\left[1-\exp(-q_t-\delta_t)\right]\delta_t l_t}{q_t +\delta_t} + \exp(-q_t-\delta_t)u_t, 
\end{equation}
where $l_t \equiv u_t + e_t$ is the size of the labour force during period $t$, which we assume to be constant since  entries or exits from the labour force are not allowed.

\subsection{Robustness check for the calculation of job finding and job exit rates}

\begin{figure}[!htbp]
	\caption{Correlation between the ratio of the estimated job finding rates and job exit rates and the ratio between employed and unemployed by worker type in Italy in the period 2004-2017.}
	\begin{subfigure}[h]{0.4\textwidth}
		\centering
		\includegraphics[width=\linewidth]{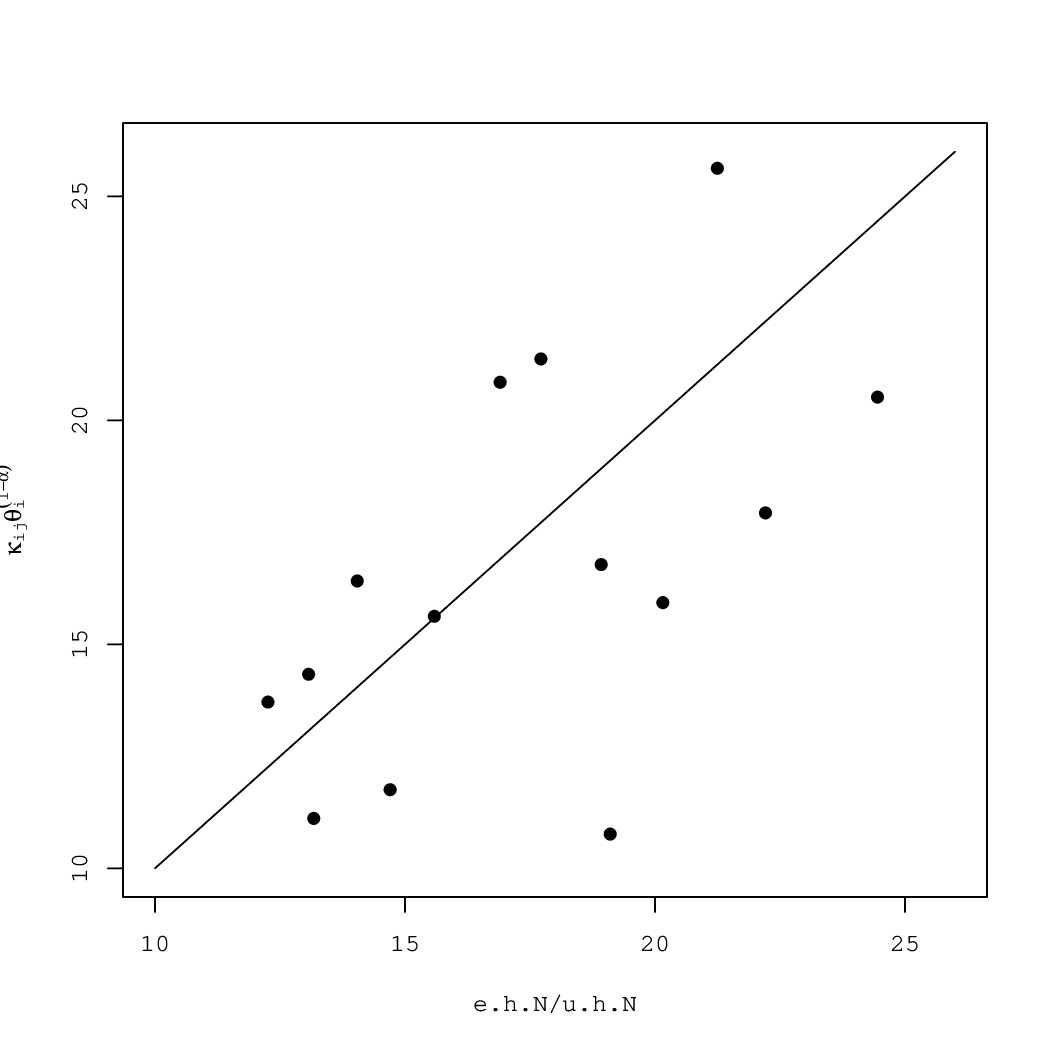}
		\caption{High-skilled natives.}
		\label{fig:checknativehighskilledemployeesrates}
	\end{subfigure}        \hfill
	\begin{subfigure}[h]{0.4\textwidth}
		\centering
		\includegraphics[width=\linewidth]{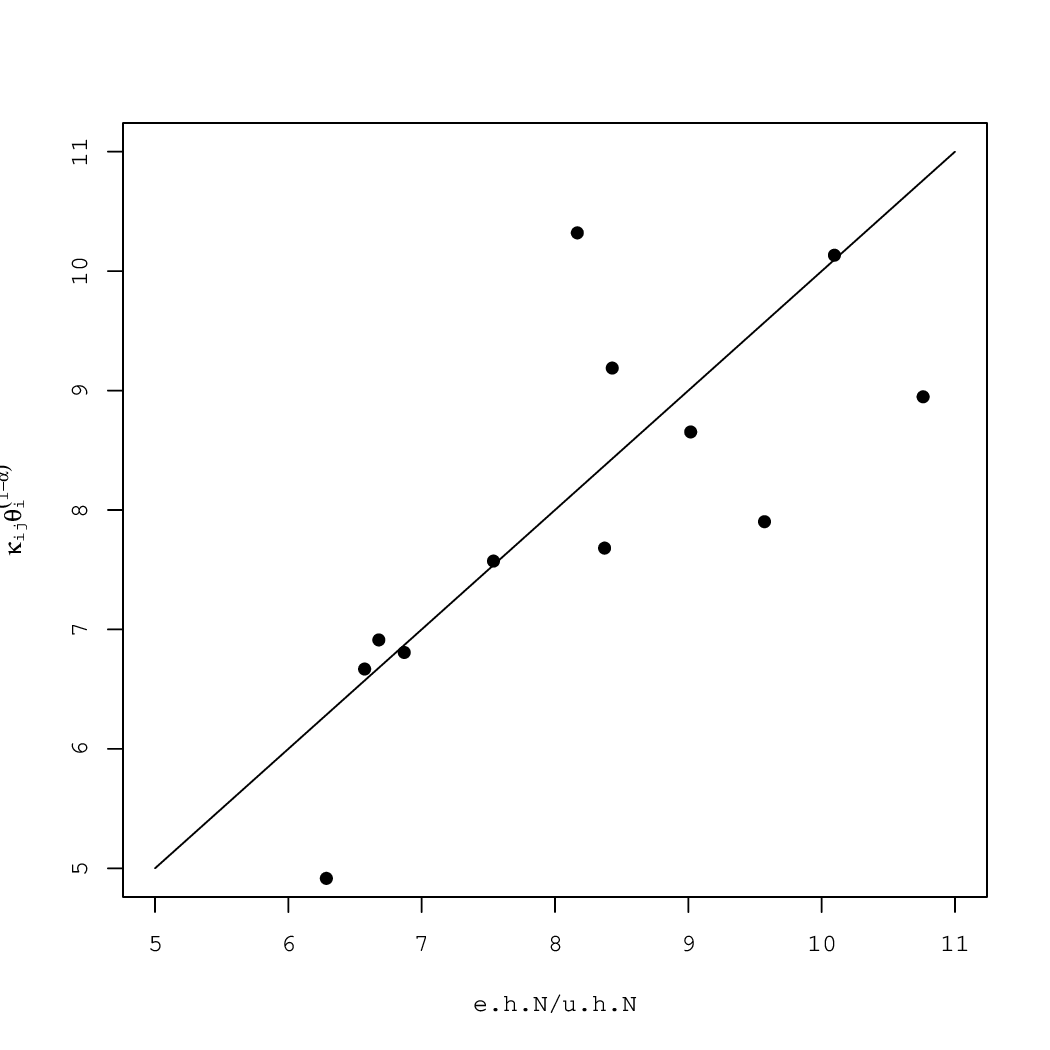}
		\caption{High-skilled non-natives.}
		\label{fig:checkimmigranthighskilledemployeesrates}
	\end{subfigure}
	\begin{subfigure}[h]{0.4\textwidth}
		\centering
		\includegraphics[width=\linewidth]{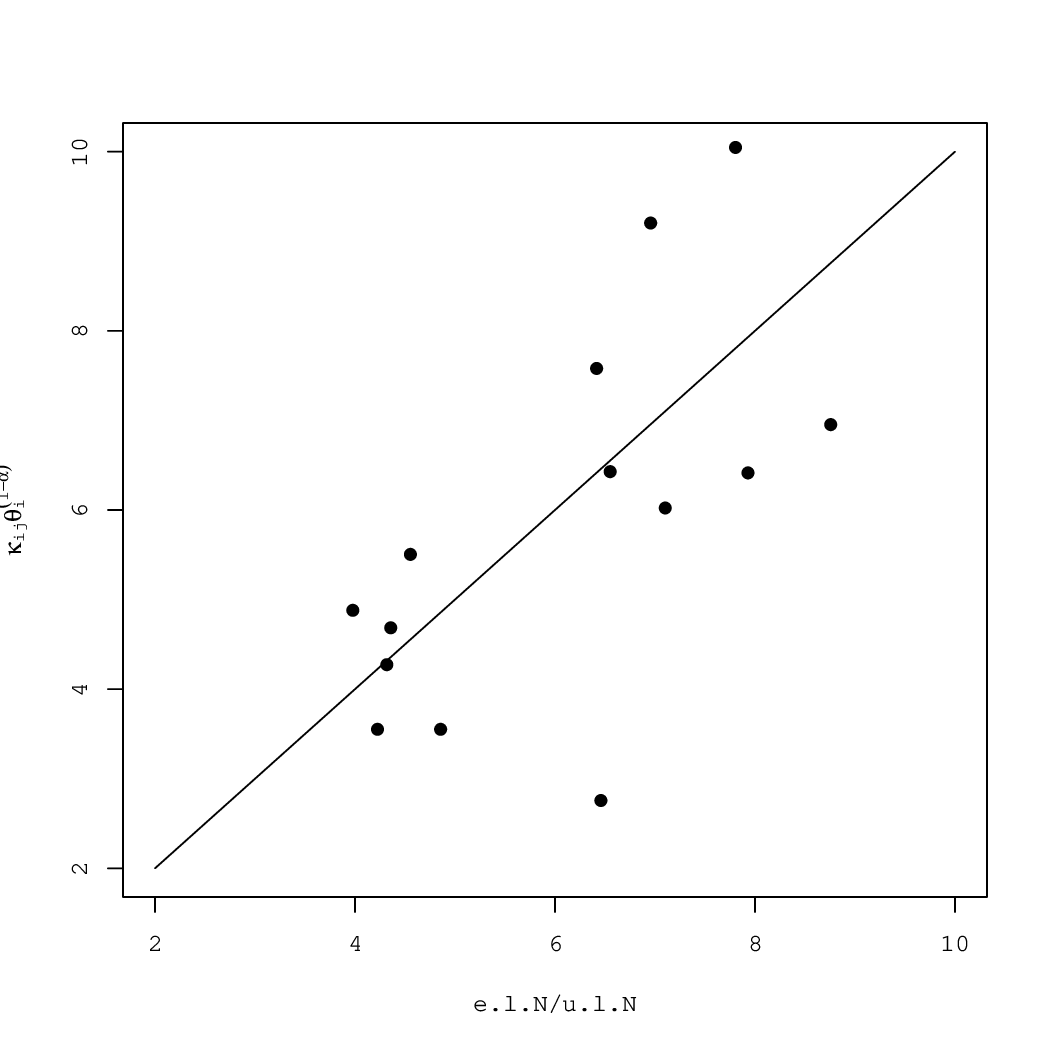}
		\caption{Low-skilled natives.}
	\end{subfigure}\hfill
	\begin{subfigure}[h]{0.4\textwidth}
		\centering
		\includegraphics[width=\linewidth]{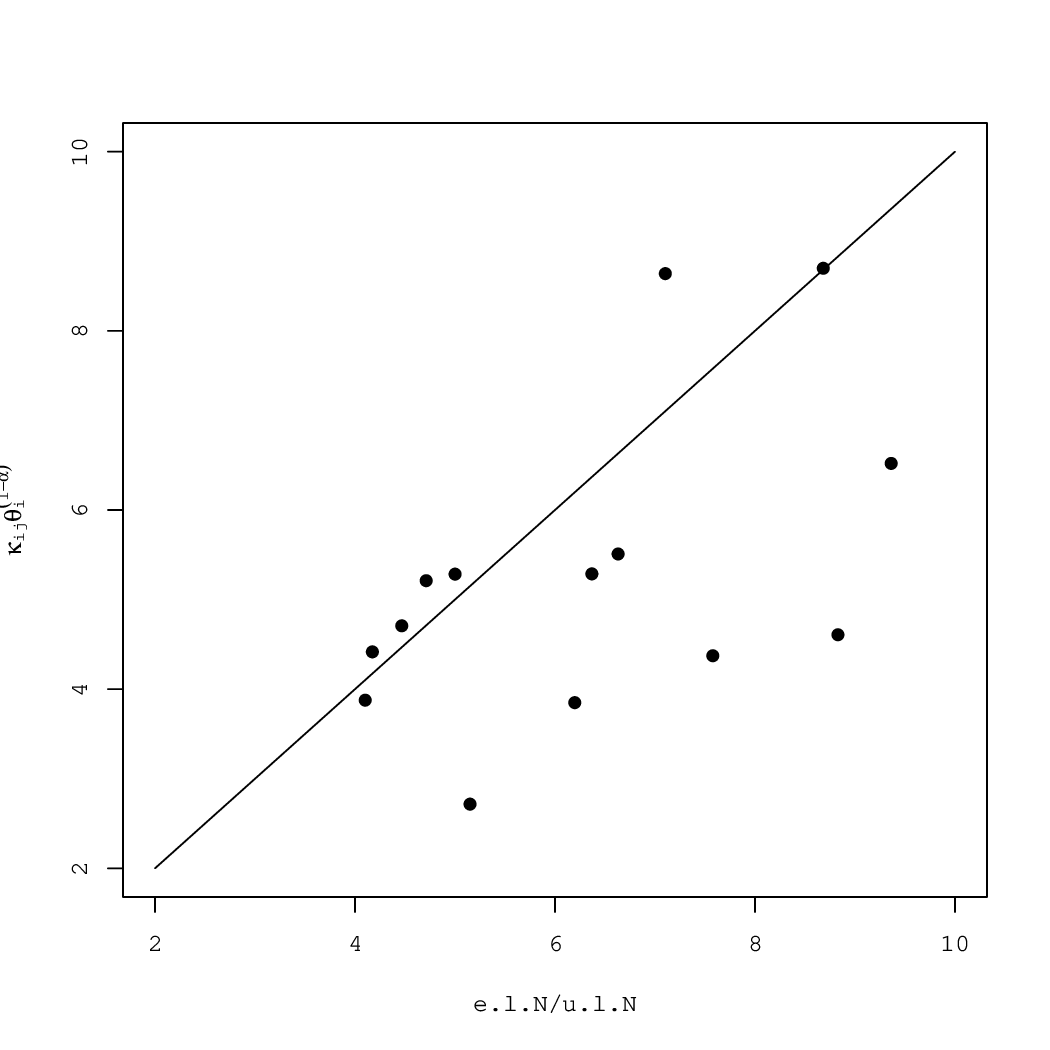}
		\caption{Low-skilled non-natives.}
		\label{fig:checknativelowskilledemployeesrates}
	\end{subfigure}\\\\[-1.8ex]
\caption*{\scriptsize{\textit{Note}: On the vertical axis we report the  ratio of the estimated job finding rates and job exit rates, while on the horizontal axis we report the ratio of employed and unemployed workers.}}
\end{figure}

From Equations (\ref{eq:equilibriumEmploymentNatives})-(\ref{eq:equilibriumUnemploymentImmigrants}), we derive the following equality:
\begin{equation}
\dfrac{e_{i,j}}{u_{i,j}} = \dfrac{\kappa_{i,j} \theta_i^{1-\alpha} }{\delta_{i,j}},
\label{testJobFindingandExitRates}
\end{equation}
where $i \in \left\{h,l\right\}$ and $j \in \left\{N,I\right\}$.
The right hand side of Equation (\ref{testJobFindingandExitRates}) reports the ratio between the job finding rate and the job exit rate per each worker type, while the left hand side is the ratio between  employed and  unemployed. If our estimates of the job finding rates and the job exit rates for employees by skill level and country of origin were approximatively correct, then their ratio should be equal to the ratio of employed and unemployed employees by skill level and country of origin. Hence, we regress the computed ratio of job finding rate and job exit rate per worker type on the ratio of employed and unemployed  by skill level and country of origin for Italy for the period 2004-2017 (Figures \ref{fig:checknativehighskilledemployeesrates}-\ref{fig:checknativelowskilledemployeesrates}). The coefficients are very close to 1, confirming the validity of our calculations (Table \ref{jfrcheck}).

\begin{table}[!htbp]
	\centering
	\caption{Check for the estimate of the job finding and exit rate for different skills and country of origin in Italy in the period 2004-2017.}
	\label{jfrcheck}
	\begin{tabular}{rrrrr}
		\hline \hline
		& Estimate & Std. Error & t value & Pr($>$$|$t$|$) \\ 
		\hline
		$e_{h,N}/u_{h,N}$ & 1.0302 & 0.0739 & 13.93 & 0.0000 \\ 
		$e_{l,N}/u_{l,N}$ & 0.9753 & 0.0857 & 11.39 & 0.0000 \\ 
		$e_{h,I}/u_{h,I}$ & 1.0949 & 0.0862 & 12.69 & 0.0000 \\ 
		$e_{l,I}/u_{l,I}$ & 1.1940 & 0.1246 & 9.58 & 0.0000 \\ 
		\hline \hline 
		\multicolumn{5}{l}{\footnotesize{\textit{Source}: our calculations using ELFS data.}}
		
	\end{tabular}
\end{table}

\clearpage

\section{The classification of employees by skill level \label{workerclassification}}

To classify employees according to their skill level is a hard task. This is because it is not clear what is a good proxy to be used to capture the skills of an individual. Possible alternatives which have been used in the literature include  educational attainment \citep{Altonji1991, dustmann2005}, occupation \citep{card2001}, or experience and education \citep{borjas2003}. All of those have pros and cons. The benefit of using the education level is that it is in general available for all employees; however, it is a rather imprecise measure of the individual skills. One of the main problems is the issue of mismatch, particularly over-education, as often employees are hired to perform a job which requires skills associated with an education level which is lower compared to the one of the individual. This phenomenon is specifically relevant for immigrants, as investigated by \cite{dustmann2012} and \cite{Eckstein2004}, who show that immigrants downgrade considerably upon arrival and therefore the allocation of immigrants according to their measured skills, such as education, would place them at different locations across the native wage distribution than where we actually find them. Alternatively, we could use as a proxy the individual's occupation, which is still an imperfect measure of the skill level, but is probably more accurate than education.

The International Labor Organization (ILO) maps the the International Standard Classification of Occupations (ISCO) in skill levels (Table \ref{OccupationtoSkill}). While the definition of \textit{skill} refers to the ability to carry out the tasks and duties of a given job \citep{ILO2012}, the definition of \textit{skill level} relates to a function of the complexity and range of tasks and duties to be performed in an occupation. 
The skill level is measured operationally by considering one or more of the following elements:
\begin{itemize}
	\item the nature of the work performed in an occupation in relation to the characteristic tasks and duties defined for each ISCO-88 skill level;
	\item the level of formal education defined in terms of the International Standard Classification of Education (ISCED-97) required for competent performance of the tasks and duties involved; and
	\item the amount of informal on-the-job training and /or previous experience in a related occupation required for competent performance of these tasks and duties.
\end{itemize}
In addition, ILO provides a mapping between skill levels and education levels, following the International Standard Classification of Education ISCED-97, as developed by UNESCO (Table \ref{OccupationtoEdu}).

Occupations at Skill Level 1 typically require the performance of simple and routine physical or manual tasks. They may require the use of hand held tools, such as shovels, or of simple electrical equipment, such as vacuum cleaners. They involve tasks such as cleaning; digging; lifting and carrying materials by hand; sorting, storing or assembling goods by hand (sometimes in the context of mechanised operations); operating non-motorised vehicles; and picking fruit and vegetables. Many occupations at Skill Level 1 may require physical strength and/or endurance. For some jobs basic skills in literacy and numeracy may be required. If required, these skills would not be a major part of the job. For competent performance in some occupations at Skill Level 1, completion of primary education or the first stage of basic education (ISCED Level 1) may be required.  Occupations classified at Skill Level 1 include office cleaners, freight handlers, garden labourers and kitchen assistants.
\begin{table}[!htbp]
	\footnotesize
	\centering
	\caption{Mapping of ISCO-08 major groups to skill levels.} 
	\label{OccupationtoSkill}
	\global\let\restorecr=\\
	\begin{tabular}{lc} 
		%\toprule 
		\hline \hline
		\global\let\oldcr=\\
		\global\def\crsng{\global\let\\=\crdbl\oldcr}
		\global\def\crdbl{\global\let\\=\crsng\oldcr[2.5pt]}
		\global\let\\=\crdbl
		{ISCO-08 major groups}&{Skill Level}\\
		\hline
		1 - Managers, senior officials and      legislators &3 + 4\\
		\hline
		2 - Professionals&4\\
		\hline
		3 - Technicians and associate  professionals&3\\
		\hline
		4 - Clerks&\\
		5 - Service and sales employees&\\
		6 - Skilled agricultural and fishery employees&\\
		7 - Craft and related trades employees&\\
		8 - Plant and machine operators, and assemblers&2\\
		\hline
		9 - Elementary occupations&1\\
		\hline
		0 - Military occupations&1 + 4\\
		%\bottomrule
		\hline \hline
	\end{tabular}
	\caption*{\footnotesize{\textit{Source}: International Labor Organization (ILO).}} 
\end{table}

Occupations at Skill Level 2 typically involve the performance of tasks such as operating machinery and electronic equipment; driving vehicles; maintenance and repair of electrical and mechanical equipment; and manipulation, ordering and storage of information.
For almost all occupations at Skill Level 2 the ability to read information such as safety instructions, to make written records of work completed, and to accurately perform simple arithmetical calculations is essential. Many occupations at this skill level require relatively advanced literacy and numeracy skills and good interpersonal communication skills. In some occupations these skills are required for a major part of the work. Many occupations at this skill level require a high level of manual dexterity.  Occupations classified at Skill Level 2 include butchers, bus drivers, secretaries, accounts clerks, sewing machinists, dressmakers, shop sales assistants, police officers, hairdressers, building electricians and motor vehicle mechanics.

Occupations at Skill Level 3 typically involve the performance of complex technical and practical tasks which require an extensive body of factual, technical and procedural knowledge in a specialised field. Occupations at this skill level generally require a high level of literacy and numeracy and well developed interpersonal communication skills. These skills may include the ability to understand complex written material, prepare factual reports and communicate with people who are distressed. The knowledge and skills required at Skill Level 3 are usually obtained as the result of study at a higher educational institution following completion of secondary education for a period of 1-3 years (ISCED Level 5b). In some cases extensive relevant work experience and prolonged on the job training may substitute for the formal education. Occupations classified at Skill Level 3 include shop managers, medical laboratory technicians, legal secretaries, commercial sales representatives, computer support technicians, and broadcasting and recording technicians.

\begin{table}[!htbp]
	\footnotesize
	\centering
	\caption{Mapping of ISCO-08 major groups to education level (ISCED-97) groups.} 
	\label{OccupationtoEdu}
	\global\let\restorecr=\\
	\begin{tabular}{cl} 
		%\toprule
		\hline \hline
		\global\let\oldcr=\\
		\global\def\crsng{\global\let\\=\crdbl\oldcr}
		\global\def\crdbl{\global\let\\=\crsng\oldcr[2.5pt]}
		\global\let\\=\crdbl
		{ISCO-08 }& { ISCED-97 groups}\\
		{Skill Level}& {Education Level}\\
		\hline
		4&
		6 - Second stage of tertiary education (leading to an advanced research qualification)\\
		&5a - First stage of tertiary education, 1st degree (medium duration)\\
		\hline
		3&
		5b - First stage of tertiary education (short or medium duration)\\
		\hline
		2&
		4 - Post-secondary, non-tertiary education \\
		&3 - Upper secondary level of education\\
		&2 - Lower secondary level of education\\
		\hline
		1&
		1 - Primary level of education\\
		%\bottomrule
		\hline \hline
	\end{tabular} 
	\global\let\\=\restorecr
	%\label{LPMOnlyTemp} 
	\caption*{\footnotesize{\textit{Source}: International Labor Organization (ILO).}}
\end{table} 

Occupations at Skill Level 4 typically involve the performance of tasks which require complex problem solving and decision making based on an extensive body of theoretical and factual knowledge in a specialised field. The tasks performed typically include analysis and research to extend the body of human knowledge in a particular field, diagnosis and treatment of disease, imparting knowledge to others, design of structures or machinery and of processes for construction and production.
Occupations at this skill level generally require extended levels of literacy and numeracy, sometimes at a very high level, and excellent interpersonal communication skills. These skills generally include the ability to understand complex written material and communicate complex ideas in media such as books, reports and oral presentations.
The knowledge and skills required at Skill Level 4 are usually obtained as the result of study at a higher educational institution for a period of 3-6 years leading to the award of a first degree or higher qualification (ISCED Level 5a or higher). In some cases experience and on the job training may substitute for the formal education. In many cases appropriate formal qualifications are an essential requirement for entry to the occupation.

We classify individuals in two categories, high-skilled and low-skilled. To do so, we follow the ILO classification, and we refer to high-skilled employees (with skill levels 3 or 4) as those individuals who work as managers, professionals or technicians. Moreover, we refer to low-skilled employees  (with skill levels 1 or 2) as those individuals  who work as clerks, sales employees, craft employees, plant and machine operators and in elementary occupations. 
For those for whom, we do not observe the occupation as they are currently unemployed, we use the occupation in their last job. For those for whom no information is available, either because they are stepping for the first time in the labour market or because they have not worked before in Italy or because they did not report the information, we use the education level. The majority of unemployed without information on previous occupation are young and their average age is below 40, both among natives and non-natives. In order to correct for the issue of mismatch, we look at the probability for high educated employees (with a tertiary level of education) and low educated employees (primary or secondary levels) under the age of 40 to work in a high-skilled occupation (with skill levels 3 or 4, as classified by ILO) versus a low skill occupation (with skill levels 1 or 2, as classified by ILO) for both natives and immigrants.  We then randomly assign the unemployed into high-skilled and low-skilled according to their education level, taking into account the probabilities of falling in the different categories reported in Table \ref{Doubletableeduocc}.

\begin{table}[!htbp]
	\centering
	\caption{Employed employees  by country of origin, occupation and education levels In Italy in the period 2008-2017.} 
	\label{Doubletableeduocc}
	\footnotesize
	\begin{tabular}{p{3cm}l cc|cc}
	%\toprule
	\hline \hline
		&&\multicolumn{4}{c}{ {ISCED-97 Education level}}\\
		\cline{3-6}
		&&\multicolumn{2}{c}{Non-natives}&\multicolumn{2}{c}{Natives}\\
		&&\multicolumn{1}{c}{Low }&\multicolumn{1}{c}{High }&\multicolumn{1}{c}{Low }&\multicolumn{1}{c}{High }\\
		\hline \\[-1.8ex]
		{Occupation level} &Low&0.825&0.335&0.685&0.1127\\
		&High&0.175&0.665&0.315&0.8873\\
		%\bottomrule
		\hline \hline
		\multicolumn{6}{l}{\footnotesize{\textit{Note}: These statistics refer to employees below the age of 40, as this    }}\\\multicolumn{6}{l}{\footnotesize{is the category which is most exposed to the issue of downgrading.}}\\\multicolumn{6}{l}{\footnotesize{ \textit{Source}: our calculations based on the Italian Labour Force }}\\\multicolumn{6}{l}{\footnotesize{Statistics (RCFL) data.}}
	\end{tabular}
\end{table}

\newpage 

\section{The Italian labour market \label{app:ItalianLabourMarket}}

\subsection{The size and type of immigration in Italy \label{sec:ilm}}

The size of non-natives in Italy becoming larger and larger, reaching approximately  15\% of the total workforce in 2017 (from approximately 6\% in 2004). Low-skilled employees are the strong majority (about 90\%) of the non-native workforce in 2017, and there exists a strong heterogeneity in wages, job creation and job exit rates among high-skilled, low-skilled, native and non-native employees.

\begin{figure}[!htbp]
	\centering
	\caption{Stocks of native and non-native population and workforce.}
	\label{stocksandgrowthratepopulation}
	\begin{subfigure}[h]{0.45\textwidth}
		\centering
		\includegraphics[width=0.8\linewidth]{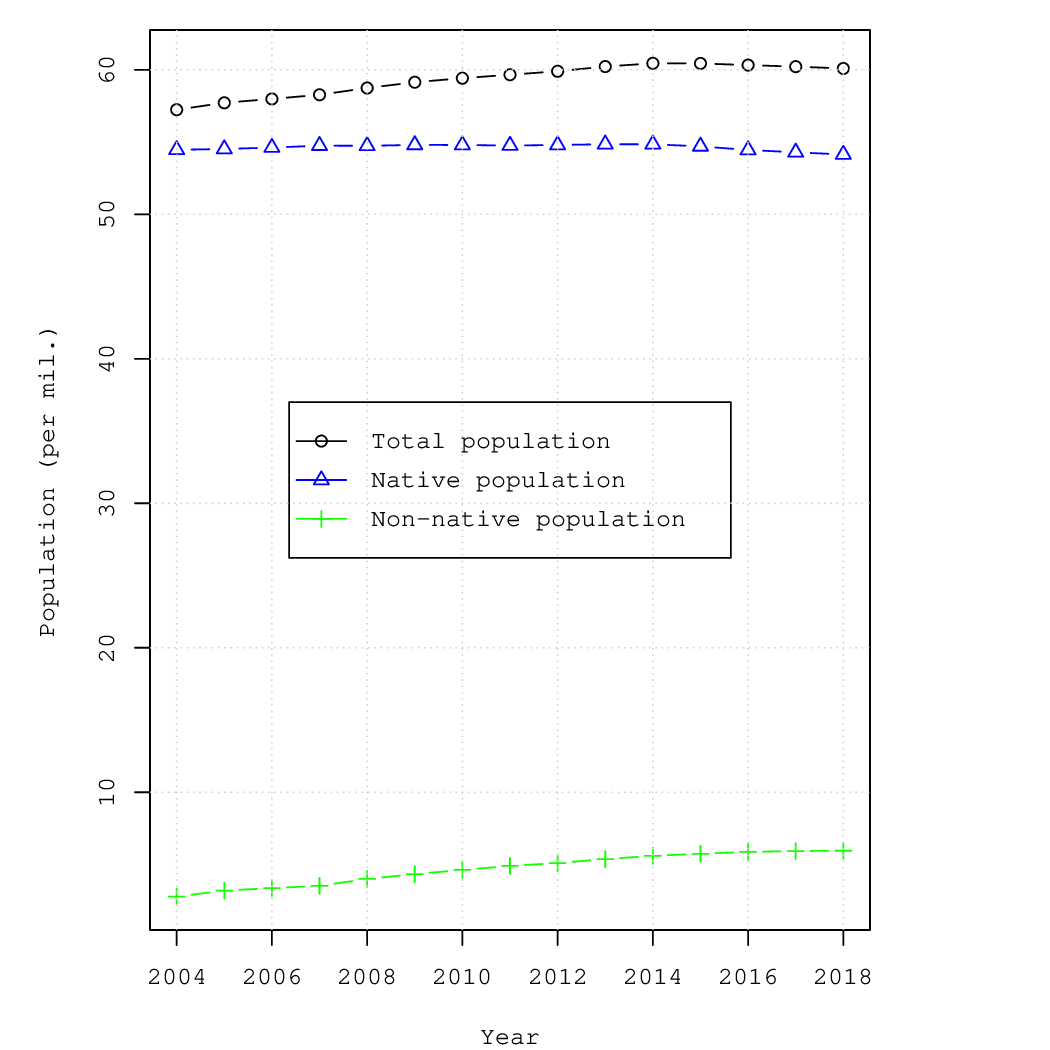}
		\caption{Stock of native and non-native population.}
		\label{fig:nativepopulation}
	\end{subfigure}
\begin{subfigure}[h]{0.45\textwidth}
	\centering
	\includegraphics[width=0.8\linewidth]{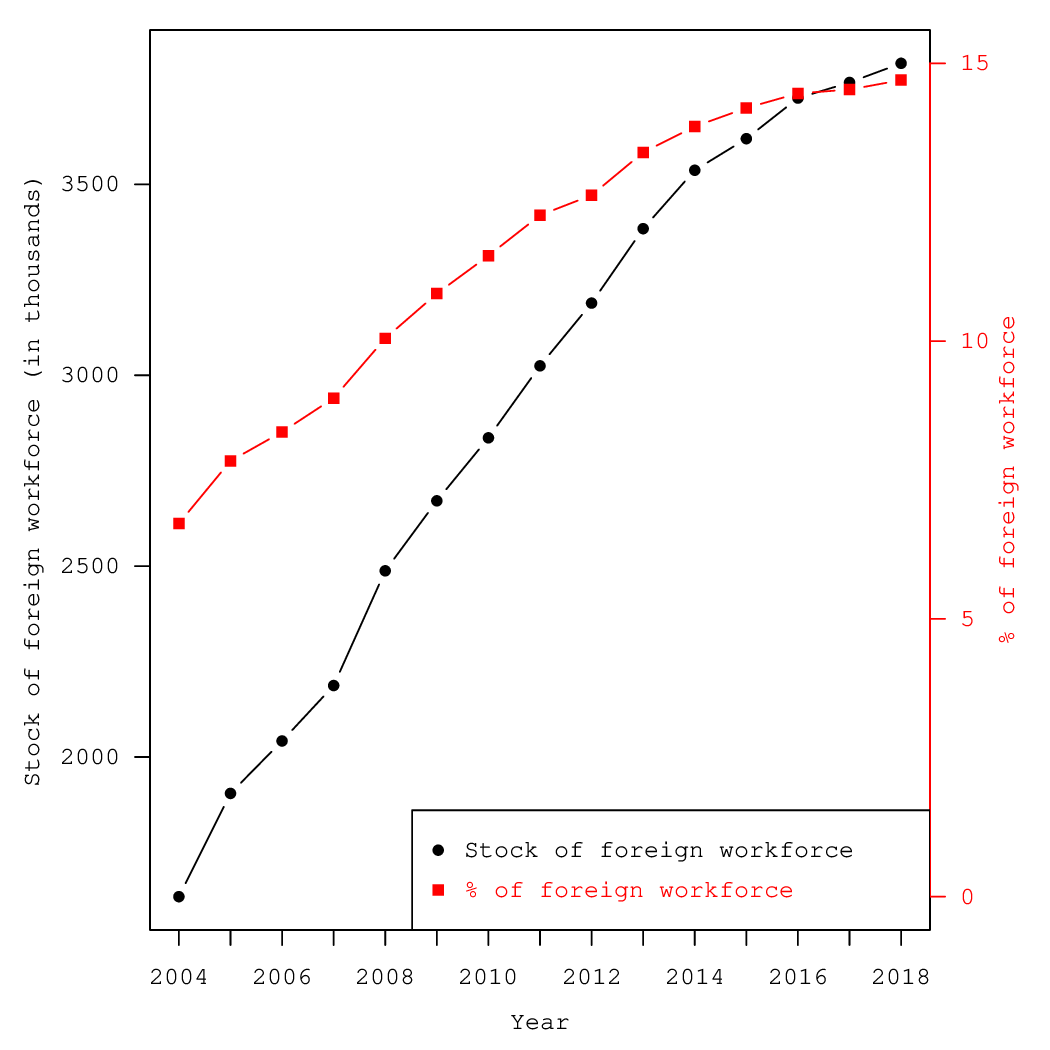}
		\caption{Stock and share of foreign-born workforce. }
	\label{fig:foreignpopulationstockandshare}
	\end{subfigure}
\vspace{0.3cm}
	\caption*{\scriptsize{\textit{Source}: Own calculations based on the Italian Labour Force Statistics (RCFL) data.}}
\end{figure}

The increase in the Italian population in the past two decades is due exclusively to the increase in the number of non-natives present on the territory (Figure \ref{stocksandgrowthratepopulation}). The stock of non-natives in the workforce in Italy increased from less than 2 millions in 2004 to approximately 4 millions in 2018, corresponding to a surge in the share in the labour force from approximately 6\% in 2004 to 15\% in 2018.

Table \ref{InflowsOutflows} shows that there were  approximately 270.000 new entries in 2005, but this number almost doubled in 2007, when the pick of the inflow was reached. Since 2008, the total number of entries slowed down, and went back to an average of 270,000 in the period 2013-2018. The total number of inflows, which shows a pick in 2007, was almost entirely ascribable to the entry of Romania and Bulgaria in the EU in 2007. %Specifically, the number of Romanians resident in Italy jumped from approximately 38.000 in 2006 to more than 261.000 in 2007, while the number of Bulgarians increased from approximately 2.000 in 2006 to more than 12.000 in 2007. The number of non-EU entries (from China, India, Middle East and Africa) has been oscillating for the whole period considered.
The total number of outflows as a percentage of unemployed (both  EU and non-EU nationals) has been quite stable around 10\% over time, with a single pick of $14\%$ in 2008.\footnote{For natives inflow and outflow rates, please refer to Table \ref{inflowoutflownatives} in the Appendix \ref{app:shareTemporaryJobs}.}

\begin{table}[!htbp]
	\centering
	\caption{Immigration inflows and outflows (in thousands). } 
	\label{InflowsOutflows}
	\scriptsize
	\begin{tabular}{@{\extracolsep{-1pt}}lccccccccccccccc} 
%\toprule 
\hline \hline 
		%		\global\let\oldcr=\\
		%		\global\def\crsng{\global\let\\=\crdbl\oldcr}
		%		\global\def\crdbl{\global\let\\=\crsng\oldcr[2pt]}
		%		\global\let\\=\crdbl
		\\[-1.8ex]
		&2004&2005&2006&2007&2008&2009&2010&2011&2012&2013&2014&2015&2016&2017&2018\\
		\midrule
		\textbf{Inflow}\\
		\hline
		\\[-1.8ex]
		Total  &	373.1&	267.6&	242.1&	490.4&	462.3 &	392.5&	419.6&	354.3&	321.3&	279.0&	248.4&	250.0&	262.9&	301.1&	285.5\\
%		Inflow - Total  	&394.8 &282.8	&254.6&	515.2&	496.5	&406.7&	424.5	&354.3&	321.3&	279.0	&248.4&250.5&262.9&301.1&\\
	%	Inflow - Non EU &	&205.1&181.2	&187.5	&277.3	&265.6&302.3&	234.6&	211.9&	197.6&	176.7&&&&\\
		 Non EU &178.4&	128.9&	115.8&	112.7&	168.3&	179.0&	205.6&	175.2&	168.6&	155.8&	145.2&	152.8&	166.3&	204.4&	190.0\\
		 Romania	&	63.4&	43.9&	38.2&	261.3&	162.3&	100.7&	90.9&	90.1&	81.7&	58.2&	50.7&	46.4&	45.2&	43.5&	40.1\\
				 Bulgaria&3.8&2.3&	2.1&	12.7&	7.8&	6.1&	5.8&	5.1&	4.8&	3.7&	2.9&	2.8&	2.8&	2.6&	2.5\\
		%Inflow - Romania&	66.1&	45.3	&	39.7	&	271.4	&	174.6	&	105.6	&	92.1	&	90.1	&	81.7	&	58.2	&	50.7&46.4&45.2&43.5&\\
		%Inflow - Bulgaria	&4.1&2.4&	2.1&	13.4&	8.4&	6.2&	5.9&	5.1&	4.8&	3.7&	2.9&2.8&2.8&2.6&\\
		\hline
		\\
		\textbf{Outflow}\\
		\hline\\[-1.8ex]
		%Outflow - Total&14.0&16.0&	17.0	&20.3&	27.0&	32.3&	32.8&	32.4&	38.2	&43.6&	48.0&44.7&&40.5&\\
		%	Outflow - Total&49.9&	53.9&	58.4&	51.1&	61.6&	64.9&	67.5&	82.4&	106.2&	125.7&	136.3&	146.9&	157.0&	155.1&	156.9\\
			%	Outflow - Italians&39.1&41.9&	46.3&	36.3&	39.5&	39.0&	39.5&	50.0&	68.0&	82.1&	88.8&	102.2&	114.5&	114.5&	116.7\\
					Total &10.8&11.9&	12.1&	14.8&	22.1&	25.9&	28.0&	32.4&	38.2&	43.6&	47.5&	44.7&	42.5&	40.5&	40.2\\
		%Unempl. &105.0&	133.0&	123.0&	131.0&	157.0&	226.0&	250.0&	278.0&	346.0&	455.0&	466.0&	456.0&	437.0&406.0&400.0\\
		%	Outflow - Non EU&&	10.3	&11.0&	10.9&	11.6&	13.7&	18.4&	17.7&	21.1&	24.1&--\\
		
		\%  Unempl. &10.2&8.9&	9.8&	11.3&	14.1&	11.5&	11.2&	11.7&	11.0&	9.6&	10.2&9.8&9.7&9.9&10.0\\
	%\bottomrule
	\hline \hline \\[-1.8ex]
		\multicolumn{15}{l}{\textit{Note}: The last row reports the ratio between the outflow of non-natives and the total  unemployed non-natives.}\\
		\multicolumn{15}{l}{\textit{Source}: Italian Institute of Statistics (ISTAT).}
	\end{tabular} 
	\global\let\\=\restorecr
	%\label{LPMOnlyTemp} 
\end{table}

The activity rate and the employment rate are much higher for non-natives than for natives (Table \ref{LabourForce}). %Among non-natives, 1 million were in the labour force in 2004 and almost 0.96 millions were employed. In 2018 2.8 millions were in the labour force and 2.4 millions were employed. 
The 2008-2009 economic crisis drove the employment rate of non-native employees down by more than 8 percentage points between 2004 and 2012, while the activity rate went down by 3 percentage points. %The number of employed non-natives has significantly declined from more than 21 millions to less than 20 millions in 2014, while picking up in the following years. 
The employment rate of Italian employees dropped by 2 percentage points between 2004 and 2014, and raised back in 2018, while the activity rate remained constant, with an increasing trend after 2013. The unemployment rate is higher among non-natives than among natives. The gap was of approximately 2 percentage points in the period 2004-2008, went up to 6 percentage points in 2013 and down again to 4 percentage points in 2018.

\begin{table}[!htbp]
	\centering
	\caption{Labour force statistics by immigration status. } 
	\label{LabourForce} 
	\scriptsize
	\setlength{\tabcolsep}{3pt}
	%	\begin{adjustbox}{angle=90}
	\begin{tabular}{lccccccccccccccc} 
		%\toprule
		\hline \hline \\[-1.8ex]
		&2004&2005&2006&2007&2008&2009&2010&2011&2012&2013&2014&2015&2016&2017&2018\\
\hline
		\\
		\textbf{Non-natives}\\
		\hline
		\\
		Labour force&1.070&	1.291&	1.422&	1.578&	1.847&	2.017&	2.162&	2.308&	2.456	&2.638&	2.760&2.815&2.838&2.829&2.855\\
		Employed  &965& 1.158	&1.299&	1.447&	1.690	&1.790&	1.912&	2.030&	2.110&	2.183&2.294&2.359&2.401&2.423&2.455\\
		\midrule
		Activity rate (\%)&74.2&73.4&	73.6&	73.2&	73.2&	72.5&	71.3&	70.9&	70.5&	70.5&	70.4&70.3&70.4&70.8&71.2\\
			Empl. rate (\%)&66.9	&65.8&	67.2&	67.1&	67.0&	64.3	&63.1&	62.3&	60.6&	58.3&	58.5&	58.9&	59.5&	60.6&	61.2\\
		Unempl. rate (\%) &9.8&		10.3&	8.6&	8.3&	8.5&	11.2&	11.6&	12.0&	14.1&	17.2&	16.8&16.2&15.4&14.3&14.0\\
		\hline
		\\
		\textbf{Italians}\\
		\hline
		\\
		Labour force&23.237&22.993&	22.990	&22.797&	22.908&	22.589&	22.420&	22.351&	22.801	&22622&	22.755&22.683&22.932&23.101&23.116\\
		Employed  &21.398& 	21.249&	21.459&	21.447&	21.400	&20.909&	20.615&	20.568&	20.456&	20.008&	19.985&20.106&20.357&20.600&20.760\\
		\midrule
		Activity rate (\%)&	62.2&	61.8&	62.1&	61.8&	62.2&	61.5&	61.2&	61.3&	62.8&	62.6&63.2&63.3&64.3&64.8&65.0\\
		Empl. rate (\%)&57.2&	57.1&	57.9&	58.1&	58.1&	56.8&	56.2&	56.3&	56.3&	55.2&	55.4&56.0&57.0&57.7&58.2\\
		Unempl. rate (\%)&7.9&		7.6&	6.7&	5.9&	6.6&	7.4&	8.1&	8.0&	10.3&	11.6&	12.2&11.4&11.2&10.8&10.2\\
		%\bottomrule
\hline \hline 
		\\[-1.8ex]	\multicolumn{15}{l}{\textit{Note}: Individuals in the labour force and employed individuals are reported in thousands.}\\
		\multicolumn{15}{l}{\textit{Source}: Italian Institute of Statistics (ISTAT).}
	\end{tabular} 
	%	\end{adjustbox}
\end{table}

\subsection{Inflow and outflow of natives \label{app:inflowOutflowNatives}} 

Table \ref{inflowoutflownatives} reports the inflow and outflow  of natives. The share of the number of natives who leave the country and the total number of unemployed natives is rather small. It ranges between 2.1\% and 3.2\% until 2014 and although increasing in the last few years it is never higher than 5\%.
\begin{table}[!htbp]
	\centering
	\caption{Inflows and outflows of natives (in thousands). } 
	\label{inflowoutflownatives}
	\scriptsize
	\begin{tabular}{lccccccccccccccccc} 
		%\toprule
		\hline \hline 
		%		\global\let\oldcr=\\
		%		\global\def\crsng{\global\let\\=\crdbl\oldcr}
		%		\global\def\crdbl{\global\let\\=\crsng\oldcr[2pt]}
		%		\global\let\\=\crdbl
		&2004&2005&2006&2007&2008&2009&2010&2011&2012&2013&2014&2015&2016&2017&2018\\
		\midrule
		
		Inflow - Total	&	41.8&	37.3&	37.7&	36.7&	32 .1&	29.3&	28.2&	31.5&	29.5&	28.4&	29.3&	30.1&	37.9&	42.4&	46.8\\
		Outflow - Total&39.2&41.9&	46.3&	36.3&	39.5&	39.0 &	39.5&	50.1&	68.0&	82.1&	88.9&	102.3&114.5&114.5&116.7\\
		\midrule
		Outflow &2.1 &2.4&3.0&	2.8&	2.6&	2.3&2.2&	2.8&	2.9&	3.1&	3.2&3.9&4.4&4.6&4.9\\
		(\%  unemployed)&&&&&&&&&&&&&&\\
		%\bottomrule
		\hline \hline 
			\multicolumn{15}{l}{\textit{Note}: The last row reports the ratio between the outflow of natives and the total unemployed natives.}\\
		\multicolumn{15}{l}{\textit{Source}: Italian Institute of statistics (ISTAT).}
	\end{tabular} 
	\global\let\\=\restorecr
	%\label{LPMOnlyTemp} 
\end{table}

\subsection{Employees' occupation, composition and unemployment}\label{workeroccupation}

Table \ref{WorkersByOccupation} provides data on the distribution of non-native  (Panel A) and native (Panel B) employees by occupation in Italy from 2004 to 2018. More than 90\% of non-natives are hired either as clerks and sales employees, craft employees and machine operators or in elementary occupations, which are occupations which require lower skill levels (1 or 2 according to the ILO classification). Looking at the trend, the share of non-natives hired in elementary occupations is roughly stable over time. However, we observe a shift away from occupations such as craft employees and machine operators and managers towards occupations such as clerks and sales employees. Among natives, approximately two third of employees are hired in occupations which require lower skill levels (levels 1 or 2 of the ILO classification), while one third of employees are hired as managers and professionals, which are occupations which require higher skill levels (levels 3 or 4 of ILO classification). Over time, we observe fewer employees who work as craft employees or machine operators  and more who are employed in occupations such as clerks and sales employees and managers.

\begin{table}[!htbp]
	\caption{Distribution of native and non-native employees by occupation (in $\%$).} 
	\label{WorkersByOccupation}
	\scriptsize
	\setlength{\tabcolsep}{3pt}
	\begin{threeparttable}
		\centering
		\vspace{0.2cm}
	
		\begin{tabular}{@{\extracolsep{3pt}}lccccccccccccccc} 
			%\toprule
			\hline \hline 
			%		\global\let\oldcr=\\
			%		\global\def\crsng{\global\let\\=\crdbl\oldcr}
			%		\global\def\crdbl{\global\let\\=\crsng\oldcr[2.5pt]}
			%		\global\let\\=\crdbl
			\\[-1.8ex]
			&2004&2005&2006&2007&2008&2009&2010&2011&2012&2013&2014&2015&2016&2017&2018\\
			\hline
			\\
				\textbf{Non-natives}\\
				\hline
			\\[-1.8ex]
			\multirow{1}{3.5cm}{Managers, etc.\tnote{1}}	&10.9&	9.2&	9.4&	10.1&	8.5&	7.3&	7.2&	6.7&	5.9&	6.0&	7.0&	6.8&	6.7&7.1&7.6\\
			%	&&&&&&&&&&&&\\
			%	\\
			\multirow{1}{3.5cm}{Clerks and  sales employees\tnote{2}}	&16.4&	17.0	&18.4	&	18.8	&18.3	&	17.3 &	16.4	&	23.6		&25.6	&	26.7		&26.9	&	27.2		&28.3&30.0&29.4\\
			%\tnote{2}	&&&&&&&&&&&&\\
			%\\
			\multirow{2}{3.5cm}{Craft employees and machine operators\tnote{3}}&39.8&	40.4&	42.7&	42.8&41.2&	39.3&	38.2&	36.4&33.8&	0.318&	30.3&	30.1&	29.3&28.3&29.6\\
			&&&&&&&&&&&&\\
			%		&&&&&&&&&&&&\\		&&&&&&&&&&&&\\		&&&&&&&&&&&&\\		&&&&&&&&&&&&\\
			%\\
			\multirow{1}{3.5cm}{Elementary occupations}	&33.0&	33.3&	29.5&	28.3&	31.9&	36.1&	38.1&	33.2&	34.7&	35.4&	35.6&	35.9&	35.7&34.4&33.3\\
			% occupations	&&&&&&&&&&&&\\
			\hline
			%Total (in thousands) &1157&	1299&	1446&	1690&	1790&	1911&	2030&	2109&	2183&	2293&	2359&	2401\\
			%	\\
			%	\\
%			\vspace{0.2cm}
%		\end{tabular} 
\\
		\textbf{Natives}\\
%		\begin{tabular}{@{\extracolsep{5pt}}lccccccccccccccc} 
%			\toprule
%			%\global\let\oldcr=\\
%			%\global\def\crsng{\global\let\\=\crdbl\oldcr}
%			%\global\def\crdbl{\global\let\\=\crsng\oldcr[2.5pt]}
%			%\global\let\\=\crdbl
%			&2004	&2005&2006&2007&2008&2009&2010&2011&2012&2013&2014&2015&2016&2017&2018\\
	\hline
	\\[-1.8ex]
			\multirow{1}{3.5cm}{Managers, etc.\tnote{1}}	&35.5&	35.5&	37.8&	38.8&	38.5&	37.5&36.9&	36.4&36.5&	37.4&	37.4&	37.6&	37.8&38.0&38.5\\
			%&&&&&&&&&&&&\\
			%&&&&&&&&&&&&\\	
			\multirow{1}{3.5cm}{Clerks and  sales employees\tnote{2}}&27.5	&	27.7&26.8&	26.7&	27.5&	28.4&	29.3&	30.0&	30.5&	30.5&	30.6&30.7&	30.8&30.7&30.6\\
		%	sales employees\tnote{2}&&&&&&&&&&&&\\
			%&&&&&&&&&&&&\\
			\multirow{1}{3.5cm}{Craft employees and  machine operators\tnote{3}}	&27.6&	27.5&26.5&	26.0&	25.7&	25.5&	25.0&	24.7&	23.7&22.9&	22.7&	22.4&	22.0&21.9&21.6\\
			&&&&&&&&&&&&\\
			%&&&&&&&&&&&&\\
			%&&&&&&&&&&&&\\
			%&&&&&&&&&&&&\\
			%&&&&&&&&&&&&\\
			%&&&&&&&&&&&&\\
			\multirow{1}{3.5cm}{Elementary occupations}	&8.3&8.2&	7.8&	7.4&	7.2&7.3&	7.5&	7.7&	7.9&	8.1&	8.1&	8.1&	8.2&8.2&8.2\\
		%	&&&&&&&&&&&&\\
			\multirow{1}{3.5cm}{Military service}	&1.2&	1.2&1.2&	1.2&1.1&	1.2&1.3&	1.2&1.3&1.2&	1.2&1.2&	1.2&1.2&1.1\\
			%\\
			%\hline
			
			%Total (in thousands) &	21249&	21458&	21447&	21400&	20909&	20615&	20567&	20457&	20008&	19986&	20107&	20357\\
%\bottomrule
\hline \hline 
			\\[-1.8ex]
			\multicolumn{8}{l}{\textit{Source}: Italian Institute if Statistics (ISTAT) and OECD.}
		\end{tabular} 
		%	\end{tabular} 
		\begin{tablenotes}
\scriptsize
			\item[1] It includes also professionals, technicians and associate professionals.
			\item[2] It includes also service employees.
			\item[3] It includes also skilled agricultural and fishery employees, plant and machine operators and assemblers.
		\end{tablenotes}
	\end{threeparttable}
	\global\let\\=\restorecr
	%\label{LPMOnlyTemp} 
\end{table}

To have a better understanding of the distribution of non-natives across occupations, in Table \ref{ImmigrantsShareByOccupation} we report the number of non-natives as a share of the total number of employees by occupation. In 2005, approximately 4.3\% of all employees were non-natives. Specifically, among employees employed in elementary occupations, approximately 15\% were non-natives, among craft employees approximately 6\%  and among clerks or sale employees less than 3\%. Among managers the percentage of non-natives was just 1.4\%. In 2018, the number of non-natives as a share of the whole pool of employees is up to 10.6\%, while in elementary occupations, the share of non-natives is up to 32\%; among all craft employees 14\% and among clerks and sales employees approximately 10\%. The share of non-natives in managerial positions is still  low (approximately 2\%).
Evidence on the share of non-natives in 3 digit occupations (129 categories) provides further support against the hypothesis of occupational segregation of non-natives. In the large majority of low-skilled occupations (85 categories) we observe the coexistence of a large share of both natives and non-natives. If we consider occupations with at least 10,000 employees (62 categories), there is no occupation with less than 20\% non-natives.\footnote{On the other hand, there are 5 low-skilled occupations in which the share of non-natives is below 5\%.} This is the case also in occupations which are perceived as dominated by non-natives \citep{di2015female}: approximately 24\% of natives and 76\% of non-natives are employed in domestic services and 20\% of natives and 80\% of non-natives are employed as blue-collars performing manual (routine) jobs.

\begin{table}[!htbp]
	\centering
	\setlength{\tabcolsep}{3pt}
	\caption{Share of non-natives by occupation (in $\%$).} 
	\label{ImmigrantsShareByOccupation}
	\scriptsize
	\begin{threeparttable}
		\begin{tabular}{@{\extracolsep{3pt}}lccccccccccccccc} 
			%\toprule
			\hline \hline 
			&2004&2005&2006&2007&2008&2009&2010&2011&2012&2013&2014&2015&2016&2017&2018\\
			\hline
			\\
			\multirow{1}{3.5cm}{Managers, etc.\tnote{1}}	&1.4&	1.4&	1.5&	1.7&	1.7&	1.6&	1.8&	1.8&	1.6&	1.7&	2.1&	2.1&	2.0&2.2&2.3\\
			%		&&&&&&&&&&&&\\
			%		&&&&&&&&&&&&\\
			%		&&&&&&&&&&&&\\
			\multirow{1}{3.5cm}{Clerks and sales employees\tnote{2}}&2.6&	3.2&	4.0&	4.5	&5.0	&5.0&	4.9&	7.2&	8.0&	8.7&	9.2&	9.4&	9.8&10.3&10.2\\
			%		&&&&&&&&&&&&\\
			%		&&&&&&&&&&&&\\
			%		&&&&&&&&&&&&\\
			\multirow{1}{3.5cm}{Craft employees and machine operators\tnote{3}}&6.1&	7.4	&8.9&	10.0&	11.3&	0.116&	12.4&	12.7	&12.8&	13.2&	13.3&	13.7&	13.6&13.2&14.0\\
			&&&&&&&&&&&&\\
			%		&&&&&&&&&&&&\\
			%		&&&&&&&&&&&&\\
			%		&&&&&&&&&&&&\\
			%		&&&&&&&&&&&&\\
			%		&&&&&&&&&&&&\\
			%		&&&&&&&&&&&&\\
			\multirow{1}{3.5cm}{Elementary occupations}&15.2	&	18.2&	18.7&	20.5&	25.9&	29.7&31.9&	29.9&31.0&	32.6&	33.6&	34.2&	33.9&33.1&32.4\\
			%		&&&&&&&&&&&&\\
			%		&&&&&&&&&&&&\\
			\hline
			Total&4.3	&	5.2&	5.7&	6.3&	7.3&	7.9&	8.5&	9.0&	9.3&	9.8&	10.3&	10.5&	10.6&10.5&10.6\\
			
			%\bottomrule 
			\hline \hline  \\[-1.8ex]
			\multicolumn{8}{l}{\textit{Source}: Italian Institute if Statistics (ISTAT).}
		\end{tabular} 
		\begin{tablenotes}
\scriptsize
			\item[1] It includes also professionals, technicians and associate professionals.
			\item[2] It includes also service employees.
			\item[3] It includes also skilled agricultural and fishery employees, plant and machine operators and assemblers.
		\end{tablenotes}
	\end{threeparttable}
	\global\let\\=\restorecr
	%\label{LPMOnlyTemp} 
\end{table}

%\begin{table}[h]
%	\centering
%	\caption{Unemployment rate by country of origin and skill level based on education } 
%	\tiny
%	\global\let\restorecr=\\
%	\begin{tabular}{@{\extracolsep{5pt}}lcccccc} 
%		\toprule
%		\global\let\oldcr=\\
%		\global\def\crsng{\global\let\\=\crdbl\oldcr}
%		\global\def\crdbl{\global\let\\=\crsng\oldcr[2.5pt]}
%		\global\let\\=\crdbl
%
%& \multicolumn{2}{c}{2005}& \multicolumn{2}{c}{2009}& \multicolumn{2}{c}{2013}\\
%\hline
%&	Italians& 	Immigrants&	Italians& 	Immigrants&	Italians& 	Immigrants\\
%\hline
%	\\	
%Low skills&	7.501&	9.408&	7.541&	10.164&	12.427&	15.762\\
%High skills & 5.967&	8.192&5.168&	9.395&6.779&	13.985\\
%
%
%		\bottomrule
%Source: Istat.
%\end{tabular} 
%\global\let\\=\restorecr
%\label{LPMOnlyTemp} 
%\end{table}

%\subsection{Workforce composition by country of origin and skill level\label{sec:Skills}}

\begin{figure}[!htbp]
\caption{Share of non-natives in the workforce and unemployment rates by skill level.}
	\centering
	\begin{subfigure}[b]{0.44\textwidth}
		\centering
		\includegraphics[width=0.8\textwidth]{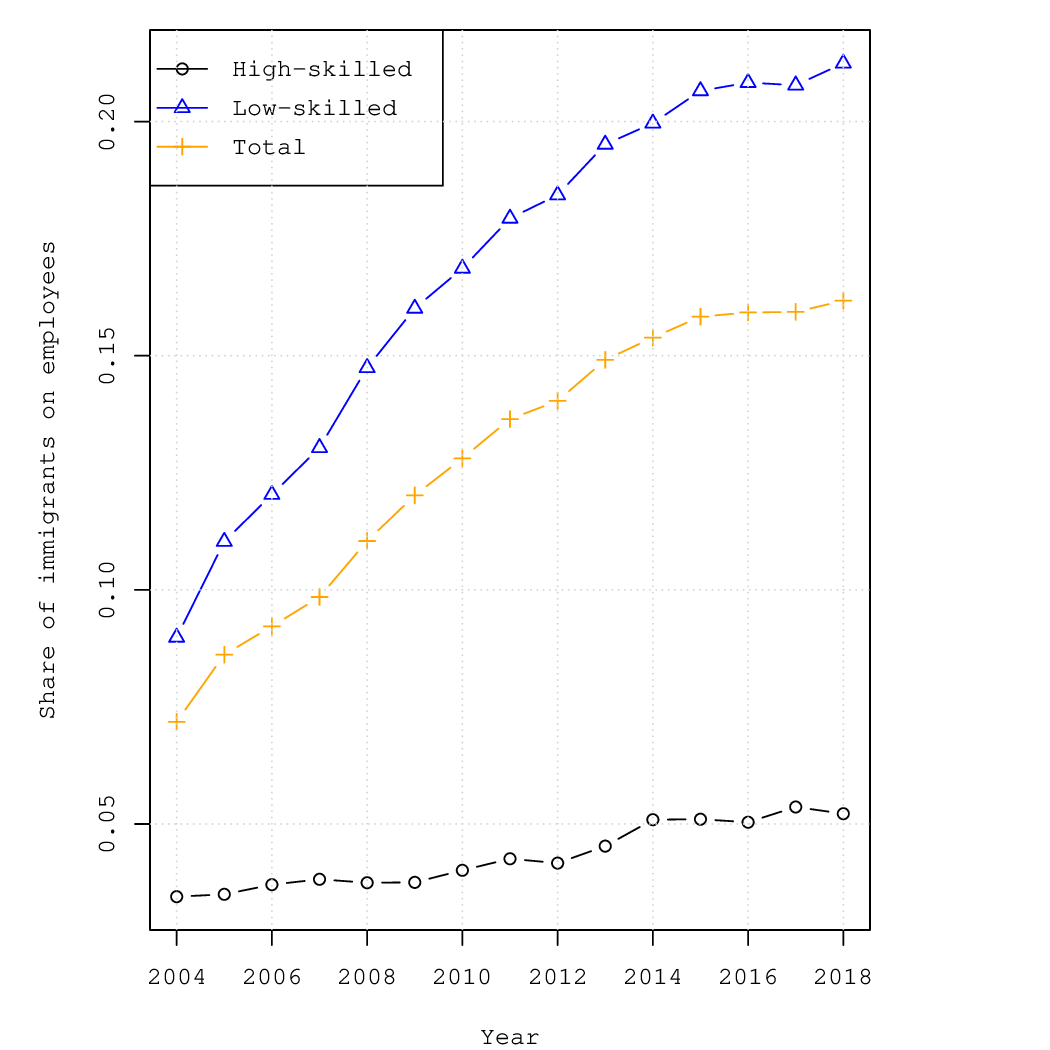}
		\caption{Share of non-natives in the \\ workforce by skill level.}
		\label{shareImmigrants}
	\end{subfigure}
	\begin{subfigure}[b]{0.44\textwidth}
		\centering
		\includegraphics[width=0.8\textwidth]{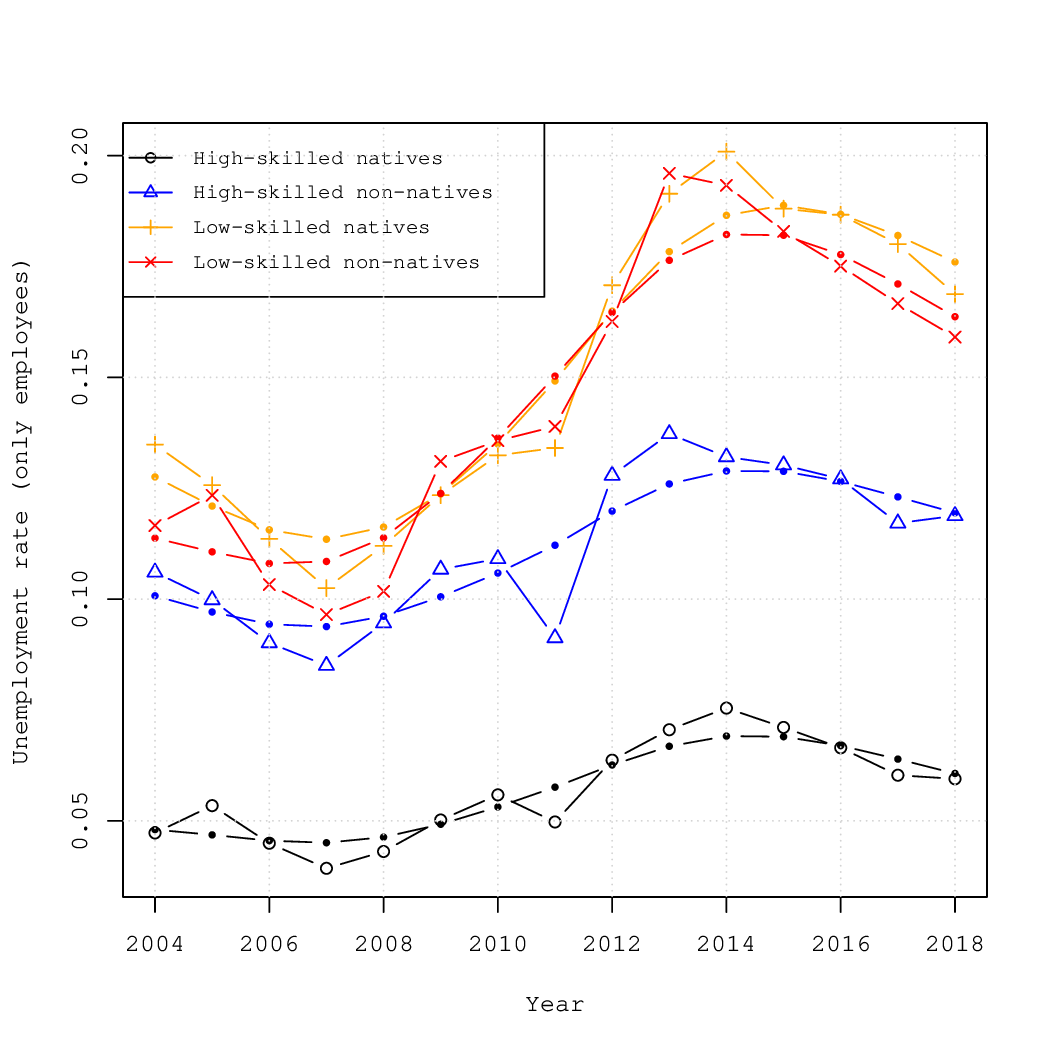}
	%	\floatfoot{\tiny{	Note: h.N=low-skilled natives; l.N=low-skill natives; h.I=low-skilled immigrants; h.I=low-skilled immigrants.}}
		\caption{Unemployment rate by country \\ of origin and skill level. }
		\label{unemploymentRatebySkill}	
	\end{subfigure}
\vspace{0.1cm}
	\caption*{\scriptsize{\textit{Note}: The left graph reports the ratio between the number of non-natives in the workforce and the total number of individuals in the workforce. The black and blue line report the same statistic when considering only high-skilled or low-skilled individuals, respectively. The right  graph reports the unemployment rate for the four categories of employees: high-skilled natives (black line), high-skilled non-natives (blue line), low-skilled natives (yellow line) and low-skilled non-natives (red line).}}
	\caption*{\scriptsize{\textit{Source}: Italian Labour Force Survey (RCFL).}}
\end{figure}

We use data from the Labour Force Survey as provided by the National Institute of Statistics (ISTAT) to compute the share of non-natives in the workforce and the unemployment rates by skill  level, according to our classification.
The great majority of non-natives is low-skilled, and their share in the workforce has increased by approximately 15\% in the period 2004-2018. The share of non-native high-skilled employees has increased by 2\% only in the period considered (Figure \ref{shareImmigrants}). Among natives and non-natives (Figure \ref{unemploymentRatebySkill}), the unemployment rate of low-skilled employees is higher compared to the unemployment rate of high-skilled employees. Among low-skilled employees, the unemployment rate is similar between natives and non-natives, while among high-skilled employees the unemployment rate of non-natives is much higher compared to the unemployment rate of natives. After the 2008/2009 crisis, the unemployment rates of low-skilled employees have increased relatively more compared to the unemployment rates of high-skilled employees, both among natives and non-natives.

\subsection{Wages \label{sec:Wages}}

\begin{figure}[!htbp]
	\caption{Monthly real mean and median net wages (in thousands).}
	\centering
	\begin{subfigure}[b]{0.44\textwidth}
		\includegraphics[width=0.8\linewidth]{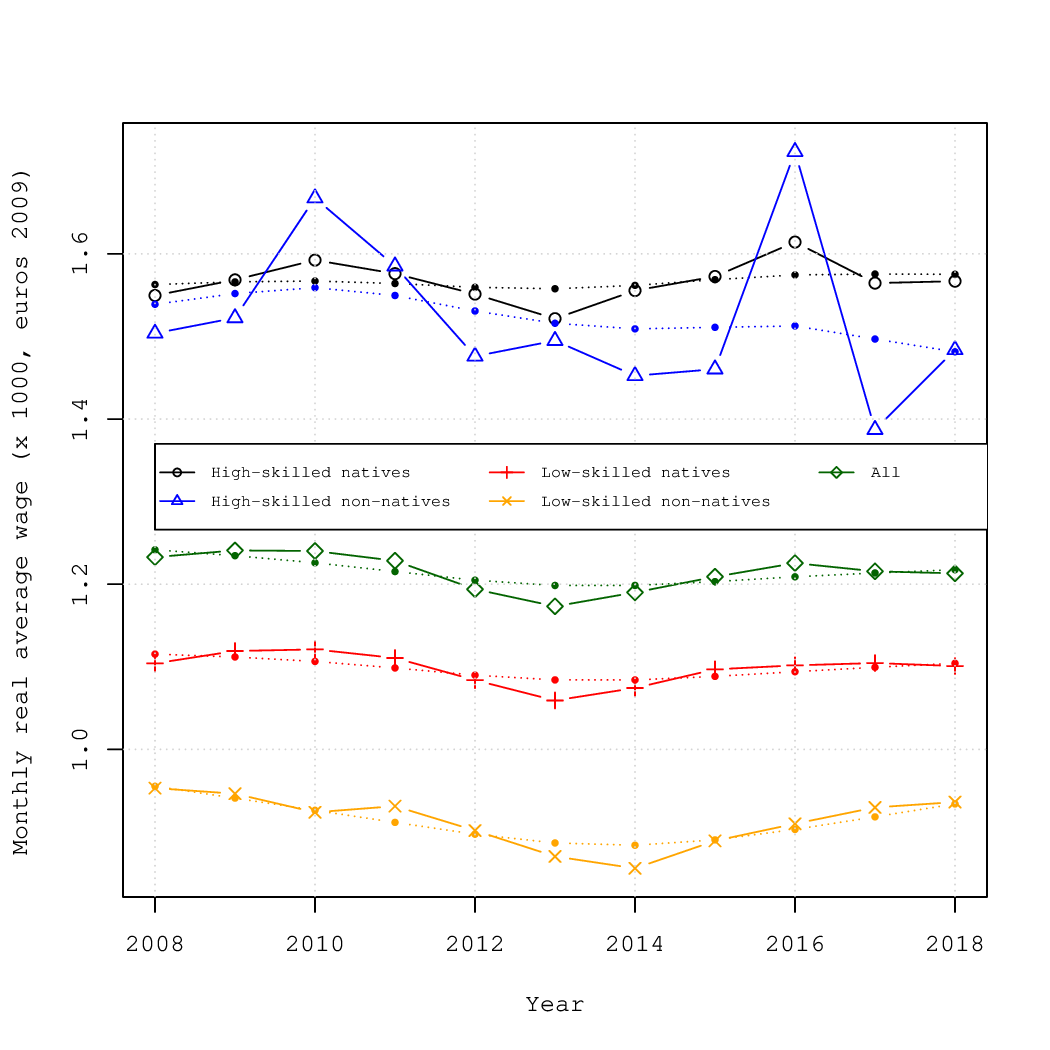}
		\caption{Monthly real average net wages. }
		\label{fig:realwagesitaly}
	\end{subfigure}
	\begin{subfigure}[b]{0.44\textwidth}
		\centering
		\includegraphics[width=0.8\linewidth]{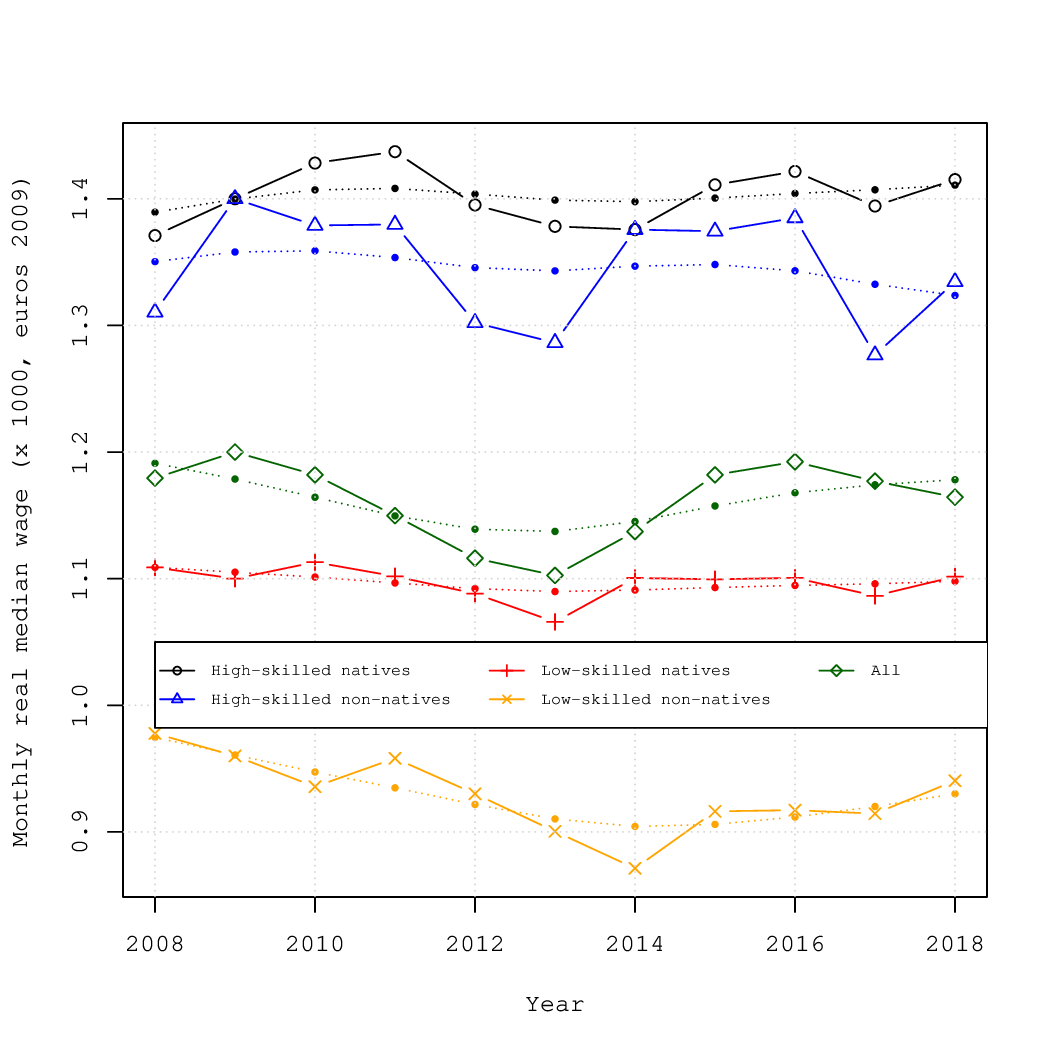}
		\caption{Monthly real median net wages. }
		\label{fig:realMedianWagesItaly}
	\end{subfigure}
	\vspace{0.3cm}
	\caption*{\scriptsize{\textit{Note}: The monthly real average wages and the monthly real median wages reported in the graphs are calculated by dividing the monthly nominal net average and median wages by the CPI index with 2009 used as the base year. The monthly real average and median wages are reported for all (green line) and the four categories of employees: high-skilled natives (black line), high-skilled non-natives (blue line), low-skilled natives (yellow line) and low-skilled non-natives (red line).\\ \textit{Source}: Italian Labour Force Survey (RCFL). }}
\end{figure}

Figures \ref{fig:realwagesitaly} and \ref{fig:realMedianWagesItaly} report the mean and median (net) real wages of employees by skill level and country of origin (natives and non-natives) for the years 2008-2018 calculated using data from  the Labour Force Survey as provided by the National Institute of Statistics (ISTAT).\footnote{Few caveats need to be mentioned here in relation to the data used. First, the information on the individual wage is not released by ISTAT before 2008. Second, the wages reported are \textit{nominal} monthly net wages, which are then converted into real monthly net wages using the price consumer index provided by ISTAT. The net wages refer to the wages earned by the employees the month before the interview, excluding additional monthly payments and thirteen and fourteen salaries. Finally, the wage distribution is left and right truncated as wages are capped in the range between 250\euro \; and  3000\euro; mean wages are therefore calculated by fitting a beta distribution  using the available data.}   
The real mean and median wages of both high-skilled and low-skilled  native employees are higher than those of non-native employees, although the gap between the two is much larger among low-skilled employees.
Specifically, employees with a low skill level who are non-natives earn 20\% less than natives. Employees with a high skill level who are non-natives earn 10\% less than natives. Native high-skilled employees earn 40\% more than native low-skilled  employees.

 Non-native high-skilled employees earn 55\% more than non-native low-skilled  employees. While the real wages of high-skilled employees both natives and non-natives, has been approximately constant during the period considered, from 2009 to 2014 the wage level of low-skilled employees has decreased in absolute terms. The wage of non-native employees has decreased more than the wage of native employees, and although it increased after 2014 it did not reach the pre-crisis level.\footnote{It is noteworthy mentioning that the wage of high-skilled non-natives may not be accurate due to their small number in the sample.} In the analysis, we use median wages (instead of mean wages) as we believe them to be more robust for the categories of employees we are considering.

\subsection{Job creation and job exit rates \label{JobCreationandJobDestruction}}

\begin{figure}[!htbp]
 	\caption{Annual job finding and job exit rates. }
 	\label{fig:JFJE}
 	\centering
 	\begin{subfigure}[b]{0.48\textwidth}
 		\centering
 		\includegraphics[width=0.8\textwidth]{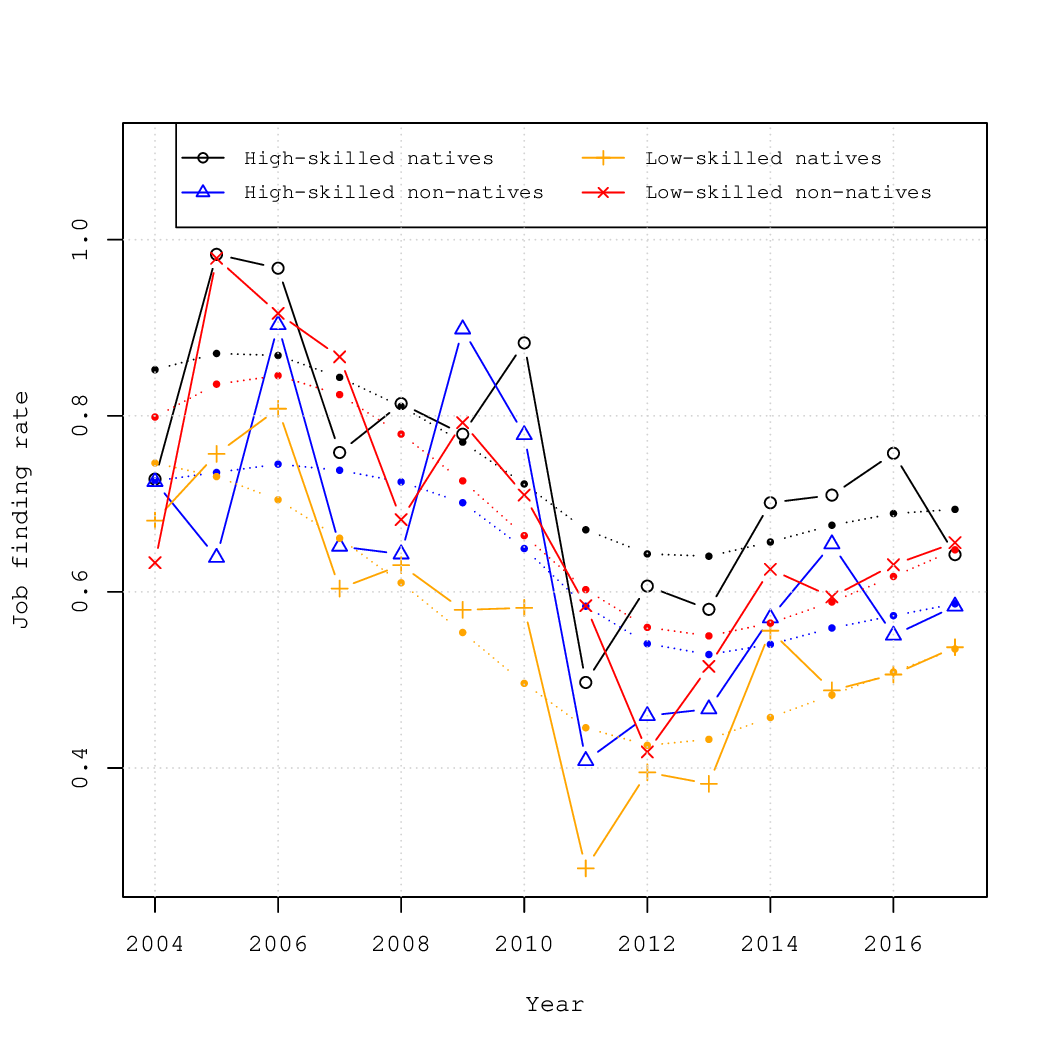}
 		\caption{Job finding rates.}
 	%	\label{tot_unem}
 	\end{subfigure}
 	\begin{subfigure}[b]{0.48\textwidth}
 		\centering
 		\includegraphics[width=0.8\textwidth]{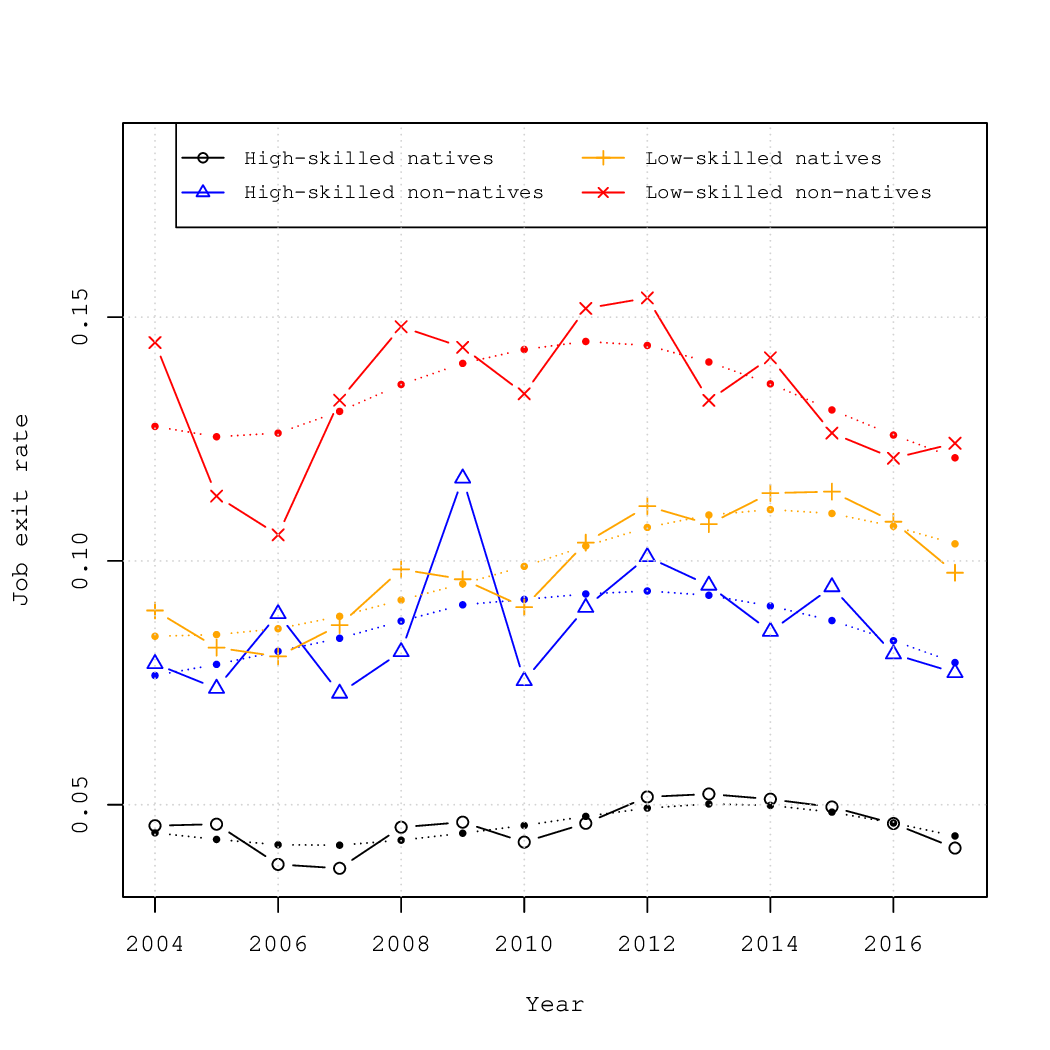}
 		\caption{Job exit rates.}
 	%	\label{tot_unem}
 	\end{subfigure}
 \vspace{0.3cm}
 	\caption*{\scriptsize{\textit{Note}: The job finding rates (left graph) and the job exit rates (right graph) are reported for  the four categories of employees: high-skilled natives (black line), high-skilled non-natives (blue line), low-skilled natives (yellow line) and low-skilled non-natives (red line).\\\textit{Source}: our calculations using data from the Italian Labour Force Survey (RCFL).}}
 \end{figure}

Figure \ref{fig:JFJE} reports the probability for a worker to find a job and the probability to lose a job, using the methodology proposed by \cite{SHIMER2012127} and  data on unemployment rates by skill level, as computed in Section \ref{workeroccupation} (see Appendix \ref{app:jfrjer}). High-skilled employees exhibit higher job finding rates and lower job exit rates compared to low-skilled employees, in agreement with the literature which provide evidence of  low educated employees having the highest gross mobility (turnover) compared to middle and high educated employees \citep{landesmaan_2015}. As in \cite{dustmann2010}, we find that  among high-skilled, natives exhibit higher job finding rates, particularly after 2011, while among low-skilled employees the opposite is true. Over time,  job finding rates across all types of employees have crashed in 2011 as a result of the crisis and while they have increased afterwards, as of 2018 they have not reached the level pre-crisis. Non-native employees tend to have higher exit rates.  This is in line with the literature, which shows that non-native employees lose their jobs more often than natives  but once being unemployed they have more probabilities of finding a job than natives \citep{barth2012immigrant, fullin_2011}. Heterogeneity between natives and non-natives may be due to the difference in the job tenure and the higher likelihood to be hired on temporary contracts.\footnote{The concentration of non-natives in the secondary labour market is the most important factor explaining their disadvantage in terms of risk of losing a job \citep{priore79, OECD2006}.  In Italy, the share of non-natives hired on temporary contracts is only slightly higher than the share of natives (Table \ref{Temps} in Appendix \ref{app:shareTemporaryJobs}), however, many non-natives hired on  a permanent job  are working  in small firms in which the risk of losing a job is  higher. Moreover, a large secondary labour market provides a great deal of poorly qualified jobs that are more suitable for non-natives, who have lower reservation wages than natives since they take the wages in their home country as a reference \citep{dustmann2000,kalter2022,kogan2007}.} The availability of financial support may also make a job search different for non-natives and natives, as poor support pushes unemployed into finding a job as soon as possible.\footnote{In Italy, non-natives are formally entitled to get unemployment benefits, but in practice they have less access, because their work history includes more spells of temporary and non-registered jobs \citep{fullin_2011} and most non-natives cannot rely on family support \citep{zimmermann2014}.}   Finally, non-native employees tend to be concentrated in industries which are more vulnerable to economic slowdowns and in low-skilled occupations.\footnote{In Italy, male non-natives are mainly employed in the construction and manufacturing sectors, which are either seasonal or very sensitive to business cycle fluctuations, whereas females non-natives are mainly employed as housekeepers and elderly caregivers, which are less sensitive to the business cycle \citep{fullinreynieri}.} 
 
\subsection{Temporary jobs among natives and non-natives \label{app:shareTemporaryJobs}}

In most countries the probability of being hired on a temporary job is significantly higher for non-natives than for natives and the more temporary work is utilized as a form of employment, the greater is be the gap between non-natives and natives \citep{OECD2006}. In Italy, surprisingly, the share of non-natives hired on temporary contracts is only slightly higher than the share of natives, as shown in Table \ref{Temps}.

\begin{table}[!htbp]
	\centering
	\scriptsize
	\caption{Share of  employees in temporary contracts among natives and non-natives (in $\%$). }
	\label{Temps}
	\begin{tabular}{lcccccccccccc}
		%\midrule
		\hline \hline 
		&2004&2005&2006&2007&2008&2009&2010&2011&2012&2013&2014&2015\\
		\midrule
		Non-natives &14.5 &	14.7&	15.3&	14.7&	15.6&	14.3&	15.1&	15.8&	16.3&	15.2&	15.8&16.4\\
		Natives & 11.8&	12.1&	12.9&	13.0&	13.1&	12.3&	12.4&	13.0&	13.5&	12.9&	13.3&14.0\\
		%\bottomrule 
		\hline \hline \\[-1.8ex]
		\multicolumn{12}{l}{\textit{Note}: in this classification non-natives are defined  by nationality and not  by country of birth.}\\	\multicolumn{8}{l}{\textit{Source}: Italian Institute of Statistics (ISTAT). }
	\end{tabular}	
\end{table}

\newpage 

\section{Details on the calibration of the model's parameters }

In this section we describe in detail the steps followed to calibrate the levels of non-native inflow rate, tax subsidy, firing cost, and vacancy cost (and some robustness checks on these).

\subsection{The vacancy cost \label{vacancycost}}

The vacancy cost in the model is equal to:
\begin{equation}
cp_ix_i=\kappa_{i,j}q(\theta) TC,
\end{equation}
where $TC$ is the total cost of opening a vacancy, $q(\theta)$ is the probability to fill a vacancy and $cp_ix_i$ is the instantaneous cost of vacancy, which is paid by the employer in each instant of time and it is proportional to the real value added. Rearranging, we that the instantaneous vacancy cost $c$ is equal to:
\begin{equation}
c=\dfrac{\kappa_{i,j}q(\theta) TC }{p_ix_i}.
\end{equation}
The next step is to move away from an instantaneous cost to a monthly cost, as per our calibration. To achieve this goal, we compute the monthly job finding rate ($\kappa_{i,j}q(\theta)^m$) and the monthly real value added ($p_ix_i^m$), to get:
\begin{equation}
c^m=\dfrac{\kappa_{i,j}q(\theta)^m TC}{p_ix_i^m}.
\end{equation}\label{monthlycost}
Since $TC=DC+OC$, where $DC$ is the direct cost and  $OC$ is the opportunity cost, we can the previous equation as:
\begin{equation}
c^m=\kappa_{i,j}q(\theta)^m \left(\dfrac{DC}{p_ix_i^m}+\dfrac{OC}{p_ix_i^m}\right).
\end{equation}
To compute the monthly cost we use the job finding rate as calculated in Appendix \ref{app:jfrjer} and data on direct cost and opportunity cost per person from The World Bank `Doing Business', which we converted into the per worker variables.

\subsection{The firing cost \label{firingcost}}

In the Italian legislation, a firing cost is not due in case of quitting, hence in this case $F$ is equal to zero.  Moreover, an employer-initiated separation is legitimate only when it satisfies a "just clause". The Italian civil law (st. n 604/1966, sect. 3) foresees that individual dismissal is legal only under the two headings: justified objective motive, i.e. "justified reasons concerning the production activity, the organization of labour in the firm and its regular functioning", and justified subjective motives, i.e. "a significantly inadequate fulfilment of the employee's tasks specified by the court". The first heading involves events which are outside the employee's control, while the second case requires misconduct on the part of the worker. The worker has always the right to appeal the firm's decision, and the final judgment ultimately depends on the court's interpretation of the case. If the separation is ruled fair, or if the worker does not appeal the firing decision, the legislation does not impose any firing cost to the firm. Conversely, when the separation is ruled unfair and illegitimate, the court imposes a specific set of transfers and "taxes" to the firm.

\begin{figure}[!htbp]
	\caption{Number of fired employees and share of fired employees on total job losses.}
	\label{trialinfo}
	\centering
	\begin{subfigure}[b]{0.40\textwidth}
		\centering
		\includegraphics[width=1\linewidth]{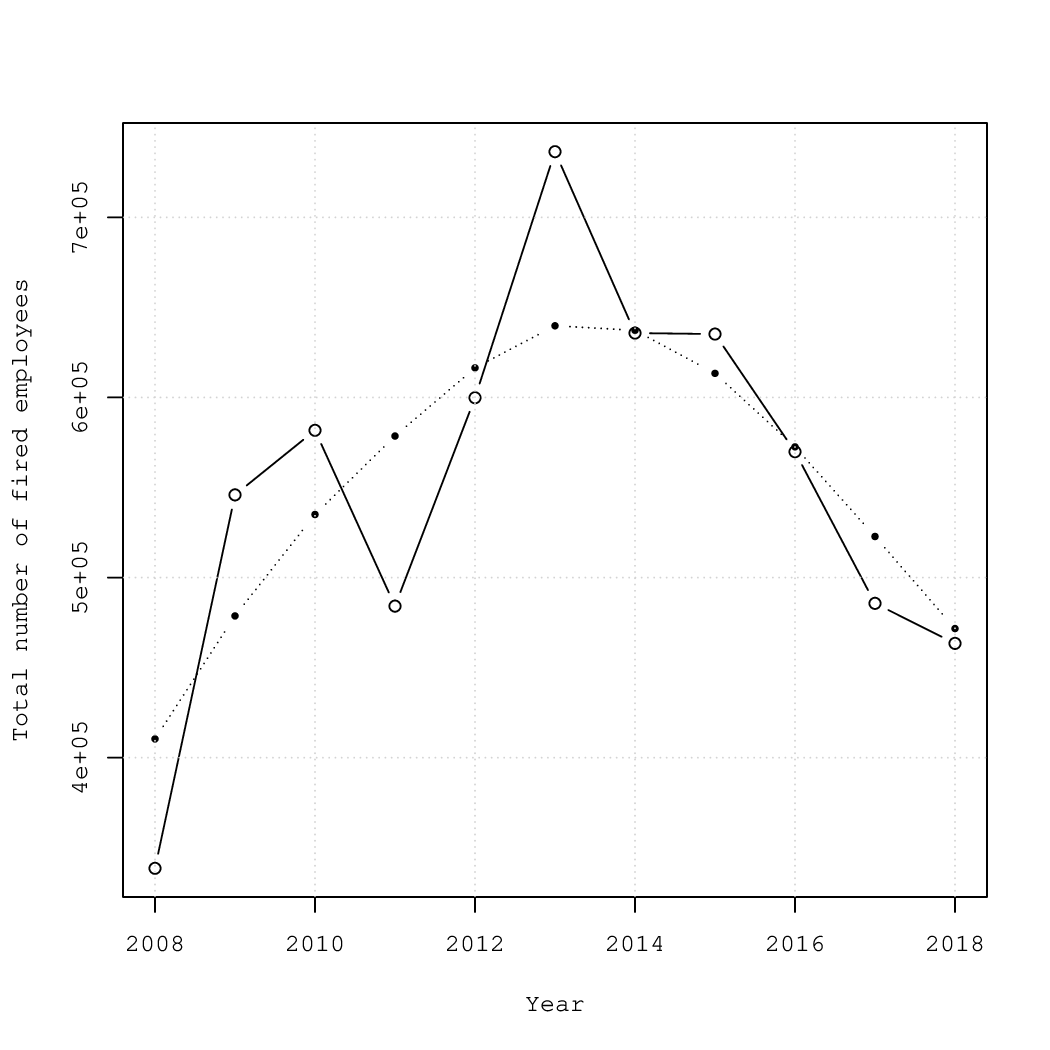}
		\caption{Total number of fired employees}
		\label{fig:numberfiredemployees}
	\end{subfigure}
	\begin{subfigure}[b]{0.40\textwidth}
		\centering
		\includegraphics[width=1\linewidth]{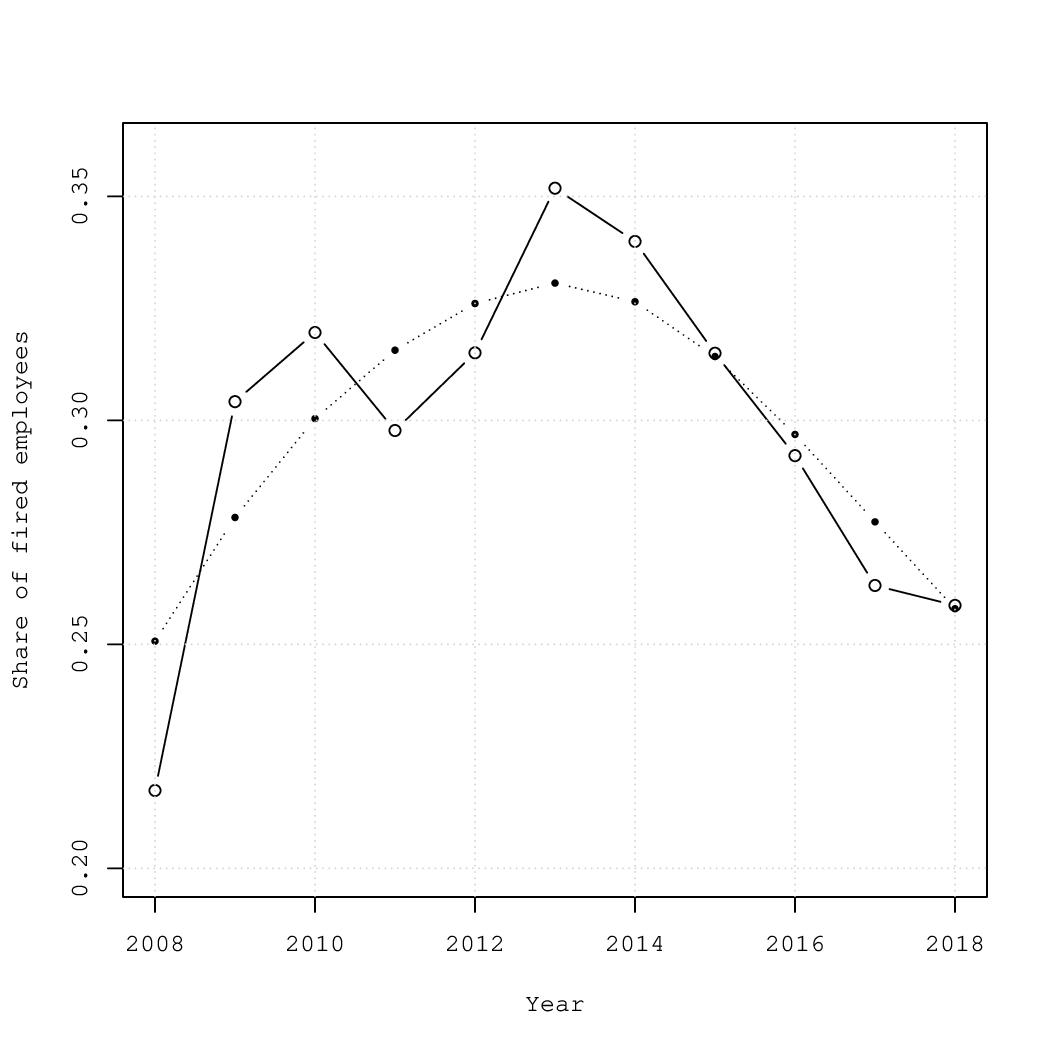}
		\caption{Share of fired employees on total job losses}
		\label{fig:sharefiredemployees}
	\end{subfigure}
	\vspace{0.1cm}
	\caption*{\scriptsize{\textit{Note}: The dotted line represents the smoothed series obtained using the Hodrick-Prescott filter.} \\ \scriptsize{\textit{Source}: Own calculation based on the Italian Labour Force Survey (RCFL).}}
\end{figure}

In Figure \ref{trialinfo}, we report the number of fired employees as the share of the total number of employees, who lost their job.

With the so called "Fornero Law" implemented in 2013 (Law n.92 del 2012), all employers who fire employees are required to contribute to the payment of the unemployment benefits, the worker is entitled to. This is also a form of firing costs. Specifically, employers need to pay 41\% of the maximum monthly unemployment benefit, per each 12 month of tenure in the previous 3 years. For instance, in 2013 the maximum monthly unemployment benefit amounted to  483,80 \euro, hence the employer is required to pay approximately 1.451,00 \euro, to the social security institute, which is approximately equal to an average  monthly salary. Moreover, the law introduced a fast-track to try to accelerate the trials in case of dismissal. The main purpose  was to create greater and faster legal certainty in the dismissal
system, especially with regard to the financial consequences of an unfair dismissal.  The proceedings in first instance were split into two phases. The first phase was initiated by means of a complaint lodged at the Labour Court, followed by a judge's order to schedule a summary hearing within forty days of the complaint, and ended with a preliminary, but enforceable, court order upholding or rejecting the claim. The second, optional, phase offered both parties the opportunity to oppose the preliminary rulings within thirty days of the initial judgment being notified; this culminated in a final, first-instance verdict. Afterwards, as under the old procedure, the parties could appeal the judgment at the \textit{Corte di Appello} (second degree) \citep{bijdevaate}.

Specifically, we consider a situation where an employer-initiated individual separation against a blue-collar worker with average tenure in a firm with more than 15 employees is ruled unfair by the judge after a trial. The computation is based on the ex-post firing cost, once the case has been taken to court and the judge has reached the verdict. Obviously, ex-ante the firm does not know with certainty whether any given individual dismissal will be appealed by the worker, and whether the separation will be ruled legitimate.

First of all, the worker should be granted the foregone wages from the separation's day up to the court ruling, while the firm should pay the foregone social insurance contributions augmented by a penalty for delayed payment. In addition, the worker may choose between a severance payments of 15 months or the right of being reinstated by the firm that unlawfully fired him. Finally, all the legal costs should be paid by the firm.

Thus, if we let $n$ be the number of months that it took to reach a court decision, $w$ the gross monthly wage, $ssc$ the social security contribution rate, $pp$ the penalty rate on foregone contributions, $sp$ the mandatory severance payment rate for unfair dismissal and $lc$ the total legal cost, the expected firing costs (EFC) when the worker opts for the severance payment over reinstatement (this happens in over 95\% of the cases), which in our model corresponds to $F\tilde{p}_ix_i$, is
\begin{equation}
EFC=\left[n+(ssc +pp)n+sp+lc\right]w.
\end{equation}
The pure transfer component paid by the firm to the worker is
\begin{equation}
S = (n + \backepsilon ssn + sp)w.
\end{equation}
where $\backepsilon$ is the share of the social security contributions that is rebated to the worker in the
form of increased future pensions. The tax component is
\begin{equation}
T =[(1-\backepsilon)ssn +ppn+lc]w.
\end{equation}

We allow the number of months that it took to reach a court decision to be different whether it is a first degree trial or an appeal: $n_{fd}$ defines the number of months in first degree trials and $n_{a}$  the number of months in case of appeal.
If we ignore discounting, and we denote as $p_f$ the probability of being fired, $p_s$ the probability of suing the company, $p_w$ the probability that the firing is ruled unfair and, $p_a$ be the  probability of appeal,the ex-ante expected firing cost is:
\begin{eqnarray}
EFC &=&p_f\left\{(1-p_s)C_{NA}+p_s\left[(1-p_w)C_L +p_w\left[n_{fd}+(ssc  +pp)n_{fd}+sp+lc\right]w\right] + \right.\\\notag &+& \left. p_s (1-p_w)p_a p_w\left[n_a+(ssc  +pp)n_a+sp+lc\right]w + p_s (1-p_w)p_a (1-p_w)C_{SD}\right\},
\end{eqnarray}
where $C_L$ is the firing costs incurred by the firm when the judge rules the firing legitimate, $C_{NA}$ is cost incurred when the worker does not appeal the firm decision, and $C_{SD}$ is the firing costs incurred by the firm when the judge rules the firing legitimate after an appeal.  Since, in the Italian legislation $C_L = C_{NA}= C_{SD} = 0$, the ex-ante expected firing cost is
\begin{eqnarray}
EFC &=&p_fp_sp_w \left[n_{fd}+(ssc  +pp)n_{fd}+sp+lc\right]w + \\\notag &+& p_fp_s (1-p_w)p_a p_w\left[n_a+(ssc +pp)n_a+sp+lc\right]w.
\end{eqnarray}
We can then compute the firing cost $F$ as a proportion of the value added of the employees (as in the model) as:
%\begin{eqnarray} \nonumber
%F &=& p_f \left\{ \frac{w}{px}\mathbbm{1}^F+\dfrac{p_sp_w \left[n_{fd}+(ss  +pp)n_{fd}+sp+lc\right]w}{\tilde{p}x}\right. + \\ 
%&+& \left. \dfrac{p_s p_a p_w \left[n_{a}+(ss  +pp)n_{a}+sp+lc\right]w}{\tilde{p}x} \right\},
%\end{eqnarray}
\begin{eqnarray}\label{eq:firingcost}
F &=& p_f\bigg\{ \mathbbm{1}^F+p_sp_w \left[n_{fd}+(ssc  +pp)n_{fd}+sp+lc\right] +\\\notag &+& p_s(1-p_w) p_a p_w\left[n_a+(ssc +pp)n_a+sp+lc\right]\bigg\}\left(\dfrac{w}{\tilde{p}x}\right),
\end{eqnarray}
where $\mathbbm{1}^F$ is the additional cost introduced by the Fornero reform in 2013.

Finally, one should recall that most employer-initiated separations do not end up in court since employers and employees may well find a satisfactory settlement before the full trial is over. In the case of an off-court agreement, the parties can save any court penalties that may eventually be imposed by a judge, and all the legal costs linked to the trial.
\begin{table}[!htbp]
	\centering
	\caption{Firing cost components.}
	\label{tab:firingcostcomponents}
	\footnotesize
	\begin{tabular}{l|rrrH}
		\hline \hline
		&Symbol& First Degree&Appeal&Total\\
		\hline
		%Trial length&$n$&12&12&24\\
		%Appeal length&$n_2$&24&&\\
		%Foregone wages&nw&\\
		%Health Insurance&sh&1/12&1/12&2/12\\
		Social Security Contributions&ssc&4/12&4/12&8/12\\
		Sanctions for Delayed Payments&pp&3/12&3/12&6/12\\
		\hline
		Legal Costs&lc&3&3&6\\
		Severance Payments &sp&15&-&15\\
		\hline \hline
	\end{tabular}
\end{table}

\begin{figure}[!htbp]
	\caption{Trial length, probabilities to sue and appeal, and estimated firing costs in Italy in the period 2008-2017.}
	\label{tab:triallengthandprob}
	
	\begin{subfigure}[b]{0.33\textwidth}
		\centering
		\includegraphics[width=\linewidth]{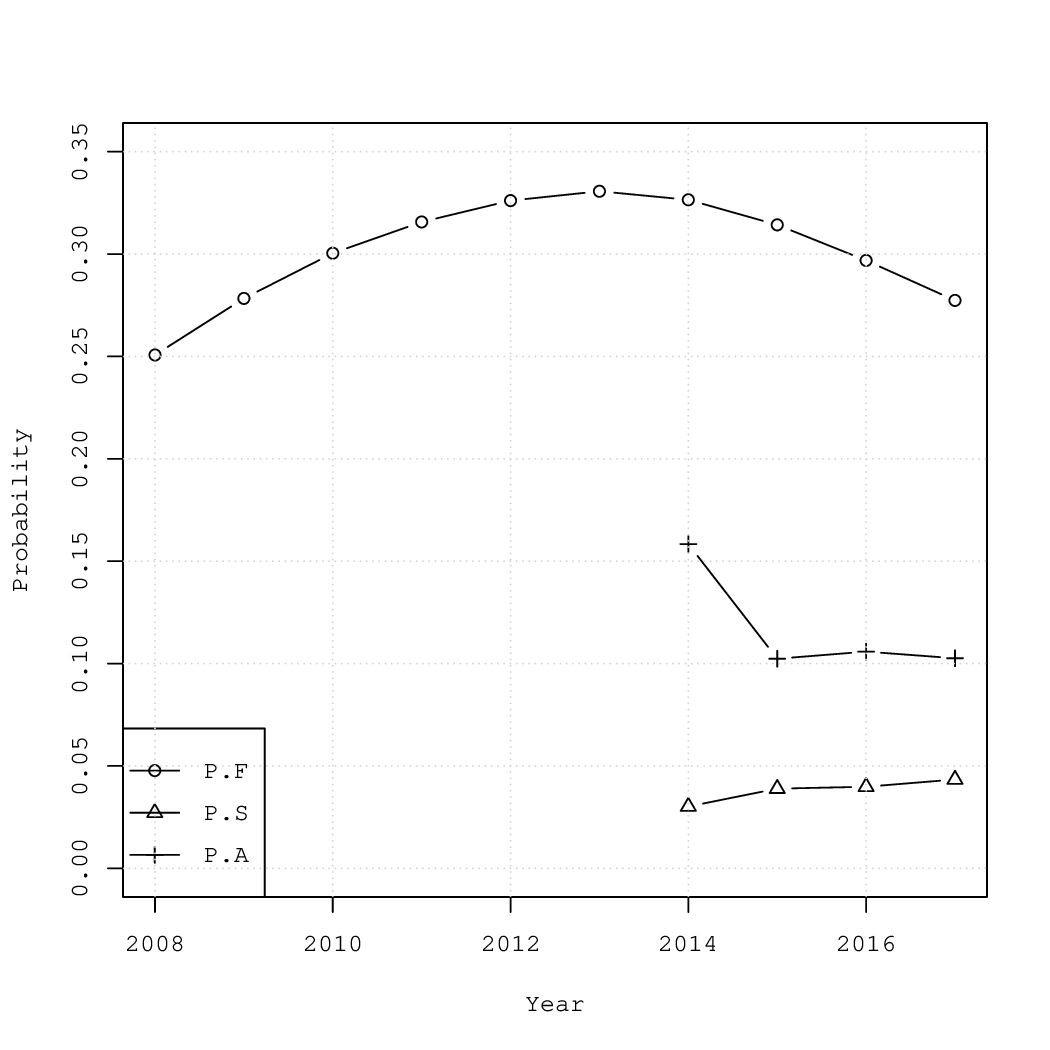}
		\caption{Probabilities to be fired (P.F), to sue (P.S.) and to appeal (P.A.).}
		\label{fig:probabilitytobefiredtosuetoappeal}
	\end{subfigure}
	\begin{subfigure}[b]{0.33\textwidth}
		\centering
		\includegraphics[width=\linewidth]{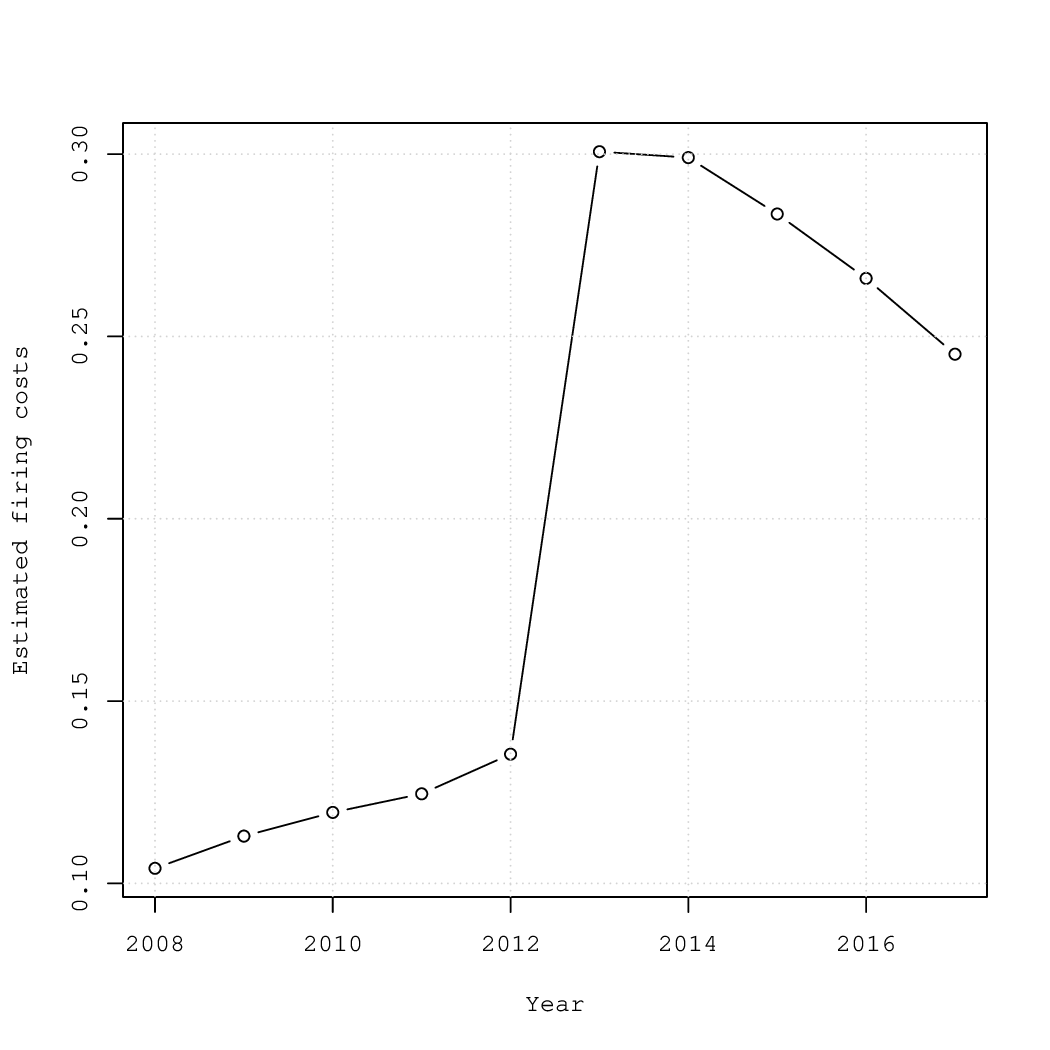}
		\caption{Estimated firing costs (F).}
		\label{fig:estimatedfiringcosts}
	\end{subfigure}
	\begin{subfigure}[b]{0.33\textwidth}
		\centering
		\includegraphics[width=\linewidth]{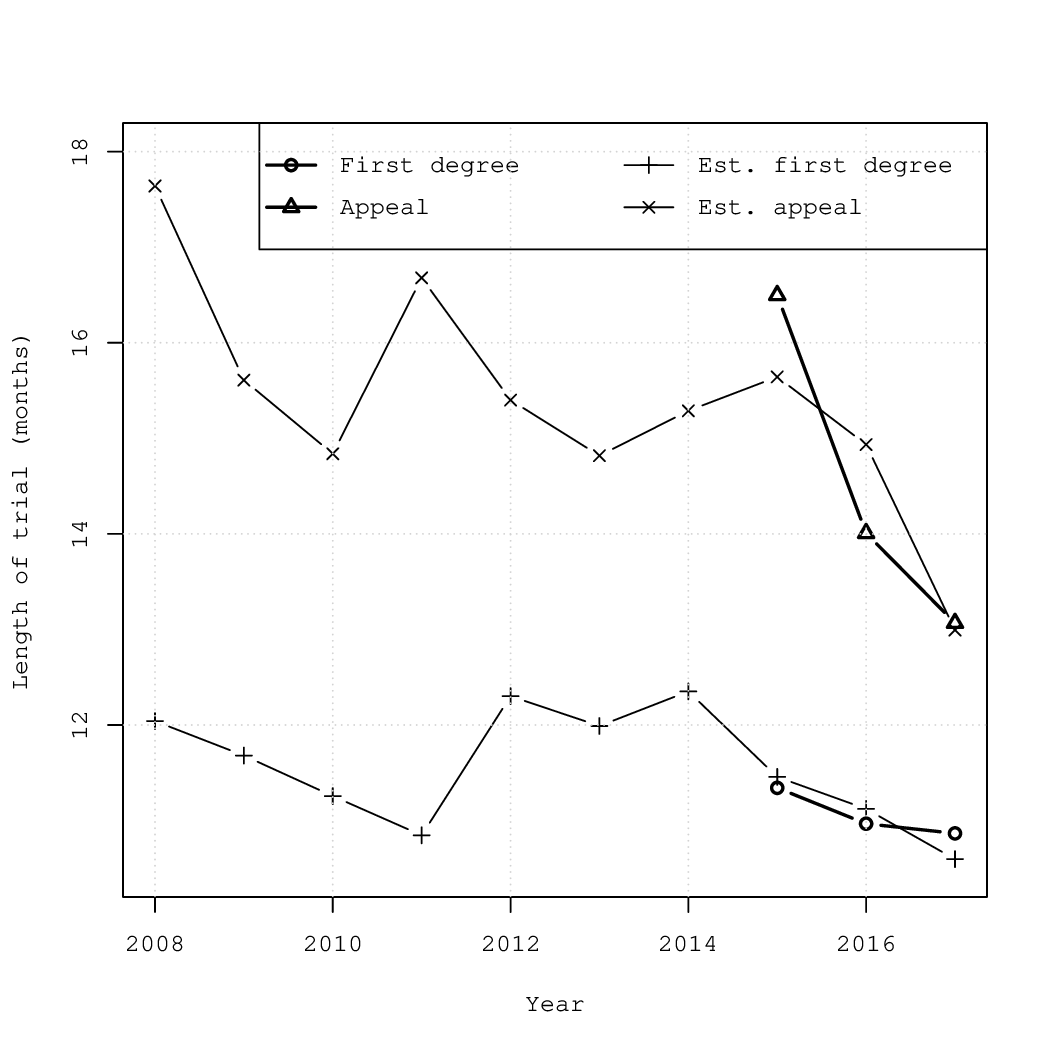}
		\caption{Trial length (in months).}
		\label{fig:trial_length}
	\end{subfigure}
	\vspace{0.1cm}
	\caption*{\textit{Note}: The trial length in Figure (c) refers to the average length of trials in first degree and appeal.\\\textit{Source}: Own calculation based on data from the Italian Department of Justice.}
\end{figure}

Following \cite{garibaldi2002firing}, we set the legal cost and the sanctions for delayed payments equal to three gross monthly salaries (Table \ref{tab:firingcostcomponents}), in line with the evidence provided by \citep{ichino_ll_1996}. The severance payments are set by law equal to 15 monthly salaries. Given the large uncertainty about the judges' decisions \citep{ichino_ll_1996, ichino2003judges}, we assume symmetry in the probability that the judges rule in favor of the worker, i.e., $p_w=0.5$. Using Equation (\ref{eq:firingcost}) and the data reported in Table \ref{tab:firingcostcomponents}, we are able to compute the estimated firing costs (Table \ref{tab:triallengthandprob}).

The average trail length is calculated using data from the Italian Ministry of Justice on ensued, pending and settled trials fro the years 2014-2017 (Figure \ref{tab:triallengthandprob}). Specifically, we compute the monthly probability of closing a trial ($P^m_{ct}$) assuming a uniform distribution of ensued and settled trials over the months, as
\begin{equation} 
P^m_{ct}=\dfrac{ST_{m}}{PT_{t-1}+ET_m},
\end{equation}
where $ST_{m}$ is the flow of settled trials in month 1 (January), $PT_{t-1}$ is the stock of pending trials at time $t-1$ and $ET_m$ is the flow of ensued trials in month 1 (January). The average length of trials is therefore computed as 
\begin{equation} 
ATL=\dfrac{1}{P^m_{ct}},
\end{equation}
assuming that the trials are distributing according to a Poisson process (Figure \ref{tab:triallengthandprob}).

Finally, in 2015 the government implement a new reform to create incentives for firms to hire employees on a permanent basis, the \textit{Jobs Act}. The reform introduced a new form of open-ended contract with firing cost increasing with tenure, reducing \textit{de facto} the firing costs to be paid by the employer in case of unfair dismissal. As part of the reform, employers and employees can also opt for negotiations right after the firing to avoid the legal route, which could be costly, long and uncertain for both parties. These negotiations allow employers to pay within two months from the event (firing of the employee) half the amount it would have had to pay in case of a judicial ruling in favor of the employee. %Starting from 2015, the number of negotiations has decreased both in first and second instance trials.
However, these new rules apply only to the new open-ended contract and hence we believe that for the years 2015-2017 the change in the average aggregate firing cost has been minimal.

\subsection{The government expenditure \label{govExp}}

\begin{figure}[h!]
	\centering
	\caption{Government expenditure for public goods and services as a share of GVA in Italy in the period 200-2017.}
	\label{fig:ratiopublicexpgoodsservicesongva}
	\includegraphics[width=0.4\linewidth]{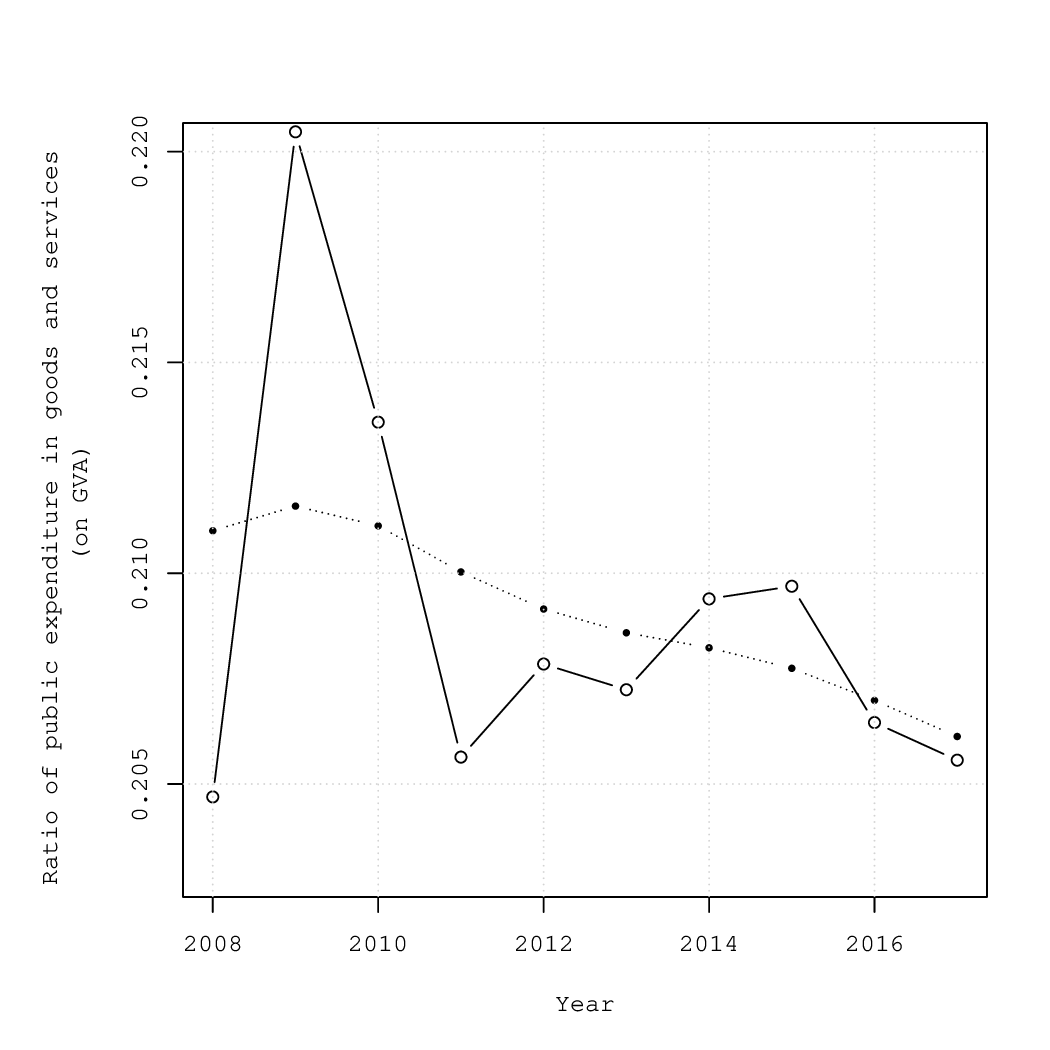}
	\caption*{\textit{Note}: The dotted line represents the smoothed series obtained using the Hodrick-Prescott filter.\\
		\textit{Source}: Eurostat.}
\end{figure}

Eurostat provides data on government expenditure by function in millions \euro. To compute the government expenditure for public goods and services we sum the relevant components reported in Table \ref{apptab:govExpPublicGoodsServ}.
\begin{table}[!htbp]
	\centering
	\caption{The taxonomy of Government expenditure for public goods and services.}
	\label{apptab:govExpPublicGoodsServ}
	\footnotesize
	\begin{tabular}{c|l}
		\hline
		\hline
		EUROSTAT Code& Function\\
		\hline
		GF02 & Defence
		\\
		GF03 & Public order and safety
		\\
		GF05 & Environmental protection
		\\
		GF06 & Housing and community amenities
		\\
		GF07 & Health
		\\
		GF08 & Recreation, culture and religion
		\\
		GF09 & Education
		\\
		GF1004 & Social protection - Family and children
		\\
		GF1006 & Social protection - Housing
		\\
		GF1007 & Social protection - Social exclusion not elsewhere classified (n.e.c.)\\
		GF1008 & Social protection - $R\&D$ Social protection
		\\
		GF1009 & Social protection not elsewhere classified (n.e.c.)\\		\hline \hline
		\multicolumn{2}{l}{\textit{Source}: Eurostat.}
	\end{tabular}
\end{table}
We then compute the real government expenditure by dividing the nominal government expenditure by the CPI index, as provided by Italian Institute of Statistics. Finally, we compute the real government expenditure for public goods and services as a share of GVA (Figure \ref{fig:ratiopublicexpgoodsservicesongva}).

\subsection{The unemployment benefits} \label{unBenefits}
\begin{figure}[!htbp]
	\centering
	\caption{Government expenditure for unemployment benefits as percentage of gross real wages in Italy in the period 2008-2017.}
	\label{fig:ratiopublicexpunbenonrealwages}
	\includegraphics[width=0.4\linewidth]{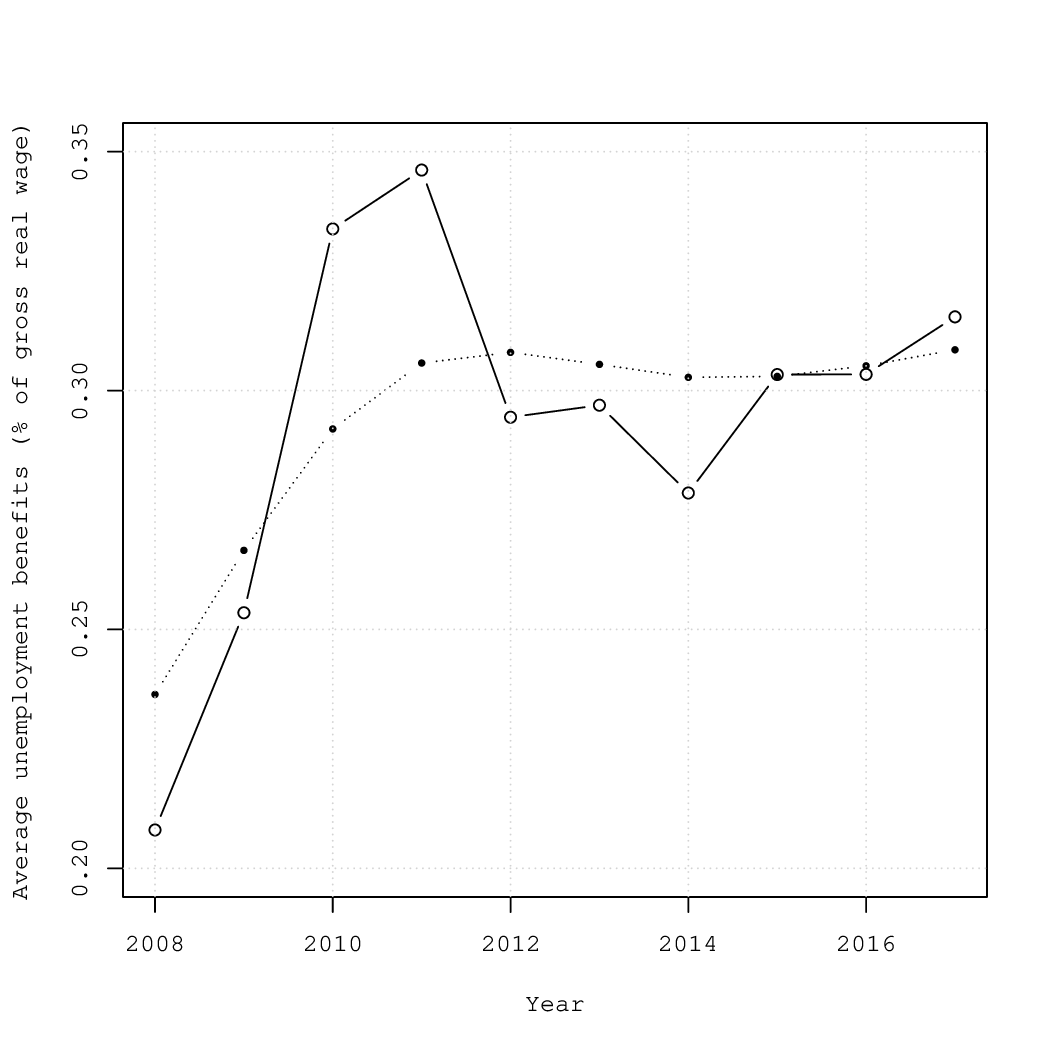}
	\caption*{\textit{Note}: The dotted line represents the smoothed series obtained using the Hodrick-Prescott filter.\\ \textit{Source}: Eurostat.}
\end{figure}

Eurostat provides data on government expenditure for unemployment benefits in millions \euro \ (EUROSTAT code GF1005). We compute the ratio of the government expenditure for unemployment benefits and the number of unemployed in the economy.
Figure \ref{fig:ratiopublicexpunbenonrealwages} reports the average government expenditure for unemployment benefits as share of gross real wages in Italy in the period 2008-2017.

\subsection{The direct and indirect taxes, and the social security contributions \label{indirectTax}}

\begin{figure}[!htbp]
	\centering
	\caption{Indirect taxes and social security contributions in Italy in the period 2008-2018.}
	\begin{subfigure}[b]{0.40\textwidth}
		\includegraphics[width=\linewidth]{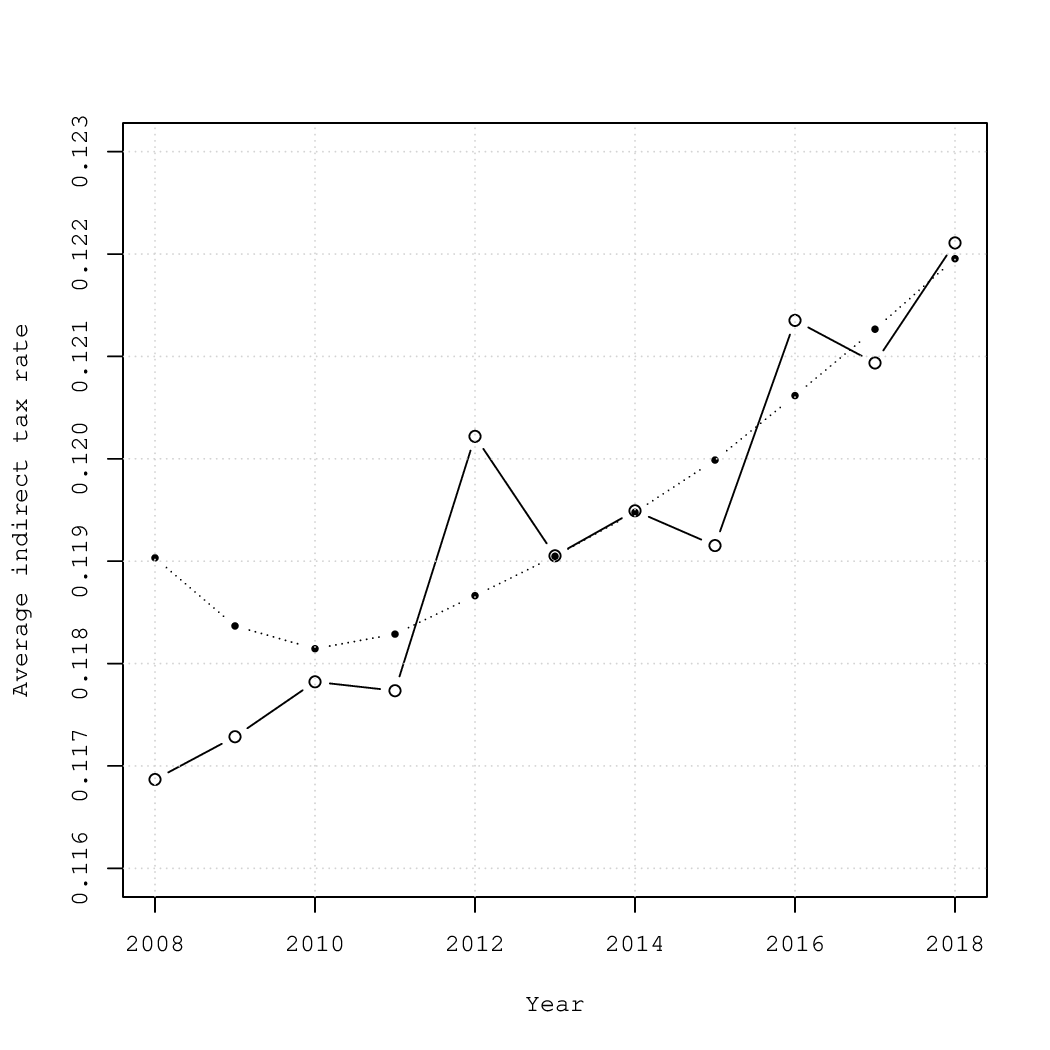}
		\caption{Indirect tax rate.}
		\label{fig:indTaxRate}
	\end{subfigure}
	\begin{subfigure}[b]{0.40\textwidth}
		\centering
		
		\includegraphics[width=\linewidth]{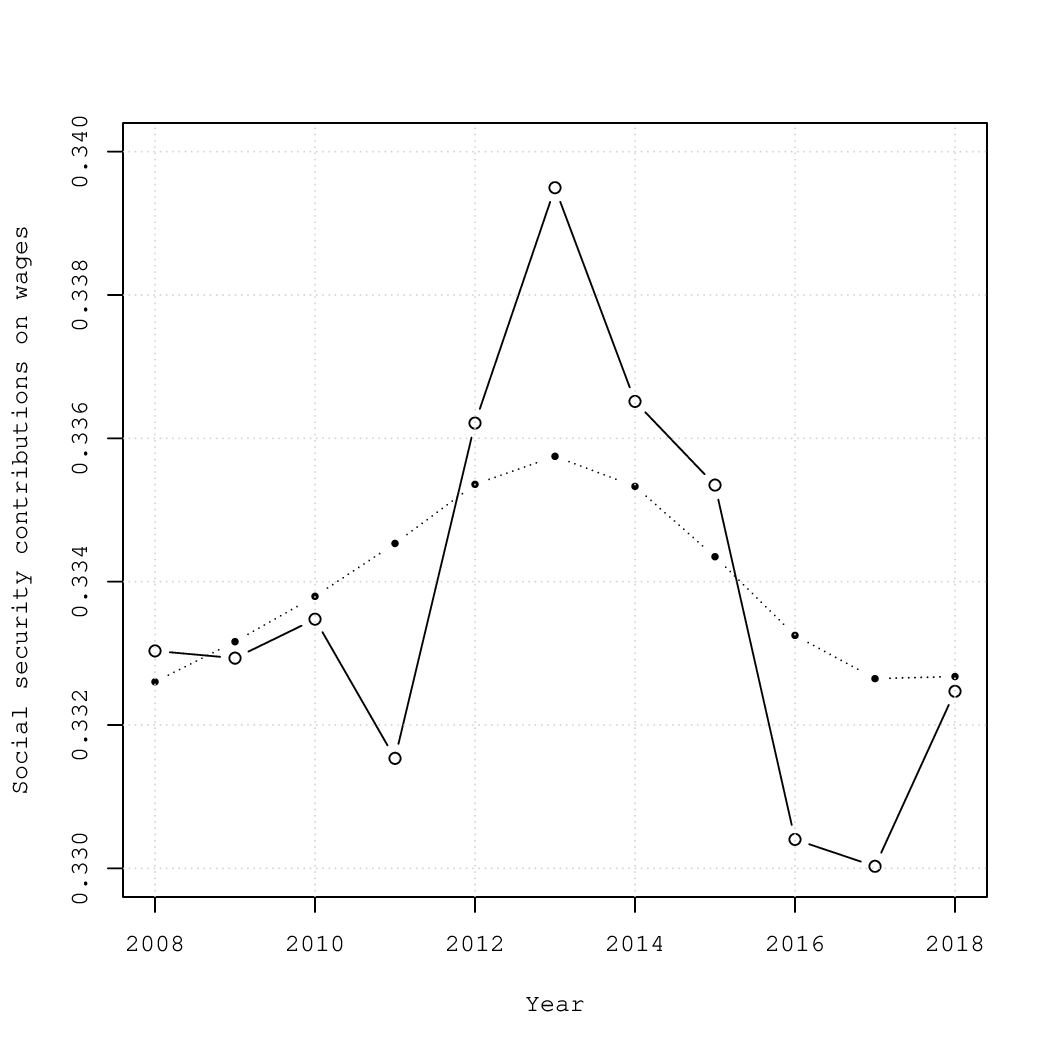}
		\caption{Social security contribution rate.}
		\label{fig:SS}
		
	\end{subfigure}
	\caption*{\textit{Note}: The dotted line represents the smoothed series obtained using the Hodrick-Prescott filter. In Figure (b) it is reported the share of social security contributions on wages. \\\textit{Source}: Italian Ministry of Economics and Finance.}
\end{figure}

In 2017, the \textit{marginal} income tax rates for the \textit{median} employee and self-employed in Italy were 43\% and  42\%, respectively \citep[Figures 5,6]{di2017effective}. Since the direct taxation system has not changed significantly in the period under consideration, in the estimates we set $t=0.43$  for all years. 
The indirect tax rate (Figure \ref{fig:indTaxRate}) is computed by  dividing the total revenues from indirect taxation by the GDP at current prices, as reported by the Italian Ministry of Economics and Finance and the  Italian  Institute of Statistics (ISTAT). Similarly, the social security contribution rate (Figure \ref{fig:SS}) is computed by  dividing the total revenues from social security contributions,  as reported by the Italian Ministry of Economics and Finance by the wage bill, as reported in the national accounts by  the  Italian  Institute of Statistics (ISTAT). 

\subsection{The inflow rate of  non-natives \label{app:ExitRateImmigrant}}
In our model the total inflow of non-natives (TII) is given by:
\begin{equation}
TII = \eta\left(\sigma_{h,I} + \sigma_{l,I}\right),
\label{eq:totInflowImmigrants}
\end{equation}
which is the product between the rate $\eta$ at which non-native employees arrive in the country and the stock of all non-native employees in the country $\sigma_{h,I} + \sigma_{l,I}$.
Using the stock of employed non-native employees (Equations (\ref{eq:equilibriumEmploymentImmigrants})) and (\ref{eq:totInflowImmigrants})), we can derive an expression for the rate $\eta$ at which non-native employees join the country:
\begin{equation}
\eta = \dfrac{TII - \lambda \left[ e_{h,I}\delta_{h,I}/\kappa_{h,I}\theta_h q\left(\theta_h\right) + e_{l,I}\delta_{l,I}/\kappa_{l,I}\theta_lq\left(\theta_l\right)\right]}{e_{h,I}\left[\kappa_{h,I}\theta_h q\left(\theta_h\right)+\delta_{h,I}\right]/\kappa_{h,I}\theta_hq\left(\theta_h\right) + e_{l,I}\left[\kappa_{l,I}\theta_l q\left(\theta_l\right)+\delta_{l,I}\right]/\kappa_{l,I}\theta_lq\left(\theta_l\right)}.
\label{eq:calEtaViaEmploymentRate}
\end{equation}

\begin{figure}[!htbp]
	\caption{Non-native inflow and outflow rates in Italy in the period 2004-2018.}
	\label{inflowoutflowrates}
	\centering
	\begin{subfigure}[b]{0.40\textwidth}
		\label{fig:differentestimateeta}
		\centering
		\includegraphics[width=\linewidth]{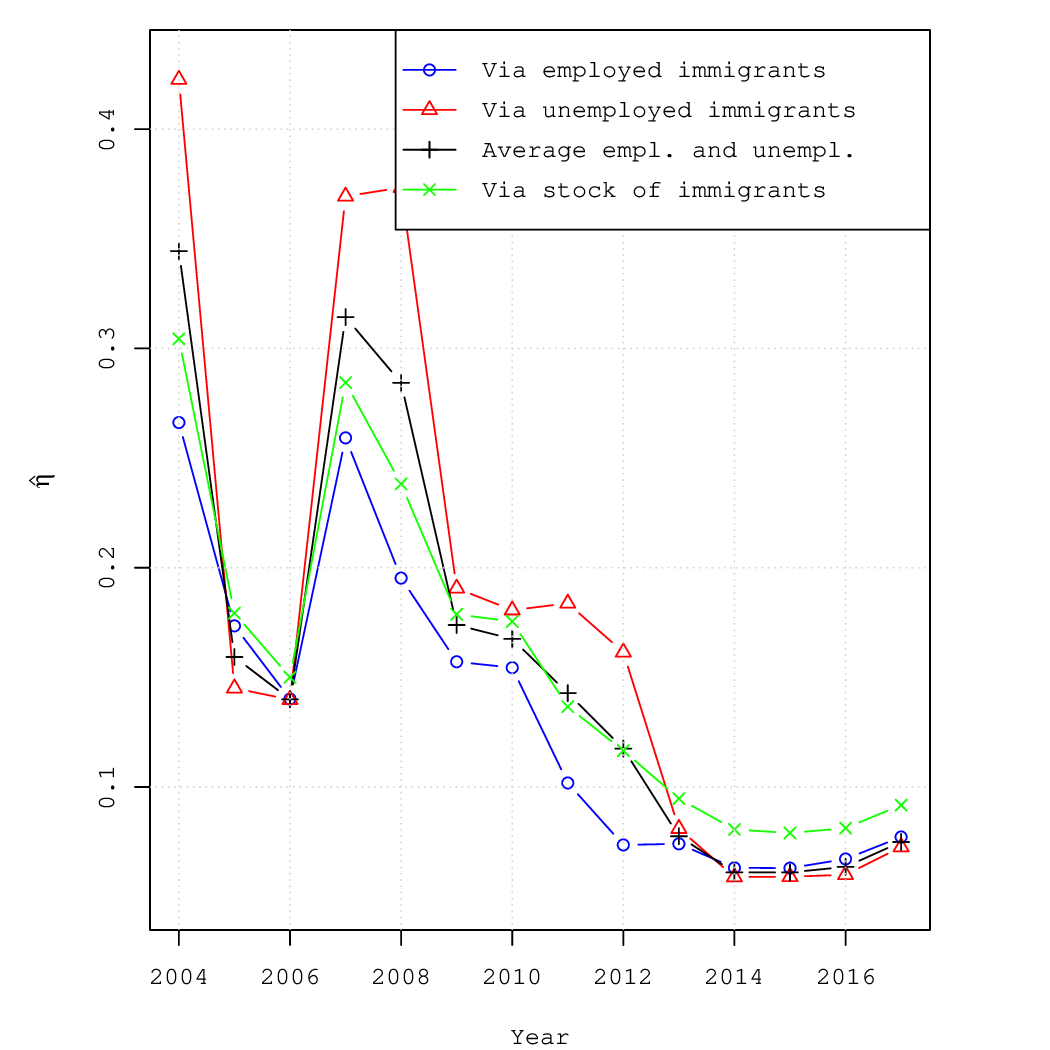}
		\caption{Non-native inflow rate \\(different approaches compared)}
	\end{subfigure}
	\begin{subfigure}[b]{0.40\textwidth}
		\centering
		\label{fig:rateinflowsoutflowsnonativeworkers}
		\includegraphics[width=\linewidth]{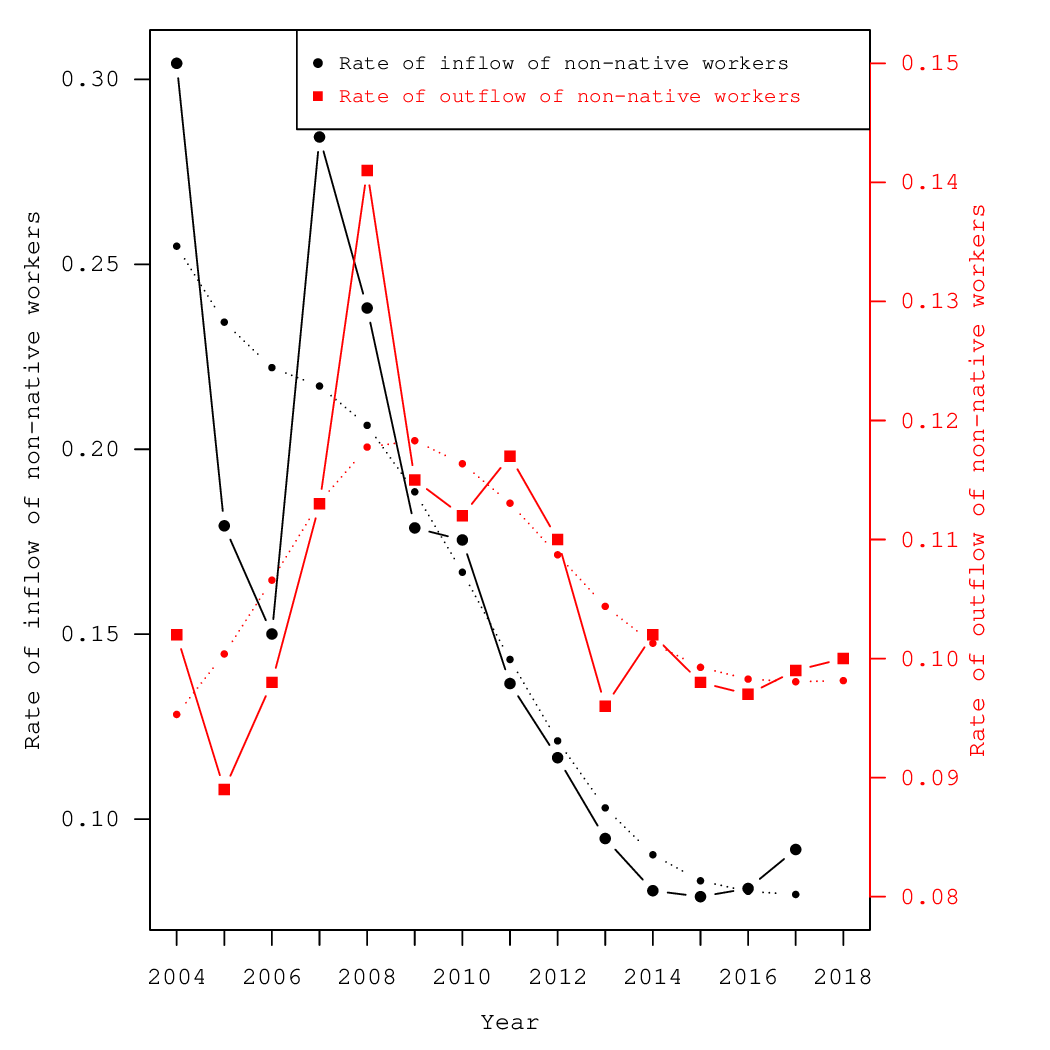}
		\caption{Inflow and outflow rates of \\non-native employees}
	\end{subfigure}
	\vspace{0.5cm}
	\caption*{\textit{Source}: Own calculations based on the model.}
\end{figure}

The same rate can be computed using the stock of unemployed non-native employees (Equations (\ref{eq:equilibriumUnemploymentImmigrants}) and (\ref{eq:totInflowImmigrants}):
\begin{equation}
\eta = \dfrac{TII - \lambda \left( u_{h,I} + u_{l,I}\right)}{u_{h,I}\left[\kappa_{h,I}\theta_h q\left(\theta_h\right)+\delta_{h,I}\right]/\delta_{h,I} + u_{l,I}\left[\kappa_{l,I}\theta_l q\left(\theta_l\right)+\delta_{l,I}\right]/\delta_{l,I}}.
\label{eq:calEtaViaUnemploymentRate}
\end{equation}

Figure \ref{inflowoutflowrates} reports the estimates of $\eta$ based on Equations (\ref{eq:equilibriumUnemploymentImmigrants}), (\ref{eq:calEtaViaEmploymentRate}) and (\ref{eq:calEtaViaUnemploymentRate}) using data as provided by the OECD statistics.

\clearpage

%\subsection{Tax subsidies and public trasfers \label{taupar}}

%In order to match the direct taxes and the tax gap as observed in the data, we estimate the tax subsidy $\tilde{\tau}$ and the lump-sum transfer $\tilde{m}$. The tax gap is determined as the sum of  three variables: i) the charges deductible from taxable income; ii) tax deductions; and, possibly, iii) tax evasion; , i.e. the difference between the theoretical total tax burden calculated on the basis of the actual tax system and the effective revenues collected by Government. Below we provide some evidence on the tax gap. 

\subsection{Tax gap in Italy}\label{app:TaxGap}

The ‘tax gap’ is defined in the literature as the difference between the amount of taxes that the government should collect based upon the laws it has in operation in an annual accounting period and the actual amount of taxes paid during that same period. The \textit{tax gap} (TG) includes two components: the \textit{compliance gap} and the \textit{policy gap}. While the former depends on the tax compliance of individuals who are required to pay taxes in accordance with the legislation (direct tax evasion), the latter is influenced by (fiscal) policy decisions, which reduce the tax revenues (tax deductions). Table \ref{tab:taxgap} reports the net direct taxes, tax deductions and direct tax evasion for the years 2010-2017. We compute the tax gap as the sum of the tax deductions and direct tax evasion as a share of net direct taxes.
The tax gap for Italy ranges from $64\%$ to $68\%$ of the net direct taxes in the period considered.

\begin{table}[!htbp]
	\centering
	\caption{Net direct taxes, tax deductions, direct tax evasion and tax gap in Italy.}
	\label{tab:taxgap}
	\scriptsize
	\begin{tabular}{lHHcccccccc}
		\hline
		\hline\\[-1.8ex]
		&2008& 2009& 2010&2011&2012&2013&2014&2015&2016&2017\\
		\hline \\[-1.8ex]
	Net direct taxes (A)&&&149,442
&	152,219&
152,270
&152,238
&151,185
&155,157
&156,047
&157,516\\
		 Tax deductions (B) &&&62,482&62,112&62,984&64,406&66,150&66,069&67,553&68,985\\
		 $\%$ Tax deductions (C=B/A) &&&41.80&
		 40.80&
		 41.36&
		 42.30&
		 43.75&
		 42.58&
		 43.29&
		 43.79\\
		 Direct tax evasion (D)&NA&NA&33,851&
		 		 35,990&
		 		 35,512&
		 		 37,523&
		 		 35,837&
		 		 37,183&
		 		 38,216
&
		 		 38,486\\
		 		 $\%$ Tax evasion (E= D/A)&&&22.65&
		 		 23.64&
		 		 23.32&
		 		 24.64&
		 		 23.70&
		 		 23.96&
		 		 24.49&
		 		 24.43\\
		 		 		 $\%$ Tax gap (TG=C+E) &&&64.46&
		 		 		 64.45&
		 		 		 64.69
&
		 		 		 66.95&
		 		 		 67.46
&
		 		 		 66.55&
		 		 		 67.78&
		 		 		 68.23\\
	\hline \hline \\[-1.8ex]
\multicolumn{11}{l}{\scriptsize{\textit{Note}: Net Direct Taxes, Tax Deductions and Direct Tax Evasion are reported in million euros.}}\\ \multicolumn{11}{l}{\scriptsize{\textit{Source}: Italian Ministry of Economics and Finance.}}
	\end{tabular}
\end{table}

\clearpage

\section{The system of equations defining the model's equilibrium (only for referees) \label{sec:systemEqqEquilibrium}}
	
The equilibrium is the solution of the following system of 24 nonlinear equations in 24 variables ($i \in \left\{h,l\right\})$.

\subsection{Equilibrium in the labour markets}
\begin{scriptsize}
	\begin{equation}
		e_{i,N}=\sigma_{i,N} \left[\dfrac{\kappa_{i,N}\theta_i q\left(\theta_i\right)}{\delta_{i,N} + \kappa_{i,N} \theta_i q\left(\theta_i\right)}\right];
	\end{equation}
	\begin{equation}
		u_{i,N}=\sigma_{i,N} \left[\dfrac{\delta_{i,N}}{\delta_{i,N} + \kappa_{i,N}\theta_i q\left(\theta_i\right)}\right];
	\end{equation}
	\begin{equation}
		e_{i,I}=\sigma_{i,I} \left\{ \dfrac{\kappa_{i,I}\theta_i q\left(\theta_i\right)}{\delta_{i,I} + \kappa_{i,I}\theta_i q\left(\theta_i\right)} \right\};
	\end{equation}
	\begin{equation}
		u_{i,I}=\sigma_{i,I} \left\{  \dfrac{ \delta_{i,I}}{\delta_{i,I} +\kappa_{i,I}\theta_i q\left(\theta_i\right)} \right\};
	\end{equation}
	
	\subsection{Wage equations}
	\begin{eqnarray}\notag
		\tilde{w}_{i,N}&=&\notag\beta_j (1-t)\left\{ \dfrac{(r+\delta_{i,N}+ \kappa_{i,N} \theta_i q\left(\theta_i\right))   }{(1-t)[(r+\delta_{i,N})(1-b(1-\beta_j))+\beta_j\kappa_{i,N} \theta_i q\left(\theta_i\right)]}\right\}\tilde{p}_i x_i + \\\notag &-& \dfrac{(1-\beta_j)(r+\delta_{i,N})\left(1-t\right)(1-b) }{(1-t)[(r+\delta_{i,N})(1-b(1-\beta_j))+\beta_j\kappa_{i,N} \theta_i q\left(\theta_i\right)]}\tilde{\tau} + \\ &+&  \left\{\dfrac{ \beta_j  r(r+\delta_{i,N}+ \kappa_{i,N} \theta_i q\left(\theta_i\right))    + \phi   (r+\delta_{i,N})(1-\beta_j)[(r+ \kappa_{i,N} \theta_i q\left(\theta_i\right))]  }{(1-t)[(r+\delta_{i,N})(1-b(1-\beta_j))+\beta_j\kappa_{i,N} \theta_i q\left(\theta_i\right)]} \right\}\tilde{p}_i x_i F;
	\end{eqnarray}
	\begin{eqnarray}\nonumber
	\tilde{w}_{i,I}\notag&=& \beta_I    (1-t)\left\{\dfrac{(r + \lambda) (r+\delta_{i,I})+ r \kappa_{i,I} \theta_i q\left(\theta_i\right)}{\left(1-t\right)\left\{ (r+\delta_{i,I})\left[(\lambda+r)-rb(1-\beta)\right]+ \beta_Ir\kappa_{i,I} \theta_i q\left(\theta_i\right)  \right\}}\right\}\tilde{p}_i x_i +\\&-&\notag \left\{ \dfrac{(1-\beta_I)(r+\delta_{i,I})\left(1-t\right)\left[(1-b)r+\lambda\right]}{\left(1-t\right)\left\{ (r+\delta_{i,I})\left[(\lambda+r)-rb(1-\beta)\right]+ \beta_Ir\kappa_{i,I} \theta_i q\left(\theta_i\right)  \right\}} \right\}  \tilde{\tau} +\\\notag&+& \left\{ \dfrac{\lambda r(1-\beta_I) (r+\delta_{i,I})}{\left(1-t\right)\left\{ (r+\delta_{i,I})\left[(\lambda+r)-rb(1-\beta)\right]+ \beta_Ir\kappa_{i,I} \theta_i q\left(\theta_i\right)  \right\}} \right\} W_{i,FC}  + \\&+&\notag\left\{\dfrac{\phi (r+\delta_{i,I})(1-\beta_I)(r + \lambda) \left( r+ \kappa_{i,I} \theta_i q\left(\theta_i\right)- \lambda\right)}{\left(1-t\right)\left\{ (r+\delta_{i,I})\left[(\lambda+r)-rb(1-\beta)\right]+ \beta_Ir\kappa_{i,I} \theta_i q\left(\theta_i\right)  \right\}}  \right.+\\&+&\notag\left. \dfrac{\beta_I  r\left[(r+\delta_{i,I})(r + \lambda)+ r\kappa_{i,I} \theta_i q\left(\theta_i\right)\right]}{\left(1-t\right)\left\{ (r+\delta_{i,I})\left[(\lambda+r)-rb(1-\beta)\right]+ \beta_Ir\kappa_{i,I} \theta_i q\left(\theta_i\right)  \right\}}\right\}\tilde{p}_i x_i F+\\
	&-& \left\{ \dfrac{\lambda(1-\beta_I)(r+\delta_{i,I})  }{\left(1-t\right)\left\{ (r+\delta_{i,I})\left[(\lambda+r)-rb(1-\beta)\right]+ \beta_Ir\kappa_{i,I} \theta_i q\left(\theta_i\right)  \right\}} \right\} \left(\iota\nu + \tilde{m} \right).\label{wageI_II}
	\end{eqnarray}
	
	\subsection{Price equations for the two intermediate goods (demand side)}
	\begin{eqnarray} \nonumber
		\tilde{p}_{i} & = &\dfrac{q(\theta_i)\pi_{i,N} \kappa_{i,N}(r+\delta_{i,I})(1-t)\tilde{w}_{i,N}}{x_i\left\{ q(\theta_i)\left[\pi_{i,N} \kappa_{i,N}(r+\delta_{i,I})(1-t-\delta_{i,N}F)+(1-\pi_{i,N})\kappa_{i,I}(r+\delta_{i,N})(1-t-\delta_{i,I}F)\right]-c(r+\delta_{i,N})(r+\delta_{i,I})\right\}} + \\ &+&\dfrac{q(\theta_i)(1-\pi_{i,N})\kappa_{i,I}(r+\delta_{i,N})(1-t)\tilde{w}_{i,I}}{x_i\left\{ q(\theta_i)\left[\pi_{i,N} \kappa_{i,N}(r+\delta_{i,I})(1-t-\delta_{i,N}F)+(1-\pi_{i,N})\kappa_{i,I}(r+\delta_{i,N})(1-t-\delta_{i,I}F)\right]-c(r+\delta_{i,N})(r+\delta_{i,I})\right\}};
	\end{eqnarray}

\subsection{Equilibrium in the intermediate good markets}
\begin{eqnarray}
x_{h}e_{h}=q_{h}&=&\left[\frac{\tilde{p}_h}{\alpha_{HS} \tilde{p}_{HS} }\right]^{1/(\rho_{HS}-1)}q_{HS}+\left[\frac{\tilde{p}_h}{\alpha_{LS} \tilde{p}_{LS} }\right]^{1/(\rho_{LS}-1)}q_{LS},\\
x_{l}e_{l}=q_{l}&=&\left[\frac{\tilde{p}_l}{\left(1-\alpha_{HS}\right)\tilde{p}_{HS} }\right]^{1/(\rho_{HS}-1)}q_{HS}+\left[\frac{\tilde{p}_l}{\left(1-\alpha_{LS}\right) \tilde{p}_{LS} }\right]^{1/(\rho_{LS}-1)}q_{LS}.
\end{eqnarray}

\begin{eqnarray}
q_{HS}=\frac{ x_h e_h \left[\frac{\tilde{p}_l}{\tilde{p}_{LS} (1-\alpha_{LS})}\right]^{1/(\rho_{LS}-1)}  - x_le_l \left[\frac{\tilde{p}_h}{\tilde{p}_{LS} \alpha_{LS}}\right]^{1/(\rho_{LS}-1)} }{\left[\frac{\tilde{p}_h}{\tilde{p}_S \alpha_{HS}}\right]^{1/(\rho_{HS}-1)}\left[\frac{\tilde{p}_l}{\tilde{p}_{LS} (1-\alpha_{LS})}\right]^{1/(\rho_{LS}-1)} -\left[\frac{\tilde{p}_h}{\tilde{p}_{LS} \alpha_{LS}}\right]^{1/(\rho_{LS}-1)}\left[\frac{\tilde{p}_l}{\tilde{p}_{HS} (1-\alpha_{HS})}\right]^{1/(\rho_{HS}-1)}}\\
q_{LS}=\frac{x_l e_l \left[\frac{\tilde{p}_h}{\tilde{p}_{HS} \alpha_{HS}}\right]^{1/(\rho_{HS}-1)} - x_h e_h\left[\frac{\tilde{p}_l}{\tilde{p}_{HS} (1-\alpha_{HS})}\right]^{1/(\rho_S-1)} }{\left[\frac{\tilde{p}_h}{\tilde{p}_{HS} \alpha_{HS}}\right]^{1/(\rho_{HS}-1)}\left[\frac{\tilde{p}_l}{\tilde{p}_{LS} (1-\alpha_{LS})}\right]^{1/(\rho_{LS}-1)} -\left[\frac{\tilde{p}_h}{\tilde{p}_{LS} \alpha_{LS}}\right]^{1/(\rho_{LS}-1)}\left[\frac{\tilde{p}_l}{\tilde{p}_{HS} (1-\alpha_{HS})}\right]^{1/(\rho_{HS}-1)}}
\end{eqnarray}

\subsection{Equilibrium in the final good markets}
\begin{eqnarray}\label{appJ:priceHpriceL}
\tilde{p}_{HS}&=&\left\{\gamma^{1/(1-\rho)}+(1-\gamma)\gamma^{\rho/(1-\rho)}\left[\frac{(1-g_{LS})q_{LS}}{(1-g_{HS})q_{HS}}\right]^{\rho}\right\}^{(1-\rho)/\rho}\\ \tilde{p}_{LS}&=&\left\{(1-\gamma)^{1/(1-\rho)}+\gamma(1-\gamma)^{\rho/(1-\rho)}\left[\frac{(1-g_{HS})q_{HS}}{(1-g_{LS})q_{LS}}\right]^{\rho}\right\}^{(1-\rho)/\rho}
\end{eqnarray}

\subsection{Government expenditure in public goods}	
\begin{equation}
		\iota \nu  =  \dfrac{\iota    [g_{LS}\tilde{p}_{LS}q_{LS} +g_{HS}\tilde{p}_{HS} q_{HS}] }{\left(\sigma_{h,N}+\sigma_{l,N}+e_{h,I}+e_{l,I}+u_{h,I}+u_{l,I} + \chi + IP  \right)^\zeta}
		\label{app:budget}
\end{equation}

\subsection{Government lump-sum transfer}	
\begin{eqnarray}
		 \tilde{m} &=&   \left(\dfrac{taxgap}{\sigma_{h,N}+\sigma_{l,N}+\sigma_{h,I}+\sigma_{l,I}+ \chi}\right)\left\{  t\left[\tilde{p}_{HS} q_{HS} +  \tilde{p}_{LS} q_{LS} + b\left( \tilde{w}_{h,N}u_{h,N} + \tilde{w}_{h,I}u_{h,I} + \tilde{w}_{l,N}u_{l,N} + \tilde{w}_{l,I}u_{l,I}\right) \right] \right. + \\ \notag
		 &-& \left. \left(1-t\right)\tilde{\tau}\left[e_{l,N}+e_{l,I}+e_{h,N}+e_{h,I} + b\left(u_{l,N}+u_{l,I}+u_{h,N}+u_{h,I}\right) \right]
		   \right\} 
		\label{app:lumpSumTrasfer}
\end{eqnarray}
	 
\end{scriptsize}

\subsection{Relationship between final and intermediate good prices}
\begin{eqnarray}
\tilde{p}_{HS} &=& \left[\alpha_{HS}^{\frac{1}{1-\rho_{HS}}} \tilde{p}_h^{\frac{\rho_{HS}}{\rho_{HS} - 1}} +\left(1-\alpha_{HS}\right)^{\frac{1}{1-\rho_{HS} }} \tilde{p}_l^{\frac{\rho_{HS}}{\rho_{HS} - 1}} \right]^{\frac{\rho_{HS}-1}{\rho_{HS}}} \\
\tilde{p}_{LS} &=& \left[\alpha_{LS}^{\frac{1}{1-\rho_{LS}}} \tilde{p}_h^{\frac{\rho_{LS}}{\rho_{LS} - 1}} +\left(1-\alpha_{LS}\right)^{\frac{1}{1-\rho_{LS} }} \tilde{p}_l^{\frac{\rho_{LS}}{\rho_{LS} - 1}} \right]^{\frac{\rho_{LS}-1}{\rho_{LS}}}
\label{app:constraintsDistributionTotalOutput}
\end{eqnarray}

\subsection{Production function of final goods}
\begin{eqnarray}
		q_{HS} &=& \left[\alpha_{HS} q_{h,HS}^{\rho_{HS}}+(1-\alpha_{HS})q_{l,HS}^{\rho_{HS}}\right]^{1/\rho_{HS}}\\
		q_{LS} &=& \left[\alpha_{LS} q_{h,LS}^{\rho_{LS}}+(1-\alpha_{LS})q_{l,LS}^{\rho_{LS}}\right]^{1/\rho_{LS}}
		\label{app:productionFinalGoods}
\end{eqnarray}

\subsection{Price equations for intermediate goods (supply side)}
 \begin{eqnarray}
	p_{h}&=&p_{HS}\alpha_{HS} q_{h,HS}^{\rho_{HS}-1} [\alpha_{HS} q_{h,HS}^{\rho_{HS}}+(1-\alpha_{HS}) q_{l,HS}^{\rho_{HS}}]^{(1-\rho_{HS})/\rho_{HS}},\\
		p_{h}&=&p_{LS}\alpha_{LS} q_{h,HS}^{\rho_{LS}-1} [\alpha_{LS} q_{h,{LS}}^{\rho_{LS}}+(1-\alpha_{LS}) q_{l,{LS}}^{\rho_{LS}}]^{(1-\rho_{LS})/\rho_{LS}}\\
	p_{l}&=&p_{HS}(1-\alpha_{HS}) q_{l,HS}^{\rho_{HS}-1} [\alpha_{HS} q_{h,HS}^{\rho_{HS}}+(1-\alpha_{HS}) q_{l,HS}^{\rho_{HS}}]^{(1-\rho_{HS})/\rho_{HS}}\\
	p_{l}&=&p_{LS}(1-\alpha_{LS}) q_{l,LS}^{\rho_{LS}-1} [\alpha_{LS} q_{h,LS}^{\rho_{LS}}+(1-\alpha_{LS}) q_{l,LS}^{\rho_{LS}}]^{(1-\rho_{LS})/\rho_{LS}}.
\end{eqnarray}
%from which, in real prices (e.g. $p_{v} \equiv P \tilde{p}_v$):\footnote{With 
%\[
%	P=\left[p_S\left(\frac{p_S}{\gamma}\right)^{1/(\rho-1)}+p_M\left(\frac{p_M}{1-\gamma}\right)^{1/(\rho-1)}\right]^{(\rho-1)/\rho} = \left[P_S^{\rho/(\rho-1)} \gamma^{1/(1-\rho)} + P_M^{\rho/(\rho-1)} \left(1-\gamma\right)^{1/(1-\rho)}  \right]^{(\rho-1)/\rho}.
%\]
%}

\subsection{Equations for the quantity of intermediate goods used in production of the two final goods}

\begin{eqnarray}
	q_{h,HS}&=&\left(\frac{\tilde{p}_{h}}{\tilde{p}_{HS} \alpha_{HS}}\right)^{1/(\rho_{HS}-1)}q_{HS};\\
	q_{h,LS}&=&\left(\frac{\tilde{p}_{h}}{\tilde{p}_{LS} \alpha_{LS}}\right)^{1/(\rho_{LS}-1)}q_{LS};\\
	q_{l,HS}&=&\left[\frac{\tilde{p}_{l}}{\tilde{p}_{HS} (1-\alpha_{HS})}\right]^{1/(\rho_{HS}-1)}q_{HS};\\
	q_{l,LS}&=&\left[\frac{\tilde{p}_{l}}{\tilde{p}_{LS} (1-\alpha_{LS})}\right]^{1/(\rho_{LS}-1)}q_{LS},
\end{eqnarray}
%and in the equilbrium{HS}:
%\begin{eqnarray}
%\tilde{p}_{S,h} &=& \tilde{p}_{M,h} = \tilde{p}_{h} \\
%\tilde{p}_{S,l} &=& \tilde{p}_{M,l} = \tilde{p}_{l}.
%\end{eqnarray}

\newpage

\section{Welfare equations of employers and employees (only for referees)}

\begin{eqnarray}
W^{u}_{i,N} &=& \dfrac{Z^u_{i,N}}{r } +\left(\dfrac{\kappa_{i,N} \theta_i q\left(\theta_i\right)}{r}\right) \left\{\left(\dfrac{\beta_N }{1-\beta_N} \right) \left[ \dfrac{(1-t)(\tilde{p}_i x_i - \tilde{w}_{i,N})}{r+\delta_{i,N}}- \dfrac{\delta_{i,N}\tilde{p}_i x_iF}{r+\delta_{i,N}}+\tilde{p}_i x_iF \right]\right. + \notag \\&+&\left. \phi \tilde{p}_i x_i F \bigg\} \right.;\\ \notag\\
W^{u}_{i,I} &=& \dfrac{Z^u_{i,I}+\lambda W_{i,FC}}{(r + \lambda)} +\left(\dfrac{\kappa_{i,I} \theta_i q\left(\theta_i\right)}{r + \lambda}\right)\left\{\left(\dfrac{\beta_I}{1-\beta_I}\right)  \left[ \dfrac{(1-t)(\tilde{p}_i x_i - \tilde{w}_{i,I})}{r+\delta_{i,I}}- \dfrac{\delta_{i,I}\tilde{p}_i x_iF}{r+\delta_{i,I}}+\tilde{p}_i x_iF \right]\right. + \notag \\&+&\left. \phi \tilde{p}_i x_i F  \bigg\} \right.;\\ \notag\\
W^{e}_{i,N}&=&W^{u}_{i,N}+\dfrac{(1-b)\left(1-t\right)(\tilde{w}_{i,j}+\tilde{\tau} + \delta_{i,N}\phi \tilde{p}_i x_i F)}{r+\delta_{i,N}+ \kappa_{i,N} \theta_i q\left(\theta_i\right)};\label{we-wu_native}\\\notag\\
W^{e}_{i,I}&=&W^{u}_{i,I}+\dfrac{(1-b)\left(1-t\right)\left(\tilde{w}_{i,j}+\tilde{\tau} \right)-\lambda\left(W_{i,FC} - W^u_{i,I}\right)+\delta_{i,I}\phi \tilde{p}_i x_i F}{r+\delta_{i,I}+ \kappa_{i,I} \theta_i q\left(\theta_i\right)} \label{we-wu_immigrant};\\ \notag\\
J_{i,N} &=& \dfrac{\left(1-t\right)\left(\tilde{p}_i x_i - \tilde{w}_{i,N}\right) + \iota \nu -  \delta_{i,N} \tilde{p}_i x_i F}{r + \delta_{i,N}};\\\notag\\
J_{i,I} &=& \dfrac{\left(1-t\right)\left(\tilde{p}_i x_i - \tilde{w}_{i,I}\right) + \iota \nu -  \delta_{i,I} \tilde{p}_i x_i F}{r + \delta_{i,I}}.
\end{eqnarray}

\newpage
\section{Calibrated parameters \label{app:calibratedParameters}}
	
	\begin{figure}[!htbp]
		\centering
		\caption{The twenty-two calibrated parameters of the model (with $g_{HS}=g_{LS}$).}
		\label{fig:calibratedParameters}
		\includegraphics[width=\linewidth]{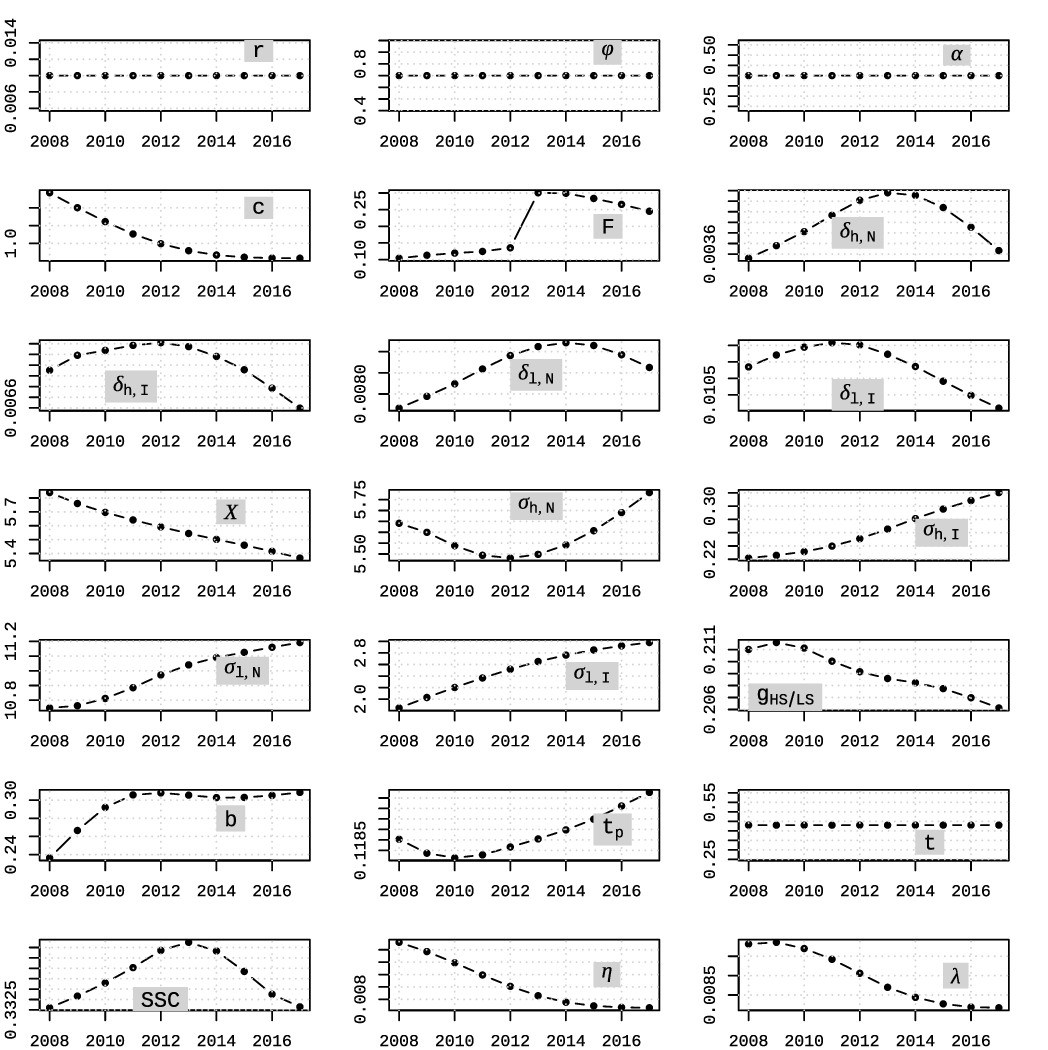}\vspace{0.2cm}
		\caption*{\textit{Note}: the list includes: the discount rate (r), the share of firing cost transferred to employees ($\phi$), the elasticity of the matching function with respect to unemployment ($\alpha$), the vacancy cost ($c$), the firing cost ($F$), the job destruction rate ($\delta$) for the four types of individuals, the mass of employers ($\chi$), the mass of employees by each type ($\sigma$), the government expenditure as $\%$ of GVA ($g$), the unemployment benefits ($b$), the tax gap ($t_p$), the direct tax rate ($t$), the social security contribution ($SSC$), the rate at which non-natives enter the country ($\eta$) and the rate at which non-natives exit the country ($\lambda$).}
	\end{figure}

\clearpage

\section{Observed versus matched moments \label{app:matchedMoments}}

\begin{figure}[!htbp]
	\centering
	\caption{The twenty-six observed versus matched moments.}
	\label{fig:momentsVsMatchedMoments}
	\includegraphics[width=\linewidth]{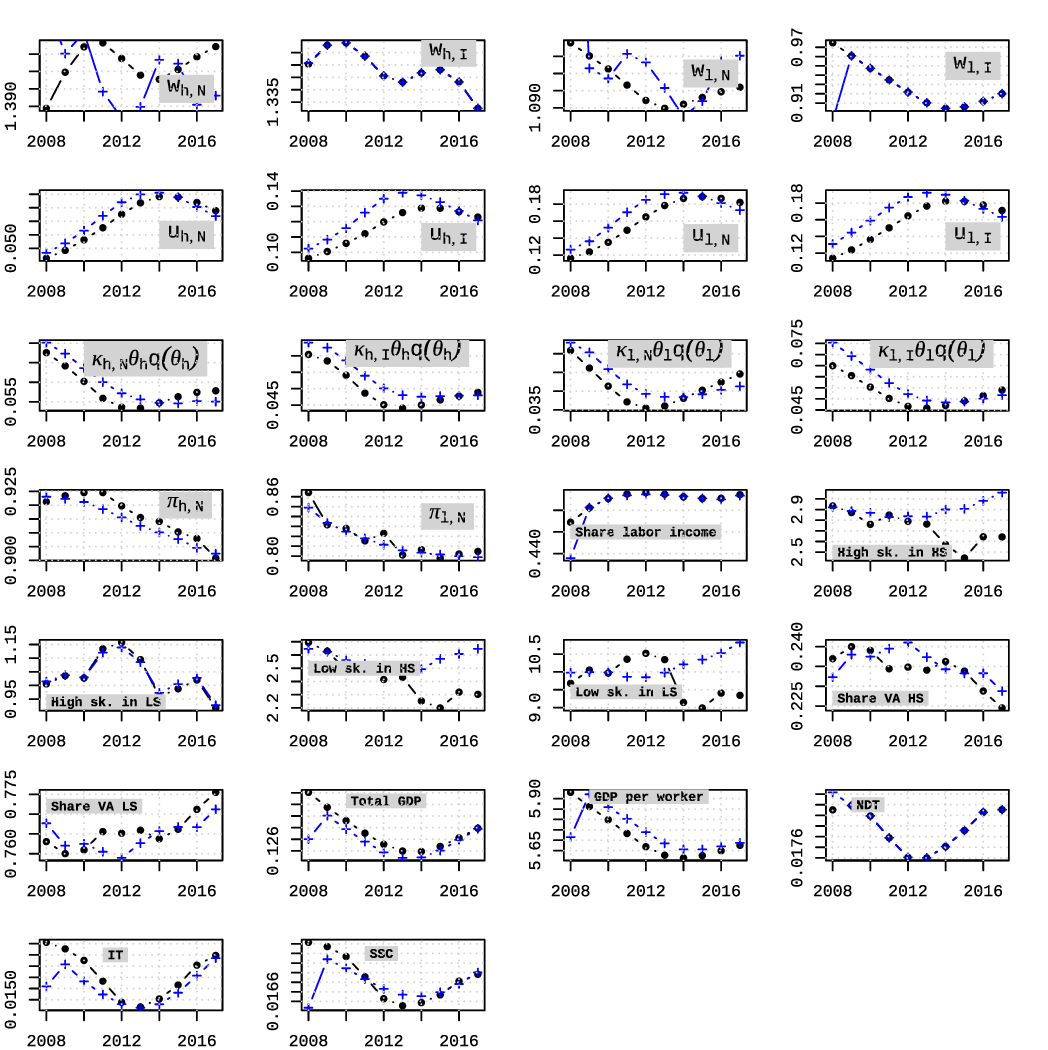}
	\vspace{0.2cm}
		\caption*{\textit{Note}: The  observed moments are in black with solid circles, while the matched moments are in blue with plus symbols. We indicate with $N$ native employees, $I$ non-native employees, $h$ high-skilled employees and $l$ low-skilled employees. The moments include net real wages ($w$),  unemployment rates ($u$) and job finding rates ($\kappa\theta q(\theta)$)  for the four categories of employees, the shares of native unemployed employees in the high(low)-skill intensive market ($\pi$), the share of labour income, the share of high(low)-skilled workers in the high(low)-skill intensive sector, the share of value added in the high(low)-skill intensive sector, the total real GDP, the  real GDP per worker, the  net direct taxes (NDT), the  indirect taxes (IT) and the  social security contributions (SSC).}
\end{figure}

\newpage

\section{Matched parameters \label{app:matchedParameters}}

\begin{figure}[!htbp]
	\centering
	\caption{The sixteen matched parameters.}
	\label{fig:matchedParameters}
	\includegraphics[width=0.9\linewidth]{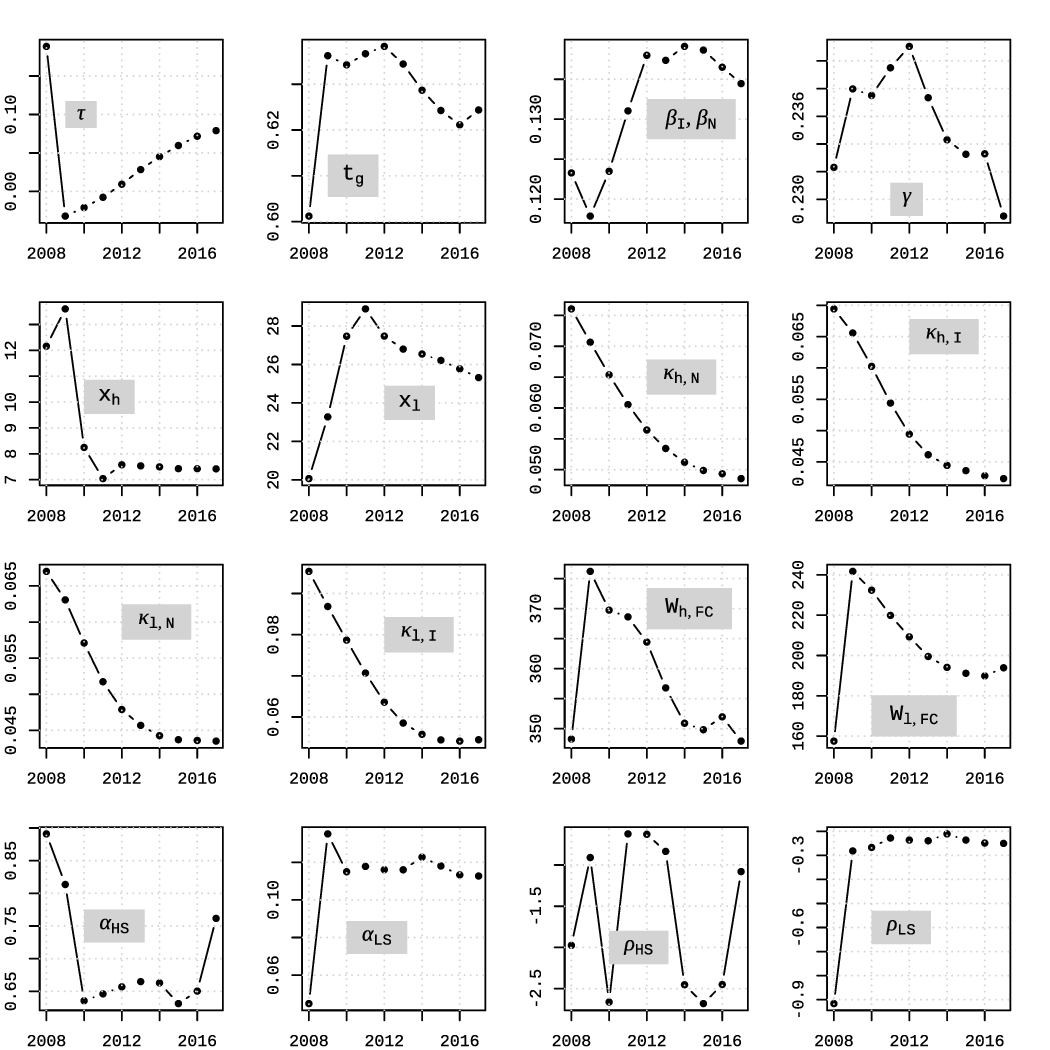}
		\vspace{0.5cm}
		\caption*{\textit{Note}: we indicate with $N$ native employees, $I$ non-native employees, $h$ high-skilled employees and $l$ low-skilled employees. The matched parameters include the tax subsidy ($\tau$), the tax gap ($t_{g}$), the bargaining power of the employees ($\beta$) (which is assumed to be the same for natives and non-natives), the elasticity of substitution between high-skill and low-skill intensive final goods ($\gamma$), the value added of high-skilled and low-skilled employees ($x_h$ and $x_l$), the hiring chances of the four types of employees ($\kappa$), the lifetime utility of high-skilled and  low-skilled employees abroad ($W_{h,FC}$ and $W_{l,FC}$) and the parameters of the CES production function of final good sectors ($\alpha_{HS}$,$\alpha_{LS}$,$\rho_{HS}$,$\rho_{LS}$).}
\end{figure}

\newpage

\section{Other endogenous variables \label{app:relevantEndoVariables}}
\begin{figure}[!htbp]
	\caption{Gross value added of firms in the high-skill (HS) intensive and low-skill (LS) intensive markets ($px$), price of the high-skill (HS) intensive good ($p_{HS}$) and price of the low-skill (LS) intensive good ($p_{LS}$), and per-capita value of the provision of public goods ($\nu$).}
\label{fig:relevantEndoVariables}
	\begin{subfigure}[t]{0.48\textwidth}
		\centering
		\includegraphics[width=0.7\textwidth]{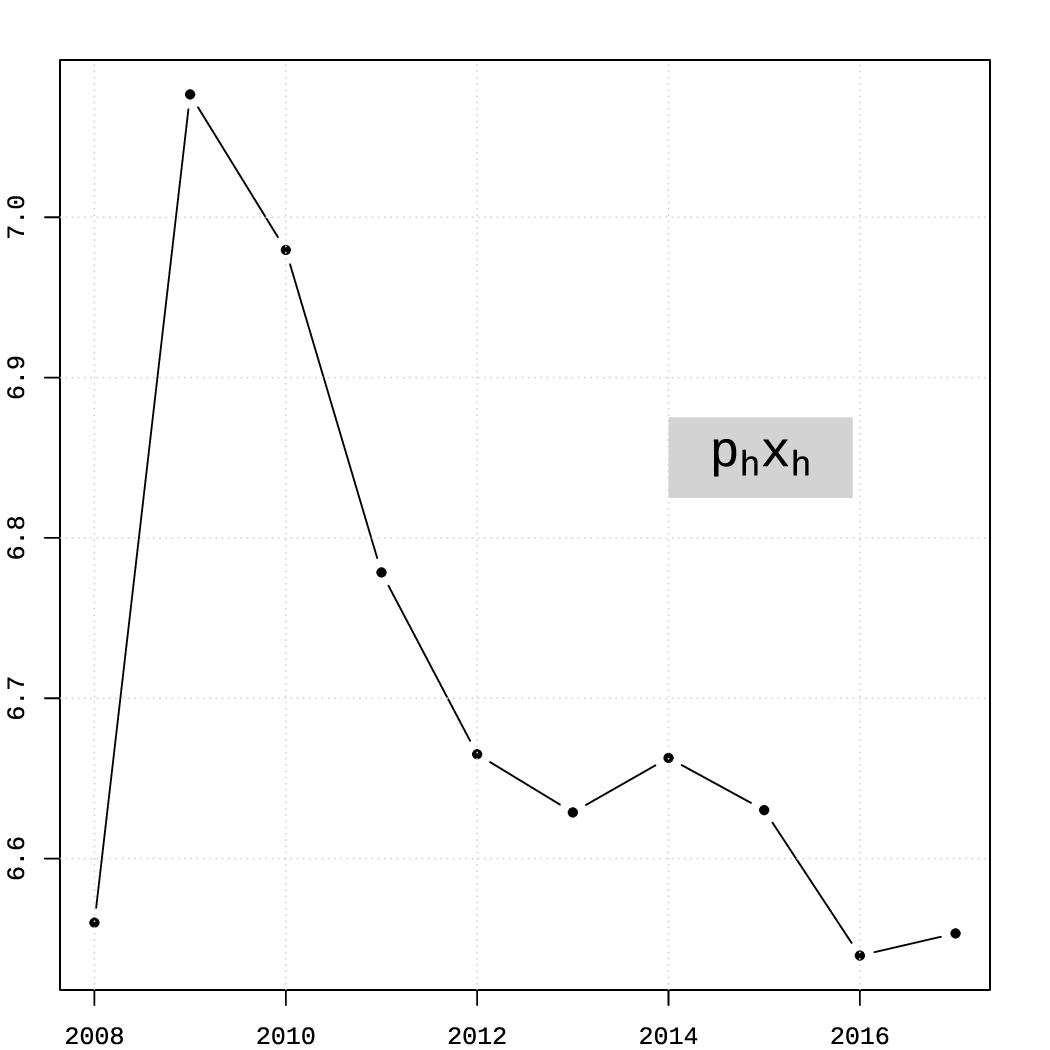}
		\caption{Gross Value Added (GVA) in the high-skill (HS) intensive market.}
	\end{subfigure}\hfill
	\begin{subfigure}[t]{0.48\textwidth}
		\centering
		\includegraphics[width=0.7\textwidth]{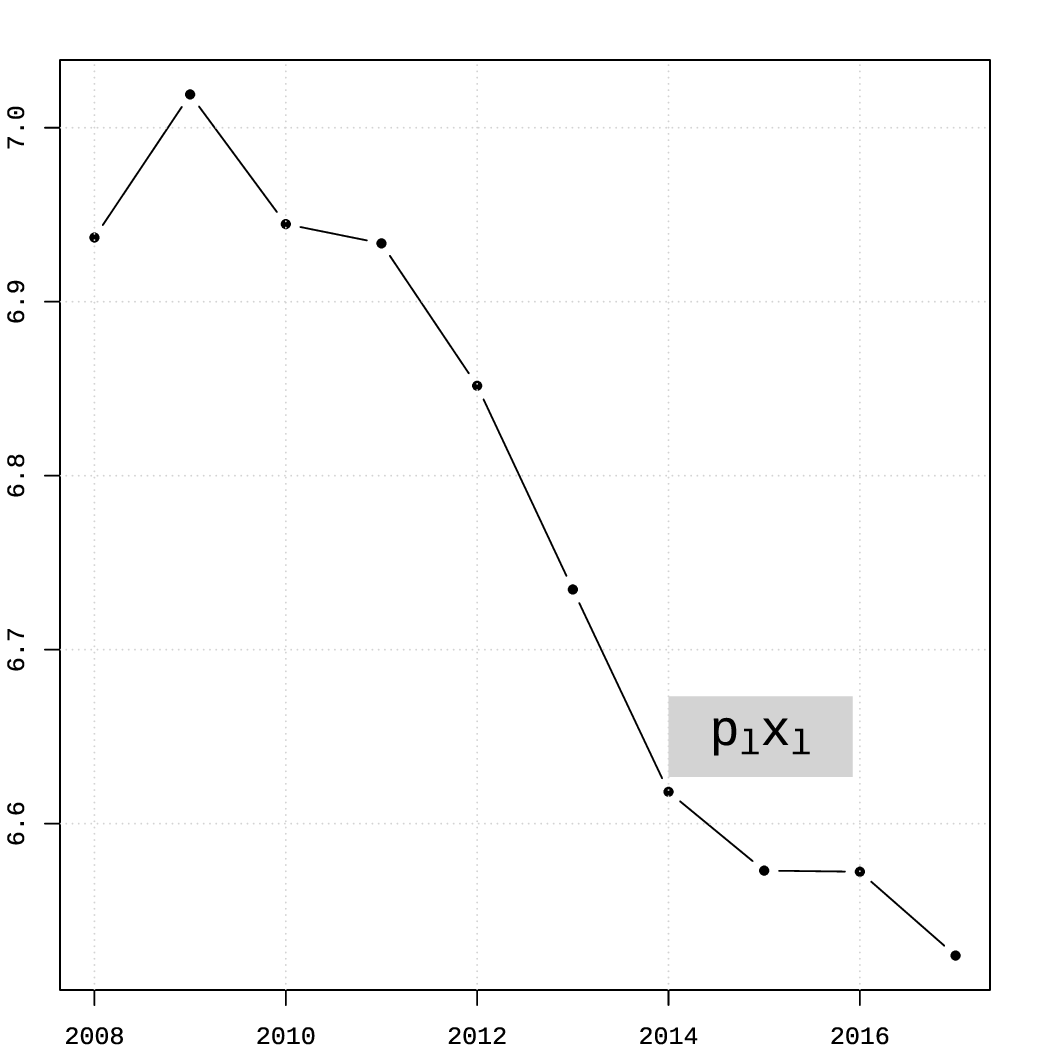}
		\caption{Gross Value Added (GVA) in the low-skill (LS) intensive market.}
	\end{subfigure}
	\begin{subfigure}[t]{0.48\textwidth}
			\centering
			\includegraphics[width=0.7\textwidth]{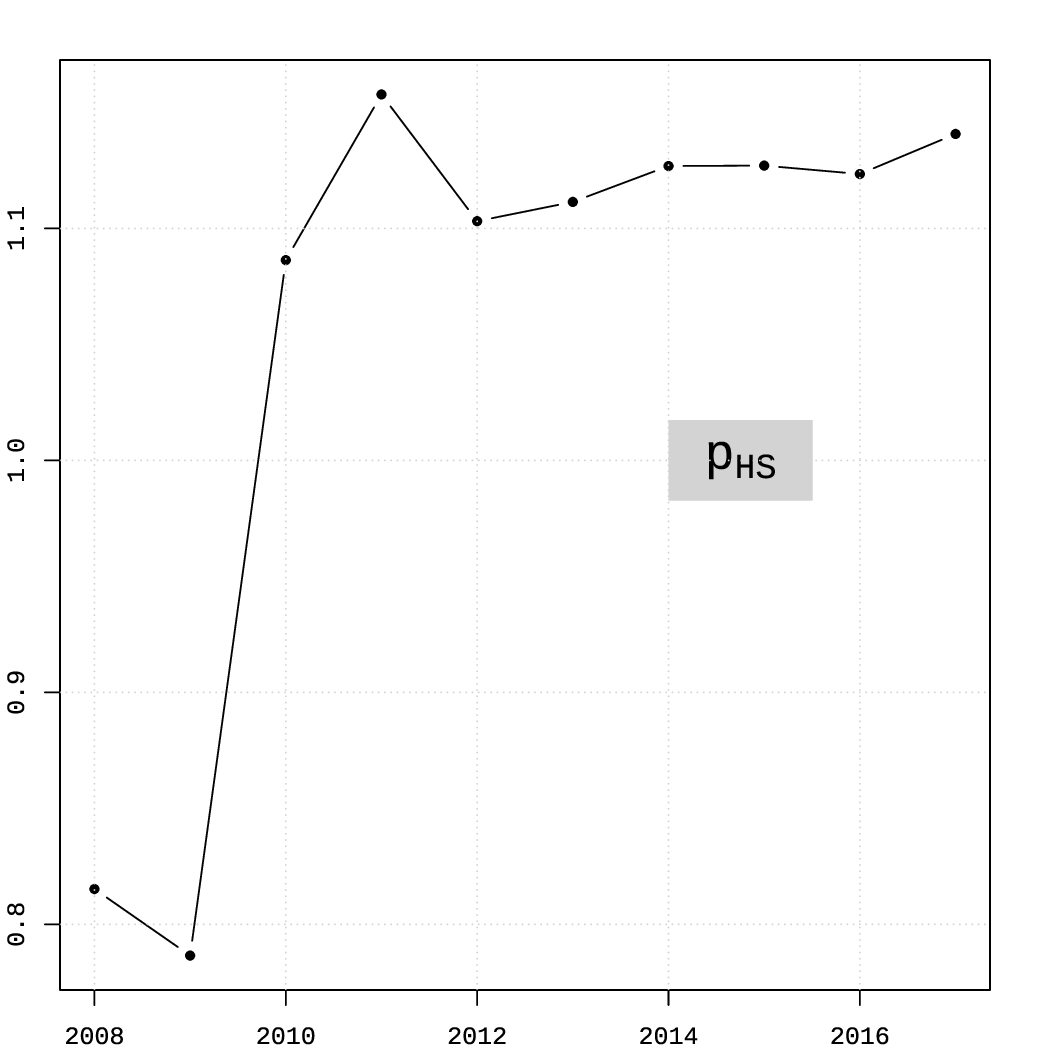}
			\caption{Price of the high-skill (HS) intensive final good.}
		\end{subfigure}
		\begin{subfigure}[t]{0.48\textwidth}
			\centering
			\includegraphics[width=0.7\textwidth]{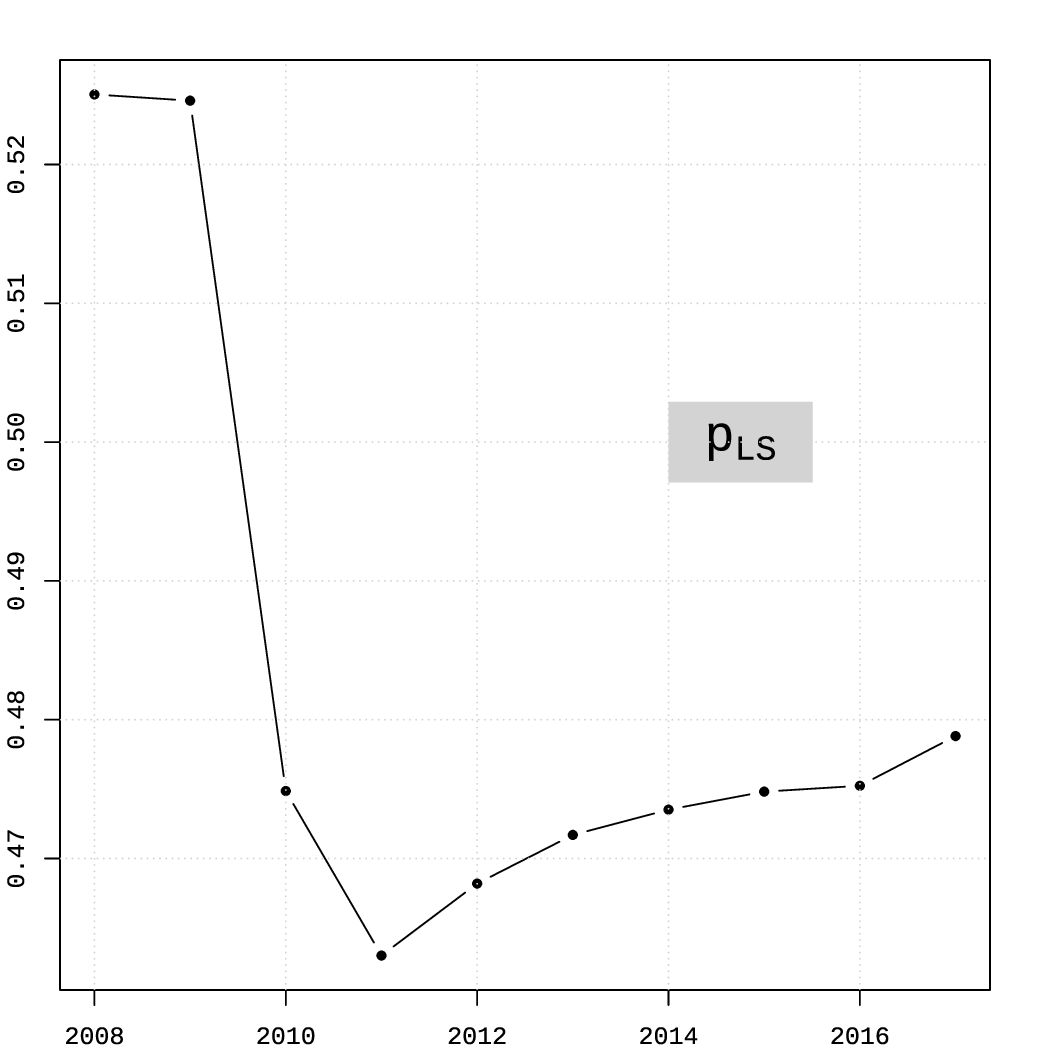}
			\caption{Price of the low-skill (LS) intensive final good.}
		\end{subfigure}
	\begin{subfigure}[t]{0.48\textwidth}
		\includegraphics[width=0.7\textwidth]{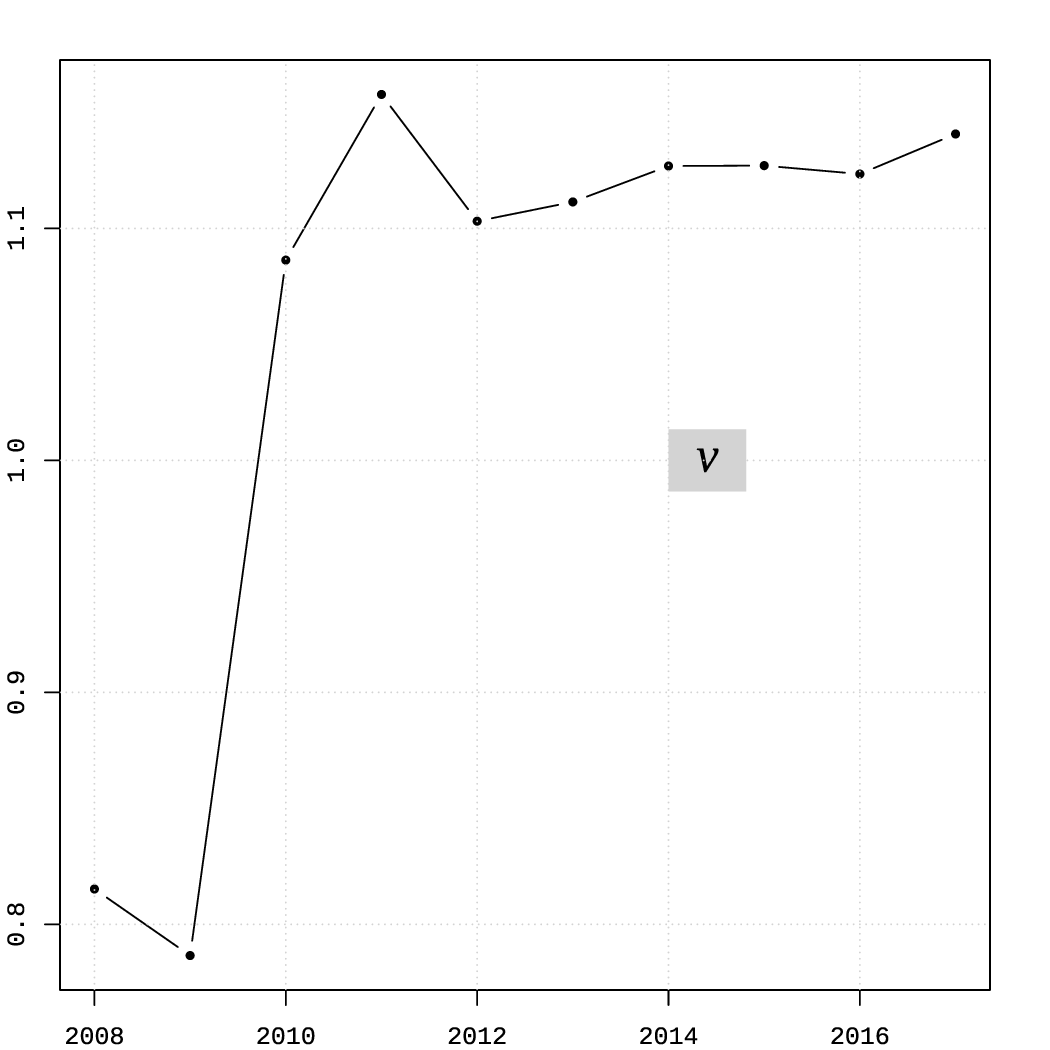}
		\caption{Per-capita provision of public goods. }
		\label{PublicGood}
	\end{subfigure}
\end{figure}

\clearpage

\section{List of LS versus HS final good sectors}\label{app:sectors}

\begin{table}[!htbp]
	\tiny
	\centering
	\caption{List of sectors with their ATECO code, high-skilled/low-skilled (H/L) employees share, classification and name.}
	\label{table:listSectorsHSLS}
	\begin{tabular}{rlcl}
		\hline \hline
		ATECO & Share & Sector & Sector's name \\
		code & H/L & classification & \\
		\hline
		01-01 & 0.05 & LS & Crop and animal production,hunting and related service activities \\ 
		02-02 & 0.07 & LS & Forestry and logging \\ 
		03-03 & 0.10 & LS &Fishing and aquaculture \\ 
		05-09 & 0.38 & LS &Mining and quarrying \\ 
		10-12 & 0.19 & LS &Manufacture of food products, beverages and tobacco products  \\ 
		13-15 & 0.18 & LS &Manufacture of textiles, wearing apparel, leather and related products \\ 
		16-16 & 0.15 & LS & Manufacture of wood, and of products of wood and cork, except furniture; articles of straw and plaiting materials  \\ 
		17-17 & 0.23 & LS & Manufacture of paper and paper products \\ 
		18-18 & 0.45 & LS & Printing and reproduction of recorded media \\ 
		19-19 & 0.55 & LS & Manufacture of coke and refined petroleum products \\ 
		20-20 & 0.62 & LS & Manufacture of chemicals and chemical products \\ 
		21-21 & 1.33 & HS & Manufacture of pharmaceutical products and pharmaceutical preparations \\ 
		22-22 & 0.28 & LS & Manufacture of rubber and plastic products\\ 
		23-23 & 0.22 & LS & Manufacture of other non metallic mineral products \\ 
		24-24 & 0.23 &  LS & Manufacture of basic metals \\ 
		25-25 & 0.19 & LS & Manufacture of fabricated metal products, except machinery and equipment  \\ 
		26-26 & 1.21 & HS & Manufacture of computer, electronic e optical products \\ 
		27-27 & 0.43 & LS & Manufacture of electrical equipment \\ 
		28-28 & 0.56 & LS & Manufacture of machinery and equipment n.c.a. \\ 
		29-29 & 0.31 & LS & Manufacture of motor vehicles, trailers and semi-trailers\\ 
		30-30 & 0.49 & LS &  Manufacture of other transport equipment \\ 
		31-32 & 0.24 & LS & Manufacture of furniture and other manufacturing  \\ 
		33-33 & 0.30 & LS &  Repair and installation of machinery and  equipment \\ 
		35-35 & 0.95 & LS &  Electricity, gas, steam and air conditioning supply \\ 
		36-36 & 0.64 & LS & Water collection, treatment and supply \\ 
		37-39 & 0.14 & LS & Sewerage, remediation activities and other waste management services; waste collection, \\
		&&&treatment and disposal activities, \\ 
		41-43 & 0.13 & LS & Construction \\ 
		45-45 & 0.14 & LS & Wholesale and retail trade; repair of motor vehicles and motorcycles \\ 
		46-46 & 0.48 & LS & Wholesale trade, except of motor vehicles and motorcycles\\ 
		47-47 & 0.14 & LS & Retail trade, except of motor vehicles and motorcycles \\ 
		49-49 & 0.12 & LS & Land transport and transport via pipelines \\ 
		50-50 & 0.70 & LS & Water transport \\ 
		51-51 & 0.89 & LS & Air transport \\ 
		52-52 & 0.34 & LS &Warehousing and support activities for transportation \\ 
		53-53 & 0.16 & LS & Postal and courier activities \\ 
		55-56 & 0.04 & LS & Accommodation, food and beverage services \\ 
		58-58 & 2.76 & HS & Publishing activities \\ 
		59-60 & 3.2 & HS & Motion picture, video and television programme production, sound recording and music \\
		&& &publishing activities, programming and broadcasting activities \\ 
		61-61 & 1.22 & HS & Telecommunications \\ 
		62-63 & 5.7 & HS & Computer programming, consultancy and related activities \\ 
		64-64 & 1.94 & HS & Financial service activities, except insurance and pension funding \\ 
		65-65 & 1.31 & HS & Insurance, reinsurance and pension funding,except compulsory social security \\ 
		66-66 & 1.00 & LS & Activities auxiliary to financial services and insurance activities \\ 
		68-68 & 0.41 & LS & Real estate activities \\ 
		69-70 & 0.91 & LS & Legal and accounting activities, Activities of head offices; management consultancy activities \\ 
		71-71 & 3.82 & HS &  Architectural and engineering activities; technical testing and analysis\\ 
		72-72 & 3.60 & HS & Scientific research and development \\ 
		73-73 & 2.13 & HS &  Advertising and market research \\ 
		74-75 & 2.84 & HS & Other professional, scientific and technical activities; veterinary activities\\ 
		77-77 & 0.29 & LS & Rental and leasing activities \\ 
		78-78 & 0.43 & LS & Employment activities \\ 
		79-79 & 1.03 & HS & Travel agency, tour operator reservation service and related activities \\ 
		80-82 & 0.06 & LS &  Security and investigation activities, services to buildings and landscape activities,\\
		&&& office administrative,office support and other business support activities \\ 
		84-84 & 0.48 & LS & Public administration and defence; compulsory social security \\ 
		85-85 & 4.04 & HS &  Education \\ 
		86-86 & 2.21 & HS & Human health activities \\ 
		87-88 & 0.79 & LS & Residential care activities, social work activities without accommodation \\ 
		90-92 & 0.78 & LS & Creative, arts and entertainment activities, libraries, archives,museums and other cultural activities,\\
		&&& gambling and betting activities \\ 
		93-93 & 0.78 &  LS & Sports activities and amusement and recreation activities \\ 
		94-94 & 0.93 & LS & Activities of membership organisations \\ 
		95-95 & 0.57 & LS & Repair of computers and personal and household goods \\ 
		96-96 & 0.04 & LS & Other personal service activities \\ 
		97-98 & 0.00 & LS & Activities of households as employers of domestic personnel; undifferentiated goods-and services-producing\\
		&&& activities of private households for own use \\ 
		\hline
		\hline
		\multicolumn{4}{l}{\scriptsize{\textit{Source}: Own calculation based on the Italian Labour Force Survey (RCFL) and ISTAT.}}
	\end{tabular}

\end{table}

\end{document}